\documentclass[structabstract]{aa}  
\usepackage{graphicx}
\usepackage{txfonts}
\usepackage{natbib}
\bibpunct{(}{)}{;}{a}{}{,}
\usepackage{amssymb}
\usepackage{amsmath}
\usepackage{xcolor}
\usepackage{lipsum}
\usepackage{textcomp}
\usepackage{longtable}
\usepackage{multirow}
\usepackage[colorlinks=true,linkcolor=blue,citecolor=blue,draft=false]{hyperref}
\usepackage{enumerate}
\usepackage{url}
\usepackage{breakurl}
\usepackage{rotating}
\usepackage{geometry}
\usepackage{ulem}
\usepackage{comment}

\usepackage{lscape}
\usepackage{diagbox}

\newcommand{\msun}{$M_{\sun}$}
\newcommand{\hour}{$^{\mathrm{h}}$}
\newcommand{\minute}{$^{\mathrm{m}}$}
\newcommand{\second}{$^{\mathrm{s}}$}
\def\HI{\hbox{H\,{\sc i}}}
\def\HII{\hbox{H\,{\sc ii}}}

\newcommand{\ccIa}{$N_{\mathrm{CC}}/N_{\mathrm{Ia}}$}
\newcommand{\abund}[1]{\makebox[2em]{#1:}}

\newcommand{\NOB}{\ensuremath{N_{{\rm OB}}}\xspace}

\newcommand{\xmm}{XMM-{\it Newton}\xspace}
\newcommand{\chandra}{{\it Chandra}\xspace}
\newcommand{\cxo}{{\it Chandra}\xspace}

\newcommand{\ROSAT}{ROSAT}

\newcommand{\eg}{\mbox{e.\,g.}\xspace}

\defcitealias{2006ApJ...652L..45R}{R06}
\defcitealias{2014ApJ...785L..27Y}{Y14}
\defcitealias{2016A&A...585A.162M}{MHK16}
\defcitealias{2004A&A...421.1031V}{vdH04}
\defcitealias{2017ApJS..230....2B}{B17}

\def\msun{$M_{\odot}$}

\def\HI{\ion{H}{I}}
\def\HII{\ion{H}{II}}
\newcommand{\SII}{[S\,{\sc ii}]}
\newcommand{\NII}{[N\,{\sc ii}]}
\newcommand{\OIII}{[O\,{\sc iii}]}
\newcommand{\Halpha}{H${\alpha}$}

\newcommand{\D}{$^\circ$}
\def\p0{\phantom{0}}

\def\it{\sl}

\def\confcount{19}
\def\candcount{4}
\def\finalconfcount{21}
\def\finalcandcount{2}

\def\totalcount{23}

\begin{document}

   \title{The supernova remnant population of the Small Magellanic Cloud}

   \subtitle{}


	\author{
	Pierre~Maggi\inst{1}
    \and Miroslav~D.~Filipovi\'c \inst{2}
	\and Branislav~Vukoti\'c \inst{3}
	\and Jean~Ballet \inst{4}
	\and Frank~Haberl \inst{5}
	\and Chandreyee~Maitra \inst{5}
	\and Patrick~Kavanagh \inst{6}
	\and Manami~Sasaki \inst{7}
	\and Milorad~Stupar \inst{2}
	}
	\authorrunning{Maggi et al.}

   \institute{
   Universit\'e de Strasbourg, CNRS, Observatoire astronomique de Strasbourg, UMR 7550, F-67000 Strasbourg, France\\\email{pierre.maggi@astro.unistra.fr}
\and
University of Western Sydney, Locked Bag 1797, Penrith South DC, NSW 2751, Australia
\and
Astronomical Observatory, Volgina 7, PO Box 74 11060 Belgrade, Serbia
\and
AIM, CEA, CNRS, Universit\'e Paris-Saclay, Universit\'e Paris Diderot, Sorbonne Paris Cit\'e, F-91191 Gif-sur-Yvette, France
\and
Max-Planck-Institut f\"ur extraterrestrische Physik, Gie{\ss}enbachstraße, 85748 Garching, Germany
\and
School of Cosmic Physics, Dublin Institute for Advanced Studies, 31 Fitzwillam Place, Dublin 2, Ireland
\and
Remeis Observatory and ECAP, Universit\"at Erlangen-N\"urnberg, Sternwartstr. 7, 96049 Bamberg, Germany
    }

   \date{Received Day Month Year\,/\,Day Month Year}

  \abstract
  {}
  {We present a comprehensive study of the supernova remnant (SNR) population of the Small Magellanic Cloud (SMC). We measure multiwavelength properties of the SMC SNRs and compare them to those of the Large Magellanic Cloud (LMC) population.}
  {This study combines the large dataset of \xmm\ observations of the SMC, archival and recent radio continuum observations, an optical line emission survey, and new optical spectroscopic observations. We can thus build a complete and clean sample of \confcount\ confirmed and \candcount\ candidate SNRs. The homogeneous X-ray spectral analysis allows to search for SN ejecta and Fe~K line emission, and to measure interstellar medium (ISM) abundances. We estimate the ratio of core-collapse to type~Ia supernova rates of the SMC based on the X-ray properties and the local stellar environment of each SNR.}
  {After the removal of unconfirmed or misclassified objects, and the addition of two newly confirmed SNRs based on multi-wavelength features, we present a final list of \finalconfcount\ confirmed SNRs and \finalcandcount\ candidates. While no Fe~K line is detected even for the brightest and youngest SNR, we find X-ray evidence of SN ejecta in 11 SNRs. We estimate a ratio of 4.7$_{-1.9} ^{+0.6}$ core-collapse supernova to every type~Ia SN, three times higher than in the LMC. The difference can be ascribed to the absence of the enhanced star formation episode in the SMC, which occurred in the LMC 0.5~--~1.5~Gyr ago. The hot-gas abundances of O, Ne, Mg, and Fe are 0.1~--~0.2 times solar. Their ratios with respect to SMC stellar abundances reflect the effects of dust depletion and partial dust destruction in SNR shocks. We find evidence that the ambient medium probed by SMC SNRs is less disturbed and less dense on average than in the LMC, consistent with the different morphologies of the two galaxies.}
  {}
  
   \keywords{ISM: supernova remnants, Magellanic Clouds, ISM: abundances, supernovae: general, stars: formation, X-rays: ISM}
   \maketitle
%

\section{Introduction}
\label{introduction}
A fraction of stars end their lives in powerful supernova (SN) explosions \cite[e.g.][ and references therein]{2017hsn..book.....A}. This is the case after core-collapse (CC) for some of the most massive stars (zero age main sequence mass $\gtrsim 8~M_{\odot}$), and through the thermonuclear disruption of the CO core of a white dwarf (the so-called type~Ia SNe), being possibly ignited when the Chandrasekhar mass is reached via accretion, or during the merger of a double white dwarf binary. Both types of SNe release large quantities of freshly-produced elements from light $\alpha$-group elements (O, Ne, Mg) to intermediate-mass elements (Si, S) and heavier Fe-group elements (Ti, Cr, Fe, Ni), produced during thermostatic nuclear burning and in the final, explosive nucleosynthesis episode. Together with stellar winds and neutron star mergers, SNe are responsible for the enrichment and chemical content of the Universe \citep{2013ARA&A..51..457N,2017ARNPS..67..253T}.

Furthermore, the tremendous energy release of a SN ($\sim 10^{51}$~erg) is transferred to the surrounding interstellar medium (ISM). The object created in the ISM by a SN is called a supernova remnant (SNR). The SN ejecta launched at velocities greater than $10^4$~km~s$^{-1}$ drive shock waves in the ambient medium, heating the ISM and ejecta up to X-ray emitting temperatures \citep{2012A&ARv..20...49V}. Cosmic rays (particles) are accelerated at the shock front where the magnetic field is turbulent, and electrons with energies up to 100~TeV (for the youngest SNRs) emit synchrotron radiation from radio to X-rays \citep{1995Natur.378..255K}. Optical line emission can arise mostly from charge exchange and collisional excitation of neutrals at fast shocks \citep{2001ApJ...547..995G}, shock-ionised material, or radiative cooling if the conditions are conducive \citep{1971ApJ...167..113C}.

Remnants remain visible for several $10^4$ years, as opposed to a few hundred days for their parent SNe. Consequently, the SNR population of a galaxy collectively holds precious information on the dozen or hundreds of SNe which exploded recently within it. For instance, the ratio between the rates of SNe of each type (CC vs. type~Ia) can in principle be recovered; the morphologies of individual SNRs can be linked to asymmetries either intrinsic to the explosion or coming from its surrounding ISM \citep{2013A&A...552A.126W,2014IAUS..296..239L}; and the abundances of newly-synthesised ejecta constrain details of both stellar evolution and explosion physics \citep{2017ApJ...836...79C,2018ApJ...861..143L}.

The many SNe exploding in a galaxy are the main source of energy of its ISM, in the form of kinetic energy, turbulence, and cosmic ray acceleration \citep{2004RvMP...76..125M}. They offer a mode of star formation regulation, as the combined shocks of many SNe can launch galactic winds which expels gas. Heated to temperatures $T > 10^6$~K, the dominant elements of the ISM (C, N, O, Ne, Mg, Si, S, Fe) emit many lines in the X-ray band which can be used to infer their abundances. Therefore, studies of populations of SNRs in a galaxy can reveal key information on the SNe themselves \textit{and} can be used to probe the host galaxy.

The Milky Way population currently contains about 300 SNRs \citep{2019arXiv190702638G}. Their study is hampered by large distance uncertainties and line-of-sight confusion/crowding, which prevent accurate comparison of objects. Even more problematic is the strong interstellar absorption towards most of these sources in the Galactic plane, particularly for X-rays as the important 0.5-2~keV energy band can be completely masked for $N_H > 10^{22}$~cm$^{-2}$. Despite larger distances, external galaxies are therefore better suited to SNR population studies. The SNRs of several galaxies in the Local Group \citep[M31, M33,][]{2012A&A...544A.144S,2010ApJS..187..495L,2014SerAJ.189...15G,2017MNRAS.472..308G} and beyond \citep{2004A&A...425..443P,2009ApJ...703..370C,2010ApJ...725..842L,2011Ap&SS.332..221M,2011AJ....142...20P,2012SerAJ.184...19M,2013Ap&SS.347..159O,2014ApJS..212...21L,2014Ap&SS.353..603G,2015AJ....150...91P,2018PASA...35...15Y} have been studied at various wavelengths. Closer to us, our Galactic neighbours the Large and Small Magellanic Clouds (LMC, SMC) provide excellent benchmarks for the study of star-forming galaxies. At only 50 and 60~kpc, respectively \citep{2019Natur.567..200P,2005MNRAS.357..304H,2014ApJ...780...59G}, they are close enough that we can detect and spatially resolve SNRs from radio to X-rays, and are located behind only a moderate Galactic foreground \citep[$N_H \lesssim$ a few $10^{20}$~cm$^{-2}$,][]{1990ARA&A..28..215D}.

In the LMC, our knowledge about the SNR population was built up over time with radio, optical, and X-ray observations. In \citet[][hereafter \citetalias{2016A&A...585A.162M}]{2016A&A...585A.162M}, a comprehensive X-ray study of the LMC SNRs was conducted, taking advantage of the coverage of a large fraction of the SNR population ($\sim 60$ objects) with the \xmm\ X-ray observatory during targeted observations and the extensive survey of the LMC (PI: F. Haberl). The radio counterpart to that study, also presenting 15 further SNR candidates, was published in \citet[][hereafter \citetalias{2017ApJS..230....2B}]{2017ApJS..230....2B}.

In this work, we attempt to provide the most comprehensive study of the SNR population of the SMC. As the sample is about three times smaller than in the LMC, we combine in a single work both the X-ray and radio-continuum analyses, together with archival optical emission line data and new optical spectroscopy. Previous similar studies were more limited in their scope. \citet{2004A&A...421.1031V} analysed \xmm\ data for 13 SMC SNRs. We expand considerably on this work, using the much larger body of \xmm\ observations accumulated since, a larger population augmented with newly-discovered objects \citep[e.g.][]{2008A&A...485...63F}, and, as already mentioned, a multiwavelength approach including deep radio and optical surveys.

The paper is organised as follows: in Sect.\,\ref{observations} we describe the \xmm, radio, and optical observations used and how the data were reduced. The X-ray imaging and spectral analyses are detailed in Sect.\,\ref{data}. Our results are presented and discussed in Sect.\,\ref{results}, starting with the final sample and rejected objects in Sect.\,\ref{results_sample} and \ref{results_rejectedSNRs}, respectively, followed by the X-ray spectral properties of SMC SNRs (Sect.\,\ref{results_spectral}). Candidates and confirmed SNRs are discussed individually in Sect.\,\ref{results_notes_candidates} and \ref{results_notesSNRs}. We then measure the gas-phase abundances of the SMC ISM and the ratio of CC to type~Ia SNe (Sect.\,\ref{results_abundances} and \ref{results_typing}), discuss the radio properties, size, and morphology of SNRs in both Magellanic Clouds (MCs; Sect.\,\ref{results_radio_properties}), and probe the 3D spatial distribution of SNRs within the SMC (Sect.\,\ref{results_3D}). Our findings are summarised in Sect.\,\ref{summary}.

\section{Observations and data reduction}
\label{observations}

\subsection{X-ray data }
\label{observations_xray}

There are upwards of 120 \xmm observations of the SMC. In this work, all observations useful for imaging and/or spectroscopic purposes were included. Lists of observation IDs (ObsIDs) sorted by off-axis angle and exposure time were compiled for each SNR. Thanks to the compactness of the SMC and its dense coverage with \xmm, all SNRs have multiple observations available. That number ranges from two (for two SNRs) to 38 for four of them. Those highly covered are those in the field of view of SNR 1E~0102.2$-$7219 (including itself), which is used as a calibration source \citep{2017A&A...597A..35P} and thus monitored frequently.

All data were processed with the ``MPE pipeline'', used for \xmm surveys of M31 \citep{2005A&A...434..483P,2011A&A...534A..55S}, M33 \citep{2004A&A...426...11P,2006A&A...448.1247M}, and the SMC \citep{2012A&A...545A.128H}. A summary of the important steps of the pipeline was given in \citetalias{2016A&A...585A.162M}. The difference with \citet{2012A&A...545A.128H} is that data were reprocessed with version 16.0.0 of the \xmm Science Analysis Software\,\footnote{SAS, \url{http://xmm.esac.esa.int/sas/}}. The resulting event lists and associated good time interval files (\texttt{gti}, one file per detector) were used as the primary source for subsequent analysis.

\begin{figure*}
  \center
    \includegraphics[height=0.31\vsize]{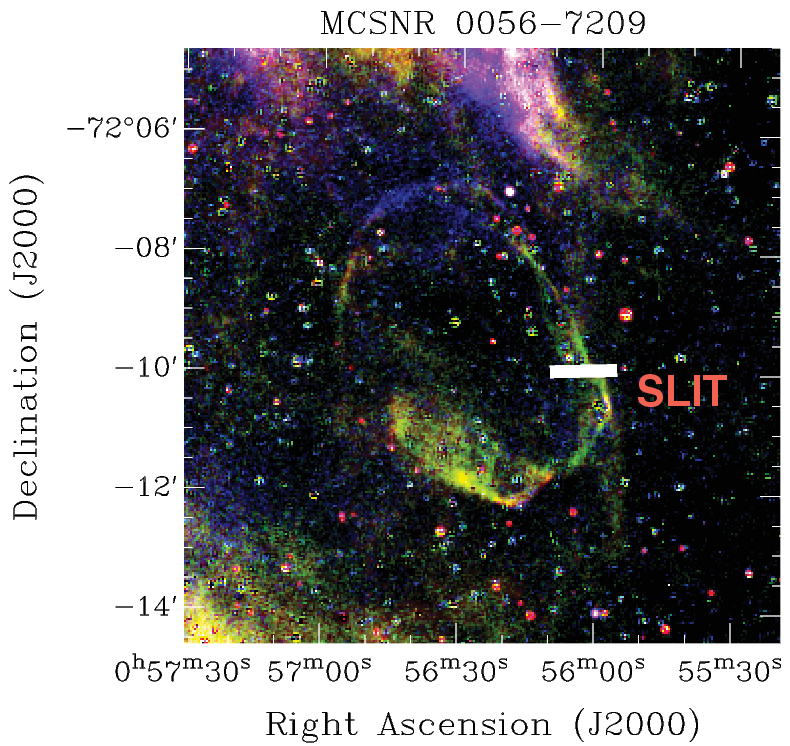}
    \includegraphics[height=0.30\vsize]{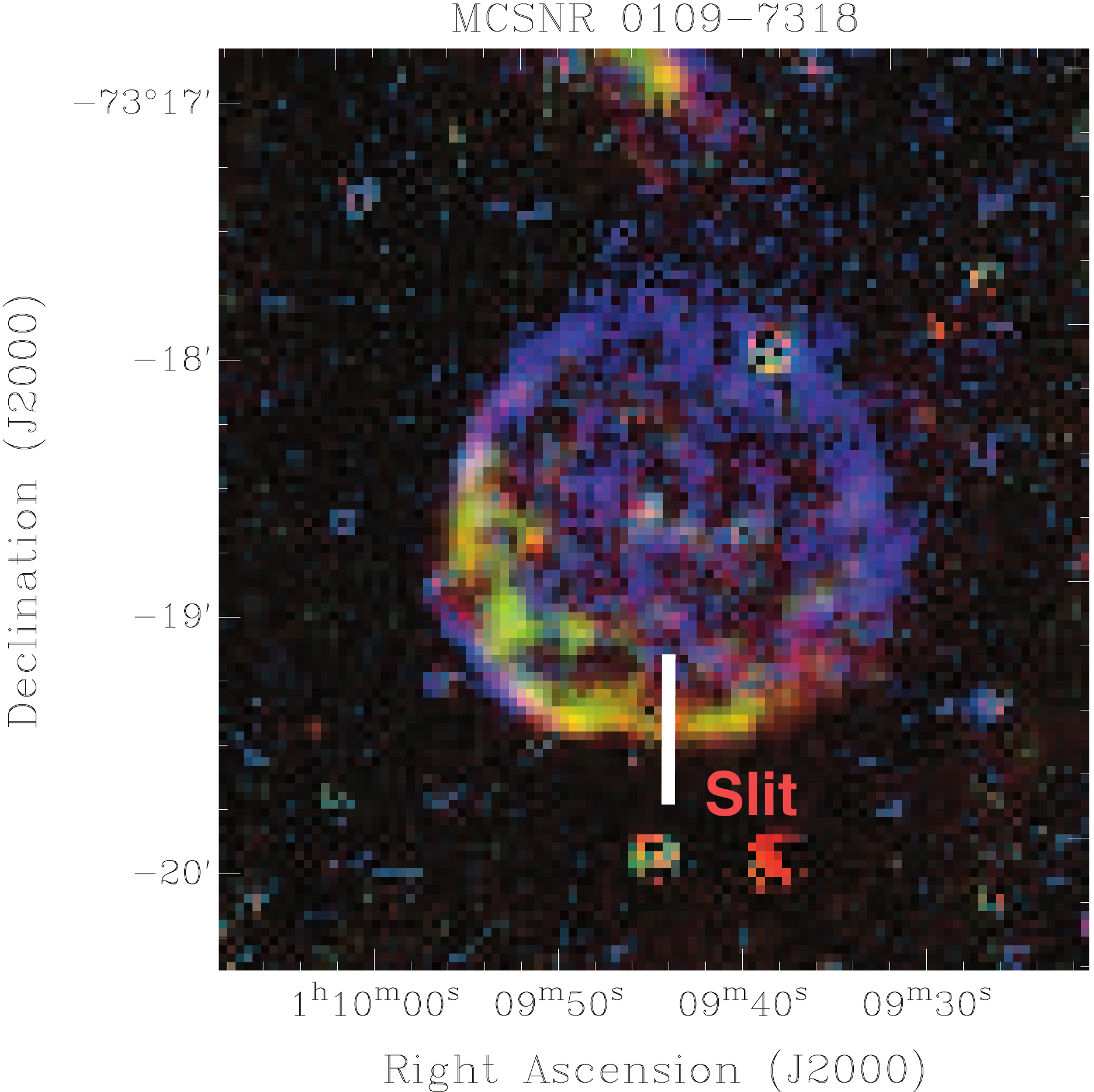}
    \caption{\textit{Left\,:} MCSNR candidate J0056$-$7209 on a composite MCELS image (R, G, B = [\ion{S}{ii}], H$\alpha$, and [\ion{O}{iii}], respectively). On the east and west side of the ellipse, fragmented filaments typical of older supernova remnants are clearly seen. The white bar shows the position of the WiFeS spectrograph slit. The spectrograph slit is actually a combination of 25 $\times$ 1\arcsec\ wide adjacent slits each, repeated 36 times to yield an effective 25$\times$36\arcsec\ field of view on the sky.
    \textit{Right\,:} Same as left for MCSNR candidate J0109--7318. Compared with J0056$-$7209, this one exhibits more fragmented filaments creating [\ion{S}{ii}] and H$\alpha$ arcs also common in morphological structures of old supernova remnants.}
    \label{MC}
\end{figure*}

\subsection{Radio continuum data}
In a similar manner as in \citetalias{2017ApJS..230....2B} where the LMC radio continuum sample were investigated, we used all available radio continuum data described in various surveys to date \citep{1997A&AS..121..321F,1998A&AS..130..421F,2002MNRAS.335.1085F,2004MNRAS.355...44P,2005MNRAS.364..217F,2006MNRAS.367.1379R,2007MNRAS.376.1793P,2011SerAJ.182...43W,2011SerAJ.183...95C,2011SerAJ.183..103W,2012SerAJ.184...93W,2012SerAJ.185...53W}. Some earlier radio continuum studies of selected SMC SNRs were shown in \cite{2008A&A...485...63F,2011A&A...530A.132O,2012A&A...537L...1H,2014AJ....148...99C,2015A&A...584A..41M}. Where possible, our study improves upon these previous SMC SNR studies. We also examine Australia Telescope Online Archive to search for in depth coverage of SNRs studied here. Apart from the ATCA (and Parkes) radio telescope, we make use of MWA observations as described in \citet{2018MNRAS.480.2743F} as well as the Australia Square Array Pathfinder (ASKAP) Early Science Project on the SMC (Joseph et al., submitted). Typical data reduction procedures as described in above papers were used. We specifically used the MIRIAD tasks \texttt{imfit} and \texttt{imsum} in order to extract flux density, extensions (diameter/axes $D$ and position angle PA) for each radio detected SNR. The radio spectral index ($\alpha$) based on at least two flux density ($S$) measurements is then estimated defined as $S_{\nu} \propto \nu^\alpha$. 


\subsection{Optical observations of SMC SNRs candidates}
\label{observations_supplementary}
In our optical search for the SMC SNRs we initially used the Magellanic Clouds Emission Line Survey (MCELS) (see \citealt{2012ApJ...755...40P}). In this survey which covered optical wavebands of \OIII\ at 5007\AA, \Halpha\ at 6563\AA\ and \SII\ at 6716/6731\AA, we found two new objects (see Fig.~\ref{MC}) whose morphological structures are typical of SNRs. This was the initial motivation for us to go further and obtain optical spectra of these objects, named candidates MCSNR~J0056$-$7209 and MCSNR~J0109$-$7318, and try to confirm their nature.

Spectral observations were undertaken on June 8, 2015 (see Table~2), using the Wide-Field Spectrograph (WiFeS) on the 2.3-m telescope of the Mount Stromlo and Siding Spring Observatory (MSSSO). The WiFeS spectrograph is an image slicer and behaves as an integral field unit (IFU) providing spatially-resolved spectroscopy \citep[see details in][]{2007Ap&SS.310..255D}. The final result, after complex data reduction of WiFeS observations, is a cube with R.A., Dec. and wavelength as third dimension. From that cube, we can generate 1D spectra. WiFeS consists of twenty-five 1\arcsec\ wide adjacent slits which are each 36\arcsec\ in length to yield an effective $25\times36$\arcsec\ on the sky. As our granted observational night definitely was not photometric, we performed observations only in the red part of the spectrum between 5700~\AA\ and 7000~\AA\ using the R7000 grating  with 1200 lines~mm$^{-1}$. In addition, due to the non-photometric night we could neither apply observations of spectrophotometric standard stars to get real line fluxes (but used simple counts) nor estimate the true extinction.

\section{Data analysis}
\label{data}

\subsection{X-ray imaging}
\label{data_imaging}
Images were created with a pixel size of 2\arcsec$\times$~2\arcsec, using single to quadruple-pixel events (\texttt{PATTERN} = 0 to 12) with \texttt{FLAG = 0} from the MOS detectors. Single and double-pixel events (\texttt{PATTERN} = 0 to 4) from the pn detector with \texttt{(FLAG \&\& 0xf0000) = 0} (i.e. including events next to bad pixels or bad columns) were used. To avoid the higher detector noise contribution from the double-pixel events below 500~eV, only single-pixel events were selected at these low energies. Exposure maps taking into account the energy-dependent telescope vignetting were produced with the task \texttt{eexpmap}. Out-of-time (OoT) images were created from the EPIC-pn OoT event lists, scaled by the corresponding OoT fraction $f_{\mathrm{OoT}}$\,\footnote{Values taken from the \xmm\ Users Handbook.}, and subtracted from the corresponding source+background images.

Images and exposure maps were extracted in various energy bands for all three cameras. The set of energy bands was tailored to the thermal spectrum of SNRs: a soft band from 0.3~keV to 0.7~keV includes strong lines from oxygen; a medium band from 0.7~keV to 1.1~keV comprises Fe L-shell lines as well as K-shell lines from \ion{Ne}{IX} and \ion{Ne}{X}; and a hard band (1.1~--~4.2~keV) includes K lines from Mg, Si, S, Ca, Ar, and possibly non-thermal continuum.

The detector background was subtracted from the images. We used filter wheel closed (hereafter FWC) data, obtained with the detectors shielded from astrophysical and soft-proton backgrounds. FWC observations are collected several times per year as part of the \xmm calibration efforts and made available by the \xmm\ Science Operations Centre\,\footnote{\url{http://www.cosmos.esa.int/web/xmm-newton/filter-closed}}. The instrumental background contribution $f_{\mathrm{FWC}}$ to the science image is estimated from the count rate in the detector corners for each instrument individually, as they are always shielded from the X-ray telescopes. The FWC images were scaled by $f_{\mathrm{FWC}}$ and removed from the science image to create the background-subtracted image. Only FWC exposures in full-frame mode are available for MOS detectors, excluding all other modes from our analysis.

We combined all suitable observations of an SNR to produce an image centred on the source. In each band, we merged the images from pn and both MOS. The smoothing of the combined images was done in adaptive mode with the SAS task \texttt{asmooth}. It calculates a library of Gaussian kernels such that the resulting images reach a minimum (Poissonian) signal-to-noise ratio of 5 everywhere. Regions of good statistics (\eg bright sources) are smoothed with a small kernel, whereas fainter regions are more thoroughly smoothed. The minimum kernel size for adaptive smoothing is either 10\arcsec\ or 20\arcsec, depending on the available data and the surface brightness of the SNR under investigation. Only MCSNR~J0104$-$7201 (1E 0102.2$-$7219) was smoothed with a smaller kernel of 6\arcsec\ owing to its small size and excellent photon statistics available. We divided the combined image by the corresponding vignetted and smoothed exposure map. The combined exposure map was produced by weighting the MOS exposure maps with a factor of 0.4 relative to pn, to account for the lower effective area. The smoothing of the exposure map is done with the same template of kernels as the for the images. The resulting composite images (soft-medium-hard bands) are shown in Appendix~\ref{appendix_images} with radio and optical features.

\subsection{X-ray spectra}
\label{data_spectra}
We follow the spectral analysis method described extensively in \citetalias{2016A&A...585A.162M}: we simultaneously fit source and background spectra (hereafter \texttt{SRC} and \texttt{BG}), where the latter is explicitly modelled rather than subtracted. This is critical for the analysis of faint \textit{extended} sources such as SNRs in the SMC. We correct the event lists for vignetting with the SAS task \texttt{evigweight} prior to extraction. This accounts for the energy-dependent effective area variation across the extents of SNRs and background regions. The redistribution matrices are produced by the SAS task \texttt{rmfgen}, and the ancillary response files by \texttt{arfgen}. The latter is used in unvignetted mode (equivalent to a flat detector map), returning the on-axis effective area, because the vignetting is already corrected event-wise.

We use the same event pattern for spectra as for imaging. We use the spectral-fitting package XSPEC \citep{1996ASPC..101...17A} version 12.9.0e, with spectra rebinned with a minimum of 25~counts to allow the use of the $\chi^2$-statistic. Interstellar absorption is reproduced by the photoelectric absorption model \texttt{phabs} in XSPEC (or \texttt{vphabs}, where the prefix ``v'' indicates that abundances can vary), using cross-sections set to those of \citet{1992ApJ...400..699B}.

The extraction regions for \texttt{SRC} and \texttt{BG} spectra are defined manually, usually guided by the X-ray contours. Simple shapes (circles, ellipses) are preferred, but an arbitrary shape (e.g. polygonal region) is also used if required. Point sources detected during the pipeline data reduction with the task \texttt{edetectchain} are excluded. Details of the definition of extraction regions are given in \citetalias{2016A&A...585A.162M}. 

When defining extraction regions, we also screen out observations not suited for spectroscopy (that might have been used for imaging). For instance, we do not use the shorter observations if many longer exposures are available, those at large off-axis angle, and those where the SNR is only partially in the field of view (i.e. over the detector edges). In the end, a variety of spectra combination can be found, from pn/MOS1/MOS2 data from a single observation (e.g. for MCSNR~J0051$-$7321), up to a combination of spectra from six observations (e.g. for MCSNR~J0058$-$7217, fitting 16 spectra simultaneously).

Spectra extracted from FWC data at the same detector positions as the \texttt{SRC} and \texttt{BG} regions are used to fit the instrumental background model. It comprises electronic noise and particle-induced background, as described in \citet{2008A&A...478..575K,2012PhDT......ppppS}; and \citetalias{2016A&A...585A.162M}. The instrumental background is not vignetted, and the vignetting-weighting process used on science data distorts its spectrum, particularly at high-energy where the vignetting effect is the strongest. We correct for this by including an ad-hoc multiplicative spline function in the model of the instrumental background. The best-fit models are used in subsequent fits (including astrophysical signal) with no free parameter, as the instrumental background averaged in the FWC dataset matches generally well with the one in the SNR spectra.

One component of the background is the SMC diffuse emission. This was studied by \citet{2012PhDT......ppppS}, who modeled the diffuse emission with a thermal model\,:
\begin{equation}
S_{\mathrm{SMC\ diff}} = 
\mathrm{\texttt{phabs}} \left(N_H ^{\mathrm{Gal}} \right)
\times 
\mathrm{\texttt{vphabs}} \left( N_H ^{\mathrm{SMC*}}, 0.2 Z_{\sun} \right)
S_{\mathrm{apec}} ^{\mathrm{SMC}}
\label{eq_SMCdiff}
\end{equation}
where $S_{\mathrm{apec}} ^{\mathrm{SMC}}$ is the emission from an \texttt{apec} model\,\footnote{Using AtomDB 3.0, \url{http://www.atomdb.org/index.php}.} at temperature $kT^{\mathrm{SMC}}$ and normalisation $N^{\mathrm{SMC}}$. The foreground column density $N_{H} ^{\mathrm{Gal}}$ at the location of each analysed source is taken from the \ion{H}{I} maps of \citet{1990ARA&A..28..215D}. $N_H ^{\mathrm{SMC*}}$ is between 0 and $N_H ^{\mathrm{SMC}}$, the total line-of-sight column density through the SMC \citep{1999MNRAS.302..417S}. Parameters of SMC diffuse emission are given by \citet{2012PhDT......ppppS} in a grid of 240 boxes (size of 9\arcmin\,$\times$\,9\arcmin), containing all but one SNR (MCSNR~J0127$-$7333). The significance of the diffuse emission component is higher than 3$\sigma$ in all the boxes hosting an SNR, except for MCSNR~J0040$-$7336, where the diffuse emission is very faint (the normalisation of the diffuse component generally correlates with its significance). For completeness, we included the diffuse SMC emission in the background model of MCSNR~J0041$-$7336, as it does not affect the rest of the fit much.

The remaining astrophysical X-ray background (AXB) comprises Galactic and extragalactic components: unabsorbed thermal emission from the local hot bubble and an absorbed two-temperature plasma emission from the halo. The cosmic X-ray background is modelled as a power law with a photon index $\Gamma$ of 1.41 \citep{2004A&A...419..837D}. The final model for the AXB is\,:
\begin{equation}
\begin{aligned}
S_{\mathrm{AXB}} =
S ^1 _{\mathrm{apec}} + & \mathrm{\texttt{phabs}} (N_H ^{\mathrm{Gal}} ) ( S ^2 _{\mathrm{apec}} + S ^3 _{\mathrm{apec}} + \\
 & \mathrm{\texttt{vphabs}} \left(N_H ^{\mathrm{SMC}}, 0.2 Z_{\sun} \right) N_{\mathrm{CXB}} E^{-\Gamma} )
\end{aligned}
\label{eq_XRB}
\end{equation}
The temperatures of the thermal components ($kT^1 = 108$~eV, $kT^2 = 36$~eV, and $kT^3 = 247$~eV) and normalisations are taken from \citet{2012PhDT......ppppS}, where this model was fit on observations around the main SMC field. Thus, it gives a fair representation of Galactic foreground and extragalactic background towards the SMC.

Another non-X-ray background component is the soft proton contamination (SPC), which we model following the prescription of \citet{2008A&A...478..575K}. The SPC parameters are highly time-variable and position-dependent, so they were different for each instrument and observation.

The final background model (FWC + $S_{\mathrm{SMC\ diff}}$ + $S_{\mathrm{AXB}}$ + SPC) is fit to the \texttt{BG} spectra prior to fitting source emission. In most cases good fits are obtained with only a constant renormalisation factor for $S_{\mathrm{SMC\ diff}}$ and $S_{\mathrm{AXB}}$. For five SNRs, the background model fit was significantly improved by varying the parameters of the SMC diffuse emission ($N_H ^{\mathrm{SMC*}}$ and $kT^{\mathrm{SMC}}$) or the normalisations of the various XRB components. This is likely due to variations of the X-ray background or diffuse emission on small angular scales.

\section{Results and discussion}
\label{results}

\subsection{Final sample}
\label{results_sample}

We searched all available optical, radio and X-ray surveys in order to secure the most complete population of the SMC SNRs. The number of confirmed SNRs in the SMC is currently at \confcount\ (see Table~\ref{tab:smcsnrs}). Sources previously classified as SNRs which were not included in the final sample are discussed in Sect.\,\ref{results_rejectedSNRs}. In addition, we list in this work two SMC SNR candidates which are presented here for the first time (Table~\ref{tbl:candsnrs}). These new SNR candidates are given the identifiers MCSNR candidate J0106$-$7242 and MCSNR candidate J0109$-$7318, and join two other candidates, MCSNR candidate J0056$-$7209 and MCSNR candidate J0057$-$7211, which were presented in \citet{2012A&A...545A.128H}. Primarily, we classified the \candcount\ SMC SNR candidates based on the well established criteria described in \citet{1998A&AS..130..421F}. For more details, see Table~\ref{tbl:candsnrs} and Sect.~\ref{results_notesSNRs}.

The extent of all \totalcount\ SNRs and SNR candidates is primarily measured using MCELS images, with some additional information obtained via our various radio images as well as \cxo,\xmm, or \ROSAT\ surveys when needed. Where possible, we determined SNR diameters from the highest resolution image available including optical and X-ray images. We estimated that the error in diameter is smaller than 2\arcsec\ or $\sim$0.58~pc. We also found that our diameters shown here could be different at different wavebands (usually within $\sim$10\% ) as it was the case in the LMC \citepalias{2017ApJS..230....2B,2016A&A...585A.162M}. All SMC SNRs and SNR candidates' radio flux density measurements are shown here for the first time and their associated errors are well below 10\%. For the sake of consistency, we assumed a common distance of 60~kpc to all sources for our measurements and derived properties. The expected dispersion along the line-of-sight due to the depth of the SMC (see Sect.\,\ref{results_3D}) is likely higher than the uncertainties of e.g. angular sizes and fluxes. Because of their very low surface brightness we could not detect radio emission from two of the \candcount\ SMC SNR candidates. Also, we could not measure the flux density of MCSNR~J0103$-$7201 (Haberl et al., in prep) as it is a very weak radio source with a very thin (but distinguishable in our high sensitivity ATCA-CABB observations) shell, which overlaps with the neighbouring massive \HII\ region DEM~S103 (see Fig.\,\ref{fig_appendix_sfh5}). Therefore, Table~\ref{tab:smcsnrs} is a compilation of our own measurements as well as those of other papers for this well established sample of the SMC SNRs.

\subsection{Objects not included}
 \label{results_rejectedSNRs}
 
We present here a list of objects previously classified as SNR or SNR candidate that, upon closer scrutiny and in light of the new datasets in radio and X-rays, can no longer be bona-fide SNRs. Most of these objects were originally suggested to be (possible) SNRs based on a single feature (e.g. radio, X-rays). None were later confirmed by a multi-wavelength detailed study, although they have been since included in SNR samples and compilations. This attempt at ``cleaning'' the literature will be beneficial for population studies, making sure they do not include such unrelated objects \citep[as e.g. in][]{2010MNRAS.407.1301B,2019ApJ...871...64A}, which is likely to introduce biases.

\paragraph{N\,S19 / [FBR2002] J004806$-$730842}: From the ATCA radio catalogue of \citet{2005MNRAS.364..217F}, this source is the very confused LHA 115-N 19 (hereafter N\,19) \ion{H}{ii} complex. Radio emission thus includes both thermal contribution from the \ion{H}{ii} region and non-thermal (synchrotron) emission by the nearby three genuine SNRs in that area (see Sect.\,\ref{results_notesSNRs} and Fig.\,\ref{fig_appendix_sfh1}). The new ASKAP data do not resolve this source into a shell, as expected for a true SNR. Very faint soft X-ray emission was noted in the SMC X-ray survey \citep{2012A&A...545A.128H}, but it is to the northeast of the radio source and from a larger region, itself surrounded by small optical features. We conclude that the radio features are likely just thermal emission from optical nebulosities within N\,19, while the large size, soft and very faint X-ray diffuse emission is akin to a superbubble, resembling e.g. LHA 120-N 51D in the LMC \citep{2003IAUS..212..637B,2010ApJ...715..412Y}.

\paragraph{SNR B0045$-$73.3}: Like N\,S19, it is a radio source in the N\,19 complex \citep{2002MNRAS.335.1085F}, but the ASKAP emission is not resolving a shell, instead highlighting optical nebulosity of the \ion{H}{ii} region with interior [\ion{O}{iii}] emission, a better indication of photoionisation. Some X-ray diffuse emission is also seen \citep{2012A&A...545A.128H}, but again it does not match the radio or optical features. This could be just hot gas seen within N\,19.
 
\paragraph{N\,S21 / [FBR2002] J004748$-$731727}: It is an unresolved radio source \citep{2002MNRAS.335.1085F} with no X-ray emission \citep[down to $\sim10^{-14}$~erg~cm$^{-2}$~s$^{-1}$~arcmin$^{-2}$,][]{2012A&A...545A.128H}. The small optical nebulosities at that position point towards the radio source being thermal emission from photoionised \ion{H}{ii} regions.
 
\paragraph{IKT7 / [HFP2000]424}: It was suggested as an SNR candidate based on \textit{Einstein} observatory hardness ratios \citep{1983IAUS..101..535I}. Later this source was confidently identified via its 172~s pulsations as the Be/X-ray binary AX J0051.6$-$7311 \citep{2004A&A...414..667H}. The absence of optical, radio, or diffuse X-ray emission proves that this is not an SNR, a conclusion already mentioned in \citet{2012A&A...545A.128H}.

\paragraph{DEM S130 / [FBR2002] J010539$-$720341}: This is a radio source around a bright emission-line star (LHA 115-N 78C). DEM S130 designates the compact \ion{H}{ii} region which likely is the source of thermal radio emission. There is no X-ray emission despite this area being the deepest covered with \xmm, in the field of view of MCSNR~J0104$-$7201.

\paragraph{LHA 115-N 83C}: There are no studies of this radio candidate \citep{2005MNRAS.364..217F} in the southeast of the SMC, which shows no X-ray emission. It is likely a compact \ion{H}{ii} region (photoionised, filled with [\ion{O}{iii}] emission) within the LHA~115-N~83 (= NGC 456) complex.


\subsection{X-ray spectral properties of SMC SNRs}
    \label{results_spectral}
    
\subsubsection{General properties}
    \label{results_spectral_general}
    
The results of the spectral analysis for the SMC sample are given in Appendix~\ref{appendix_tables} (Table~\ref{appendix_table_spectra_all}). Only MCSNR~J0103$-$7201 and J0104$-$7201 have not been included (see Sect.\,\ref{results_notesSNRs}). 

\begin{figure}
  \center
    \includegraphics[width=0.99\hsize]{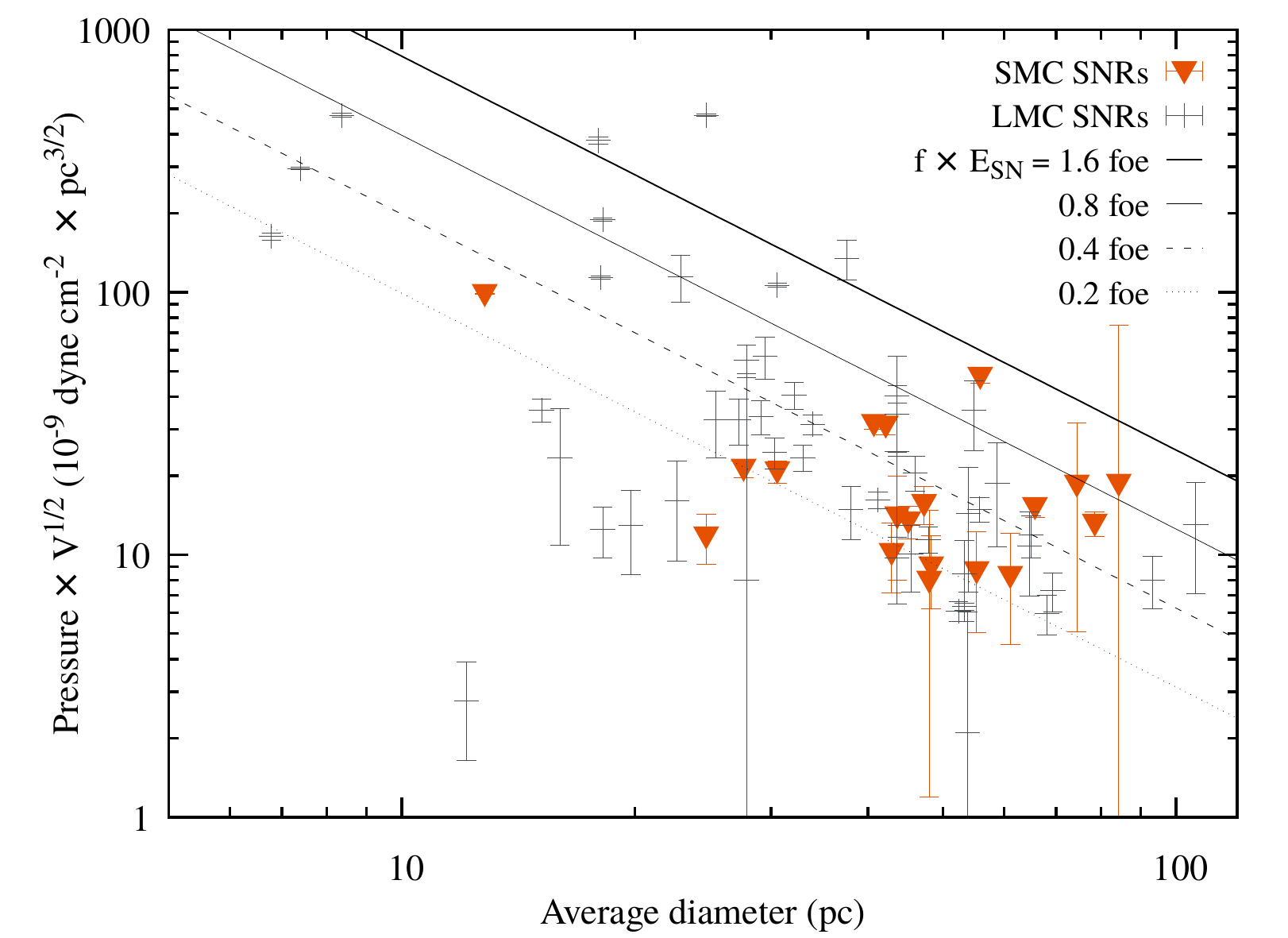}
   \caption{$P \times V^{1/2}$ of SNRs in SMC (orange triangles, this work) and LMC (grey plus signs, \citetalias{2016A&A...585A.162M}) as a function of their size. Lines of constant SN energy (times filling factor $f$, see text in Sect.\,\ref{results_spectral_general}) are overplotted in units of $10^{51}$~erg~$\equiv$1~foe.}
  \label{fig_PV}
\end{figure}

The X-ray analysis opens the possibility to follow the evolution of thermal energy ($P = n \, kT$) of the SNR, whose volume integral we expect to be at most 0.47 $E_{\mathrm{SN}}$ in the Sedov phase, as function of the size $R$. A proxy for the density is obtained from the emission measure as $n_H = f^{-1} 0\sqrt{EM \, / \,1.2\,V}$, with a factor 1.2 for a fully ionised plasma, and a filling factor $f < 1$. The volume was calculated assuming an ellipsoid shape with minor and major axes, and a third axis assumed to be in between these values. We propagate the uncertainty of this assumption in our calculation. To avoid having the size $R$ entering both $x-$ and $y-$axes (because $n \propto R^{-1/3}$), we plot in Fig.\,\ref{fig_PV} the product $P\times\sqrt{V} \equiv f^{-1} \sqrt{EM} \times kT$ as a function of average diameter in pc. For multiple-component spectra, we used the sum of all $EM$, and the $EM$-averaged temperature. Lines of constant energy $f \times E_{\mathrm{SN}} \propto R^{-3/2}$ are overplotted. We show the results for the SMC population (this work) and the LMC SNRs \citepalias{2016A&A...585A.162M}. There is no tight downward correlation. The scatter reflects at least some intrinsic variation in explosion energy, but most likely it is due to part of these SNRs having entered the radiative phase (in many cases indicated by prominent optical emission), which lowered their internal energy.

\subsubsection{Fe~K emission}
    \label{results_spectral_Fe}

Fe~K lines were shown to be a valuable tool to distinguish type~Ia from CC SNRs \citep[][but see caveats in \citealt{2017A&A...597A..65M}]{2014ApJ...785L..27Y}. However, no Fe~K emission has been reported from SMC SNRs in the literature or found in our analysis. We examined the high-energy emission of MCSNR~J0104$-$7201 (1E 0102.2$-$7219), the brightest SNRs in our sample with the hottest plasma, to assess the presence of faint Fe~K emission or derive upper limits. With a simple Bremsstrahlung continuum, we fit the spectrum above 3~keV, a band devoid of strong line besides Fe K, as Ar and Ca lines ($\sim3-4.5$~keV) have much smaller equivalent widths. Then, a Gaussian line with zero width is added at a centroid energy ranging from 6.4~keV to 6.7~keV and the 3$\sigma$ uncertainty on the line normalisation is calculated. Over the tested centroid range the error bars always cross zero, consistent with no detection.

\begin{table}[t]
\caption{SNRs used for measurements of ISM composition (top part), 
and with detected ejecta (bottom part).}
\begin{center}
\label{table_ejecta}
\begin{tabular}{l c c c c | c c c}
\hline\hline
\noalign{\smallskip}
  \multicolumn{1}{c}{\multirow{2}{*}{MCSNR}} &
\multicolumn{4}{c}{High X/Fe flags} &
\multicolumn{3}{c}{Low X/Fe flags}  \\
& \multicolumn{1}{c}{O} &
  \multicolumn{1}{c}{Ne} &
  \multicolumn{1}{c}{Mg} &
  \multicolumn{1}{c}{Si} &
  \multicolumn{1}{c}{O} &
  \multicolumn{1}{c}{Ne} &
  \multicolumn{1}{c}{Mg} \\
\noalign{\smallskip}
\hline
\noalign{\smallskip}
J0046$-$7308 & Y & Y & --- & Y & --- & --- & ---\\
J0047$-$7308 & --- & Y & Y & --- & --- & --- & ---\\
J0047$-$7309 & Y & Y & Y & --- & --- & --- & ---\\
J0049$-$7314 & --- & --- & --- & --- & Y & Y & Y \\
J0051$-$7321 & Y & Y & Y & Y & --- & --- & ---\\
J0059$-$7210 & Y & Y & --- & --- & --- & --- & ---\\
J0103$-$7209 & --- & Y & --- & --- & --- & --- & ---\\
J0104$-$7201 & Y & Y & Y & Y & --- & --- & ---\\
J0105$-$7223 & Y & Y & Y & Y & --- & --- & ---\\
J0105$-$7210 & --- & --- & --- & --- & Y & Y & Y \\
J0106$-$7205 & --- & --- & --- & --- & Y & Y & Y \\
\noalign{\smallskip}
\hline
\noalign{\medskip}
& \multicolumn{7}{c}{ISM abundances} \\
& \multicolumn{2}{c}{O} & \multicolumn{2}{c}{Ne} & 
\multicolumn{2}{c}{Mg} & \multicolumn{1}{c}{Fe} \\
\noalign{\smallskip}
\hline
\noalign{\smallskip}
J0047$-$7308 & \multicolumn{2}{c}{---} & \multicolumn{2}{c}{---} & 
\multicolumn{2}{c}{---} & \multicolumn{1}{c}{Y}\\
J0047$-$7309 & \multicolumn{2}{c}{---} & \multicolumn{2}{c}{---} & 
\multicolumn{2}{c}{---} & \multicolumn{1}{c}{Y}\\
J0051$-$7321 & \multicolumn{2}{c}{Y} & \multicolumn{2}{c}{Y} & 
\multicolumn{2}{c}{Y} & \multicolumn{1}{c}{Y}\\
J0052$-$7236 & \multicolumn{2}{c}{Y} & \multicolumn{2}{c}{Y} & 
\multicolumn{2}{c}{Y} & \multicolumn{1}{c}{Y}\\
J0056$-$7209 & \multicolumn{2}{c}{Y} & \multicolumn{2}{c}{Y} & 
\multicolumn{2}{c}{Y} & \multicolumn{1}{c}{Y}\\
J0058$-$7217 & \multicolumn{2}{c}{Y} & \multicolumn{2}{c}{Y} & 
\multicolumn{2}{c}{Y} & \multicolumn{1}{c}{Y}\\
J0059$-$7210 & \multicolumn{2}{c}{---} & \multicolumn{2}{c}{---} & 
\multicolumn{2}{c}{Y} & \multicolumn{1}{c}{Y}\\
J0105$-$7223 & \multicolumn{2}{c}{Y} & \multicolumn{2}{c}{Y} & 
\multicolumn{2}{c}{Y} & \multicolumn{1}{c}{Y}\\
\noalign{\smallskip}
\hline
\end{tabular}
\end{center}
\end{table}

\begin{figure*}
  \center
    \includegraphics[width=0.98\columnwidth,angle=0,trim=0 0 0 0mm,scale=1.01,clip]{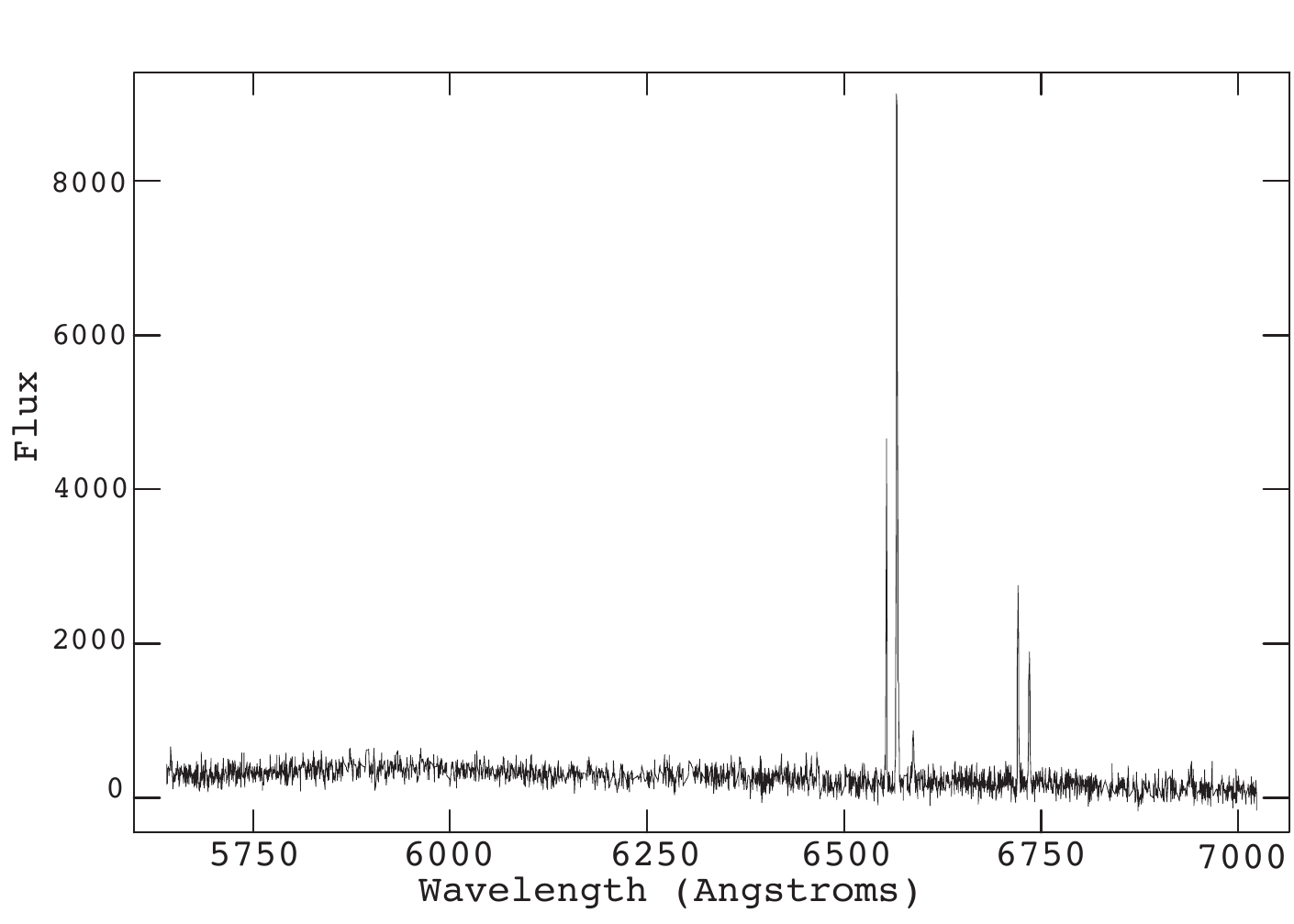}
    \hspace{0.01\columnwidth}
    \includegraphics[width=0.98\columnwidth,trim=0 0 0 7mm, scale=0.98, clip]{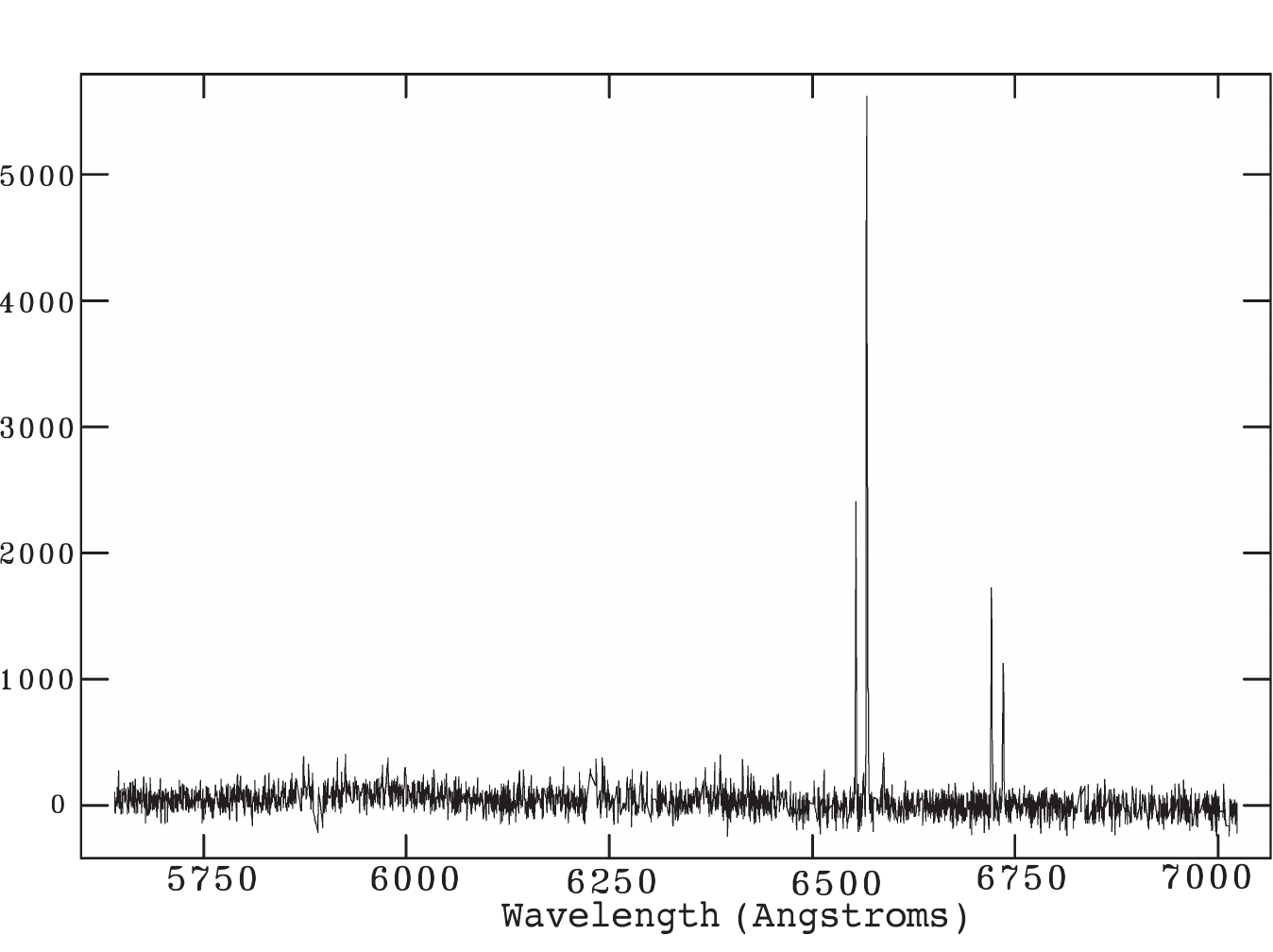}
   \caption{Red part of optical spectra of SNR candidates J0056--7209 (left) and J0109--7318 (right) as seen by the WiFeS spectrograph. All main lines characteristics of old SNRs are seen: \NII$\lambda\lambda 6548,6583$\AA, \Halpha\ and \SII$\lambda\lambda 6717, 6731$\AA.}
  \label{candidate_spectra}
\end{figure*}

\begin{table*}
    \centering 
  \small
  \caption{Emission line intensities$^\dagger$ and ratios for two SNR candidates observed with the WiFeS spectrograph, taking \Halpha=100}
  \setlength{\tabcolsep}{1pt}
  \begin{tabular}{@{\extracolsep{4pt}}ccccccccccccccccc}    
  \hline\noalign{\smallskip}
  Date&Object Name&\multicolumn{2}{c}{Slit position (J2000)}&\NII &\Halpha&\NII&\SII&\SII&\NII/\Halpha&\SII/\Halpha&\SII &Electron Density\\
   &MCSNR&RA&DEC& 6548\AA& &6583\AA &6717\AA &6731\AA & & &6717/6731\AA &(cm$^{-3}$) \\
  \hline\noalign{\smallskip}
08/11/2015&J0056--7209& 00 55 59 &--72 10 04 & 26.7 & 100$^a$ & 14.8 & 27 & 20.4 & 0.42 & 0.47 & 1.32 & $\sim$10$^{2}$\\
08/11/2015&J0109--7318& 01 09 47 &--73 19 27 & 29.5 & 100$^b$ & 9.1  & 28.1 & 18 & 0.39 & 0.46 & 1.56 & LDL \\
   \hline
  \end{tabular}
%
\tablefoot{
\tablefoottext{a}{\Halpha\ flux = 19154 counts;}
\tablefoottext{a}{\Halpha\ flux = 8837 counts.}
\tablefoottext{$^\dagger$} {The rms wavelength dispersion error from the arc calibrations was 0.09\AA\ while the relative percentage error in the flux determination from the calibration using the brightest lines was estimated as $\sim$13\%.}
  }
  \label{table_spectro}
\end{table*}

The upper limit was at most $0.9 \times 10^{-6}$~ph\,s$^{-1}$\,cm$^{-2}$. In the whole sample of Fe~K emitting SNRs, only two Galactic sources (Cas~A and W49B) and two LMC sources (N103B and N132D) would have fluxes above this limit at the SMC distance and would potentially be detectable. At an age of about 2000~yr \citep[1700 to 2600~yr,][]{2006ApJ...641..919F,2019ApJ...874...14X}, the expansion measurement and explosion modelling suggest that the remaining unshocked ejecta in MCSNR~J0104$-$7201 include most of the iron produced in the SN \citep{2019ApJ...874...14X}, explaining the lack of Fe emission. All other SMC SNRs are more evolved and their X-ray temperatures are likely too low to promote Fe~K line emission, in addition to being fainter overall, further preventing detection with existing instruments.

\subsubsection{Detection of SN ejecta}
    \label{results_spectral_ejecta}
Supernova ejecta can be revealed in X-ray spectra by high abundances of metals, significantly above the average SMC abundance (0.1 to 0.2 times solar) or even 
super-solar. Since there are stark contrasts in the nucleosynthesis yields of thermonuclear and CC SNe, ratios of O, Ne, and Mg to Fe abundances can provide 
valuable information as to the type of progenitor of a given SNR.

Using the results of our spectral analysis, we flag in Table~\ref{table_ejecta} the SNRs where the X/Fe abundance ratios (where X is O, Ne, Mg, or Si) are significantly higher or lower than the average SMC value \citep{1992ApJ...384..508R}. The three SNRs with \textit{low} X/Fe ratios are those akin to several evolved LMC SNRs with iron-rich, centrally-bright emission \citepalias{2016A&A...585A.162M}. In some cases an elevated Fe abundance or sometimes a pure Fe component are needed to fit the spectra, leading to their classification as ``low X/Fe'' cases, even if the low-Z elements were not left free. We make use of these flags in the typing of SMC SNRs (Sect.\,\ref{results_typing}).

\subsection{New SMC SNR candidates}
 \label{CAND}

\subsubsection{Optical spectroscopy}
\label{CAND_spectro}
Examination of WiFeS 1D spectra (see Fig.~\ref{candidate_spectra}) shows all main lines typical of old SNRs. They show lines of \NII\ at 6548 and 6583\AA, \Halpha\ and \SII\ at 6717 and 6731\AA. The latter are shock-sensitive line: the ratio of \SII\ lines to \Halpha\ should be $\gtrsim$~0.40 for an object to be classified as an SNR shock from the point of view of optical spectra. Our MCSNR candidates have values of 0.47 and 0.46 (see also Table ~\ref{table_spectro}) so we can classify them as SNRs. Also, the ratio between the individual \SII\ lines (6717~\AA / 6731~\AA) of 1.32 and 1.56 is fully in line with optical SNR spectra. The only exception are \NII\ lines whose ratios to \Halpha\ have values of 0.39 and 0.42, which is somewhat low. In Milky Way SNRs, this (mostly) would not be accepted as an SNR, however in the SMC it is different. From the early optical spectral observations of the MCs \citep[see examples in][]{1979AuJPh..32..123D} it is well known that \NII\ lines are very weak due to an abundance effect: nitrogen, particularly in the SMC, is even more underabundant (by 0.3~--~0.5~dex) than other elements \citep{1992ApJ...384..508R,2019AJ....157...50D}.
So, if we exclude the value and comparison of nitrogen lines with \Halpha, all other spectral characteristics of these two candidates match SNRs.



\subsubsection{Notes on individual SNR candidates}
\label{results_notes_candidates}
\paragraph{MCSNR candidate J0056$-$7209} is a large optical shell of 99~pc by 65~pc in size (Fig.~\ref{MC}, left), among the largest SNRs in the SMC. The weak diffuse X-ray emission, first identified in \citet{2012A&A...545A.128H}, is however confined to the northern region of the optical loop only, centred at RA(J2000)=00\hour56\minute33.0\second\ and Dec(J2000)=--72\D08\arcmin00\arcsec\ and with extent of about 48~pc. Using  WiFeS spectroscopic data, we found a strong indication of shock excitation with \SII/\Halpha\ ratio of 0.46 (Fig.\,\ref{candidate_spectra}, left), lending further support to a true SNR nature for this source.  We also found that the weak \OIII\ emission is coinciding with the X-ray emission of MCSNR candidate J0056$-$7209, which we confirm as of thermal nature with low metal abundances (Fig.\,\ref{fig_appendix_candidates}, Table~\ref{appendix_table_spectra_all}). The lack of radio continuum detection is surprising but not unheard of \citep[see Venn diagrams of ][\citetalias{2017ApJS..230....2B}]{2010ApJ...725..842L}, although this issue arises more commonly for galaxies beyond the Local Group where sources are often not spatially resolved.
Based on optica and X-ray features, we can nevertheless confidently confirm this source as a \textit{bona-fide} SNRs, attributing it the identifier MCSNR~J0056$-$7209.

\paragraph{MCSNR candidate J0057$-$7211 (aka N\,S66D)}: it was first suggested as an SNR candidate based on the \xmm\ mosaic of the SMC \citep{2012A&A...545A.128H}. The faint and extended soft X-ray emission is very close to the listed position, which is based on our new ASKAP EMU (SMC Early Science Project) radio continuum images (Joseph et al., submitted). The radio emission forms a partial shell at the north, while the diffuse thermal X-ray emission completes a shell in the south and south-western quadrant (Fig.\,\ref{fig_appendix_candidates}; note a likely unrelated point source at the south-western edge of the SNR). The low absorption measured in X-rays towards this SNR (Table~\ref{appendix_table_spectra_all}) suggests a position on the near side of the SMC. Its radio detection is clear, although it is a low surface brightness SNR (S$_{1 \rm{GHz}} =$~0.031~Jy). Its radio SED is quite steep ($\alpha=-0.75\pm0.04$). There is no obvious optical emission associated to that source (Fig.\,\ref{fig_appendix_candidates}), but the combination of radio and X-ray evidence leads us to conclude as a true SNR nature for it, to which we assign the identifier MCSNR~J0057$-$7211. However, it is an optically ``quiet'' SNR, similar to some Galactic SNRs as described by \citet{2008MNRAS.390.1037S}.

\paragraph{MCSNR candidate J0106$-$7242}: this is a newly suggested SMC SNR candidate based solely on its radio-continuum detection and morphology. We discovered this candidate in our new ASKAP radio continuum images (Joseph et al., submitted). Although a low surface brightness SNR candidate (S$_{1 \rm{GHz}} =$~0.0236~Jy), its radio SED is typical for an SNR ($\alpha=-0.55\pm0.02$, Fig.\,\ref{fig_appendix_candidates}). We do not detect any optical emission from this object -- similarly to SMC SNR [HFP2000] 334. Also, we found no significant X-ray detection, but this region is poorly covered with low exposure time (20~ks only, combining all EPIC detectors). From the soft and medium fluxed mosaic image in and around the radio SNR position (Table~\ref{tbl:candsnrs}), we estimate a 3~$\sigma$ upper limit of $1.9\times10^{34}$~erg s$^{-1}$ for the 0.3-8~keV luminosity (or $8.3\times 10^{-15}$~erg~cm$^{-2}$~s$^{-1}$~arcmin$^{-2}$ for surface brightness) of the MCSNR candidate J0106$-$7242. As several SMC SNRs are similarly faint (Table~\ref{tab:smcsnrs}), there is still comfortable room for a subsequent X-ray detection with deeper observations. Until then this object remains a good SNR candidate.

\paragraph{MCSNR candidate J0109$-$7318}: we suggest this shell-like object (Fig.~\ref{MC}, right) as an SNR candidate because of its strong \SII\ emission. Namely, its \SII/\Halpha\ ratio is 0.46 (Fig.~\ref{candidate_spectra}, right), which indicates shock emission that could be attributed to an SNR nature. However, none of our present generation radio images show any signs of the object due to its projected proximity to a bright radio source \citep[the background AGN XMMU J011053.5-731415,][]{2013A&A...558A.101S}, whose sidelobes distort any nearby emission. The X-ray coverage is mediocre, and in addition it was covered at high off-axis angle in the two overlapping observations of that region, further decreasing the possibility of detecting associated emission. Like MCSNR candidate J0106$-$7242, this object remains a candidate awaiting X-ray confirmation.

\subsection{Notes on individual SMC SNRs}
 \label{results_notesSNRs}

\paragraph{MCSNR~J0041$-$7336\,/\,DEM~S5\,:} This SNR is a particularly large optical shell around a central X-ray emission (Fig.\,\ref{fig_appendix_sfh0}), first studied with ROSAT and \xmm\ in \citet{2000A&AS..142...41H} and \citet{2008A&A...485...63F}, respectively. The presence of an X-ray point-like source within the remnant associated to resolved radio emission recently led \citet{2019MNRAS.486.2507A} to the discovery of a candidate pulsar wind nebula (PWN). The X-ray analysis of the diffuse emission (SNR component) in this work, which was also presented in \citet{2019MNRAS.486.2507A}, reveals that the emission arises from shocked ambient medium, as no abundance enhancement is found.

\paragraph{The "triumvirate" of LHA 115-N 19\,:} In this large optical emission nebula lie the three SNRs MCSNR~J0046$-$7308, J0047$-$7308, and J0047$-$7309. Although it is a confused region in the optical due to the bright emission and the lack of well-defined borders, these three sources have strongly different X-ray colours that allow us to distinguish them (Fig.\,\ref{fig_appendix_sfh1}). The spectral fits indicate a similar temperature for these three SNRs ($\approx 0.6$~keV). The variety in X-ray colours is instead due to variations in $N_H$ (by up to an order of magnitude), ionisation age, and abundances. For instance, MCSNR~J0047$-$7309 has highly elevated O, Ne, and Mg abundances (several times solar), while the strongest abundance enhancement of J0046$-$7308 is Si. Ne and Mg are also higher than solar in J0047$-$7308, while its high absorption ($N_H = 1.4 \pm 0.2 \times 10^{22}$~cm$^{-2}$)  prevents a meaningful measurement of its oxygen abundance. Finally, we note that the large variation of $N_H$ between these three objects indicates that even though close in \textit{projected} position, they might be at different distances and \textit{not} associated to the same star-forming event/region, as implicitly assumed in \citet{2019ApJ...871...64A}.

\paragraph{MCSNR~J0048$-$7319\,/\,IKT~4\,:} This faint SNR has an irregular X-ray emission, filling a well-defined optical shell (Fig.\,\ref{fig_appendix_sfh0}) and peaking in the Fe L-shell band, which led to the suggestion \citepalias{2004A&A...421.1031V} that this was a type~Ia SNR, similar to those discovered later by \citet{2006ApJ...652.1259B,2014MNRAS.439.1110B,2014A&A...561A..76M}. We have doubled the exposure time compared to the first \xmm\ analysis of \citetalias{2004A&A...421.1031V}, enabling us to confirm enhanced iron abundances. However, the Mg abundance is also formally enhanced in our spectral fits, making a conclusion as to the type of progenitor indecisive for IKT~4.

\paragraph{MCSNR~J0049$-$7314\,/\,IKT~5\,:} We found the interior X-ray emission to be enriched in iron (Fig.\,\ref{fig_appendix_sfh1}), which we fit with a supplementary Fe-only component. The first component possibly shows enhanced Mg abundance, but not as markedly as in IKT~4. This iron-rich core inside an [\ion{S}{ii}] shell and radio dimness make it very similar to other evolved type~Ia SNRs found in the LMC.

\paragraph{MCSNR~J0051$-$7321\,/\,IKT~6\,:} The third brightest among SMC SNRs in X-rays, IKT~6 has two components, with ejecta-dominated emission in the centre, surrounded by a soft X-ray shell of shocked SMC-abundance ISM (Fig.\,\ref{fig_appendix_sfh2}). The abundance pattern of the ejecta (elevated Ne, Mg, and Si) betray a core-collapse SN origin. The shell can be used to measure SMC ISM abundances (Sect.\,\ref{results_abundances}).

\paragraph{MCSNR~J0052$-$7236\,:} Only the south-west part of this structure was suggested as an SNR, before \citet{2012A&A...545A.128H} suggested a possible close connection with other X-ray knots further north, linked by very faint emission. This could either be two close SNRs or a single large one. The slightly brighter SW X-ray knots correlate with the strongest optical emission, while the N part exhibits small filaments in [\ion{S}{ii}], possibly part of the remnant (Fig.\,\ref{fig_appendix_sfh2}). Our X-ray spectral analysis reveals that the N and SW knots have strikingly similar spectra ($N_H$, $kT$, abundances). Combined with the morphology, we propose that these knots indeed form a single, large SNR, actually the largest SNR of the SMC.

\paragraph{MCSNR~J0058$-$7217\,/\,IKT~16\,:} This SNR is atypical in the SMC because of the hard extended source near its centre, suggested as the first PWN of the SMC \citep{2011A&A...530A.132O} and then confirmed with high-resolution \chandra\ observations \citep{2015A&A...584A..41M}. To characterise the extended soft X-ray emission from the underlying SNR, we included the pulsar and PWN components obtained in the \chandra\ analysis with fixed parameters in our analysis of the integrated emission. The X-ray size is roughly circular (1.2\arcmin\ radius) and matches some fait optical filamentary structure (Fig.\,\ref{fig_appendix_sfh3}), albeit over a confused larger nebula.

\paragraph{MCSNR~J0059$-$7210\,/\,IKT~18\,:} Just 10\arcmin\ north-east of IKT~16, IKT~18 has an irregular centre-filled X-ray morphology (Fig.\,\ref{fig_appendix_sfh3}). Its location in a large optical nebula (N66) makes an optical identification difficult, but analysis of radio to \Halpha\ emission ratio allowed \citet{1991MNRAS.249..722Y} to separate the larger \ion{H}{ii} region from the SNR emission which matches the detected X-ray SNR fairly well. The abundances measured in the X-ray spectrum are low, except for O and Ne that are slightly above the SMC average values. This possibly points to ejecta contamination from regions we cannot pinpoint with the available data and spatial resolution of \xmm.

\paragraph{MCSNR~J0100$-$7133\,/\,DEM~S108\,:} The northernmost SNR of the SMC sample (by half a degree) is detected in radio, optical, and X-rays (Fig.\,\ref{fig_appendix_sfh4}). The low surface brightness of the latter does not allow for a definite conclusion regarding elemental abundances. The diffuse interior X-ray emission is well outlined by an optical shell with enhanced [\ion{S}{ii}] and strong [\ion{O}{iii}] emission typical of radiative shocks \citep{1971ApJ...167..113C}, indicating an evolved SNR.

\paragraph{MCSNR~J0103$-$7209\,/\,IKT~21\,:} Analysis of this SNR is complicated by the bright point-source AX~J0103$-$722 within it, first identified in \citet{1994AJ....107.1363H}, and confirmed as a Be/X-ray binary by \citet{2000ApJ...531L.131I}. A compact radio and optical shell ($\approx$90\arcsec\ diameter) is seen around the X-ray binary (Fig.\,\ref{fig_appendix_sfh4}), suggested as an SNR by \citet{1984ApJS...55..189M}. No X-rays were found in this region \citep{1995MNRAS.275.1218Y} until the \xmm analysis of \citetalias{2004A&A...421.1031V}, who modelled the faint thermal X-ray emission simultaneously with that of the binary. Much like for MCSNR~J0052$-$7236, the \xmm mosaic \citep{2012A&A...545A.128H} revealed a much larger diffuse emission than the former compact nebula. We used that 270\arcsec\ diameter region for our X-ray spectral analysis. We could thus afford to excise the X-ray binary point-source contribution by excluding a circle of 50\arcsec\ radius (i.e. 90\,\% encircled energy fraction at 8\arcmin\ off-axis angle). This removes about 15\,\% of the total SNR area. The emission was best fit with an NEI model, with only neon having a higher abundance than SMC ISM, which we take as a marginal indication of a core-collapse origin (Sect.\,\ref{results_spectral_ejecta}). Finally, we note that the best-fit $N_H$ of the large thermal SNR is about half that towards the X-ray binary \citep[$3.9\times10^{21}$~cm$^{-2}$;][]{2000ApJ...531L.131I}. In addition, the binary is far from the centre of our larger SNR, making an SNR-binary physical association far less likely than e.g. in SXP~1062 \citep{2012A&A...537L...1H}.

\paragraph{MCSNR~J0103$-$7247\,/\,[HFP2000]~334\,:} Without any detected optical emission (Fig.\,\ref{fig_appendix_sfh5}), this object, discovered via its faint radio and X-ray emission \citep{2008A&A...485...63F}, was thought to host a putative PWN because of a central radio and X-ray point-like source. Later resolved with \chandra \citep{2014AJ....148...99C}, it was however attributed to a background object. We included the spectral parameters of that source obtained with \chandra \citep{2014AJ....148...99C} in our analysis of the integrated X-ray emission. The extent of the SNR was measured using both radio and X-ray contours.

\paragraph{MCSNR~J0103$-$7201\,:} This faint SNR is detected as an H$\alpha$ circular shell (Fig.\,\ref{fig_appendix_sfh5}) centered on the long-period X-ray pulsar SXP~1323 \citep{2019MNRAS.485L...6G}. Deep radio and co-added \chandra\ data reveal faint radio and X-ray emission of the shell, thus confirming its SNR status (Haberl et al., in prep.). Interestingly, this is the second case in the SMC of an SNR containing a Be X-ray binary after SXP~1062 (see below), both objects harbouring long-period pulsars ($> 1000$~s).

\paragraph{MCSNR~J0104$-$7201\,/\,IKT~22\,:} Most commonly known as 1E 0102.2$-$7219, this is the brightest X-ray and radio SNR in the SMC (Fig.\,\ref{fig_appendix_sfh5}). It has been and continues to be extensively studied. Not wanting to expand on the bulk of past detailed studies, our approach was merely to include it in our analysis for completeness, and use a multi-component spectral model to derive consistently its X-ray luminosity. We needed three NEI components with variable abundances to satisfactorily reproduce the integrated emission. For pn spectra, a redshift parameter was added to account for small gain variations \citep{2017A&A...597A..35P} that shift the line centroid between observations and would otherwise cause large residuals. We have independently verified the lack of detected Fe~K emission at high energies (Sect.\,\ref{results_spectral_Fe}).

\paragraph{MCSNR~J0105$-$7223\,/\,IKT~23\,:} This is the second brightest X-ray SNR of the SMC, although the radio flux density is about the median value of the whole sample. It is also very similar to IKT~6, with a soft ISM-abundance shell enclosing an ejecta-enhanced hotter plasma. The abundances of O and Ne are clearly super-solar, and in particular are enhanced relative to iron (Table~\ref{appendix_table_spectra_all}), yielding a clear CC SN origin for this remnant. In the optical, the remnant is not seen in [\ion{S}{ii}] or H$\alpha$. However, a faint thin shell of [\ion{O}{iii}] emission delineates clearly the remnant on the \textit{outer} side of the soft X-ray shell (Fig.\,\ref{fig_appendix_sfh6}). This is strongly suggestive of the radiative part of the outer blast wave, from regions where the plasma cooled down below X-ray emitting temperatures. Furthermore, this transition is very recent, as [\ion{O}{iii}] is emitted before H$\alpha$ in cooling order.

\paragraph{MCSNR~J0105$-$7210\,/\,DEM~S128\,:} This SNR is slightly elongated and consists mainly of diffuse interior X-ray emission and a faint radio shell encasing it at the northern and southern ends (Fig.\,\ref{fig_appendix_sfh6}). In the optical only faint emission in the north can be associated to this object. The interior X-ray emission exhibits a strong iron enhancement, which was already seen with \xmm \citepalias{2004A&A...421.1031V} and \chandra \citep{2015ApJ...803..106R}, that we interpret as a strong indicator of a type~Ia SN origin (Sect.\,\ref{results_typing}).

\begin{table*}[t]
\caption{Abundances of the SMC ISM.}
\begin{center}
\label{table_abundances}
\begin{tabular}{@{\hspace{0.05em}}c @{\hspace{0.05em}} c @{\hspace{0.0em}} c @{\hspace{0.0em}} c @{\hspace{0.0em}} c @{\hspace{0.0em}} c @{\hspace{0.0em}} c @{\hspace{0.0em}} c @{\hspace{0.05em}}}
\hline\hline
\noalign{\smallskip}
\noalign{\smallskip}
  \multicolumn{1}{c}{} & 
  \multicolumn{1}{c}{SMC SNRs} &
  \multicolumn{1}{c}{weighted SNRs} &
  \multicolumn{1}{c}{RD92 (\ion{H}{II} + SNRs)} &
  \multicolumn{1}{c}{vdH04 (SNRs)} &
  \multicolumn{1}{c}{Radiative shocks} &
  \multicolumn{1}{c}{B stars} &
  \multicolumn{1}{c}{\ion{H}{II} regions}
  \\
\noalign{\smallskip}
 & (1) & (2) & (3) & (4) & (5) & (6) & (7) \\
\noalign{\smallskip}
\noalign{\smallskip}
\hline
\noalign{\smallskip}
O  & 7.80$_{-0.10}^{+0.38}$ & 7.61$_{-0.16}^{+0.59}$ & 8.03$\pm0.10$ & $8.13_{-0.16}^{+0.11}$ & 8.02$\pm$0.06 & 7.99$\pm$0.21 & 7.99$\pm$0.04\\
\noalign{\smallskip}
Ne & 7.17$_{-0.11}^{+0.39}$ & 7.02$_{-0.15}^{+0.54}$ & 7.27$\pm0.20$ & 7.50$\pm0.15$ & 7.04$\pm0.10$ & --- & 7.22$\pm$0.04 \\
\noalign{\smallskip}
Mg & 6.76$_{-0.13}^{+0.41}$ & 6.76$_{-0.13}^{+0.42}$ & 6.98$\pm0.12$ & 7.07$\pm0.13$ & (6.72)$^{\dagger}$ & 6.72$\pm$0.18 & --- \\
\noalign{\smallskip}
Fe & 6.60$_{-0.14}^{+0.32}$ & 6.35$_{-0.26}^{+0.58}$ & 6.84$\pm0.13$ & 6.75$\pm0.18$ & (6.77)$^{\dagger}$ & --- & --- \\
\noalign{\medskip}
$[$Fe/H$]$&-0.83$_{-0.14}^{+0.32}$&-1.08$_{-0.26}^{+0.58}$&-0.59$\pm0.13$&-0.68$\pm0.18$ & -0.66 & --- & --- \\
\noalign{\smallskip}
$[$O/Fe$]$&-0.06$_{-0.18}^{+0.49}$&-0.01$_{-0.60}^{+0.64}$&-0.07$\pm0.16$&0.12$_{-0.24}^{+0.21}$ &  -0.01 & --- & --- \\
\noalign{\smallskip}
$[\alpha$/Fe$]$&-0.03$_{-0.24}^{+0.75}$& 0.06$_{-0.63}^{+0.94}$ & -0.06$\pm0.15$&0.15$_{-0.31}^{+0.29}$ & -0.04 & --- & --- \\
\noalign{\smallskip}
\hline
\end{tabular}
\end{center}
\vspace{-0.3em}
\tablefoot{Elemental abundances are given as 12+log(X/H); abundance ratios follow the convention $[$X/Y$]$ = log(X/Y) - log (X/Y)$_{\sun}$. Columns (1) to (4) are listing \textit{gas-phase} abundance values, while columns (5) to (7) are total gas+dust abundances. References for (1) and (2)\,: this work; (3)\,: \citet{1992ApJ...384..508R} ; (4)\,: \citetalias{2004A&A...421.1031V} ; (5)\,: \citet{2019AJ....157...50D} ; (6)\,: \citet{2009A&A...496..841H} ; (7)\,: \citet{2001A&A...372..667T}. $^{\dagger}$ Values assumed in their model.}
\end{table*}

\paragraph{MCSNR~J0106$-$7205\,/\,IKT~25\,:} Bright in X-rays and optical but relatively radio-dim (Fig.\,\ref{fig_appendix_sfh7}), J0106$-$7205 has a debated type. The elevated iron abundances measured with \xmm \citepalias{2004A&A...421.1031V} and \chandra \citep{2011ApJ...731L...8L} led to the suggestions that it was a type~Ia SNR, like IKT~5 and DEM~S128. It was argued in \citet{2014ApJ...788....5L} that the SMC abundance pattern used in the spectral analysis of \citet{2011ApJ...731L...8L} was erroneous. While true, this does not affect the conclusion that the Ne/Fe ratio was clearly skewed towards iron. We found a similar result in our re-analysis of the \xmm\ data. On the contrary, \citet{2014ApJ...788....5L} or \citet{2016PASJ...68S...9T}, using \textit{Suzaku} data, measured a Ne/Fe $\gtrsim 1$ ratio, albeit without resolving an \ion{Ne}{X} line that would be a tell-tale sign of enhanced neon enhancement. Meanwhile, the argument of the disrupted morphology being against a type~Ia origin \citep{2014ApJ...788....5L} remains weak, as other SNRs have been found with elongated iron-rich cores where the optical emission does not follow the diffuse X-ray emission \citep[e.g. DEM~L238, DEM~L249, or MCSNR~J0527$-$7104,][]{2006ApJ...652.1259B,2016A&A...586A...4K}, as is the case in IKT~25.

\paragraph{MCSNR~J0127$-$7333\,/\,SXP~1062\,:} Located far off to the south-east, in the Wing of the SMC, this SNR is observed as an optical shell \citep{2012MNRAS.420L..13H}, and radio and X-ray shell \citep[][Fig.\,\ref{fig_appendix_sfh7}]{2012A&A...537L...1H}. The central source is an associated Be/X-ray binary, harbouring a long-period pulsar (SXP~1062). In this work we analysed the diffuse X-ray shell, including more observations that were obtained subsequently for the monitoring of the central binary \citep{2013A&A...556A.139S}.

\subsection{Abundances of the SMC ISM}
    \label{results_abundances}

The current elemental abundances of the SMC ISM have been measured using several methods:
\textit{i)} Spectrophotometric observations of photospheric abundances of B stars \citep{2009A&A...496..841H} or O-type dwarfs \citep{2013A&A...555A...1B}, as these are young, short-lived stars and thus still presenting ISM abundances at their surfaces. Complications arise from the modelling of non-local thermodynamical equilibrium effects \citep[e.g.][]{2010PASJ...62.1239T} and the amount of rotational mixing \citep{2006ApJ...638..409H}.
\textit{ii)} Photoionisation modelling of \ion{H}{ii} nebular spectra \citep{1999ApJ...518..246K,2000ApJ...541..688P,2001A&A...372..667T,2002ApJ...564..704R,2012ApJ...746..115P,2015RMxAA..51..135C} remains affected by uncertainties of available atomic data, escape fraction, and incident spectra.
\textit{iii)} Spectral modelling of radiative shocks in dense ISM clouds, as found within some SNRs \citep{1990ApJS...74...93R,2019AJ....157...50D}.

Finally, in cases where the X-ray emission of SNRs is solely comprised of, or dominated by swept-up ISM, we can use the fitted abundances as measurements of the chemical composition of the ISM gas phase. This was used previously for the LMC \citep{1998ApJ...505..732H,2016A&A...585A.162M,2016AJ....151..161S}. One advantage is that it constrains directly the set of elements most relevant to X-ray observations, those which have emission lines and absorption edges in the 0.3-10~keV band. SMC SNRs have been used previously for that purpose by \citetalias{2004A&A...421.1031V}, but only three objects were used in their study. Here, we attempt to improve this result, taking advantage of the higher number of SNRs known and observed. 

Three SNRs (MCSNR~J0052$-$7237, J0058$-$7217, and J0059$-$7210) had their abundances already measured in the fitting procedure of Sect.\,\ref{results_spectral}. To increase that number, we re-analyse the sample using their previous best-fit model and thawing the abundances of O, Ne, Mg, and Fe. The fit improvements are not statistically significant, since by construction we would have identified these cases in Sect.\,\ref{results_spectral}. If the true abundances in an SNR are very close to the starting values of \citet{1992ApJ...384..508R}, there will be no strong improvement of the $\chi^2$-statistic. Therefore, we look instead at the \textit{uncertainties}, that is, how well the abundance of a given element is constrained. Often the abundances are severely unconstrained (i.e. X/Fe between a small fraction and hundred times the solar value) and we easily discard these objects as unsuitable. We add three SNRs (MCSNR~J0047$-$7308, J0047$-$7309, and J0056$-$7209) to the sample from which (some) abundances can be measured. Although the latter is first presented here as an SNR candidate, we provided strong evidence to its confirmation as an SNR (Sect.\,\ref{results_notesSNRs}), and at any rate its thermal emission is probing the gas-phase abundance of the ISM and can be used for that purpose.

\begin{figure*}[ht]
  \begin{center}
    \includegraphics[height=0.23\textheight]{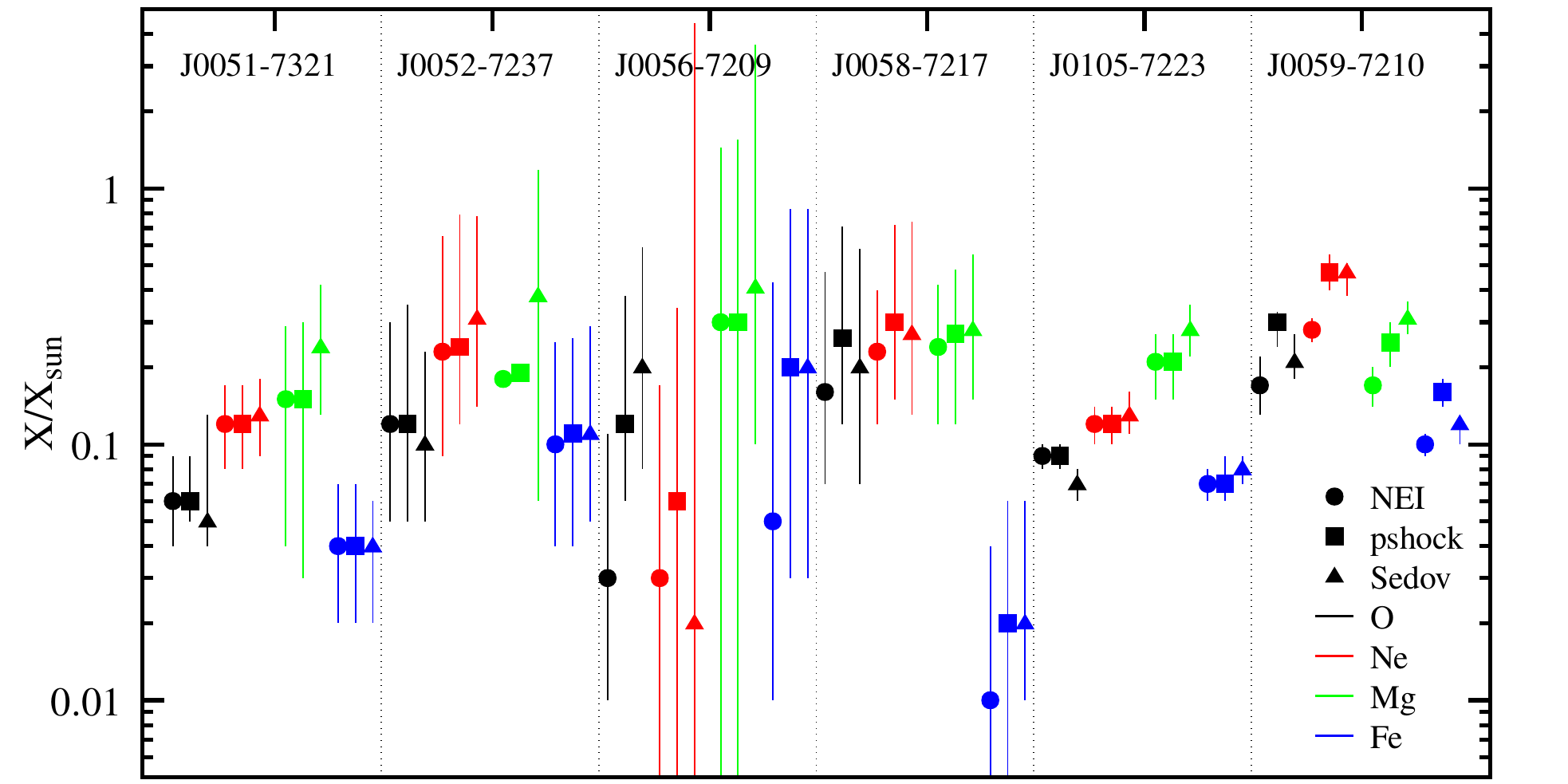}
    \includegraphics[height=0.23\textheight]{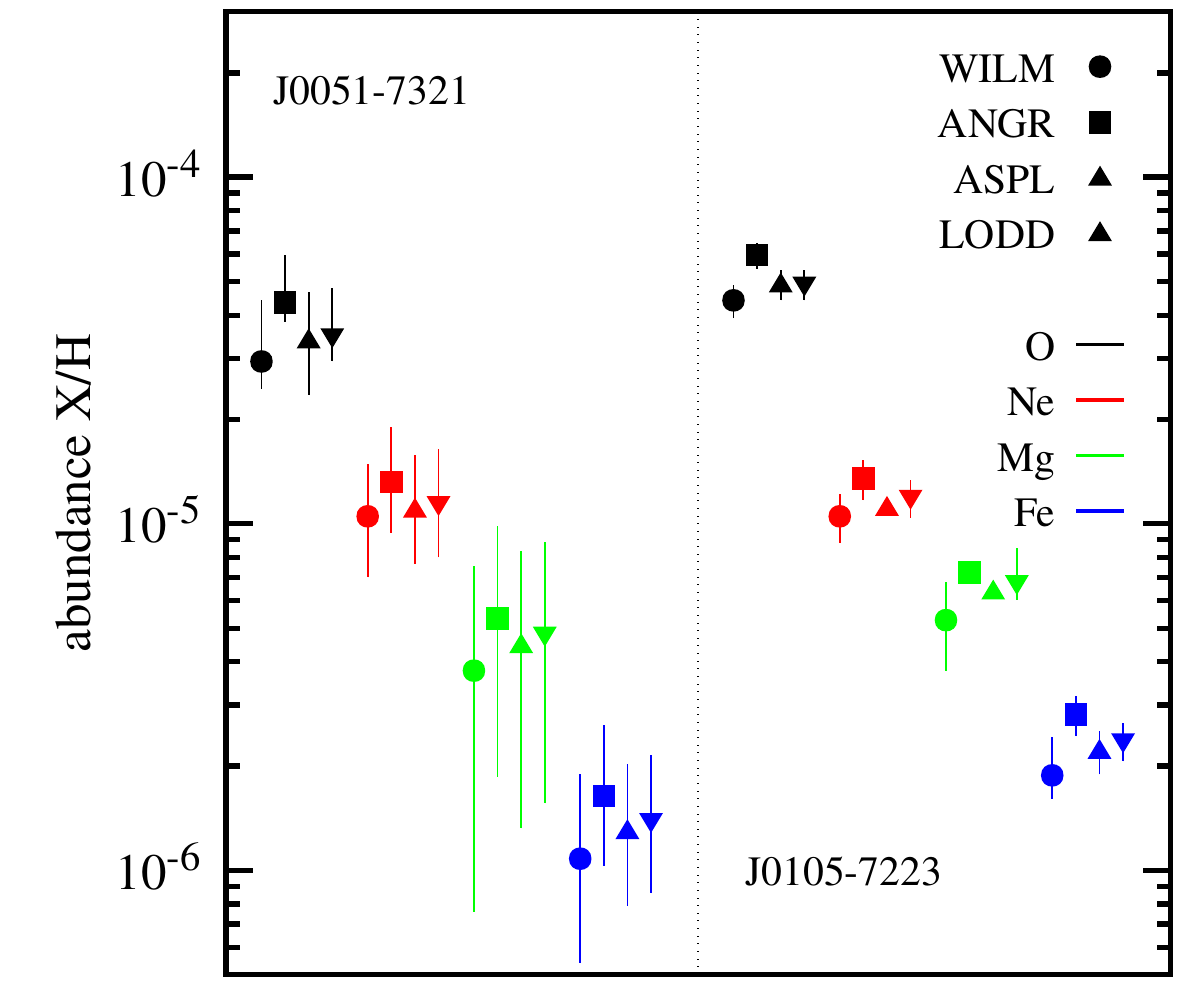}
    \caption{Comparison of the effect of type of spectral models (left) and input abundance tables (right) on abundance derived with X-ray spectra of SNRs.
    \textit{Left\,:} O, Ne, Mg, and Fe abundances (by group of three, from left to right, respectively) relative to the reference value of \citet{2000ApJ...542..914W}, in five SMC SNRs (labelled on top). Different symbols are used for the three types of spectral models used.
    \textit{Right\,:} For two SNRs, derived number abundances relative to hydrogen as function of input abundance tables (reference in Sect.\,\ref{results_abundances}), each coded by different symbols.}
  \label{fig_abund_compare}
  \end{center}
\end{figure*}

Finally, the two bright objects MCSNR~J0051$-$7321 and J0105$-$7223 clearly show two morphological components (see images in \ref{appendix_images}), a central region of shocked ejecta surrounded by a shell of shocked ISM \citep[\citetalias{2004A&A...421.1031V};][]{2005ApJ...622L.117H,2014ApJ...791...50S, 2003ApJ...598L..95P}. For these sources we go beyond the spatially-integrated analysis of Sect.\,\ref{data_spectra}. Using X-ray contours in soft and hard bands, we define an ``interior'' region (ejecta-rich) and a ``shell'' region (ISM), which is the whole SNR minus the interior region and a small buffer zone. This minimises cross-talk between regions and thus contamination of the shell with shocked ejecta emission. The shell emission is well fit by a \texttt{vpshock} model with low abundances ($\sim$ 10~\% solar). We thus obtain eight SNRs in which ISM abundances of various elements can be measured (Table~\ref{table_ejecta}). We list the mean abundances and uncertainties in Table~\ref{table_abundances}, using the simple arithmetic mean for column (1) and a mean weighted in inverse proportion to each SNR uncertainty in column (2).

Before comparing with previous studies, we investigate two potentially important sources of systematic errors: What are the effects of \textit{i)} the chosen NEI models, and \textit{ii)} the abundance table used on the derived SMC abundances? Firstly, we replace the \texttt{vpshock} model by a \texttt{vnei} (single ionisation timescale) or \texttt{vsedov} \citepalias[used in][]{2004A&A...421.1031V} in the analysis of the six SNRs where abundances other than just Fe were measured (Table~\ref{table_ejecta}). The spectral model chosen has no strong impact on the abundances, as shown on Fig.\,\ref{fig_abund_compare} (left). If anything, there is a tendency for the  \texttt{vnei} model to yield slightly lower abundances. Results of a Sedov model, as used in \citetalias{2004A&A...421.1031V} (SMC) and in \citet{1998ApJ...505..732H} to measure LMC abundances, are essentially indistinguishable from those of the \texttt{vpshock} model.

Secondly, we used other abundance tables available in XSPEC (\texttt{ANGR}\,: \citealt[][]{1989GeCoA..53..197A}, \texttt{LODD}\,: \citealt[][]{2003ApJ...591.1220L}, \texttt{ASPL}\,: \citealt[][]{2009ARA&A..47..481A}) to fit the spectra of MCSNR~J0051$-$7321 and J0105$-$7223. The fitted abundances should be insensitive to the different starting points of these tables. However, they also differ in the abundances of trace elements (e.g. odd-Z nuclei) or other non-fitted elements, such as Si and S. This affects the free electron balance ($n_e/n_H$) and thus the emission continuum. Furthermore, in some cases rare elements have lines in similar energy bands as the fitted elements, e.g. N, Ar, Ca in the 0.5-0.6 keV band dominated by oxygen. We show the absolute abundances obtained in Fig.\,\ref{fig_abund_compare} (right). There are no discernible differences. The abundance ratios, e.g. O/Fe, should be the least affected by the choice of abundance tables. Indeed, the scatter is very small, with less than 5~\% scatter between the four input tables.

Our results are best compared to those of \citetalias{2004A&A...421.1031V} since they come from the same environment. In absolute abundances we find values lower by about 0.3 to 0.5 dex. This could be ascribed to our larger sample, including several more evolved SNRs\,: As pointed in \citetalias{2004A&A...421.1031V}, larger remnants tend to have lower abundances\,\footnote{We found a similar trend in our larger sample.} as they swept up more ISM and thus further dilute the effect of potential SN ejecta contamination. On that topic, we note that our derived abundance ratios $[$O/Fe$]$ or $[\alpha$/Fe$]$ are well consistent (within 0.1 dex) with other studies, indicating that we have efficiently vetoed contamination by the more frequent CC SNRs ejecta, as we have shown for the LMC ISM as well \citep[][\citetalias{2016A&A...585A.162M}]{1998ApJ...505..732H}.

\begin{table*}[t]
\caption{``Hint-spec'' attributed to SNRs as function of spectral results.}
\begin{center}
\label{table_hints}
\begin{tabular}{@{}c c @{}}
\hline\hline
\noalign{\smallskip}
\noalign{\smallskip}
  \multicolumn{1}{c}{Hint-spec} &
  \multicolumn{1}{c}{Criteria}
  \\
\noalign{\smallskip}
\noalign{\smallskip}
\hline
\noalign{\smallskip}
1   & at least three ``low X/Fe'' flags AND no ``high X/Fe'' flag \\
1.5 & (two ``low X/Fe'' flags OR low O/Fe) AND no ``high X/Fe'' flag \\
2   & one ``low X/Fe'' flag (except O/Fe) AND ``high X/Fe'' flag \\
2.5 & low Si/Fe AND no ``high X/Fe'' flag \\
3   & ISM abundances, unfitted abundances \\
3.5 & high Si/Fe AND no ``low X/Fe'' flag \\
4   & one `high X/Fe'' flag (except O/Fe) AND no ``low X/Fe'' flag \\
4.5 & (two ``high X/Fe'' flags OR high O/Fe) AND no ``low X/Fe'' flag \\
5   & (at least three ``high X/Fe'' flags AND no ``low X/Fe'' flag) OR 
pulsar/PWN detected \\ 
\noalign{\smallskip}
\hline
\end{tabular}
\end{center}
\end{table*}

We remind that our abundances are those of the (hot) gas-phase. Compared to stellar abundance measurements \citep{2009A&A...496..841H,2013A&A...555A...1B}, or dust depletion-corrected measurements \cite[][Table~\ref{table_abundances}]{2019AJ....157...50D}, we found similar Ne abundance but lower O, Mg, Fe abundances, on average by $\approx$0.1-0.3 dex, which reflects partial depletion of these elements onto dust. However, the depletion factors $D_X = \log(N_X/N_H) - \log(N_X/N_H)_{\mathrm{stellar}}$ are less (closer to 0) than in other ISM phases (e.g. $D_{Fe} \approx -1$ in the warm ionised medium of \ion{H}{ii} regions). This can be explained by the (partial) destruction of dust by the SNR shocks \citep{2006ApJ...642L.141B,2006ApJ...652L..33W,2016ApJ...821...20K}. At least a fraction of these elements are released into the gas phase and are contributing to the observed X-ray emission. Although we do not attempt to quantify this further, the average SMC SNR depletion is less than measured directly in radiative shocks in e.g. LMC SNRs \citep{2016ApJ...826..150D,2018ApJS..237...10D}, probably an effect of the faster shocks probed in X-rays, that increase the intensity of dust grain destruction \citep{2015ApJ...803....7S}.

\subsection{The ratio of CC to type~Ia SNe in the SMC}
  \label{results_typing}
Here, we aim to establish the type (CC or Ia) of all SMC SNRs to measure $N_{\mathrm{CC}}/N_{\mathrm{Ia}}$, the ratio of CC to Ia SNe rates. We covered the various methods of SNR typing in \citetalias{2016A&A...585A.162M}. We mostly use our X-ray spectral results (i.e. the measurement of nucleosynthesis products in the ejecta), or the detection of an associated (NS) or PWN. We then add secondary evidence based on the local stellar environment of SMC SNRs to tentatively type the rest of the sample, a method we explain in detail in \citetalias{2016A&A...585A.162M}.

In Table~\ref{table_ejecta} we flagged the detection of ejecta in 11 SNRs. As in \citetalias{2016A&A...585A.162M}, we assign a number ``hint-spec'' ranging from 1 (strongly favouring a type~Ia origin) to 5 (strongly favouring a CC SN origin) depending on the flags raised, as summarised in Table~\ref{table_hints}. This leads to a (relatively) secure typing for 13 SNRs, including MCSNR~J0058$-$7217 and J0127$-$7333, which host a PWN and a BeXRB, respectively. The remaining six SNRs can only be tentatively typed using the local stellar environment, which we characterised by two metrics as described in the following paragraphs.

\begin{table}[t]
\caption{``Hint--SF'' attributed to SNRs as function of \NOB and $r$.}
\begin{center}
\label{table_hint_SF}
\begin{tabular}{l | c c c }
\hline\hline
\noalign{\smallskip}
 \backslashbox{\NOB}{$r$-value} & $r < 0.6$ & $0.6 < r < 1.5$ & $r > 1.5$  \\
\noalign{\smallskip}
\hline
\noalign{\smallskip}
\NOB $< 80$ & 1 & 1.5 & 2 \\
$80 \leq $\NOB $\leq 115$ & 2.5 & 3 & 3.5 \\
\NOB $>115$ & 4 & 4.5 & 5 \\
\noalign{\smallskip}
\hline\end{tabular}
\end{center}
\end{table}

First, we construct a $V$ vs. $(B-V)$ colour-magnitude diagram (CMD) of all stars within a projected distance of 100~pc ($\sim 5.7$ \arcmin) of each SNR, using the photometric catalogue of \citet[][hereafter MCPS]{2002AJ....123..855Z}. We add stellar evolutionary tracks from \citet{2001A&A...366..538L} to identify the upper main sequence of stars in the SMC, using initial masses from 3 to 40~\msun\ and a metallicity Z $= 0.004 = 0.1$~Z$_{\sun}$. A distance modulus of $\mu = 18.89$  is assumed, and the average extinction for ``hot'' SMC stars is taken as $A_V = 0.6$ \citep{2002AJ....123..855Z}. We use the criteria $V < 16.4$ and $B - V < 0.03$ to select blue early-type stars. The CMDs are shown in Appendix~\ref{appendix_images}. We denote \NOB the number of massive stars ($\gtrsim 8$~\msun) in the vicinity of the remnant identified this way.

Second, we plot for each SNR the star formation rate (SFR) of its surroundings as a function of lookback time (Appendix~\ref{appendix_images}), obtained from the reconstructed SMC star formation history (SFH) of \citet{2004AJ....127.1531H}. Since the SMC SFH is noisier than in the LMC \citep{2009AJ....138.1243H}, we take the average SFH in a grid of $3\times3$ cells centered on each SNR. We then compute $r=N_{\rm CC}/N_{\rm Ia}$, the ratio of CC SNe to thermonuclear SNe expected from the observed distribution of stellar ages in the neighbourhood 
of the remnants, as:
\begin{equation}
\label{eq_r}
    r = \frac{\Psi_1 M_1}{\Psi_2 M_2 + \Psi_3 M_3}
\end{equation}
where $\Psi _{i}$ is the delay-time distribution, the SN rate following a star formation event, as measured by \citet{2010MNRAS.407.1314M} in the MCs, in time intervals $i=1$, 2, and 3 corresponding to $t <$ 35~Myr, 35~Myr $< t <$ 330~Myr, and 330~Myr $< t < 14$~Gyr, respectively. This $r$ provides us with a measure of the relative size of the pool of possible progenitors of both SN types, taking into account their delay-time distributions.

As massive stars are rarely formed in isolation, high values of \NOB and the CC-to-Type Ia SN ratio $r$ in a region hosting an SNR would strongly suggest a CC SN origin, while low values favour type~Ia. We showed it to be the case in the LMC, where SNRs with well-established types (i.e. based on other methods) have bimodal distributions of \NOB and $r$ \citepalias{2016A&A...585A.162M}. In the SMC we have the additional difficulty that there is significant extent along the line of sight, such that \NOB and $r$ might not reflect the correct environment of an SNR. For instance, an SNR might be located in front or behind a star forming region, without its progenitor drawn from that stellar population. If \NOB and $r$ are low, however, it is still a solid indication that no recent star formation occurred along the line of sight, as there is not enough internal extinction to mask the bright young stars that would have been created. Therefore, we can be relatively confident for typing SNRs with low \NOB and $r$ as type~Ia, while classifying the high \NOB--$r$ SNRs as CC should be done with caution.

\begin{figure*}
   \begin{center}
        \includegraphics[width=0.48\hsize]{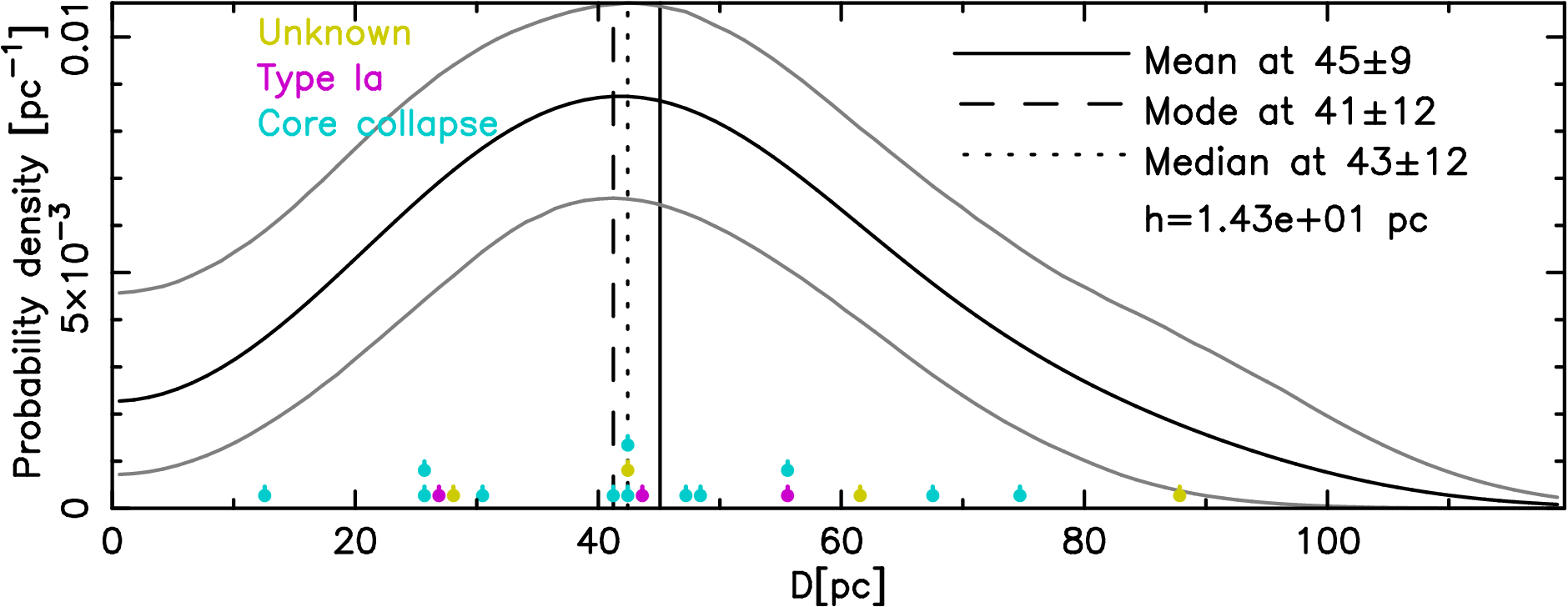}
        \hspace{1em}
        \includegraphics[width=0.48\hsize]{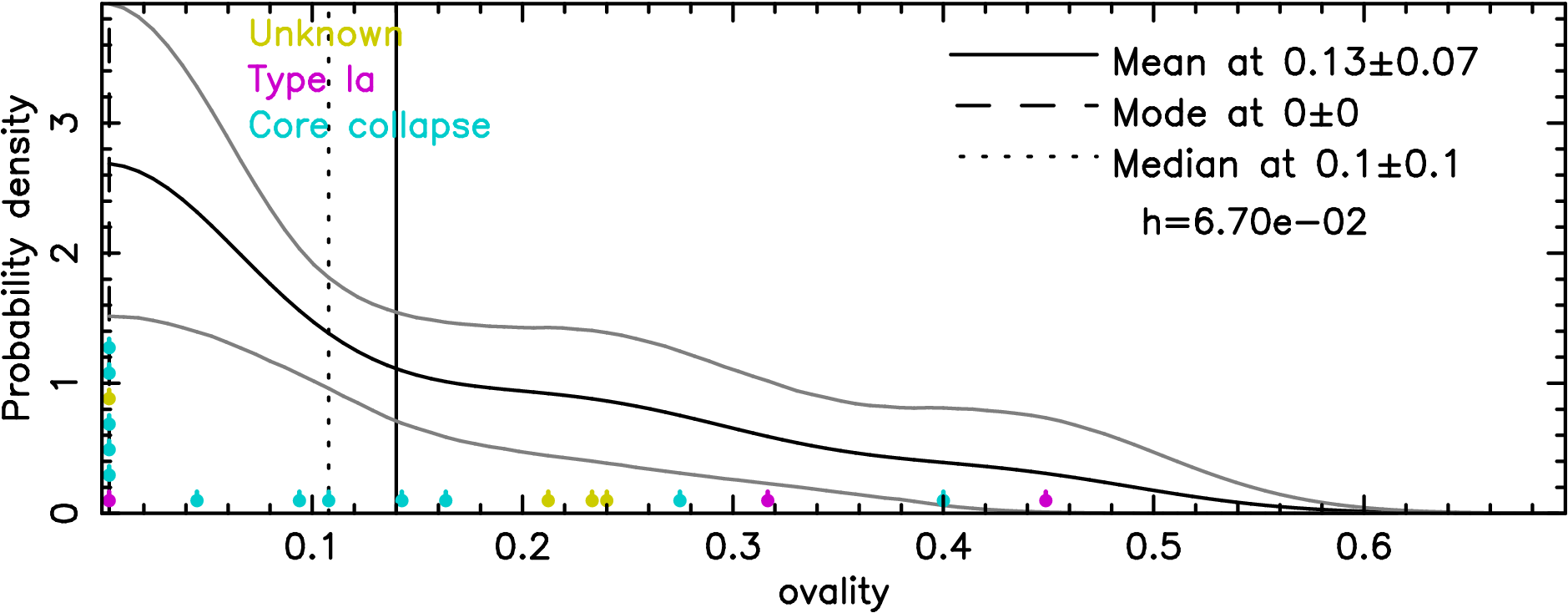}
        
        \vspace{1em}
        \includegraphics[width=0.48\hsize]{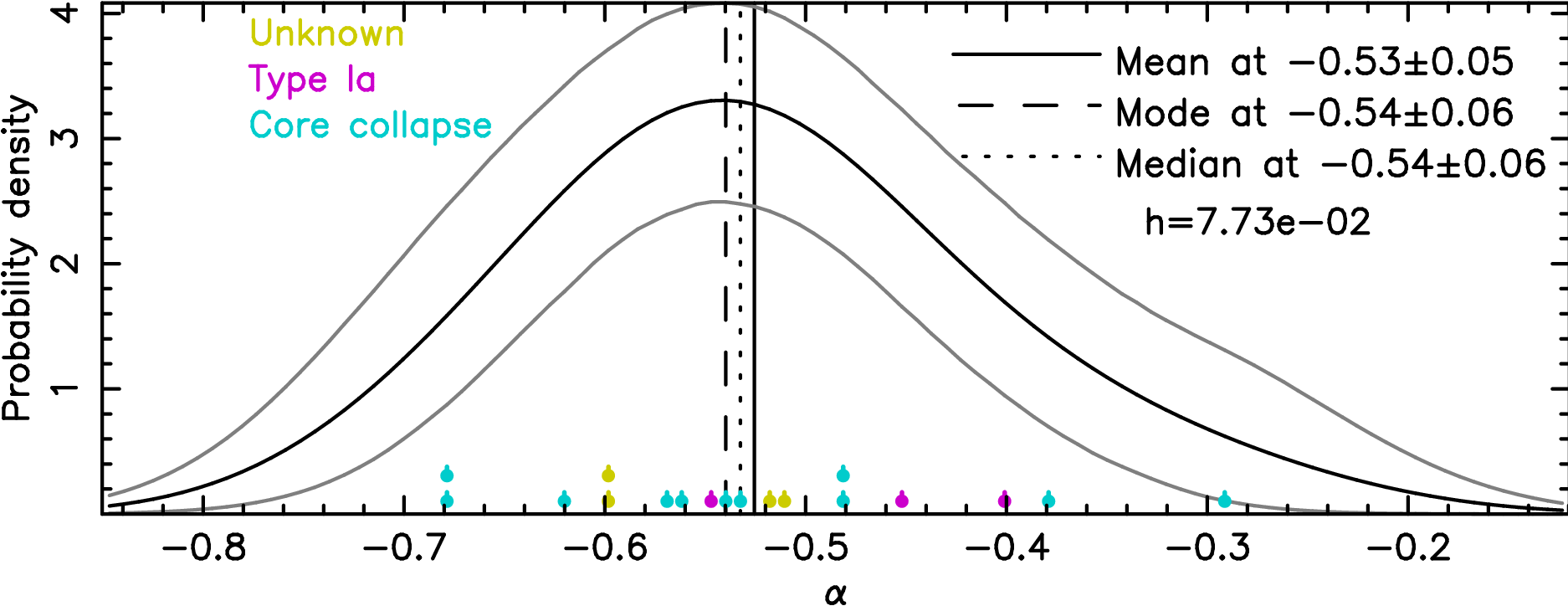}
        \hspace{1em}
        \includegraphics[width=0.48\hsize]{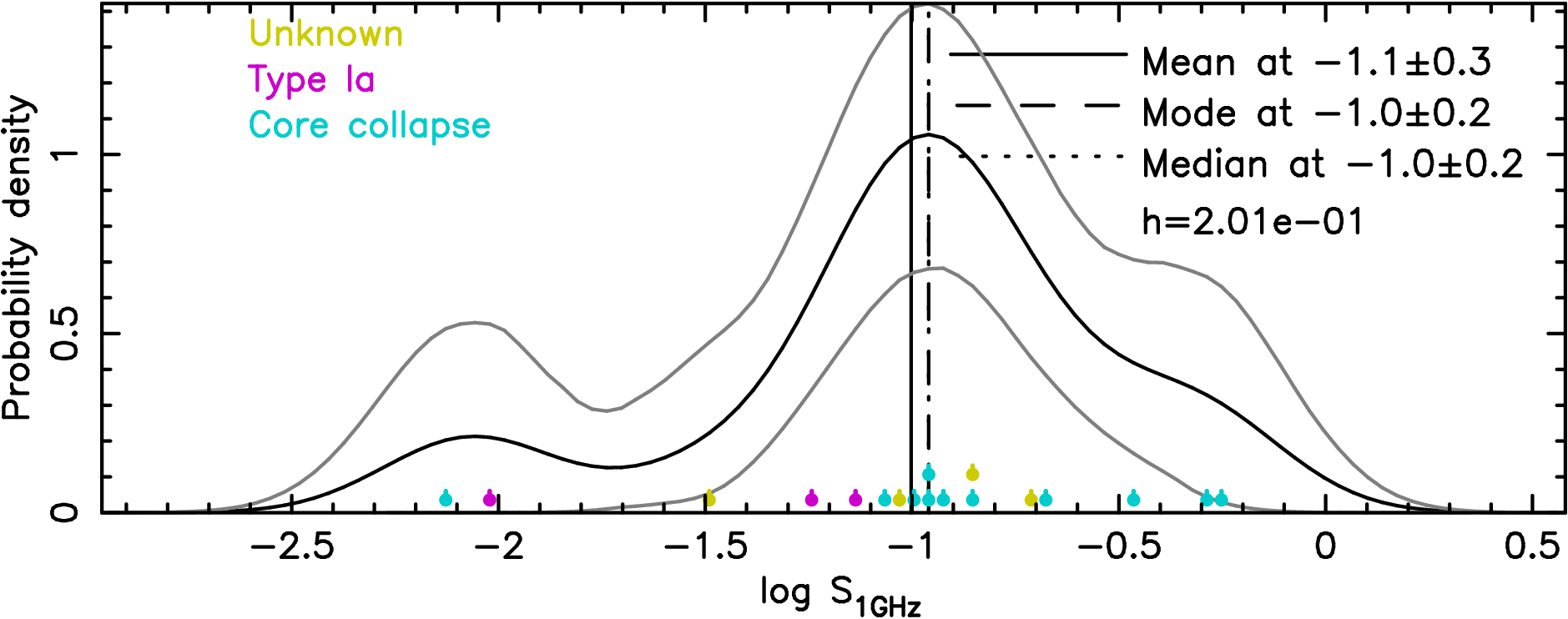}
   \end{center}
   \vspace{-0.8em}
    \caption{Distribution of parameters of 19 confirmed SMC SNRs using kernel smoothing. The colour of symbols indicate progenitor type for a particular SNR. Data points that fall in the same bin on the $x$-axis are plotted within a vertical column with equidistant spacing. The upper and lower gray curves delineate the 95\,\% uncertainty level. The corresponding kernel bandwidth $h$ and distribution parameters are shown on each panel.
    \textit{Top left\,:} Average diameter distribution.
    \textit{Top right\,:} Distribution of ovality, which is defined as $2 \, ( D_{\mathrm{maj}} - D_{\mathrm{min}} ) / ( D_{\mathrm{maj}} + D_{\mathrm{min}} )$, where $D_{\mathrm{maj}}$ and  $D_{\mathrm{min}}$ are the major and minor axes, respectively.
    \textit{Bottom left\,:} Radio spectral index distribution.
    \textit{Bottom right\,:} Estimate of the 1~GHz flux density distribution for the 18 SNRs with available 1~GHz flux density estimates from Table~\ref{tab:smcsnrs}.}
   \label{fig:kernel}
\end{figure*}

The average \NOB\ is $128 \pm 67$ for 18 SNRs\,\footnote{MCSNR~J0127$-$7333 is too far east to be in the area covered by the MCPS and  will not be included in this discussion. It is however classified as a secure CC SNR based on its associated Be/X-ray binary.}. It is higher ($160 \pm 63$) for the ten secure CC SNRs than for the (only) three likely type~Ia ($92 \pm 22$), confirming that this dagnostic has some discriminatory power, even in the SMC. The average \NOB for the remaining six SNRs of uncertain type is 86, with a very large scatter. It includes the two lowest occurrences, 18 around MCSNR~J0040$-$7336 and 32 around J0100$-$7132. MCSNR~J0056$-$7209 has \NOB$= 76$, less than all likely CC SNRs. The three other uncertain SNRs have \NOB between 117 and 142, higher than those of type~Ia SNRs and consistent with many CC SNRs.

Even when averaging over several cells, the SMC SFH is noisy, especially at recent times, which are critical. The distribution of $r$-values of all regions (not just those hosting an SNR) is less bimodal than in the LMC, without a prominent high-$r$ peak. This is again due to the elongated shape of the SMC along the line of sight. Older episodes of star-formation permeate most of the Cloud \citep{2004AJ....127.1531H} and are seen in projection in all the cells, thus lowering $r$ and blurring its peak in regions with recent star formation. Consequently, we put more emphasis on \NOB than $r$. We choose three intervals for \NOB, each split in three intervals depending on $r$, to assign a number ``hint-SF'' (for star formation) to our SNR, following the criteria from Table~\ref{table_hint_SF}. As with ``hint-spec'', values close to 1 favour type~Ia, and those close to 5 favour a CC origin.

We combined the two hints by taking their weighted mean, with a coefficient of two for the "hint-spec" which is deemed to be more critical, since it does not have the projection effect of the star-formation hint. We take a slightly more conservative approach than in the LMC, classifying sources as ``likely-Ia'' when the final hint is $< 2.5$, and ``likely-CC'' if it is $> 3.5$. Objects between 2.5 and 3.5 (inclusive) remain undecided. There are two SNRs in that category (MCSNR~J0048$-$7319 and J0100$-$7133). Therefore, we estimate that in the SMC \ccIa~$=$~4.7 (14/3), with lower and upper limits of 2.8 (14/5) and 5.3 (16/3), respectively. The limits are obtained if the undecided SNRs are assigned to either types.

The value in the LMC, measured by a similar method but with a sample three times larger, is 1.35 (1.11--1.46) \citepalias{2016A&A...585A.162M}. Hence, the ratio \ccIa\ appears to be higher in the SMC. We argued in \citetalias{2016A&A...585A.162M} that the apparent excess of type~Ia SNe in the LMC, as compared to direct SN search in the local universe or \ccIa\ measurements from intracluster medium abundances, was due to the specific recent and intermediate age SFH of the LMC. Several studies found enhanced star formation episodes at 1.5~--~2 Gyr ago and 250~--~500 Myr ago, based on both CMD fitting of field stars \citep[at various limiting magnitudes,][]{2009AJ....138.1243H,2013MNRAS.431..364W,2012A&A...537A.106R, 2014MNRAS.438.1067M} and star cluster formation history \citep[e.g.][]{2013MNRAS.430..676B}, which mostly agree with field star formation at recent times \citep{2011MNRAS.411.1495M}. Combined with the type~Ia delay-time distribution, peaking below 1~Gyr \citep{2012PASA...29..447M}, there is a large pool of possible progenitors for type~Ia SNRs.

In the SMC, several studies point to a major SFR enhancement about 5~Gyr ago \citep{2015MNRAS.449..639R,2013MNRAS.431..364W,2012ApJ...754..130C, 2009ApJ...705.1260N}, possibly related to early LMC~--~SMC interaction, with only some evidence for a secondary peak at 1.5~Gyr ago \citep{2015MNRAS.449..639R,2012ApJ...754..130C}. \citet{2004AJ....127.1531H} found the most significant intermediate star formation episode 2~--~3~Gyr ago. In recent times, SFR peaked again 200~--~400 Myr ago, most notably on the LMC side with the formation of the SMC Wing by tidal interaction. The smaller SMC SFR 0.5~--~1.5~Gyr ago as compared to the LMC could explain the currently smaller number of type~Ia SNRs.

An important caveat, however, is that while recent star formation is strong in the Bar and Wing regions which are well covered with \xmm, the outskirts (at galactocentric radius larger than 1.5\textdegree) are poorly known, and might host more type~Ia SNRs owing to the ancient SFH of these regions \citep[e.g.][]{2015MNRAS.449..639R}. The detection of low surface brightness SNRs in these areas might be possible in the near future with the \textit{eROSITA} all-sky survey and subsequent pointed surveys \citep{2012arXiv1209.3114M}.

Our suggested classification of three SNRs as type~Ia is mostly driven by their X-ray spectral features, most notably the large Fe abundance of ejecta origin, and are considered robust. In no case is an SNR classified as type~Ia based on local (projected) star-formation alone. MCSNR~J0041-7336 (DEM S5) is far off to the south-west of the main SMC Bar and thus has the lowest recent star formation of the whole sample, but a PWN candidate was recently identified in it, strongly suggesting a CC SN origin. This explains the discrepancy with \citet{2019ApJ...871...64A} who list only one type~Ia candidate SNR, but whose CC SNR classification is based on the projected star-formation history alone. Given the significant extent/depth of the SMC, such an interpretation is not warranted. A similar study for the LMC population, however, would be much more significant because the recent star forming regions and other regions potentially hosting type~Ia SNRs are better segregated thanks to the thinness of the LMC and a favourable viewing angle.

\subsection{Radio properties, size, and morphology of SMC SNRs}
\label{results_radio_properties}
As in \citetalias{2017ApJS..230....2B} we estimate the distributions of the radio parameters for the sample of 19 confirmed SMC SNRs using kernel smoothing with a Gaussian kernel. The maximum likelihood method with ``leave one out'' cross-validation is applied \citep{duin1976choice} in order to calculate the optimal smoothing kernel bandwidth ($h$). Confidence bands are calculated from $10^4$ bootstrap \citep{efron1994introduction} resamples. The kernel bandwidth, optimal for the original data sample, is used to calculate the resulting distributions of the bootstrap resamples. All distributions are calculated at $100$ equidistant points along the plotted interval (on the x-axis, Fig.\,\ref{fig:kernel}). At each $x$-axis coordinate we calculate the median value and the confidence bands as the $95\%$ confidence interval around the median value. The same procedure and re-sampled data is also used to estimate uncertainties of the distribution mean, mode and median (Fig.\,\ref{fig:kernel}). The boundary correction for the smoothed distributions that cannot have negative values (diameter and ovality) is  done using the reflection method \citep{silverman1986density}. Note that in \citetalias{2017ApJS..230....2B} a different method (smooth bootstrap resampling) was used to estimate the optimal kernel bandwidth, but the method in this work is less computationally intensive and better suited to apply data reflection. For the flux density distribution we used a log scale.

Figure~\ref{fig:kernel} shows estimates of the distributions for average diameter, ovality, radio spectral index and 1~GHz flux density. The diameter and radio spectral index distributions appear to be symmetric. The ovality shows significant asymmetry with many data points consistent with zero (circular morphology).
The median ovality is the same within the uncertainties for SMC and LMC, although the SMC distribution contains a higher fraction of circular SNRs, pointing to a less disturbed ambient medium in that galaxy.

\begin{figure}[t]
  \centering
    \includegraphics[width=0.99\hsize]{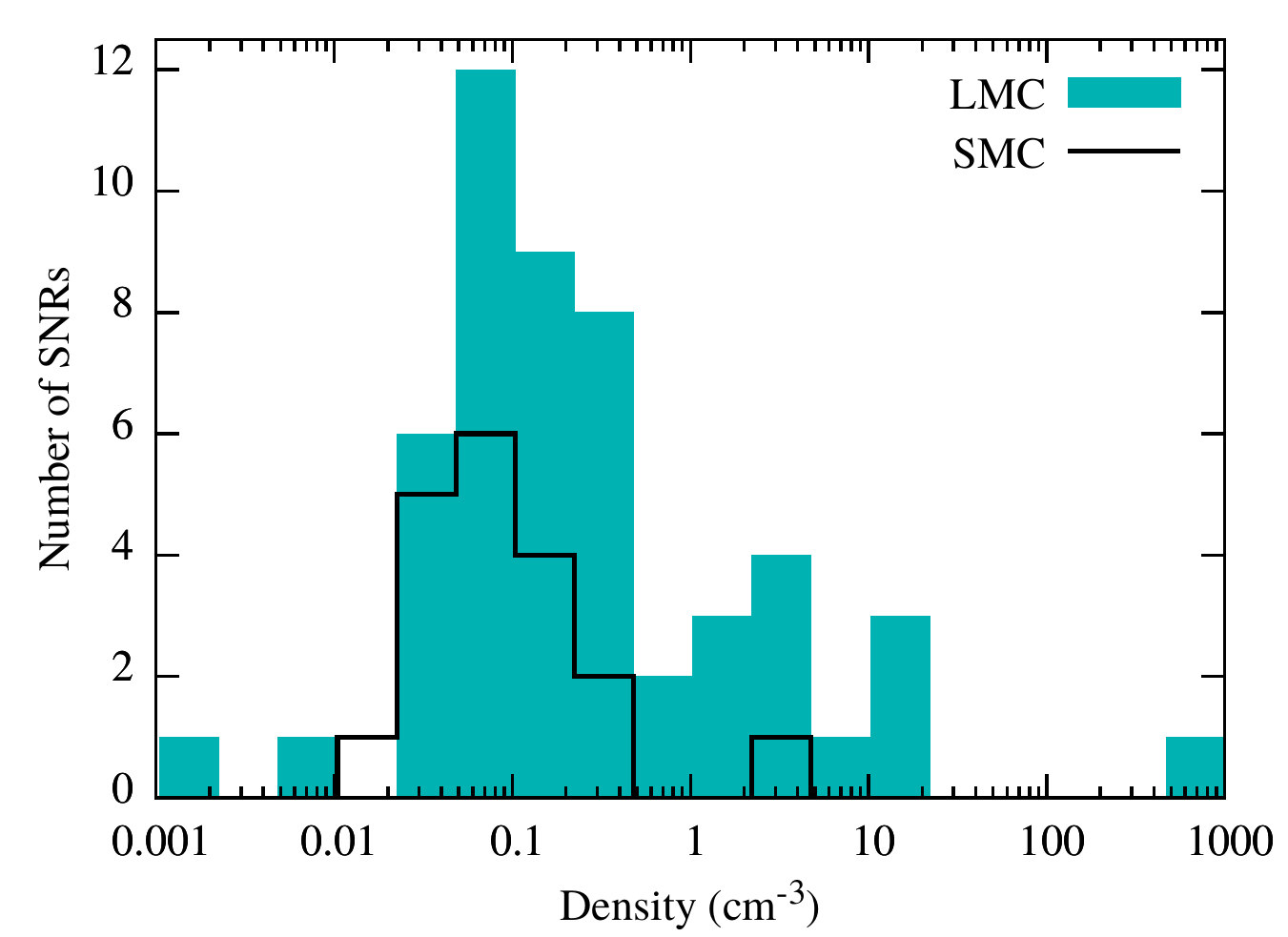}
    \caption{Histogram of ambient density around LMC and SMC SNRs, estimated from the X-ray spectrum as $n \propto \sqrt{EM / V}$ (see Sect.\,\ref{results_spectral_general}).}
  \label{fig_density}
\end{figure}

SNRs of the LMC population are slightly smaller (median of 33~pc vs. 43~pc in the SMC\,\footnote{In this work we used LMC values updated since \citetalias{2017ApJS..230....2B}, recalculating the distributions and their parameters with the same optimal kernel bandwidth and reflection methods used for the SMC.}).
The observed difference might be due to the lower completeness of the LMC SNR population compared to that of the SMC because of the lower coverage fraction e.g. in X-rays (only central areas have been surveyed by \xmm), if the outer area hosts on average larger, older SNRs, that could have been missed by previous surveys (for instance ROSAT all-sky survey and LMC pointed survey). Such incompleteness of the faint, large LMC SNR population was already suggested based on the X-ray luminosity function \citepalias{2016A&A...585A.162M}. Another plausible factor for the smaller size of LMC remnants is an ambient medium denser on average than in the SMC, which is expected given the concentration of gas and star formation in a disk in the LMC. This explanation is supported by the distribution of ambient densities shown in Fig.\,\ref{fig_density}, where the density is estimated from the X-ray derived emission measure as $n \propto \sqrt{EM / V}$ (see Sect.\,\ref{results_spectral_general}). The LMC exhibits a bimodal behaviour with about 25\,\% of SNRs studied in X-rays interacting in a denser environment ($n > 1$~cm$^{-3}$) than the rest of the population, which clusters around $n \sim 0.1$~cm$^{-3}$. Only the latter, lower density mode ($n \lesssim 0.1$~cm$^{-3}$) is seen for the SMC SNR population, the sole ``high-density'' SNR being J0104$-$7201 (IKT~22).

\begin{figure}[ht]
    \centering
    \includegraphics[width=0.90\columnwidth]{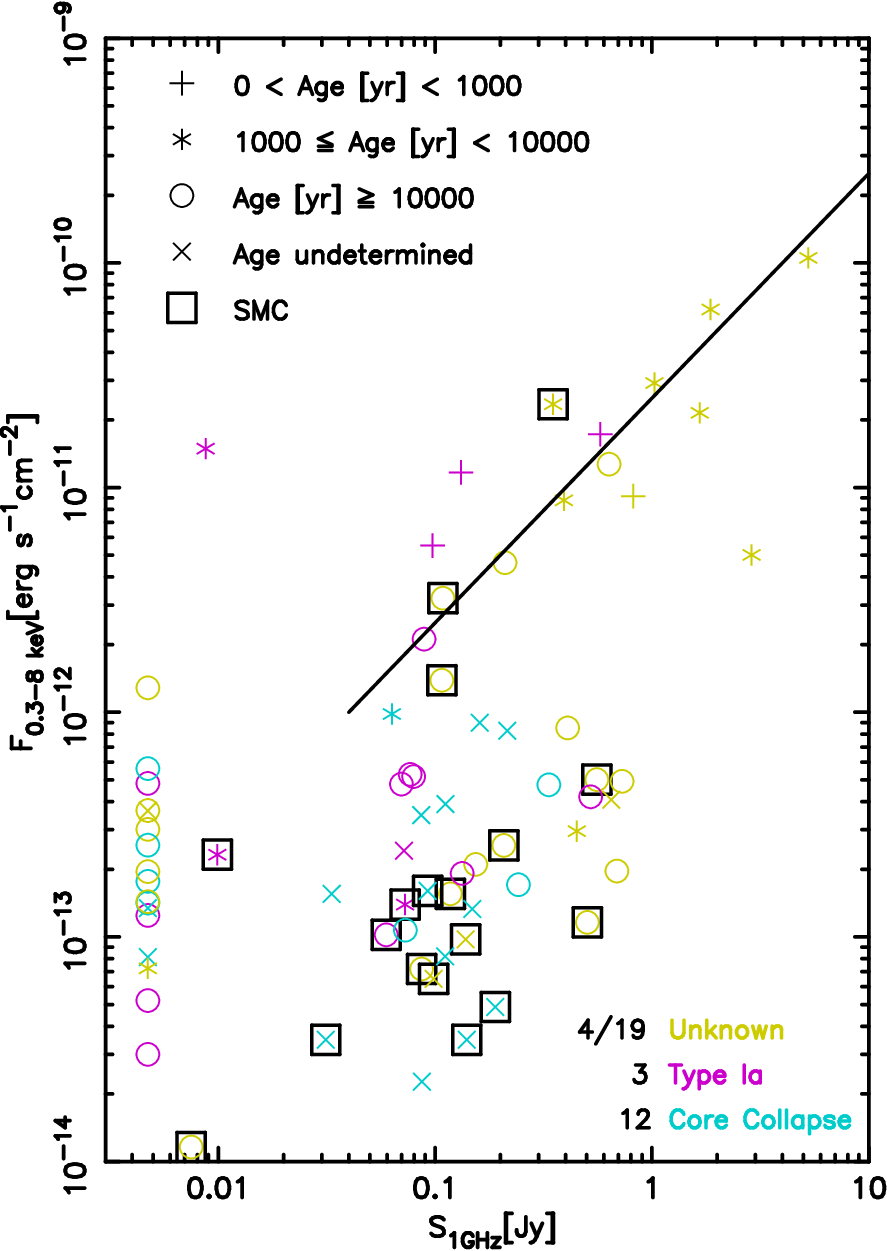}
    
    \caption{The broad-band X-ray flux vs. 1~GHz flux density for the sample of SMC and LMC SNRs with available data for age and explosion type (Table \ref{tab:smcsnrs}, \citetalias{2017ApJS..230....2B}). The position of the colour-coded symbols along the axis with no measured radio flux is offset by -0.2 dex from the faintest detection. The solid line marks the high flux correlation from \citetalias{2017ApJS..230....2B}.}
    \label{fig:s1ghz_fx_smc_lmc}
\end{figure}

The distribution of spectral indices is the same in the SMC and LMC. This is likely due to the marginal dependency of $\alpha$ with age of the SNR \citepalias{2017ApJS..230....2B}, and a similar contribution of PWN-contaminated SNRs, which if not properly resolved, would drive the radio spectra to flatter indices \cite[e.g.][]{2011A&A...530A.132O,2012A&A...543A.154H}, in both galaxies.

Finally, the radio flux densities of SMC and LMC SNRs have consistent values, with the bulk of the population around 0.1~Jy.
\citet{2009ApJ...703..370C} already noted the similar radio luminosities across extragalactic SNR populations, which can be explained because the radio luminosity, i.e. synchrotron emission, is mostly controlled by the magnetic field strength. As SNR shocks amplify $B \propto \rho_0 v_s^2$ (with $\rho_0$ the ambient density and $v_s$ the shock speed), the shock speeds and thus hydrodynamical states of the SNRs are more critical than the ambient density. Since most of the LMC and SMC are in the Sedov state, we can expect their radio luminosities to be similar.

\begin{figure*}[t]
  \centering
    \includegraphics[height=0.255\textheight]{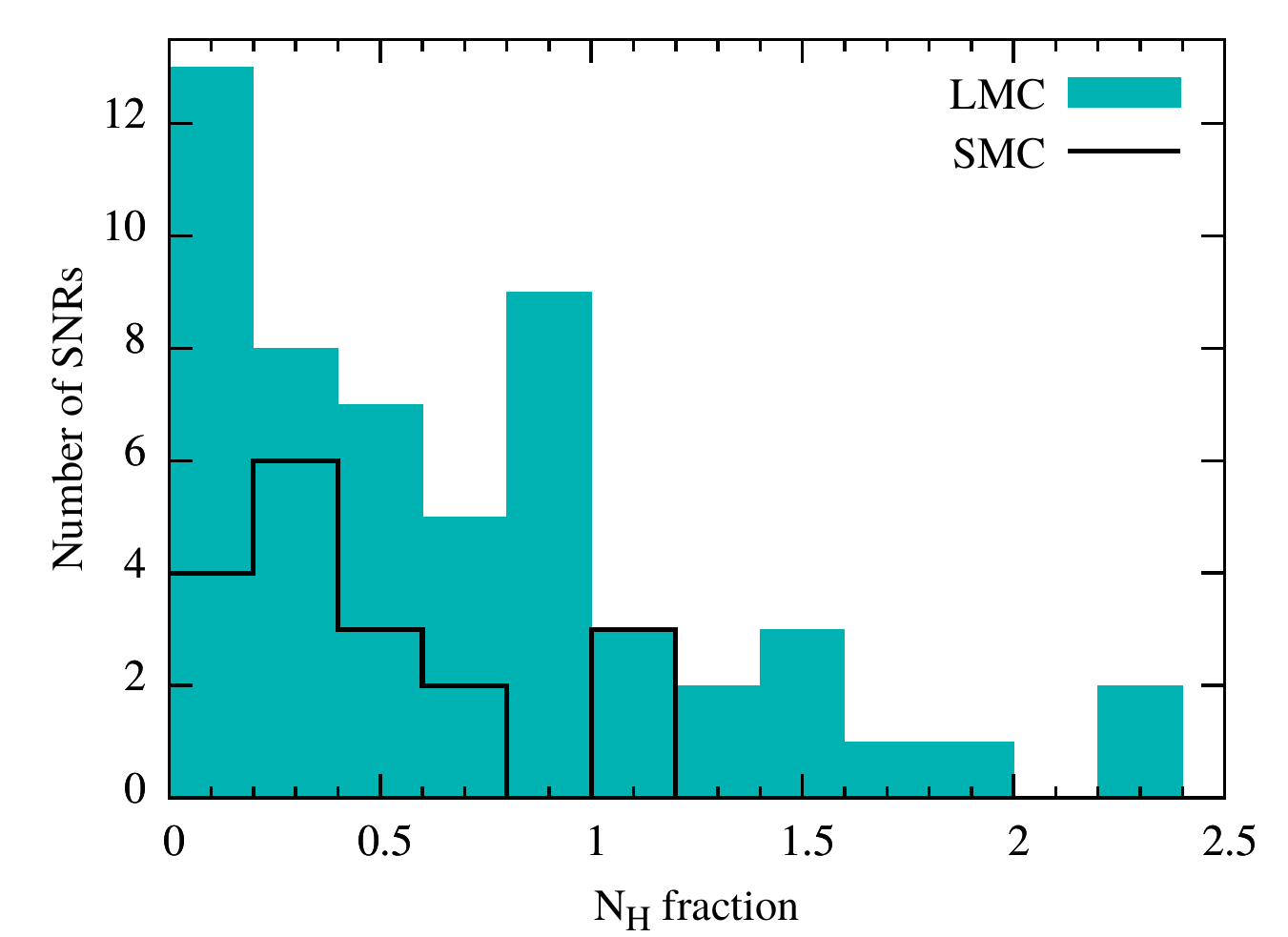}
    \hspace{1em}
    \includegraphics[height=0.255\textheight]{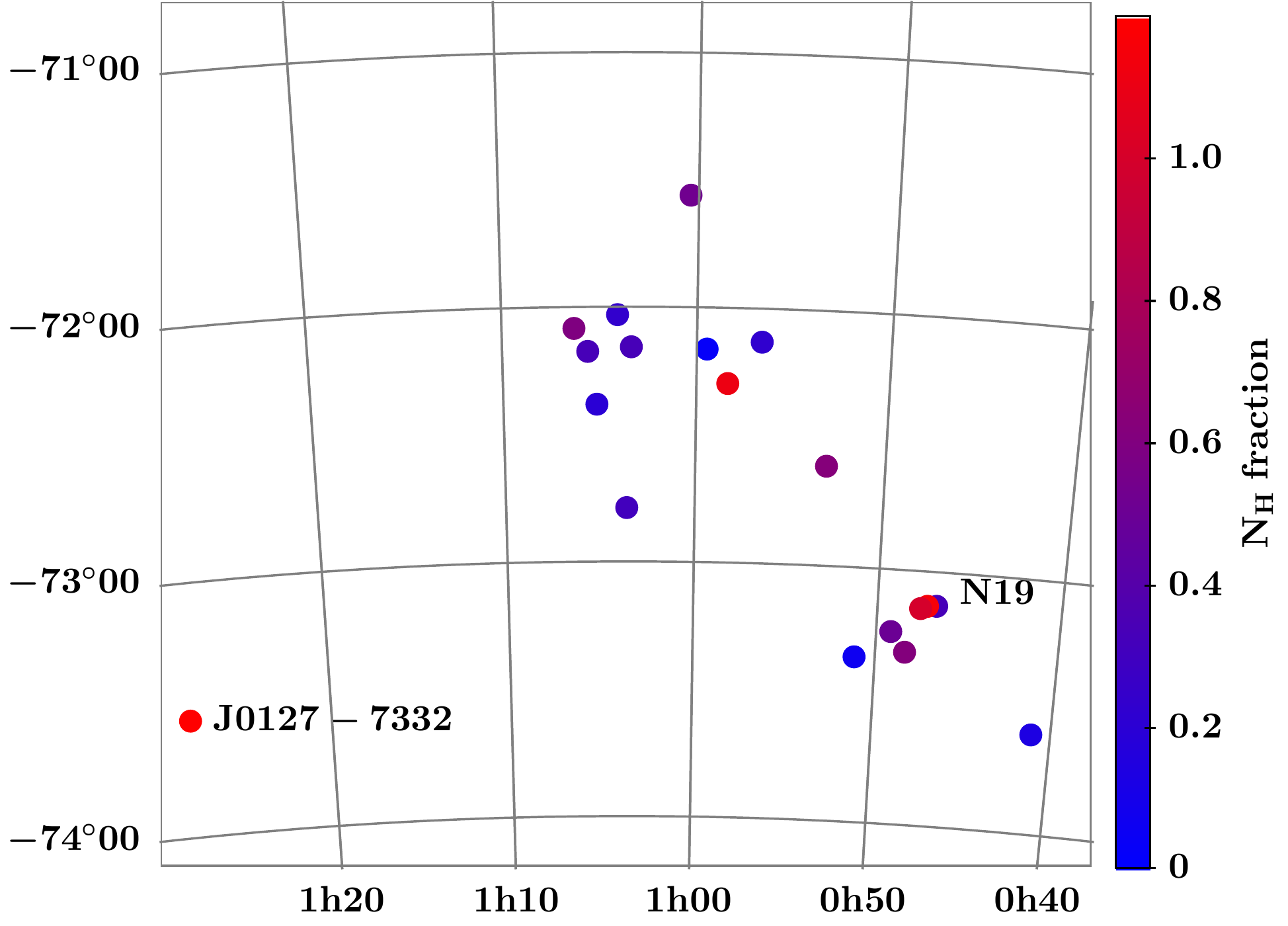}
    \caption{\textit{Left}\,: Comparison of the distribution of $N_H$ fraction for LMC and SMC SNRs. \textit{Right}\,: Spatial distribution of SMC SNRs, with $N_H$ fraction used as a proxy for the line-of-sight depth within the neutral gas.}
  \label{fig_nHfraction}
\end{figure*}

We add on Fig.\,\ref{fig:s1ghz_fx_smc_lmc} the SMC objects to the $F_X - S_{\mathrm{1\ GHz}}$ diagram of \citetalias{2017ApJS..230....2B}. SMC SNRs appear to be fainter X-ray emitters than the LMC objects (median observed 0.3~--~8~keV flux of $1.2 \times 10^{-13}$~erg~cm$^{-2}$~s$^{-1}$ vs. $4.1 \times 10^{-13}$~erg~cm$^{-2}$~s$^{-1}$). This difference is likely again due to the lower average density in which SMC SNRs explode. The X-ray luminosity scales with the density \textit{squared} and is thus much more affected by it than the radio luminosity (see above). The solid black line is plotted to guide the eye and indicates a linear correlation between X-ray and radio fluxes, which was noted for young ($\lesssim 10^4$~yr) SNRs in \citetalias{2017ApJS..230....2B}. IKT~6 and IKT~23, two of the SMC SNRs close to that line, are however estimated to be older than $10^4$~yr \citep{2005ApJ...622L.117H,2014ApJ...791...50S,2003ApJ...598L..95P}, and are probably representing the extrapolation of that multi-wavelength correlation down to lower fluxes. This highlights the need for further, both theoretical and empirical, investigation in this direction to further examine the possible origin of this correlation.

\subsection{Three-dimensional spatial distribution}
\label{results_3D}
X-ray emitting objects in a galaxy are subjected to absorption by hydrogen and metals of that galaxy's ISM. These are obviously only sensitive to the amount of material between the source and the observer, while e.g. 21~cm observations can measure the total \ion{H}{I} column density through a galaxy, e.g. for the SMC. Therefore, combining the equivalent $N_H$ measured in X-rays ($= N_{H} ^{\ X}$) with $N_{H} ^{\mathrm{\,21\,cm}}$, the line-of-sight integrated column density derived from \HI\ surveys, gives us a proxy for the location of sources within the SMC along the line of sight.

As in \citetalias{2016A&A...585A.162M}, we define the ``$N_H$ fraction'' as the ratio $N_{H} ^{\ X} / N_{H} ^{\mathrm{\,21\,cm}}$. It is a measurement of how deep an SNR is with respect to the \ion{H}{I} structure. Such a \textit{relative} line-of-sight proxy is particularly useful in the case of the SMC, because the main body of this galaxy has been shown to be inclined, with the north-eastern tip of the Bar closer than the south-western one by up to 10 kpc \citep[][]{2012ApJ...744..128S,2016ApJ...816...49S}.

The distribution of $N_H$ fraction for 19 SMC SNRs observed in X-rays is shown in Fig.\,\ref{fig_nHfraction} (left) and compared to that of the LMC. The SMC distribution is flatter than in the LMC, where there is a strong mode at 0 and a second, fainter mode around 1. This reflects the SMC neutral gas structure which has a large depth of 3-7.5 kpc \citep{2009A&A...496..399S,2010A&A...520A..74N,2011MNRAS.415.1366K,2012AJ....144..107H}, while that of the LMC has a well-defined thin disc distribution \citep{1999AJ....118.2797K}. Furthermore, there are no SMC SNRs with $N_H$ fraction $> 1$, while in the LMC this betrays the presence along the line of sight of foreground molecular clouds traced by CO emission \citepalias{2016A&A...585A.162M}. Only MCSNR J0103$-$7201 (which we did not study with \xmm) and the SNRs in the LHA 115-N 19 complex (see Sect.\,\ref{results_notesSNRs}) lie close in projection to some known giant molecular clouds (GMC) in the SMC \citep{2001PASJ...53L..45M,2007ApJ...658.1027L,2010ApJ...712.1248M}. In the latter case, MCSNR J0046$-$7308 is the best candidate to be physically associated with molecular clouds as evidenced by the detection of shocked CO emission in higher-resolution ALMA observations \citep{2019ApJ...881...85S}.

We find no obvious correlation of depth with spatial location (Fig.\,\ref{fig_nHfraction}, right). SNRs clustered closely in projected position might be at widely different $N_H$ fraction, and thus line-of-sight depth, in particular in the N19 region (south-west of SMC). This probably reflects the complex \ion{H}{I} structure of that area, with two "sheets" of neutral gas \citep[e.g.][Fig. 6]{2004ApJ...604..176S}.

\section{Summary}
\label{summary}
We summarise below our work and findings:
\begin{itemize}
    \item By combining deep, large scale \xmm\ and radio surveys of the SMC, we presented a clean list of \confcount\ bona-fide SNRs and identified \candcount\ more candidates. Upon new optical spectroscopic observations and based on multi-wavelength features, we confirm the two candidates MCSNR J0056$-$7209 and MCSNR J0057$-$7211 as \textit{bona-fide} SNRs. we also argued against the SNR nature of six poorly studied objects which were previously classified as SNRs. This leads to a final list of \finalconfcount\ SNRs and \finalcandcount\ candidates in the SMC.

    \item We characterised the SNRs using a multiwavelength approach to best capture their size and morphology. 

    \item The homogeneous X-ray spectral analysis allowed us to measure the hot-gas abundance of O, Ne, Mg, and Fe to be between 0.1 and 0.2 times their solar values. O, Mg, and Fe are only partially depleted onto dust grains, as some of the grains have been destroyed by the fast shocks producing the X-ray emission through which we are measuring these abundances.

    \item We constrained the ratio of type~Ia to core-collapse SNRs in the SMC by using both intrinsic properties (detection of SN ejecta, presence of compact remnant) and extrinsic properties (local stellar population from which the SN progenitor is taken) to infer the type of each SNR. The ratio \ccIa = 4.7 (2.8 to 5.3) is larger than that obtained from the same method in the LMC. This difference can potentially be attributed to an enhanced SFR episode 0.5~--~1.5~Gyr in the LMC which is not found in the SMC. Characterising the so far poorly-known SNR population on the outskirts of both Clouds, which is likely to preferentially contain type~Ia SNRs, is needed to provide a more definitive answer.

    \item Radio properties like the spectral index and median flux density at 1~GHz are remarkably consistent between the LMC and SMC population. This stems from the fact that such properties are governed by the ISM magnetic field and hydrodynamical states of the SNRs which are similar in both galaxies.
    
    \item LMC remnants are slightly smaller and more elongated than their SMC counterparts. A plausible explanation is a more disturbed and denser ambient medium in the LMC, as expected given the concentration of gas and star formation in the LMC disk or giant star forming complex (e.g. 30~Dor), where many SNRs explode.

    \item The SMC is inclined with respect to the plane of the sky and has significant depth, as opposed to the LMC. The line-of-sight proxy that is obtained by X-ray absorption reflects that fact. Although SNRs cannot be used as probes of absolute distances within the Cloud, we can show that several SNRs close in projection are likely to be at a different line-of-sight location. This should serve as an important caveat for studies that rely solely on the projected positions of these objects.
\end{itemize}

\begin{acknowledgements}
P.\,M. and J.\,B. acknowledge support by the Centre National d'\'Etudes Spatiales (CNES).
M.S.\ acknowledges support by the Deutsche Forschungsgemeinschaft through the Heisenberg professor grants SA 2131/5-1 and 12-1.
The Australian Compact Array and the Australian SKA Pathfinder (ASKAP) are part of the Australian Telescope which is funded by the Commonwealth of Australia for operation as National Facility managed by CSIRO. This paper includes archived data obtained through the Australia Telescope Online Archive (http://atoa.atnf.csiro.au). We used the \textsc{karma} and \textsc{miriad} software packages developed by the ATNF. Operation of ASKAP is funded by the Australian Government with support from the National Collaborative Research Infrastructure Strategy. ASKAP uses the resources of the Pawsey Supercomputing Centre. Establishment of ASKAP, the Murchison Radio-astronomy Observatory and the Pawsey Supercomputing Centre are initiatives of the Australian Government, with support from the Government of Western Australia and the Science and Industry Endowment Fund. We acknowledge the Wajarri Yamatji people as the traditional owners of the Observatory site.
This research has made use of Aladin, SIMBAD and VizieR, operated at the CDS, Strasbourg, France.
\end{acknowledgements}



\appendix

\begin{landscape}

\section{List and properties of SMC SNRs}
\label{appendix_tables}
\begin{flushleft}
We list here the confirmed (Table~\ref{tab:smcsnrs}) and candidate (Table~\ref{tbl:candsnrs}) SNRs of the SMC, with their names, position and size, X-ray, optical, radio main properties, and hints regarding their type. In Table~\ref{appendix_table_spectra_all} we give the spectral parameters of confirmed and candidate SNRs with existing \xmm\ analysis. All relevant parameters are listed with their 90\,\%~C.\.L. uncertainties: the fitted SMC absorption column density (column 2), plasma temperature $kT$ (3), ionisation age $\tau$ (4), emission measure EM $=n_d\, n_H\, V$ (5), and abundances (6). When a second component is used (in five SNRs), its parameters are given in columns (7)~--~(11). The first component is the one with higher EM. The quality of the fits are evaluated by the $\chi^2 / \nu \ (=\chi ^2 _{\mathrm{red}})$ of column (12), where $\nu$ is the number of degrees of freedom. The median $\chi ^2 _{\mathrm{red}}$ is 1.16. 90\,\% of the fitted objects have a reduced $\chi^2$ less than 1.24. For the best-fit model, we derive the total \textit{observed} (not corrected for abosrption) 0.3-8~keV X-ray flux. We convert it to an X-ray luminosity assuming a common distance of 60~kpc to all SNRs, which we list with other multiwavelength properties in the Table~\ref{tab:smcsnrs}. 
\end{flushleft}

\begin{table}[hb]
\caption{SNRs in the SMC.}
\small
\begin{tabular}{c @{\hspace{0.8em}} c @{\hspace{0.8em}} c @{\hspace{0.8em}} c @{\hspace{0.8em}} c @{\hspace{0.4em}} c @{\hspace{0.8em}} c @{\hspace{0.4em}} c @{\hspace{0.4em}} c @{\hspace{0.8em}} c @{\hspace{0.8em}} c @{\hspace{0.2em}} c @{\hspace{0.2em}} c @{\hspace{0.8em}} c @{\hspace{0.8em}} c @{\hspace{0.2em}} r @{\hspace{0.08em}/\hspace{0.08em}} c @{\hspace{0.08em}/\hspace{0.08em}} l}
\hline\hline
\noalign{\smallskip}
\noalign{\smallskip}
Name & Other name & RA & DEC & $D_{maj} \times D_{min}$ & PA & D$_{av}$ & $L_X$ & $N_H$ fraction & S$_{1 \rm{GHz}}$ & $\alpha \pm \Delta \alpha$ &  $\Sigma_{1 \rm{GHz}} (\times 10^{-20}$) & [\ion{S}{ii}]/H$\alpha$ & $N_{\mathrm{OB}}$ & $r$ & \multicolumn{3}{c}{Hints} \\
MCSNR & & (J2000)  & (J2000)  & (\arcsec) & (\degr) & (pc)  & ($10^{35}$~erg\,s$^{-1}$) & & (Jy) & & (W\,m$^{-2}$Hz$^{-1}$sr$^{-1}$) & & & & (Spec. & SF & Final)\\
(1) & (2) & (3) & (4) & (5) & (6) & (7) & (8) & (9) & (10) & (11) & (12) & (13) & (14) & (15) & (16) \\
\noalign{\smallskip}
\hline
\noalign{\smallskip}
  J0041-7336 & DEM S5       & 00:41:01.7 & --73:36:30 & 245$\times$219 & 105 & 67.6 & 0.42 & 0.14$_{-0.14}^{+0.31}$ & 0.1383 & $-$0.29$\pm$0.01 & 0.1389 & 0.8    & 18  & 0.44$_{-0.18}^{+2.55}$ & 5 & 1 & 3.67\\
  J0046-7308 & [HFP2000] 414& 00:46:40.6 & --73:08:15 & 185$\times$140 & 35  & 46.8 & 0.67 & 0.33$_{-0.09}^{+0.15}$ & 0.1176 & $-$0.38$\pm$0.05 & 0.2460 & 0.4    & 135 & 0.94$_{-0.56}^{+0.48}$ & 5 & 4.5 & 4.83\\
  J0047-7308 & IKT2         & 00:47:16.6 & --73:08:36 & 110$\times$100 & 45  & 30.5 & 0.50  & 1.16$_{-0.18}^{+0.18}$ & 0.5037 & $-$0.54$\pm$0.04 & 2.4740 & 0.6    & 139 & 0.94$_{-0.56}^{+0.48}$ & 4.5 & 4.5 & 4.5\\
  J0047-7309 &  & 00:47:36.5 & --73:09:20 & 180$\times$120 & 75  & 42.8 & 1.10  & 1.00$_{-0.20}^{+0.10}$ & 0.2074 & -0.56$\pm$0.04 & 0.5199 & 0.6    & 133 & 1.00$_{-0.60}^{+0.35}$ & 5&4.5&4.83\\
  J0048-7319 & IKT 4        & 00:48:19.6 & --73:19:40 & 165$\times$130 & 90  & 42.6 & 0.15 & 0.62$_{-0.28}^{+0.35}$ & 0.1399 & $-$0.60$\pm$0.04 & 0.3531 & 0.4    & 142 & 0.85$_{-0.50}^{+0.43}$ & 2.5&4.5&3.17\\
\noalign{\smallskip}
  J0049-7314 & IKT 5        & 00:49:07.7 & --73:14:45 & 190$\times$190 & 0   & 55.3 & 0.60  & 0.51$_{-0.23}^{+0.25}$ & 0.0727 & $-$0.45$\pm$0.04 & 0.1091 & 0.7    & 104 & 0.87$_{-0.51}^{+0.41}$ & 1&3&1.67\\
  J0051-7321 & IKT 6        & 00:51:06.7 & --73:21:26 & 145$\times$145 & 0   & 42.2 & 5.98 & 0.07$_{-0.01}^{+0.02}$ & 0.1074 & $-$0.57$\pm$0.04 & 0.2767 & 0.5    & 128 & 0.83$_{-0.48}^{+0.44}$ & 5&4.5&4.83\\
  J0052-7236 & DEM S68      & 00:52:59.9 & --72:36:47 & 340$\times$270 & 135 & 88.2 & 0.69 & 0.63$_{-0.27}^{+0.57}$ & 0.0924 & $-$0.52$\pm$0.02 & 0.0545 & 0.4    & 130 & 1.60$_{-1.16}^{+0.31}$ & 3&5&3.67\\
  J0058-7217 & IKT 16       & 00:58:22.4 & --72:17:52 & 256$\times$256 & 0   & 74.5 & 0.31 & 1.12$_{-0.33}^{+0.36}$ & 0.0866 & $-$0.53$\pm$0.03 & 0.0715 & 0.25   & 134 & 0.92$_{-0.64}^{+0.92}$ & 5&4.5&4.83\\
  J0059-7210 & IKT 18       & 00:59:27.7 & --72:10:10 & 140$\times$140 & 0   & 40.7 & 2.16 & 0.03$_{-0.03}^{+0.02}$ & 0.5569 & $-$0.48$\pm$0.03 & 1.5389 & 1.7    & 204 & 1.37$_{-0.95}^{+0.72}$ & 4.5&4.5&4.5\\
\noalign{\smallskip}
  J0100-7133 & DEM S108     & 01:00:23.9 & --71:33:41 & 210$\times$210 & 0   & 61.1 & 0.21 & 0.53$_{-0.38}^{+0.85}$ & 0.1895 & $-$0.51$\pm$0.02 & 0.2327 & \OIII & 32  & 4.48$_{-2.18}^{+3.21}$ & 3&2&2.67\\
  J0103-7209 & IKT 21       & 01:03:17.0 & --72:09:42 & 270$\times$270 & 0   & 78.5 & 0.28 & 0.33$_{-0.16}^{+0.24}$ & 0.0984 & $-$0.68$\pm$0.03 & 0.6802 & 0.5    & 260 & 2.82$_{-1.95}^{+0.09}$ & 4&5&4.33\\
  J0103-7247 & [HFP2000] 334& 01:03:29.1 & --72:47:33 & 105$\times$85  & 0   & 27.5 & 0.15 & 0.31$_{-0.19}^{+0.26}$ & 0.0313 & $-$0.60$\pm$0.05 & 0.1896 & ---    & 117 & 1.76$_{-1.00}^{+0.51}$ & 3&5&3.67\\
  J0103-7201 &              & 01:03:36.5 & --72:01:35 & 98$\times$83   & 90  & 26.2 & ---  & ---                    & ---    & ---            & ---    & 0.5    & 296 & 2.29$_{-1.56}^{+0.50}$ & 5&5&5\\
  J0104-7201 & 1E 0102.2-7219  & 01:04:01.2 & --72:01:52 & 45$\times$43   & 0   & 12.8 & 101.2&0.23$_{-0.001}^{+0.001}$& 0.3500 & $-$0.68$\pm$0.02 & 9.7926 & \OIII & 291 & 2.67$_{-1.80}^{+0.36}$ & 5&5&5\\
\noalign{\smallskip}
  J0105-7223 & IKT 23       & 01:05:04.2 & --72:23:10 & 192$\times$192 & 0   & 55.9 & 13.86& 0.20$_{-0.03}^{+0.03}$ & 0.1086 & $-$0.62$\pm$0.02 & 0.1595 & \OIII?& 96  & 2.82$_{-1.87}^{+0.94}$ & 5&3.5&4.5\\
  J0105-7210 & DEM S128     & 01:05:30.5 & --72:10:40 & 190$\times$120 & 150 & 43.9 & 0.44 & 0.33$_{-0.19}^{+0.27}$ & 0.0595 & $-$0.55$\pm$0.03 & 0.1413 & 0.6    & 110 & 2.77$_{-1.86}^{+0.20}$ & 1&3.5&1.83\\
  J0106-7205 & IKT 25       & 01:06:17.5 & --72:05:34 & 110$\times$80  & 25  & 27.3 & 1.00  & 0.60$_{-0.07}^{+0.17}$ & 0.0099 & $-$0.40$\pm$0.03 & 0.0609 & 0.4    & 61  & 2.68$_{-1.79}^{+0.24}$ & 1&2&1.33\\
  J0127-7333 & SXP1062      & 01:27:44.1 & --73:33:01 & 166$\times$166 & 0   & 48.3 & 0.05 & 3.40$_{-0.70}^{+0.63}$ & 0.0075 & $-$0.48$\pm$0.05 & 0.0148 & \OIII & --- & ---                    & 5&---&5\\
\noalign{\smallskip}
\hline
\end{tabular}
\tablefoot{The columns are the following:
(1) MCSNR identifier, in the form ``JHHMM$-$DDMM''.
(2) Old ``common'' name used in the literature.
(3), (4) Position of the remnant, in J2000 equinox.
(5) SNR extent using radio, optical, and X-ray emission (See Sect.\,\ref{results_sample}). Major and minor axes are given in arcsecond. The error in diameter is usually smaller than 2\arcsec.
(6) Position angle of the SNR, in degrees, increasing in the north-east-south direction. A 0 value is given in circular cases.
(7) Average size in parsec, assuming a common distance to all sources of 60~kpc.
(8) $L_X$, the X-ray luminosity in the 0.3~keV~--~8~keV band, in units of $10^{35}$~erg~s$^{-1}$, obtained as described in Sect.\,\ref{results_spectral_general}, at 60~kpc.
(9) $N_H$ fraction, as defined in Sect.\,\ref{results_3D}. Uncertainties are given at the 90\,\% C.L.
(10) 1~GHz flux density in Jansky. Errors are typically less than 10~\%.
(11) Radio spectral index with corresponding uncertainties.
(12) 1~GHz surface brightness.
(13) Optical [\ion{S}{ii}]/\Halpha\ ratio measured in MCELS images. \OIII\ indicates that the given SNR is dominated by \OIII$\lambda 5007\AA$\ emission.
(14) \NOB , the number of blue early-type stars within 100~pc of the remnant, and (15) $r$, the ratio of CC SNe to thermonuclear SNe expected from the observed distribution of stellar ages in the neighbourhood of the remnant, as obtained by Eq.~\ref{eq_r} (see Sect.\,\ref{results_typing}).
(16) ``Hints'' to the type of SN progenitor, based on spectral properties, local star formation, and a combination of both, as described in Sect.\,\ref{results_typing} and Tables~\ref{table_hints} \& \ref{table_hint_SF}.
}
\label{tab:smcsnrs}
\end{table}

\end{landscape}

\onecolumn
\begin{landscape}
\begin{table}
\caption{\candcount\ candidate SNRs in the SMC. These are our own measurements shown here for the first time.}
\begin{center}
\label{tbl:candsnrs}
\small
\begin{tabular}{cccccccccl}
\hline\hline
\noalign{\smallskip}
\noalign{\smallskip}
Name     & Other & RA      & DEC      & D$_{maj} \times D_{min}$ & PA  & D$_{av}$ & $L_X$ & [\ion{S}{ii}]/H$\alpha$ & Notes\\
MCSNR & Name  & (J2000)  & (J2000) & (\arcsec) & (\degr) &  (pc) & ($10^{35}$~erg\,s$^{-1}$) &  &  \\
(1) & (2) & (3) & (4) & (5) & (6) & (7) & (8) & (9) & (10) \\
\noalign{\smallskip}
\hline
\noalign{\smallskip}
J0056-7209 & SNRC1  & 00 56 28.1 & --72 09 42.2 & 340$\times$225 & 30 & 80.5 & 0.14 & 0.47 & X-ray \& Optical; \textbf{confirmed as SNR} \\
J0057-7211 & N S66D & 00 57 49.9 & --72 11 47.1 & 180$\times$145 & 45 & 47.0 & 0.17 & ---  & X-ray \& Radio ($\alpha=-0.75\pm0.04$; S$_{1 \rm{GHz}}$=0.0308~Jy); \textbf{confirmed as SNR}\\
J0106-7242 &        & 01 06 32.1 & --72 42 17.0 & 140$\times$170 &  0 & 44.9 & --- & ---  & Radio ($\alpha=-0.55\pm0.02$; S$_{1 \rm{GHz}}$=0.0236~Jy); \textbf{SNR candidate}\\
J0109-7318 & SNRC4  & 01 09 43.6 & --73 18 46.0 & 105$\times$105 &  0 & 30.5 & --- & 0.46 & Optical; \textbf{SNR candidate}\\
\noalign{\smallskip}
\hline
\end{tabular}
\end{center}
\tablefoot{
Columns (1) to (8) are as in Table~\ref{tab:smcsnrs}. Column (9) is the [\ion{S}{ii}]/\Halpha\ measured with the WiFeS spectrograph. In Notes (Col.~10) we indicate the waveband in which SNR features are detected and the final status of the source (confirmed or candidate).
}
\end{table}

\begin{longtable}[c]{@{}l @{\hspace{0.0cm}} @{\hspace{0.0cm}} c @{\hspace{0.0cm}} @{\hspace{0.0cm}} c @{\hspace{0.0cm}} @{\hspace{0.0cm}} c @{\hspace{0.3em}} @{\hspace{0.3em}} c @{\hspace{0.0cm}} @{\hspace{0.0cm}} l @{\hspace{-0.20cm}} @{\hspace{-0.20cm}} c @{\hspace{0.0cm}} @{\hspace{0.0cm}} c @{\hspace{0.0cm}} @{\hspace{0.0cm}} c @{\hspace{0.3em}} @{\hspace{0.3em}} c @{\hspace{-0.00cm}} @{\hspace{-0.0cm}} l @{\hspace{0.3em}} @{\hspace{0.3em}} c @{\hspace{0.1cm}}}
\caption{X-ray spectral results of SMC SNRs.}
\label{appendix_table_spectra_all}\\
\hline\hline
\noalign{\medskip}
  & \multicolumn{5}{c}{Component 1:} &
\multicolumn{5}{c}{Component 2:} & \multirow{2}{*}{$\chi ^2 / \nu$ \ 
$\left(\chi ^2 _{\mathrm{red}}\right)$} \\
\noalign{\medskip}
\cline{2-6} \cline{7-11}
\noalign{\medskip}
 \multicolumn{1}{c}{MCSNR} & $N_{H\mathrm{\ LMC}}$ & $kT$ & $\tau$ & EM &
\multicolumn{1}{c}{Abundances}
       & $N_{H\mathrm{\ LMC}}$ & $kT$ & $\tau$ & EM &
\multicolumn{1}{c}{Abundances} & \\
  & ($10^{21}$~cm$^{-2}$) & (keV) & ($10^{11}$ s\,cm$^{-3}$) & ($10^{58}$
cm$^{-3}$) &  & ($10^{21}$~cm$^{-2}$) & (keV) & ($10^{11}$ s\,cm$^{-3}$) &
($10^{58}$ cm$^{-3}$) &  \\
\multicolumn{1}{c}{(1)} & (2) & (3) & (4) & (5) & \multicolumn{1}{c}{(6)} & (7) & (8) & (9) & (10) &
 \multicolumn{1}{c}{(11)} & (12)\\
\noalign{\medskip}
\hline
\endfirsthead
%
%
\caption[]{(continued)}\\
\hline
\hline
\noalign{\medskip}
  & \multicolumn{5}{c}{Component 1:} &
\multicolumn{5}{c}{Component 2:} & \multirow{2}{*}{$\chi ^2 / \nu$ \ 
$\left(\chi ^2 _{\mathrm{red}}\right)$} \\
\noalign{\medskip}
\cline{2-6} \cline{7-11}
\noalign{\medskip}
 MCSNR & $N_{H\mathrm{\ LMC}}$ & $kT$ & $\tau$ & EM &
\multicolumn{1}{c}{Abundances}
       & $N_{H\mathrm{\ LMC}}$ & $kT$ & $\tau$ & EM &
\multicolumn{1}{c}{Abundances} & \\
  & ($10^{21}$~cm$^{-2}$) & (keV) & ($10^{11}$ s\,cm$^{-3}$) & ($10^{57}$
cm$^{-3}$) &  & ($10^{21}$~cm$^{-2}$) & (keV) & ($10^{11}$ s\,cm$^{-3}$) &
($10^{57}$ cm$^{-3}$) &  \\
(1) & (2) & (3) & (4) & (5) & \multicolumn{1}{c}{(6)} & (7) & (8) & (9) & (10) &
\multicolumn{1}{c}{(11)} & (12)\\
\noalign{\medskip}
\hline
\endhead
%
\hline
\endfoot
%
\hline
\endlastfoot
\noalign{\medskip}
J0041$-$7336 & 1.09$_{-0.80}^{+1.04}$ & 0.65$_{-0.12} ^{+0.09}$ & 2.24$_{-0.93} ^{+1.30}$ &
4.01$_{-0.39} ^{+0.96}$ & \multicolumn{1}{c}{RD92} & \multicolumn{5}{c}{---} & 5788.5/5562 (1.04) \\
\noalign{\medskip}
\multirow{3}{*}{J0046$-$7308} & \multirow{3}{*}{4.06$_{-1.13}
^{+1.71}$} & \multirow{3}{*}{0.60$\pm0.11$} & \multirow{3}{*}{1.01$_{-0.40}^{+0.80}$} &
\multirow{3}{*}{7.71$_{-1.64} ^{+3.45}$} & \abund{O}0.42$_{-0.10}^{+0.15}$ &
\multicolumn{5}{c}{\multirow{3}{*}{---}} & \multirow{3}{*}{12392.8/11672 (1.06)} \\*
 & & & & & \abund{Ne}0.45$_{-0.10} ^{+0.13}$ & & & & & & \\*
 & & & & & \abund{Si}2.24$_{-0.78} ^{+1.15}$ & & & & & & \\
\noalign{\medskip}
\multirow{2}{*}{J0047$-$7308} & \multirow{2}{*}{14.1$\pm2.20$} & \multirow{2}{*}{0.60$_{-0.05}^{+0.04}$} & \multirow{2}{*}{75.8$_{-50.1}^{+424}$} &
\multirow{2}{*}{13.7$_{-0.45} ^{+4.87}$} & \abund{Ne}3.11$_{-0.78}^{+1.10}$ &
\multicolumn{5}{c}{\multirow{2}{*}{---}} & \multirow{2}{*}{7554.32/6219 (1.21)} \\*
 & & & & & \abund{Mg}1.5$_{-0.42} ^{+0.58}$ & & & & & & \\*
\noalign{\medskip}
\multirow{3}{*}{J0047$-$7309} & \multirow{3}{*}{12.2$_{-2.40}
^{+1.10}$} & \multirow{3}{*}{0.63$_{-0.13} ^{+0.25}$} & \multirow{3}{*}{3.30$_{-1.69}^{+5.43}$} &
\multirow{3}{*}{5.64$_{-4.78} ^{+4.87}$} & \abund{O}8.25$_{-4.89}^{+67}$ &
\multicolumn{5}{c}{\multirow{3}{*}{---}} & \multirow{3}{*}{10224.9/8801 (1.16)} \\*
 & & & & & \abund{Ne}7.71$_{-3.63} ^{+52.3}$ & & & & & & \\*
 & & & & & \abund{Mg}3.56$_{-1.35} ^{+4.75}$ & & & & & & \\
\noalign{\medskip}
\multirow{2}{*}{J0048$-$7319} & \multirow{2}{*}{6.31$_{-2.84}^{+3.54}$} & \multirow{2}{*}{1.07$_{-0.18}^{+0.76}$} & \multirow{2}{*}{2.03$_{-0.80}^{+4.49}$} &
\multirow{2}{*}{1.03$_{-0.58} ^{+0.65}$} & \abund{Mg}2.39$_{-0.92}^{+1.82}$ &
\multicolumn{5}{c}{\multirow{2}{*}{---}} & \multirow{2}{*}{4596.9/3702 (1.24)} \\*
 & & & & & \abund{Fe}1.17$_{-0.49} ^{+1.25}$ & & & & & & \\
\noalign{\medskip}
J0049$-$7314 \tablefootmark{a} & 4.39$_{-2.02} ^{+2.13}$ & 0.52$_{-0.14} ^{+0.15}$  & 
CIE  & 3.16$_{-1.64} ^{+3.63}$ & \multicolumn{1}{c}{RD92} & 4.39 & 0.91$\pm 0.03$ & CIE &
7.28$_{-0.99} ^{+1.25}$ & \multicolumn{1}{c}{pure Fe} & 2337.0/2148 (1.09) \\
\noalign{\medskip}
\multirow{3}{*}{J0051$-$7321 \tablefootmark{a,b}} & \multirow{3}{*}{0.56$_{-0.09} ^{+0.17}$} & 
\multirow{3}{*}{0.21$\pm0.01$} & \multirow{3}{*}{CIE} & \multirow{3}{*}{171$_{-33.6} ^{+22.4}$} &
\multicolumn{1}{c}{\multirow{3}{*}{RD92}} &
\multirow{3}{*}{0.56} & \multirow{3}{*}{0.73$_{-0.10} ^{+0.16}$} &
\multirow{3}{*}{1.95$_{-0.69} ^{+0.93}$} & \multirow{3}{*}{12.4$_{-2.15}^{+5.26}$} &
\abund{Ne}1.74$_{-0.52}^{+0.77}$ & \multirow{3}{*}{2124.4/1576 (1.35)} \\*
&&&&&&&&&& \abund{Mg}0.93$_{-0.32}^{+0.46}$& \\*
&&&&&&&&&& \abund{Si}2.13$_{-0.77}^{+1.14}$& \\
\noalign{\medskip}
\multirow{4}{*}{J0052$-$7236} & \multirow{4}{*}{3.72$_{-1.60} ^{+3.35}$} &
\multirow{4}{*}{0.38$_{-0.16} ^{+0.66}$} & \multirow{4}{*}{0.71$_{-0.44}^{+3.79}$} &
\multirow{4}{*}{27.4$_{-24.6} ^{+308}$} & \abund{O}0.12$_{-0.07}^{+0.23}$ &
\multicolumn{5}{c}{\multirow{4}{*}{---}} & \multirow{4}{*}{ 9101.2/8265 (1.10)} \\*
 & & & & & \abund{Ne}0.24$_{-0.12} ^{+0.55}$ & & & & & & \\*
 & & & & & \abund{Mg}0.19 ($<0.45$) & & & & & & \\*
 & & & & & \abund{Fe}0.11$_{-0.07} ^{+0.15}$ & & & & & & \\
\noalign{\medskip}
\multirow{4}{*}{J0058$-$7217 \tablefootmark{c}} & \multirow{4}{*}{8.49$_{-2.49} ^{+2.75}$} &
\multirow{4}{*}{0.37$_{-0.09} ^{+0.26}$} & \multirow{4}{*}{6.31$_{-4.61}^{+4.94}$} &
\multirow{4}{*}{28.5$_{-22.1} ^{+60.5}$} & \abund{O}0.26$_{-0.14}^{+0.45}$ &
\multicolumn{5}{c}{\multirow{4}{*}{---}} & \multirow{4}{*}{ 41890.5/35193 (1.19)} \\*
 & & & & & \abund{Ne}0.30$_{-0.15} ^{+0.42}$ & & & & & & \\*
 & & & & & \abund{Mg}0.27$_{-0.15} ^{+0.21}$ & & & & & & \\*
 & & & & & \abund{Fe}0.02$_{-0.02} ^{+0.04}$ & & & & & & \\
\noalign{\medskip}
\multirow{4}{*}{J0059$-$7210} & \multirow{4}{*}{0.12($<0.21$)} &
\multirow{4}{*}{0.66$\pm0.03$} & \multirow{4}{*}{7.08$_{-1.48}^{+1.99}$} &
\multirow{4}{*}{25.9$_{-2.15} ^{+2.37}$} & \abund{O}0.30$_{-0.06}^{+0.03}$ &
\multicolumn{5}{c}{\multirow{4}{*}{---}} & \multirow{4}{*}{ 12096.5/10380 (1.17)} \\*
 & & & & & \abund{Ne}0.47$_{-0.07} ^{+0.08}$ & & & & & & \\*
 & & & & & \abund{Mg}0.25$\pm0.05$ & & & & & & \\*
 & & & & & \abund{Fe}0.16$\pm0.02$ & & & & & & \\
\noalign{\medskip}
J0100$-$7133 & 1.96$_{-1.40}^{+3.16}$ & 0.49$_{-0.16} ^{+0.14}$ & 3.58$_{-1.80} ^{+11.3}$ &
3.28$_{-0.99} ^{+4.96}$ & \abund{Ne}0.43$\pm0.21$ & \multicolumn{5}{c}{---} & 3722.0/3610 (1.03) \\
\noalign{\medskip}
J0103$-$7209  \tablefootmark{d} & 1.89$_{-0.91}^{+1.35}$ & 1.0$_{-0.20} ^{+0.35}$ & 0.66$_{-0.35} ^{+0.58}$ &
1.97$_{-0.40} ^{+0.45}$ & \abund{Ne}0.50$_{-0.13} ^{+0.15}$ & \multicolumn{5}{c}{---} & 12282.7/10266 (1.20) \\
\noalign{\medskip}
&&&&&&&&&&& \\
\noalign{\medskip}
J0103$-$7247 \tablefootmark{e} & 1.60$_{-0.97}^{+1.33}$ & 1.3 & 0.07 &
0.93$_{-0.35} ^{+0.45}$ & \multicolumn{1}{c}{RD92} & \multicolumn{5}{c}{---} & 1877.4/1642 (1.14) \\
\multirow{4}{*}{J0105$-$7223  \tablefootmark{a}} & \multirow{4}{*}{0.82$\pm0.01$} & 
\multirow{4}{*}{0.19$\pm0.01$} & \multirow{4}{*}{CIE} & \multirow{4}{*}{448$\pm51.7$} &
\multicolumn{1}{c}{\multirow{4}{*}{RD92}} &
\multirow{4}{*}{0.82} & \multirow{4}{*}{0.38$\pm0.01$} &
\multirow{4}{*}{CIE} & \multirow{4}{*}{85.3$_{-20.7}^{+26.3}$} &
\abund{O}1.23$_{-0.25}^{+0.35}$ & \multirow{4}{*}{8606.8/7123 (1.21)} \\*
&&&&&&&&&& \abund{Ne}1.64$_{-0.57}^{+0.53}$& \\*
&&&&&&&&&& \abund{Mg}0.94$_{-0.20}^{+0.28}$& \\*
&&&&&&&&&& \abund{Fe}0.13$\pm0.06$ & \\
\noalign{\medskip}
J0105$-$7210 \tablefootmark{a} & 1.74$_{-0.98} ^{+1.41}$ & 0.68$_{-0.11} ^{+0.20}$ & 
CIE  & 4.44$_{-0.78} ^{+1.72}$ & \multicolumn{1}{c}{RD92} & 1.74 & 0.78$_{-0.10} ^{+0.07}$ & CIE &
1.88$_{-0.58} ^{+0.63}$ & \multicolumn{1}{c}{pure Fe} & 3722.6/3610 (1.03) \\
\noalign{\medskip}
J0106$-$7205 \tablefootmark{a} & 3.76$_{-0.45} ^{+1.10}$ & 0.72$_{-0.08} ^{+0.07}$ & 
CIE  & 9.95$_{-1.51} ^{+1.59}$ & \multicolumn{1}{c}{RD92} & 3.76 & 1.0$\pm0.03$ & CIE &
7.88$_{-0.95} ^{+1.03}$ & \multicolumn{1}{c}{pure Fe} & 3816.96/311 (1.23) \\
\noalign{\medskip}
J0127$-$7333 \tablefootmark{f} & 10.2$_{-8.10} ^{+12.1}$ & 0.19$_{-0.01} ^{+0.02}$ & CIE &
25.8$_{-12.7} ^{+19.1}$ & \multicolumn{1}{c}{RD92} & \multicolumn{5}{c}{---} & 1918.9/1749 (1.10) \\
\noalign{\medskip}
\hline
\noalign{\medskip}
\noalign{\smallskip}
\multicolumn{12}{c}{Candidate SNRs with X-ray emission}\\
\noalign{\medskip}
\noalign{\smallskip}
\hline
\noalign{\medskip}
\multirow{4}{*}{J0056$-$7209} & \multirow{4}{*}{0.97($<3.10$)} &
\multirow{4}{*}{0.7$_{-0.46}^{+0.89}$} & \multirow{4}{*}{0.28$_{-0.25}^{+0.79}$} &
\multirow{4}{*}{1.48$_{-0.81} ^{+4.21}$} & \abund{O}0.12$_{-0.06}^{+0.38}$ &
\multicolumn{5}{c}{\multirow{4}{*}{---}} & \multirow{4}{*}{ 5204.2/4743 (1.10)} \\*
 & & & & & \abund{Ne}0.06$(<0.34)$ & & & & & & \\*
 & & & & & \abund{Mg}0.30$(<1.55)$ & & & & & & \\*
 & & & & & \abund{Fe}0.20$_{-0.17} ^{+0.63}$ & & & & & & \\
\noalign{\medskip}
J0057$-$7211 & 0$(<0.24)$ & 0.88$_{-0.23} ^{+0.41}$ &  1.04$_{-0.61} ^{+0.66}$ &
1.10$(<0.28)$ & \multicolumn{1}{c}{RD92} & \multicolumn{5}{c}{---} & 26563.4/22410 (1.19) \\
%
\end{longtable}
\tablefoot{Columns are described in Sect.\,\ref{results_spectral_general}.\\
\tablefoottext{a}{Same absorption column in the two components;\ }
\tablefoottext{b}{Only MOS data used;\ }
\tablefoottext{c}{Fit includes the model from \citet{2015A&A...584A..41M} for the interior PWN;\ }
\tablefoottext{d}{Bright point source AX~J0103$-$722 excised, see Sect.\,\ref{results_notesSNRs};\ }
\tablefoottext{e}{Include a power-law ($\Gamma =2.8$) and thermal parameters $kT$ and $\tau$ fixed to the values from joint \xmm/\chandra analysis of \citet{2014AJ....148...99C};\ }
\tablefoottext{f}{Central point source SXP~1062 excised;\ }
}

\end{landscape}

\section{Multiwavelength images, spectra, and local star formation history of SMC SNRs}
\label{appendix_images}

In this Appendix we show for each SNR information about their stellar environment in the form of CMD and SFH plots (see Sect.\,\ref{results_typing} for details), radio and X-ray spectra, and multiwavelength images. We show ASKAP 1320 MHz contours on colour-coded X-ray images, where the red, green, and blue components are the images in the soft, medium, and hard X-ray band as described in Sect.\,\ref{observations_xray}. The ASKAP beam size of 16.3\arcsec$\times$15.1\arcsec\ is indicated by the thick magenta ellipse. We also show X-ray contours on optical emission-line images (MCELS), with bands as on Fig.\,\ref{MC}. Radio flux density contours increase in the order white-cyan-magenta-red, with the levels used for the contours tailored for each case and given in the captions. The red, green, and blue X-ray contours are taken from the corresponding X-ray band. On the images a spatial scale of 1\arcmin\ is shown by the white bar.
For clarity, we only show up to one pn (black) and one MOS spectrum (blue), although much more might have been used for spectral analysis (see Sect.\,\ref{observations_xray}). The sum of all background components is shown has the grey dotted line. The SNR model is shown in magenta by the solid (pn) or dashed (MOS) line. For cases with a second SNR component it is displayed in green. When a contributing point source (X-ray binary, PWN), related to the SNR or not, is included, we show its emission model in cyan. Fit statistics ($\chi^2/\nu$) are given on the plot. Residuals are shown in units of standard deviation.

\begin{figure*}[ht]
    \centering
    \includegraphics[height=0.17\vsize]{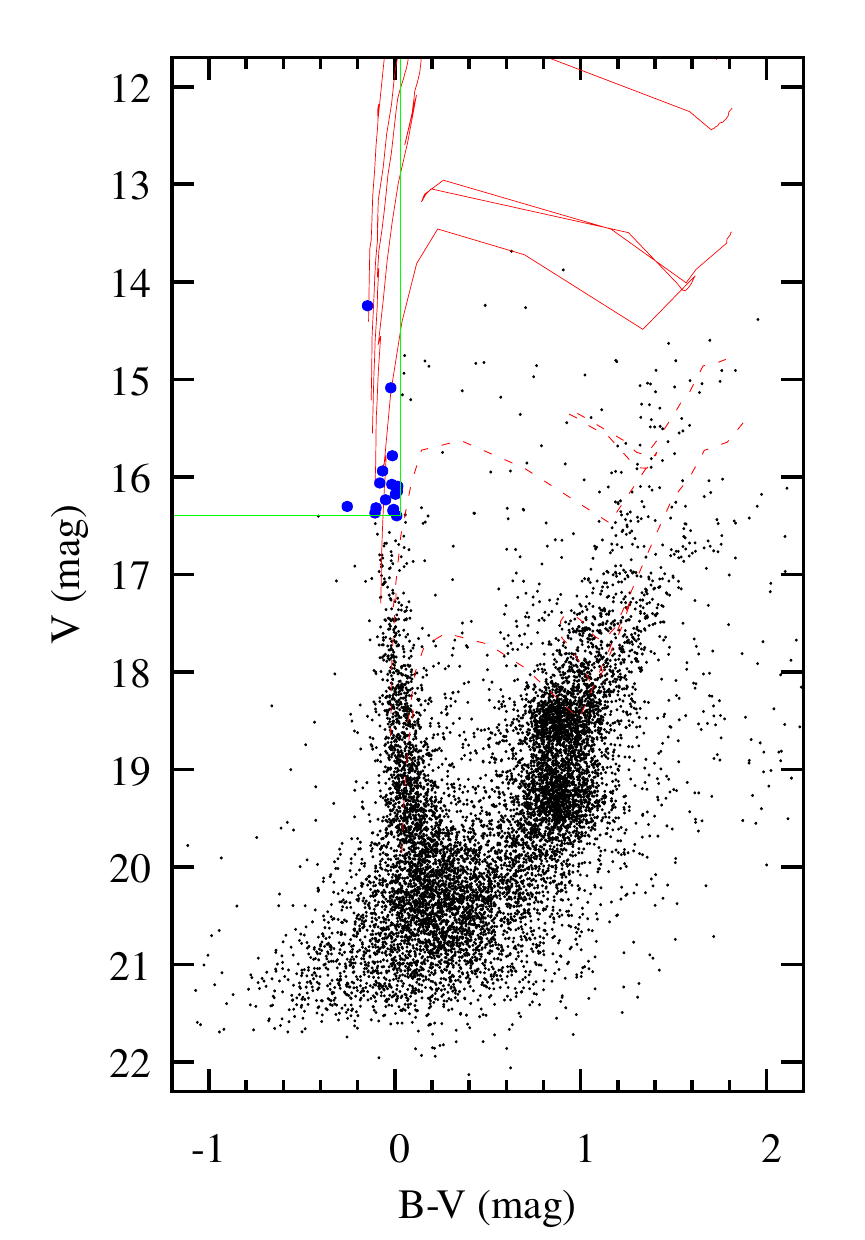}
    \includegraphics[height=0.17\vsize]{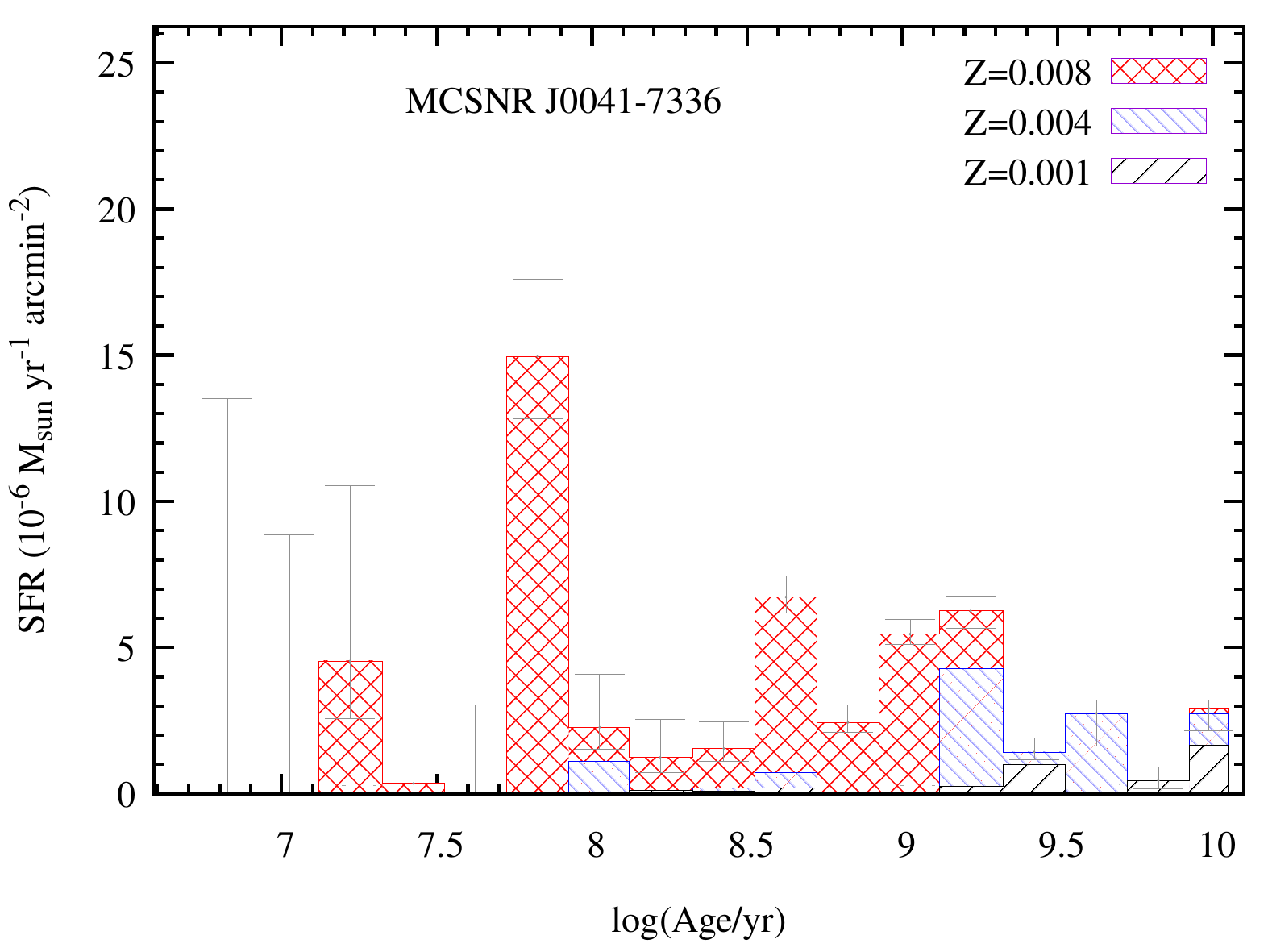}
    \includegraphics[height=0.17\vsize]{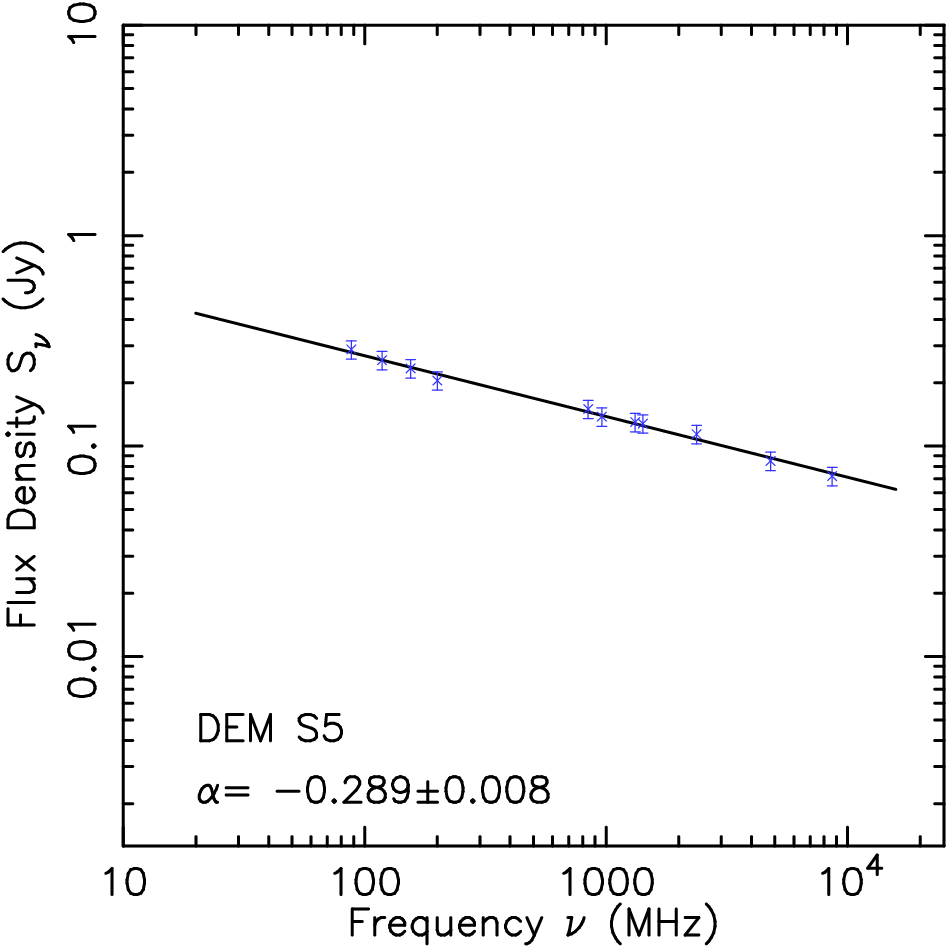}

    \includegraphics[height=0.175\vsize]{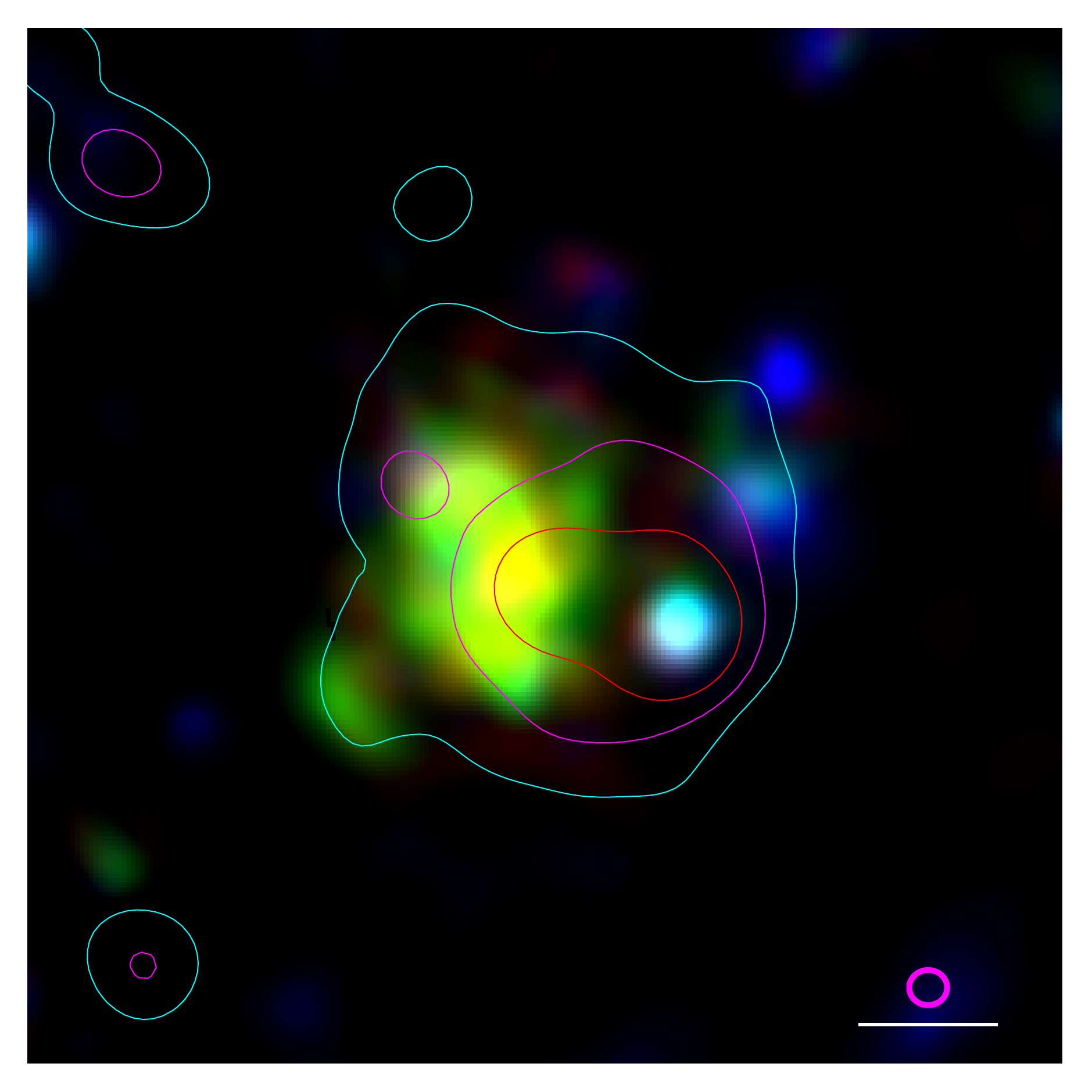}
    \includegraphics[height=0.175\vsize]{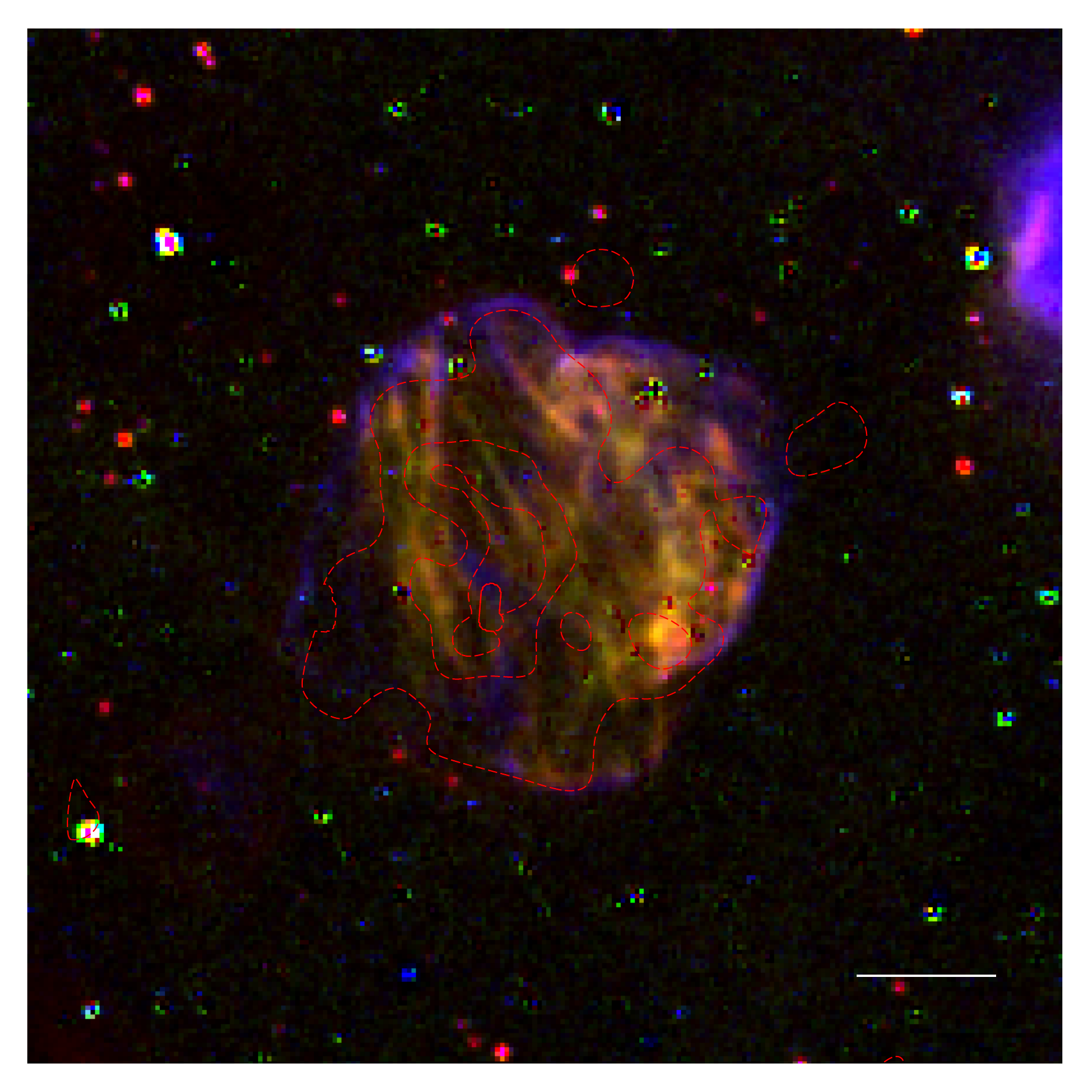}
    \includegraphics[height=0.175\vsize]{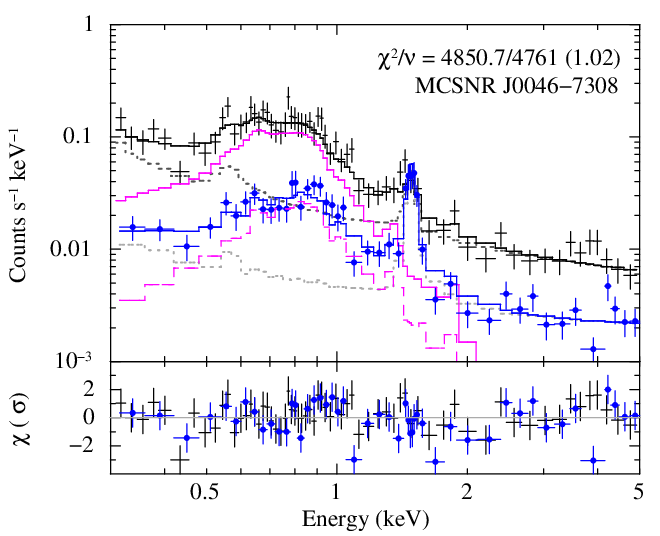}

    \vspace{0.5em}
    \hrule
    \vspace{0.5em}

    \includegraphics[height=0.17\vsize]{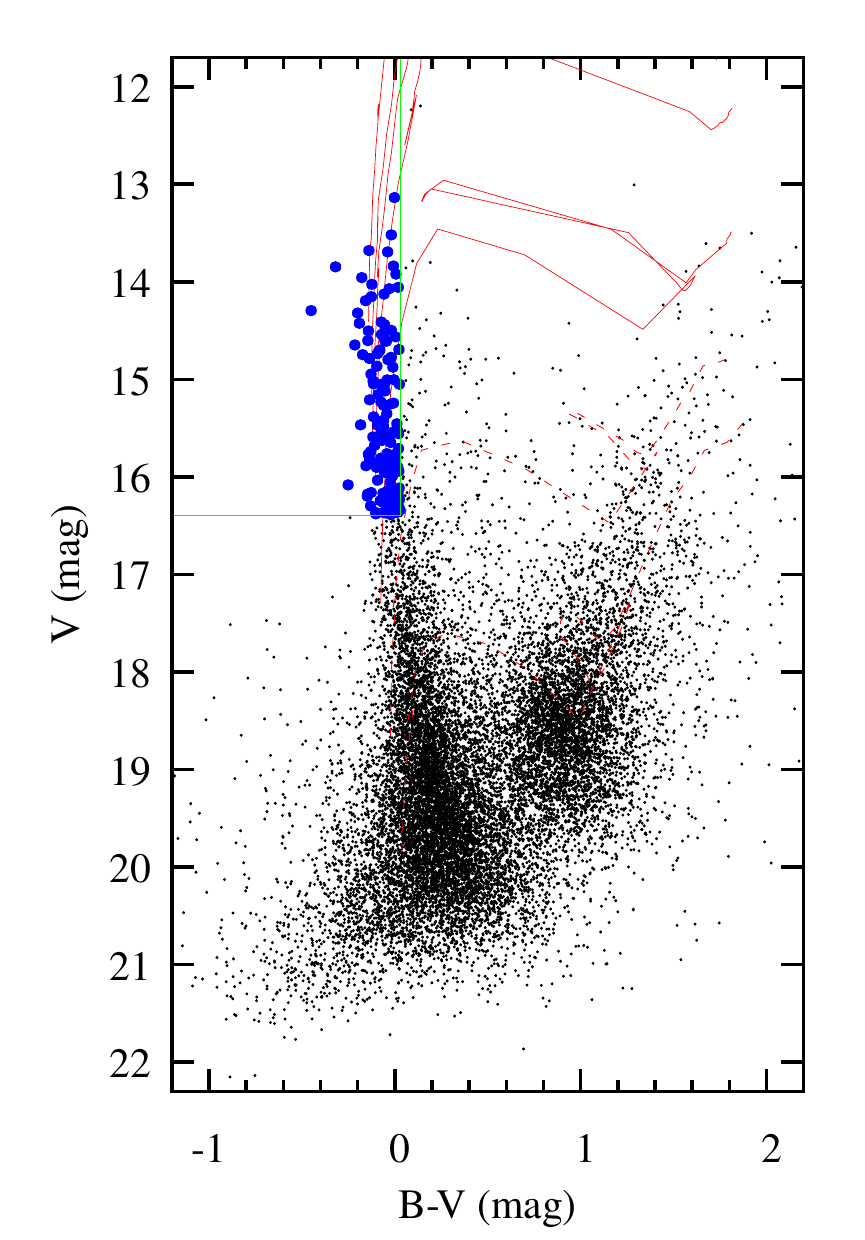}
    \includegraphics[height=0.17\vsize]{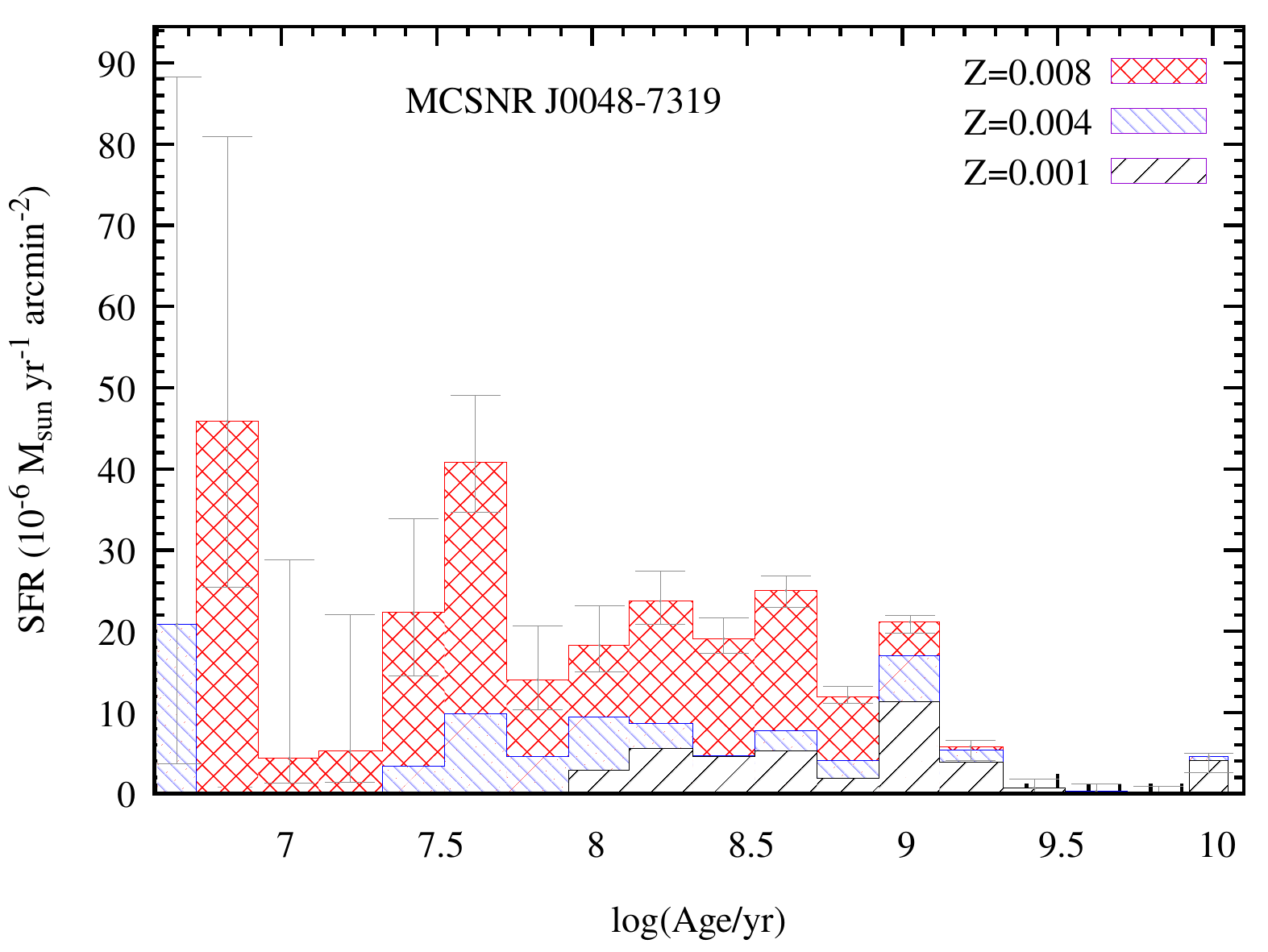}
    \includegraphics[height=0.17\vsize]{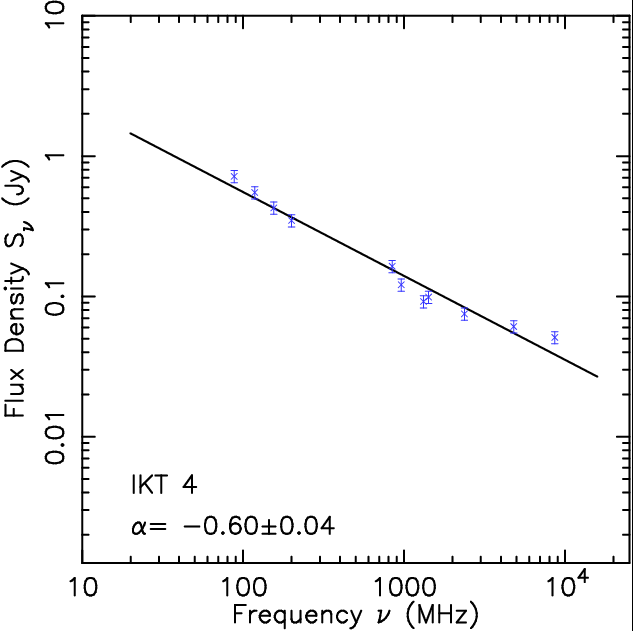}

    \includegraphics[height=0.175\vsize]{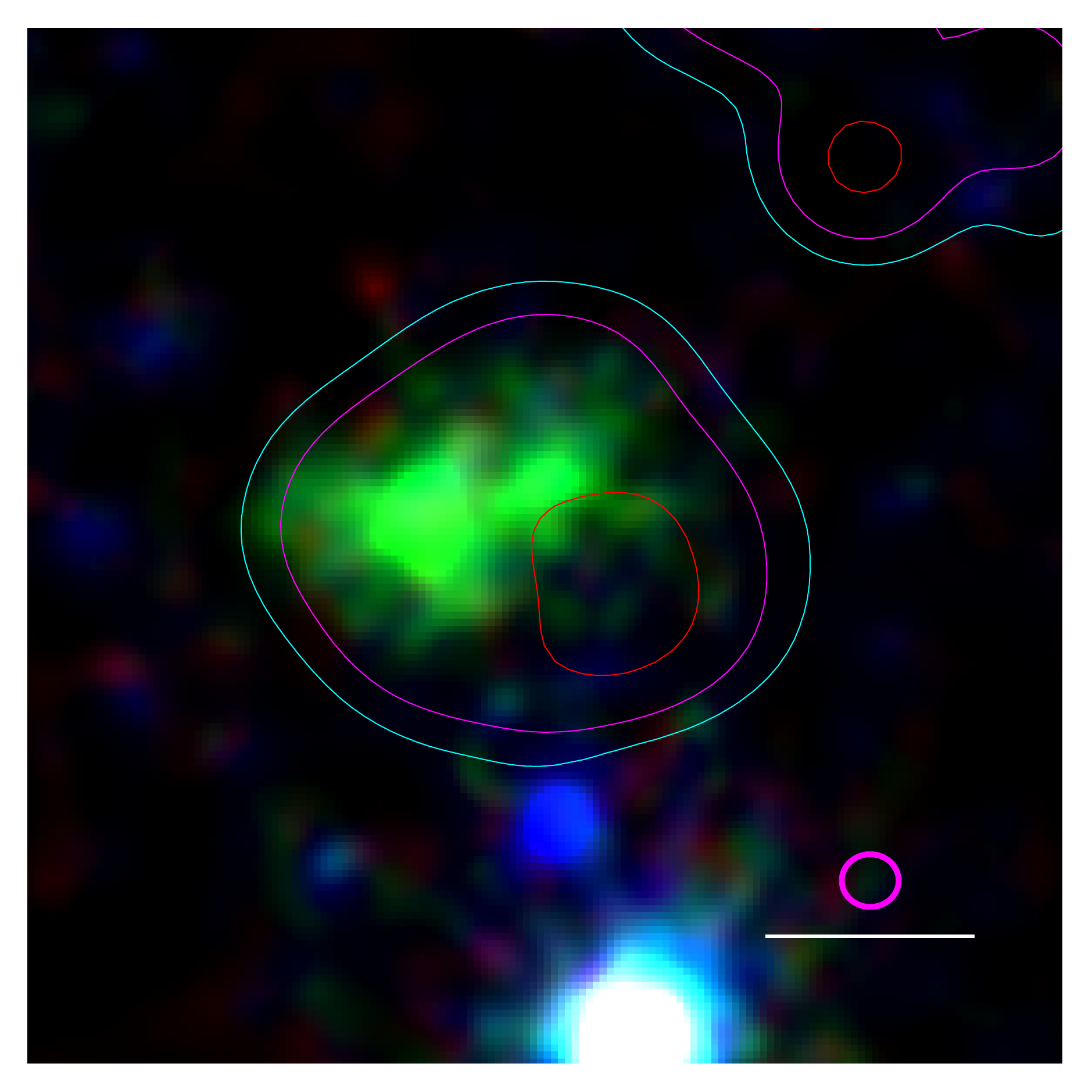}
    \includegraphics[height=0.175\vsize]{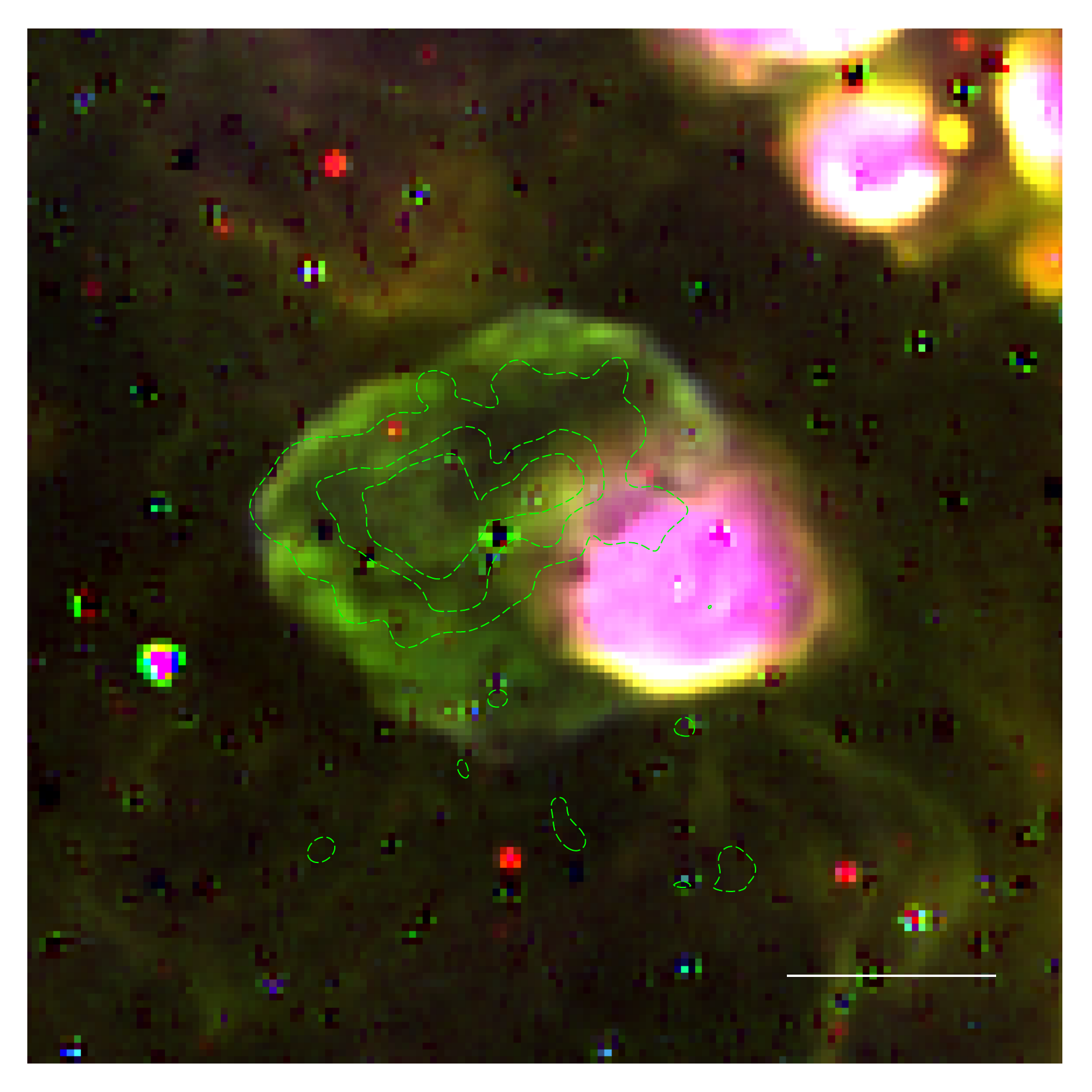}
    \includegraphics[height=0.175\vsize]{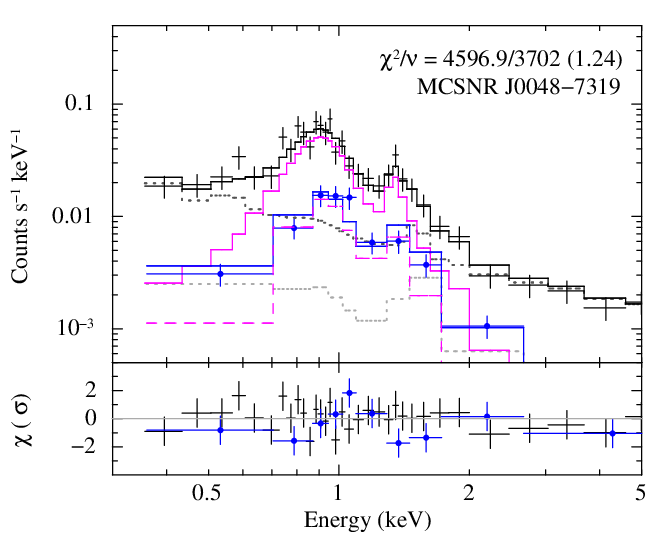}

    \caption{\textit{Top part\,:} CMD, SFH, and radio-continuum spectrum of MCSNR~J0041$-$7336 (first line), X-ray image with radio contours, optical image with X-ray contours, and X-ray spectrum (left, middle, and right panel of the second line, respectively). The radio flux density levels shown are at 0.1~mJy/beam, 0.5~mJy/beam, and 2~mJy/beam. \textit{Bottom part\,:} Same as above for MCSNR~J048$-$7319, with same radio contour levels.}

    \label{fig_appendix_sfh0}
\end{figure*}



\begin{figure*}[t]

    \centering
    \includegraphics[height=0.16\vsize]{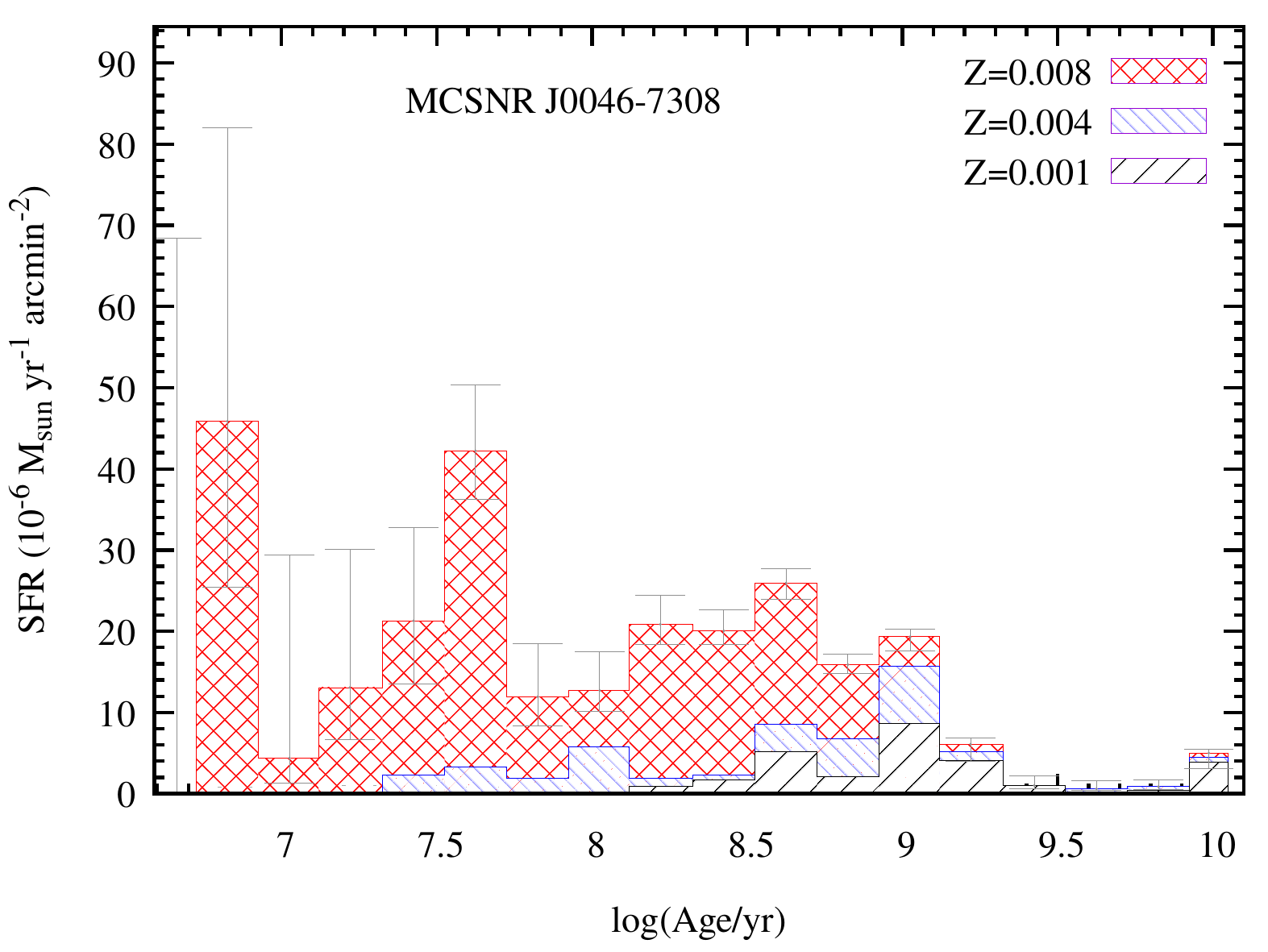}
    \includegraphics[height=0.16\vsize]{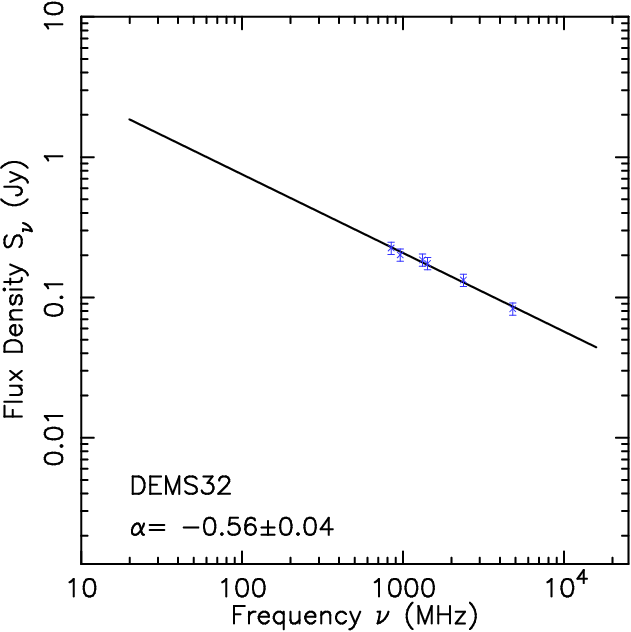}
    \includegraphics[height=0.16\vsize]{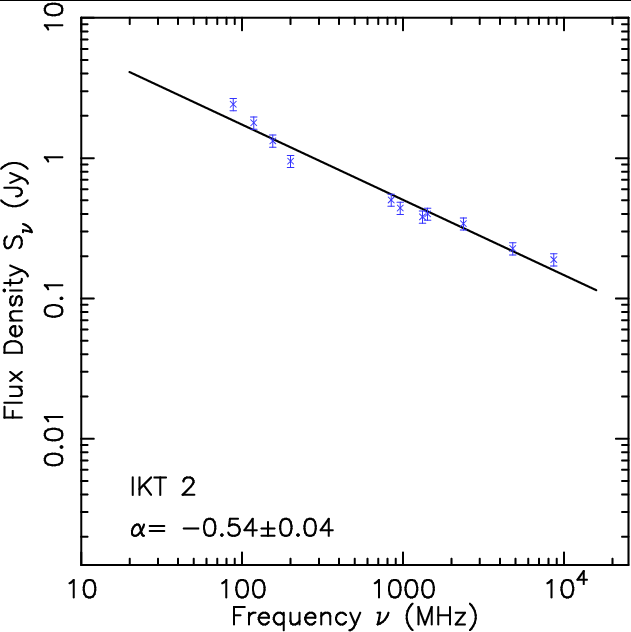}
    \includegraphics[height=0.16\vsize]{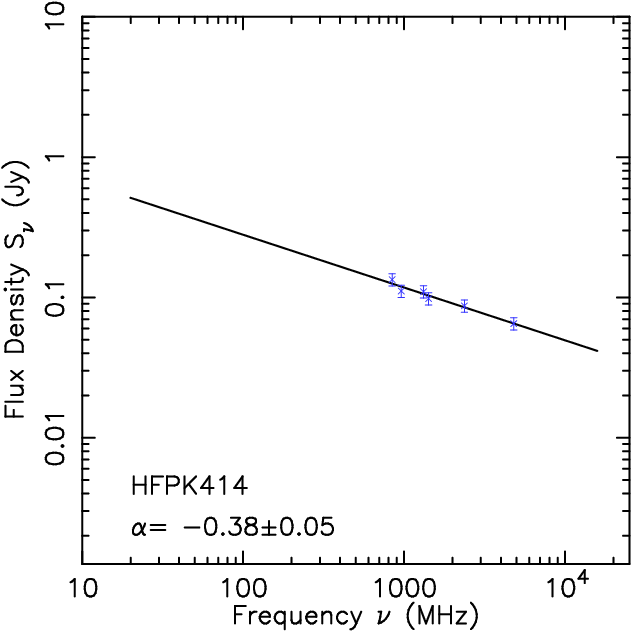}

    \includegraphics[height=0.185\vsize]{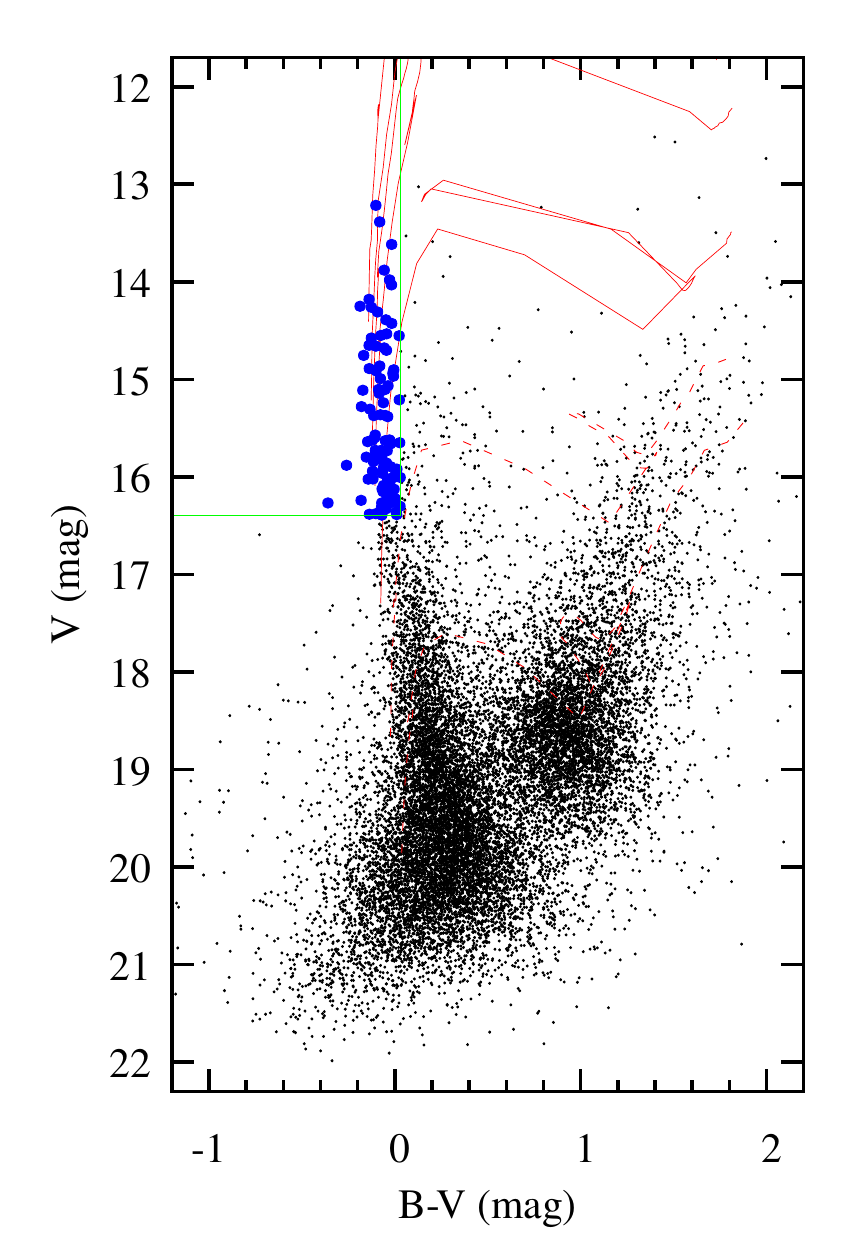}
    \includegraphics[height=0.185\vsize]{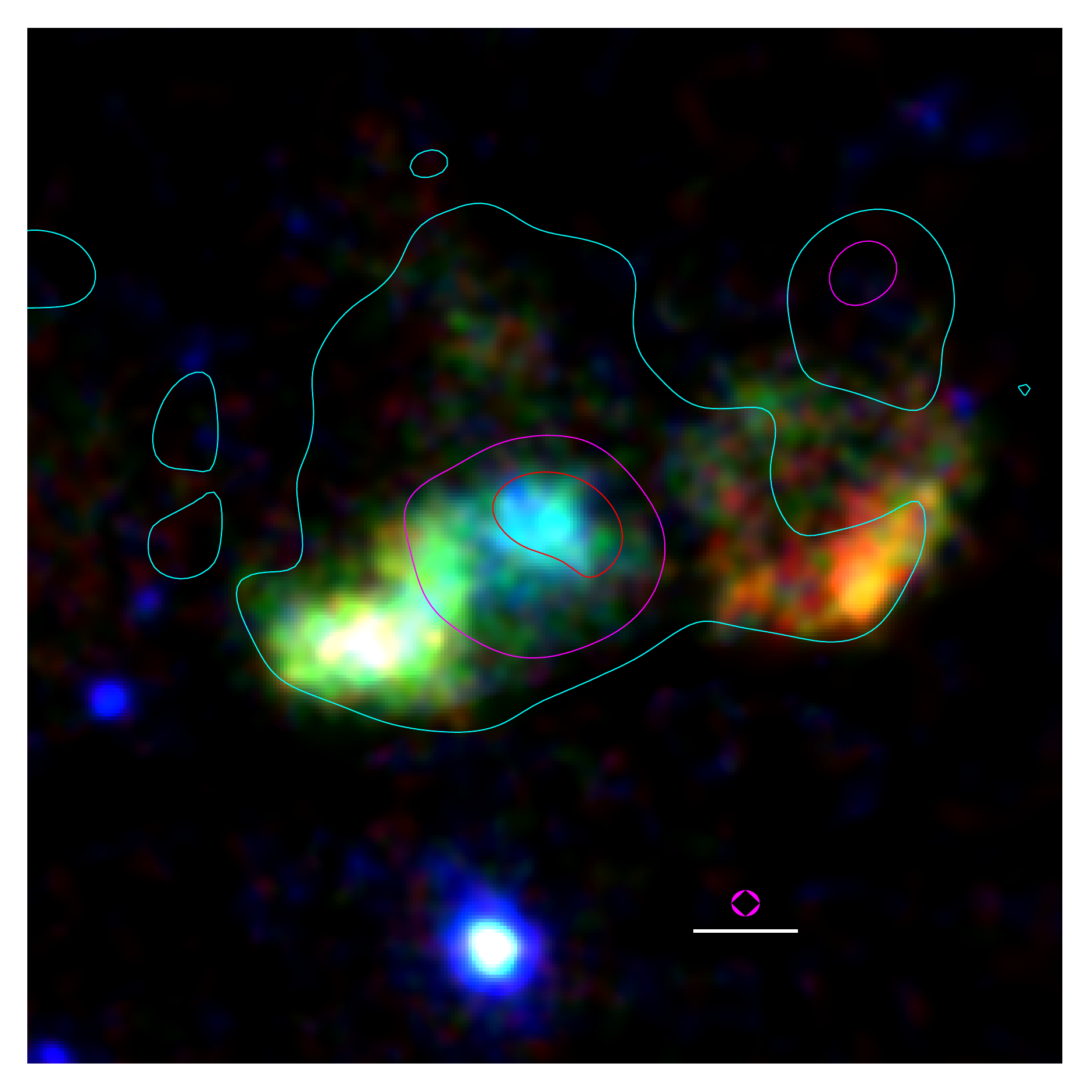}
    \includegraphics[height=0.185\vsize]{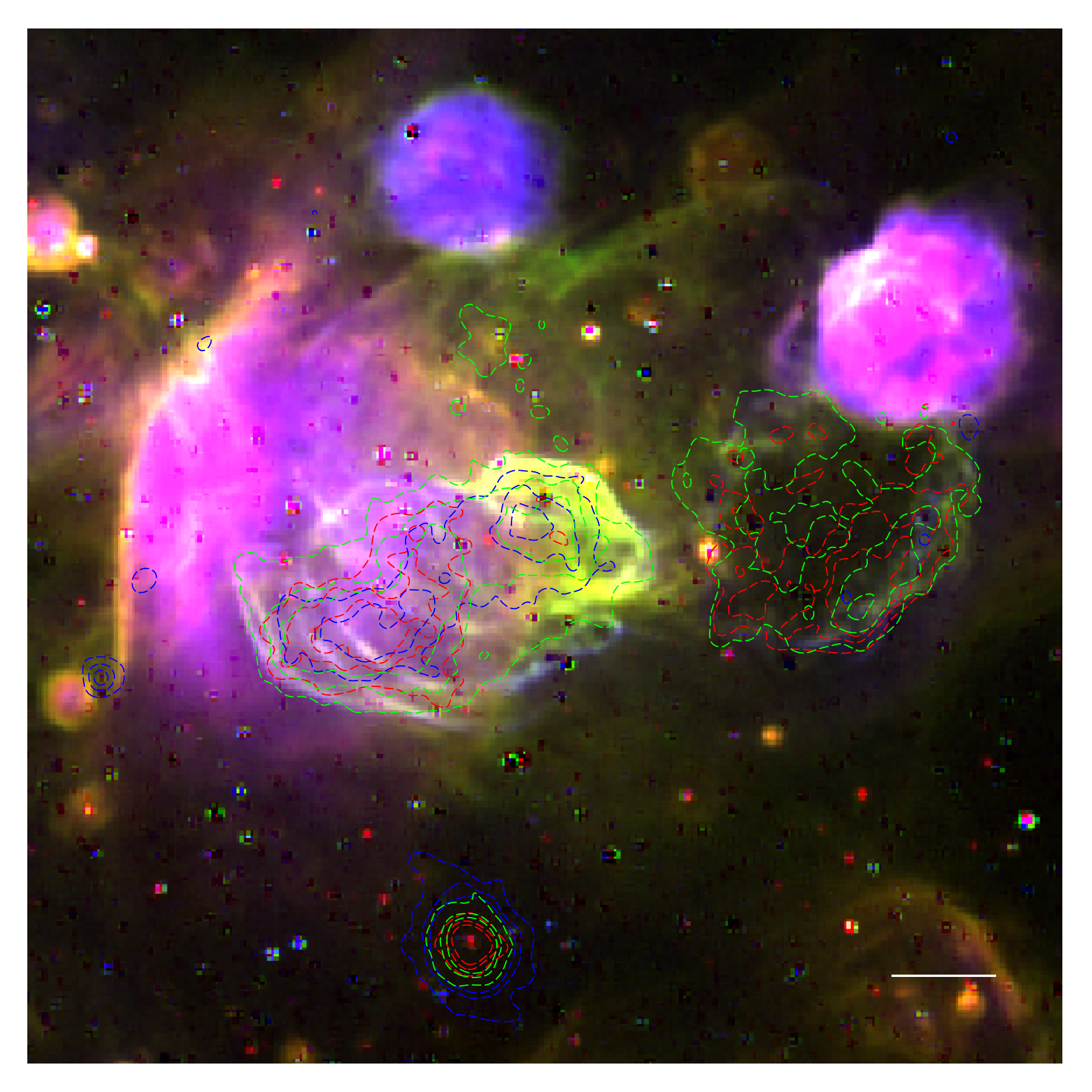}
    
    \includegraphics[height=0.19\vsize]{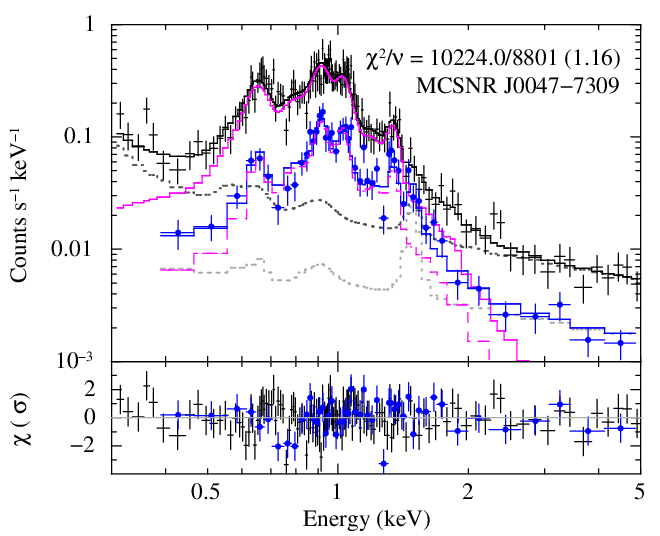}
    \includegraphics[height=0.19\vsize]{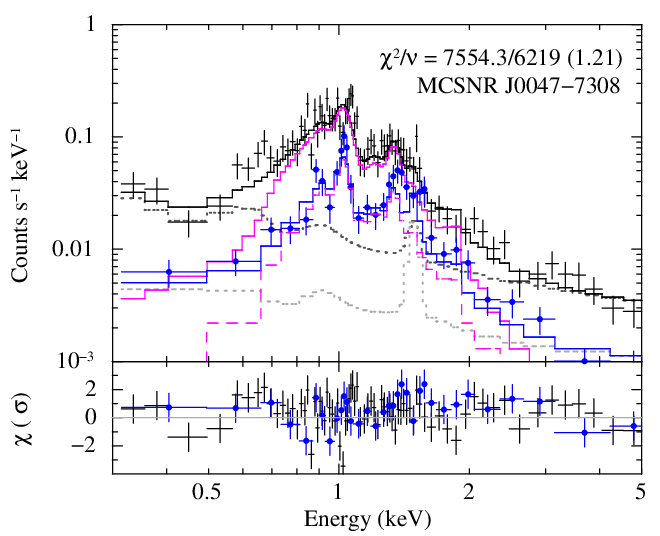}
    \includegraphics[height=0.19\vsize]{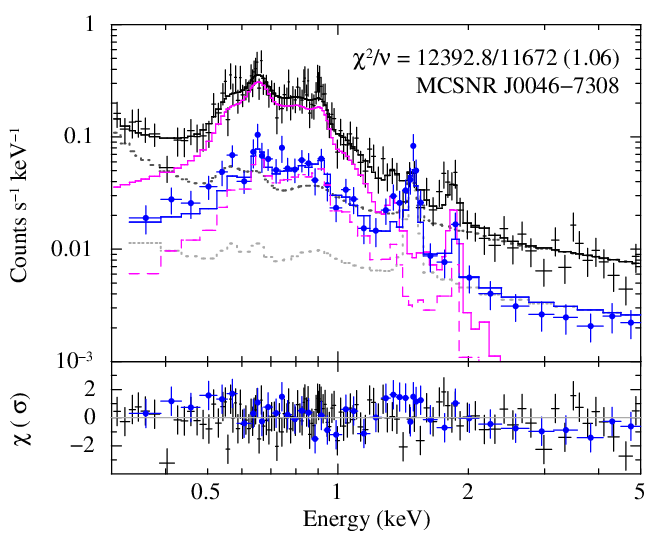}

    \vspace{1em}
    \hrule
    \vspace{1em}

    \includegraphics[height=0.18\vsize]{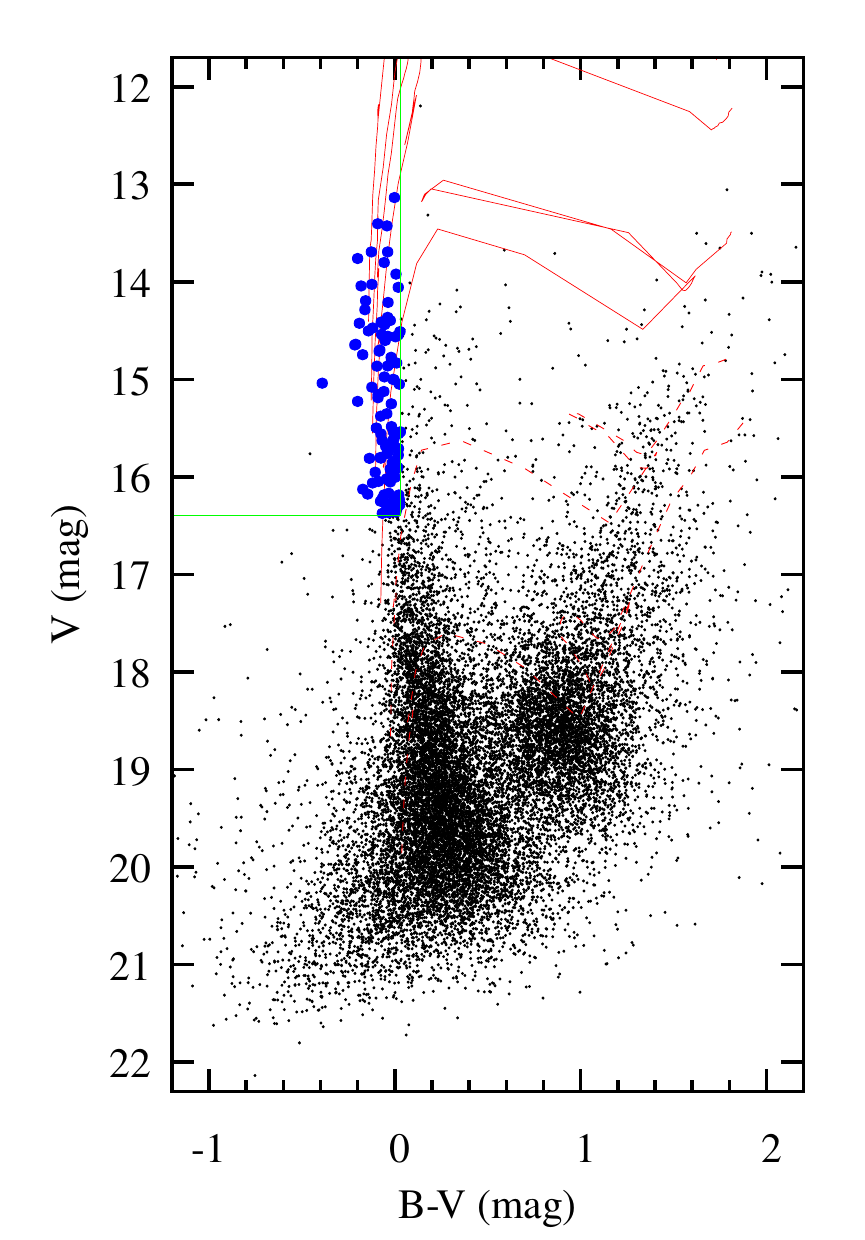}
    \includegraphics[height=0.18\vsize]{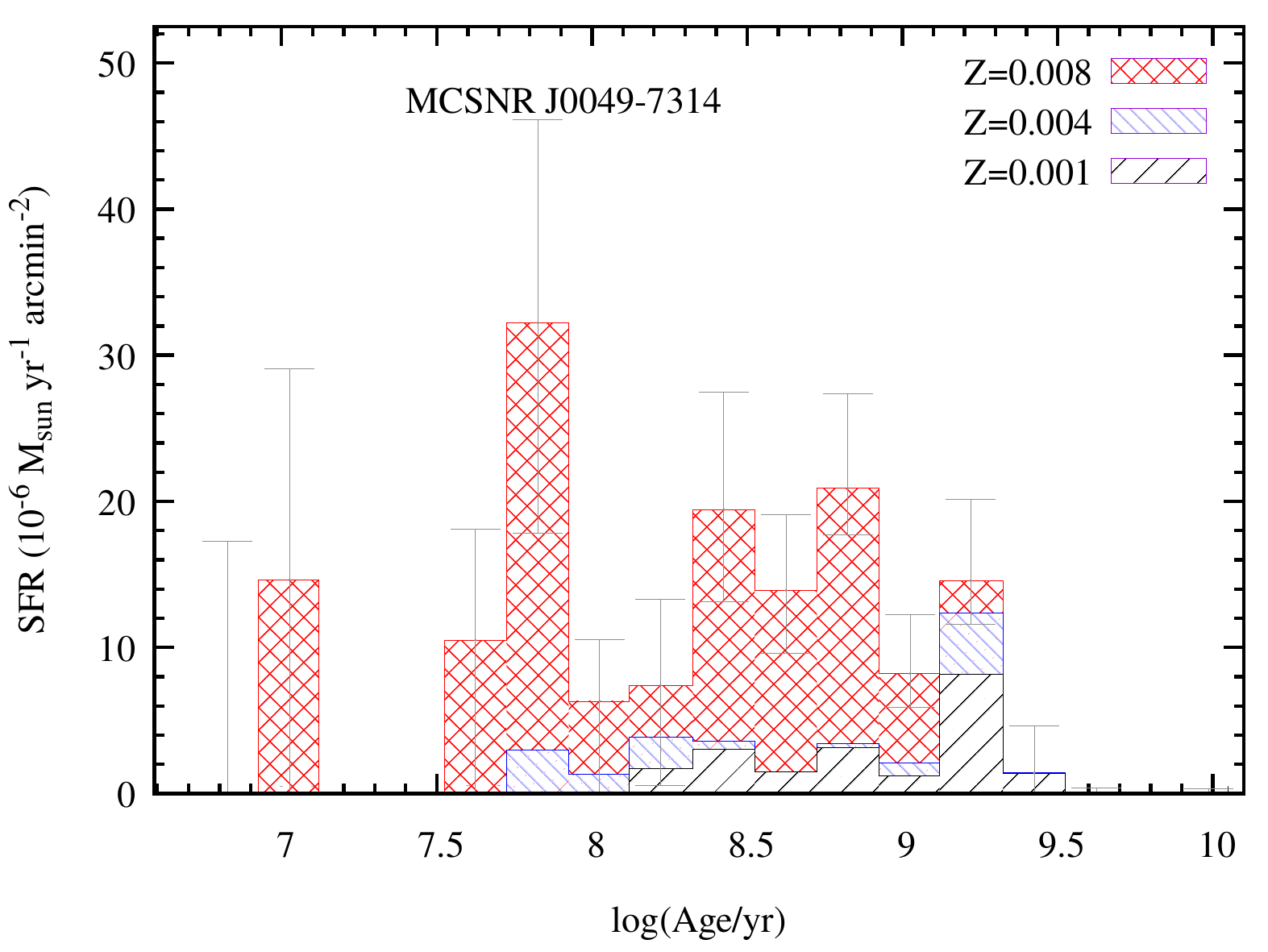}
    \includegraphics[height=0.18\vsize]{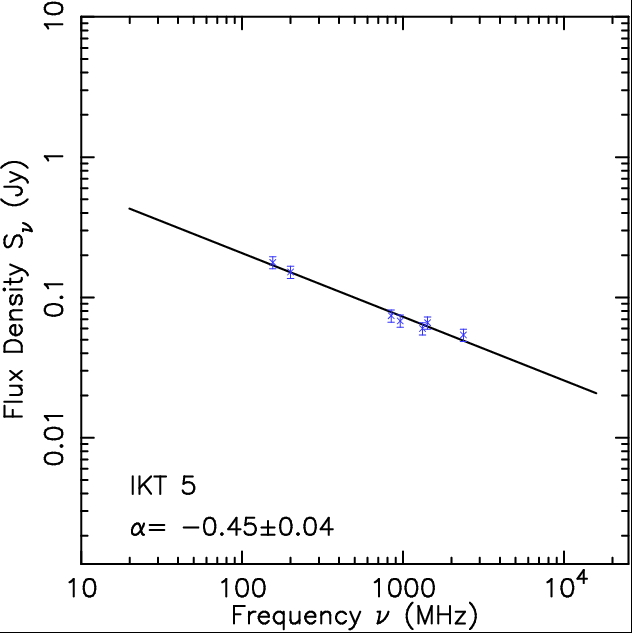}

    \includegraphics[height=0.185\vsize]{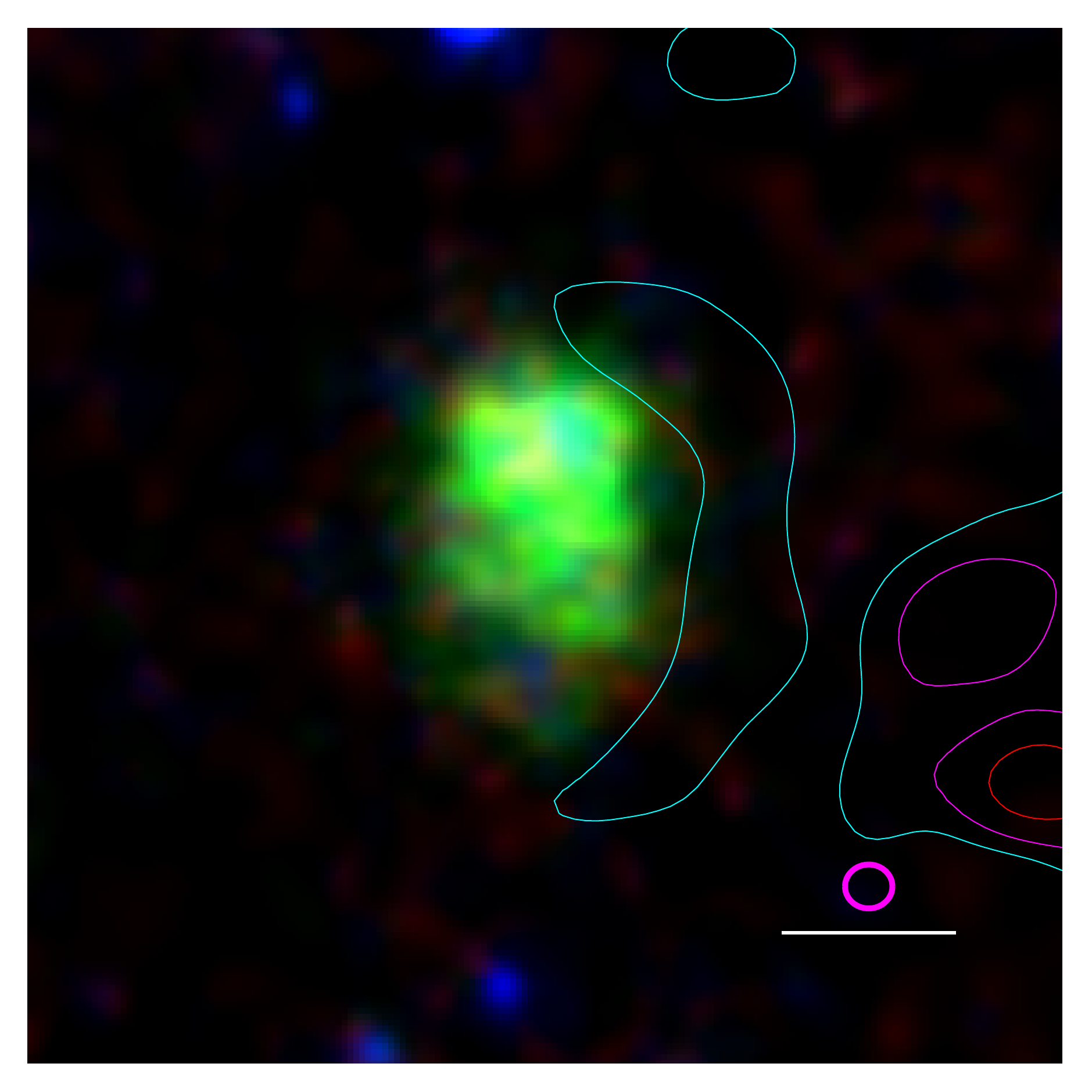}
    \includegraphics[height=0.185\vsize]{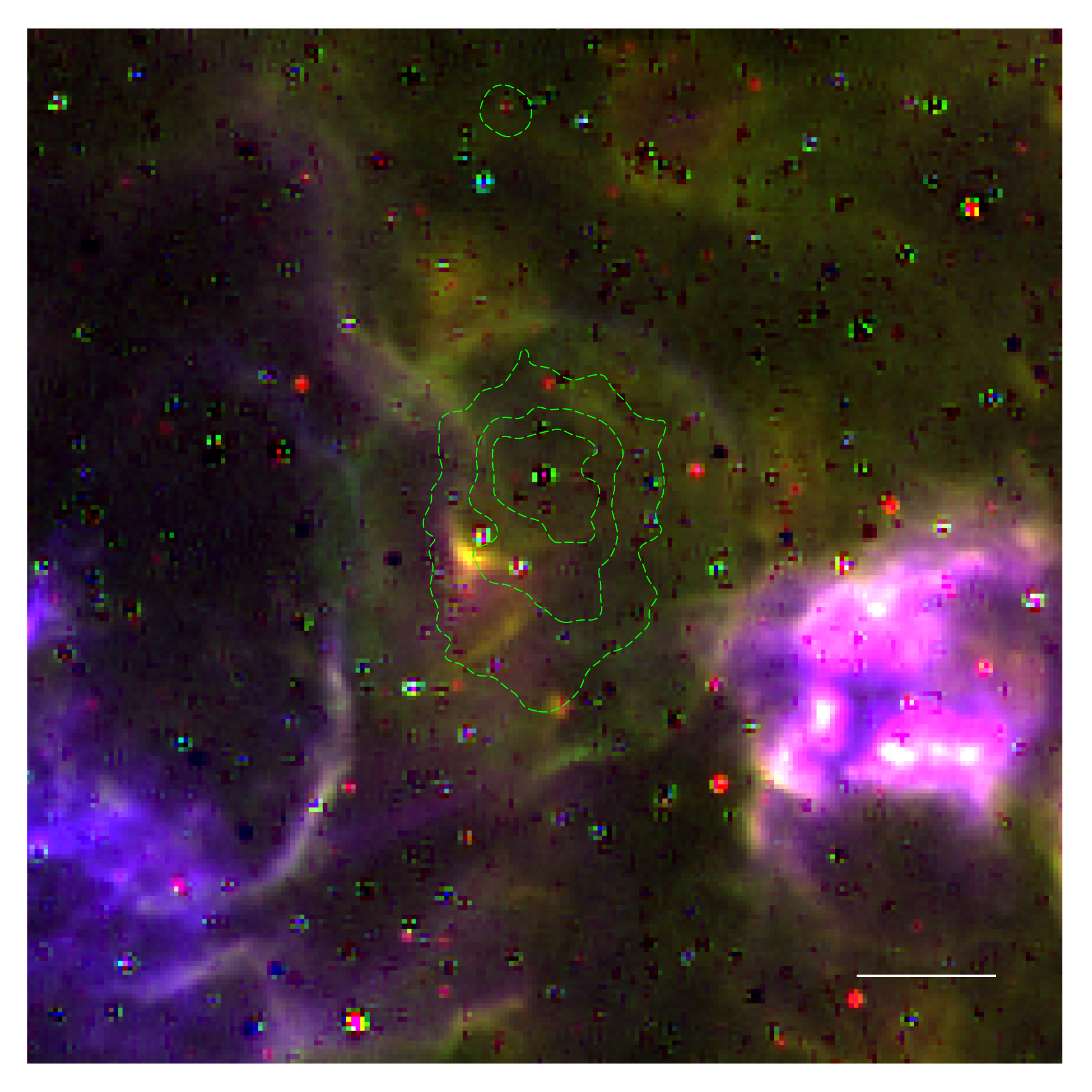}
    \includegraphics[height=0.185\vsize]{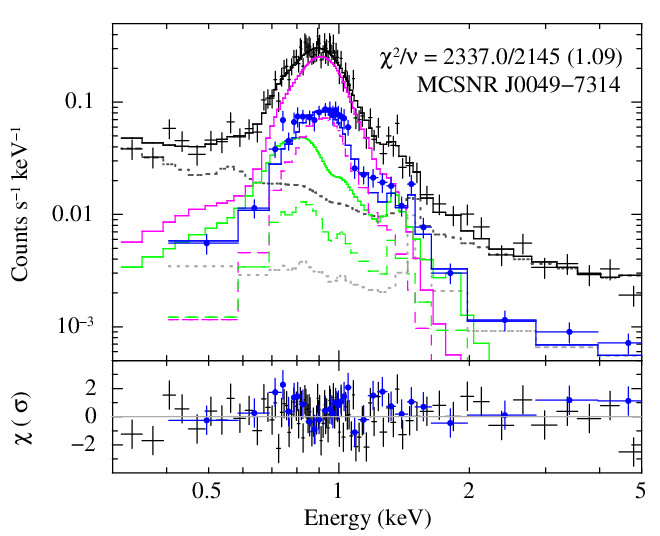}

    \caption{\textit{Top part\,:} For the three SNRs in the N19 complex (Sect.\,\ref{results_notesSNRs}) we show a single CMD and SFH as they are in the same photometric grid of \citet{2004AJ....127.1531H}. The three SNRs are shown on the same images, with from left to right (decreasing RA)\,: MCNR J0047$-$7309, J0047$-$7308, J0046$-$7308; the radio and X-ray spectra are shown in the same order. The radio contour levels are at 1, 3, and 8~mJy/beam. \textit{Bottom part\,:} Same as Fig.\,\ref{fig_appendix_sfh0} for MCSNR~J0049$-$7314, with radio contour levels at 0.5, 1, and 2~mJy/beam.}

    \label{fig_appendix_sfh1}
\end{figure*}

\begin{figure*}[t]
    \centering
    \includegraphics[height=0.23\vsize]{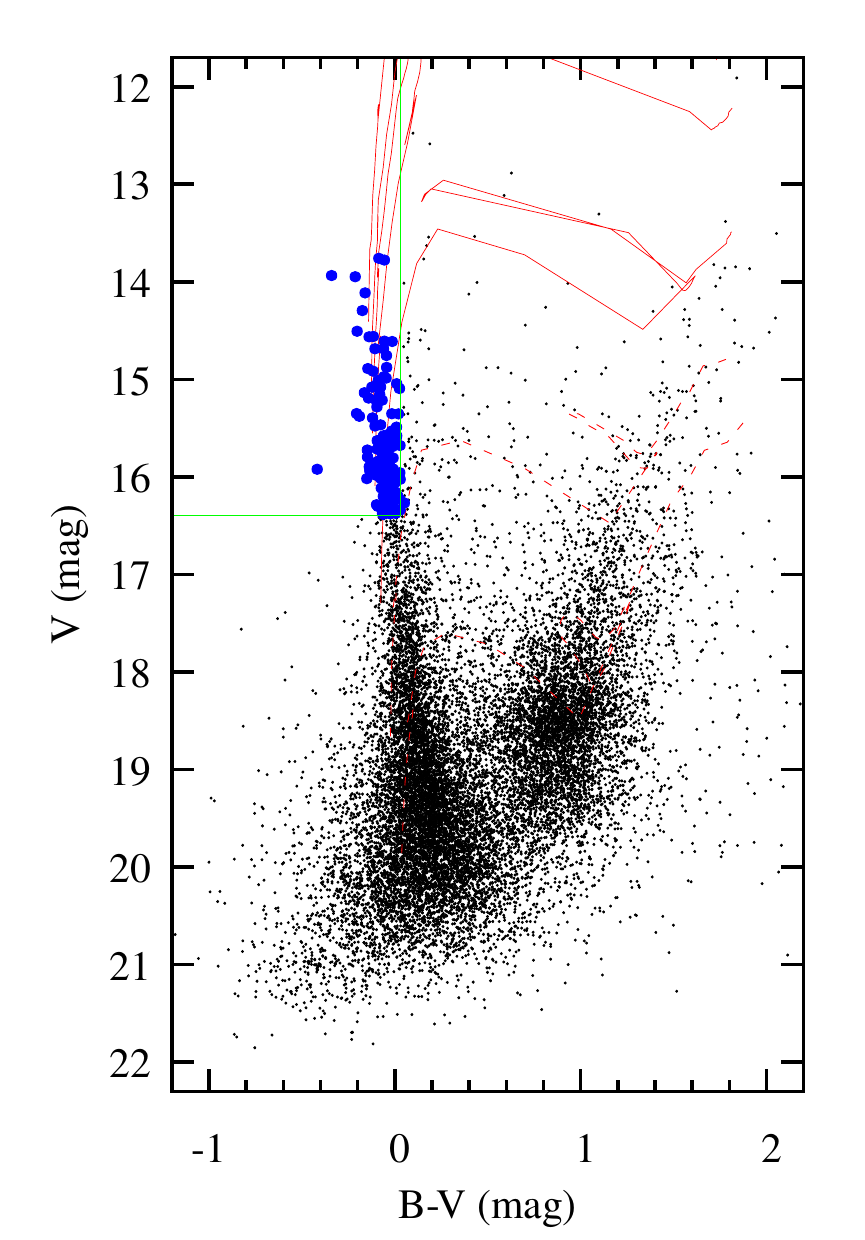}
    \includegraphics[height=0.23\vsize]{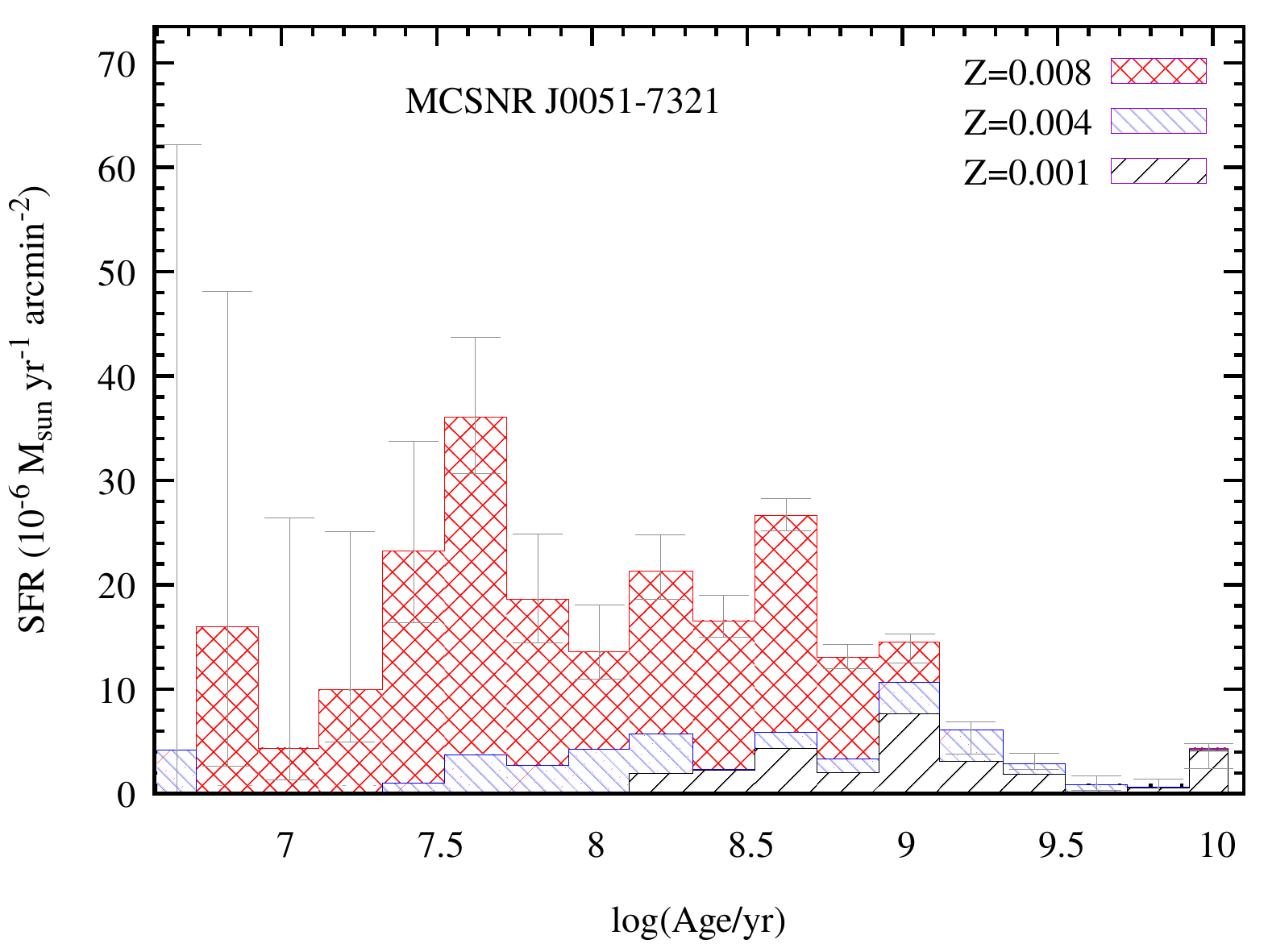}
    \includegraphics[height=0.23\vsize]{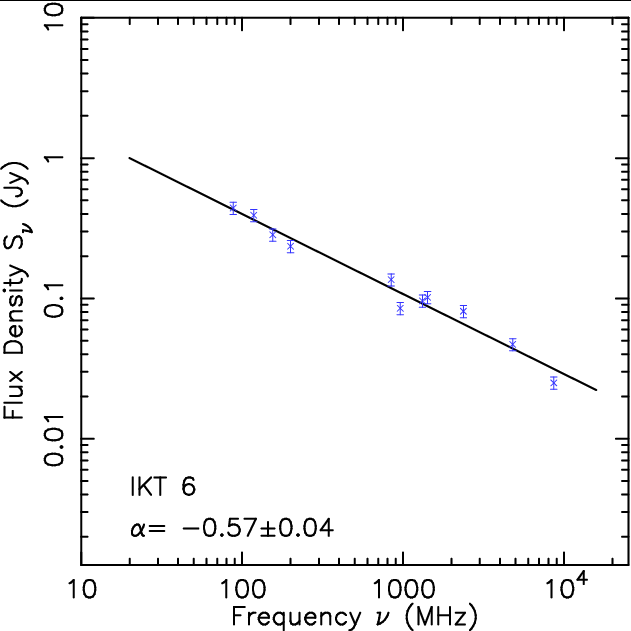}

    \includegraphics[height=0.225\vsize]{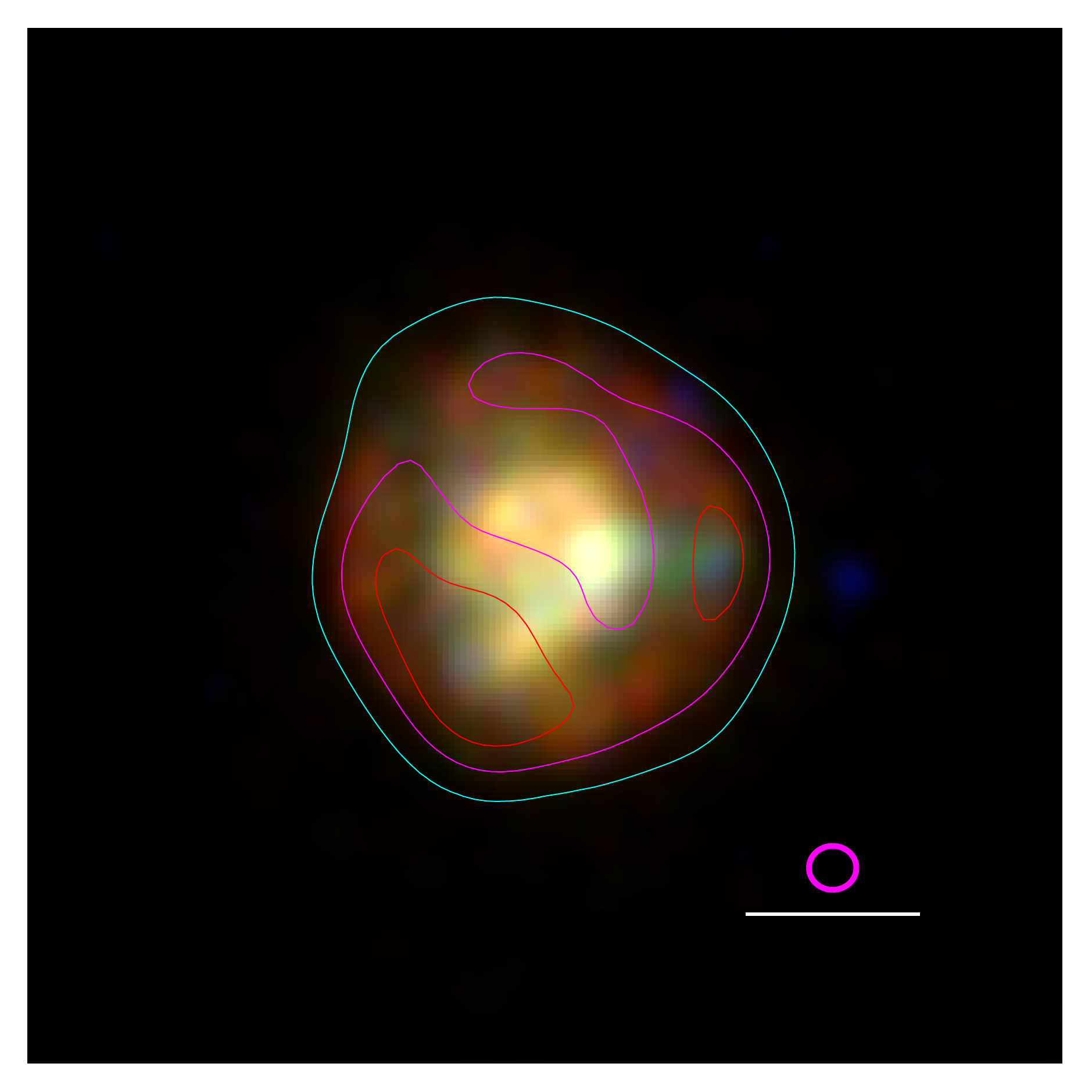}
    \includegraphics[height=0.225\vsize]{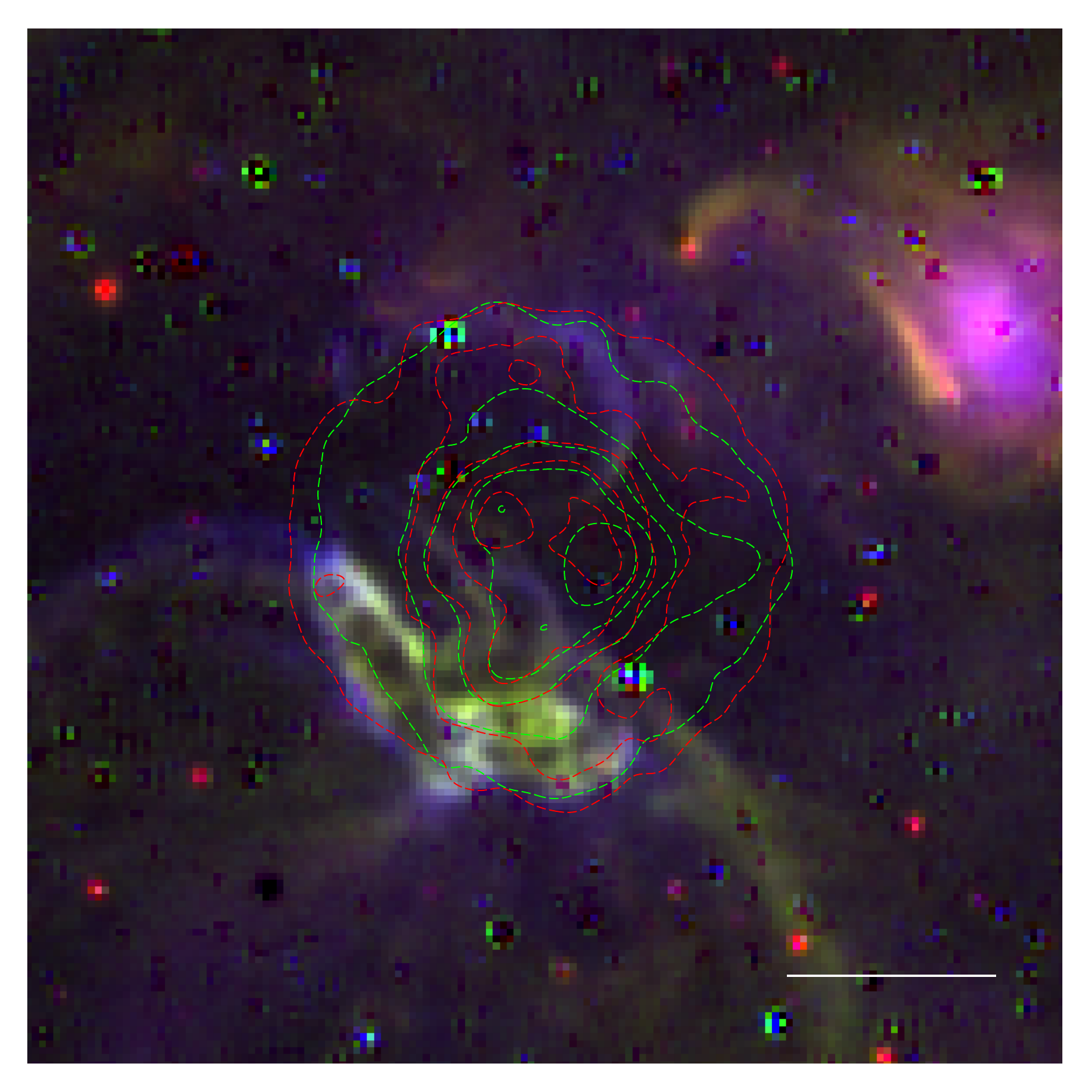}
    \includegraphics[height=0.225\vsize]{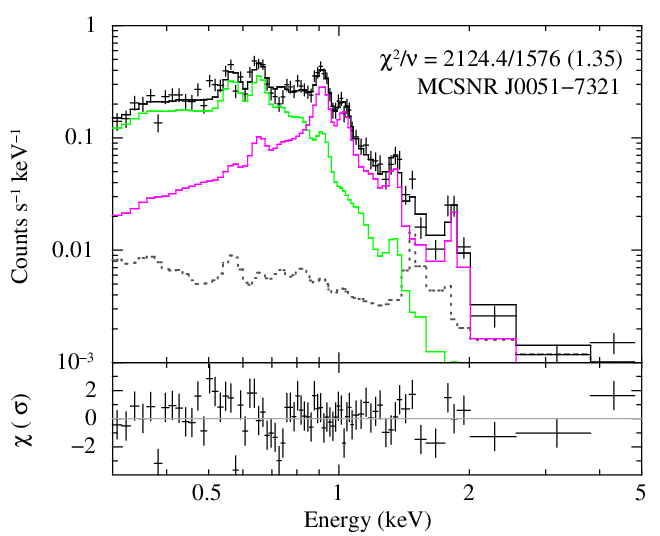}

    \vspace{1em}
    \hrule
    \vspace{1em}

    \includegraphics[height=0.23\vsize]{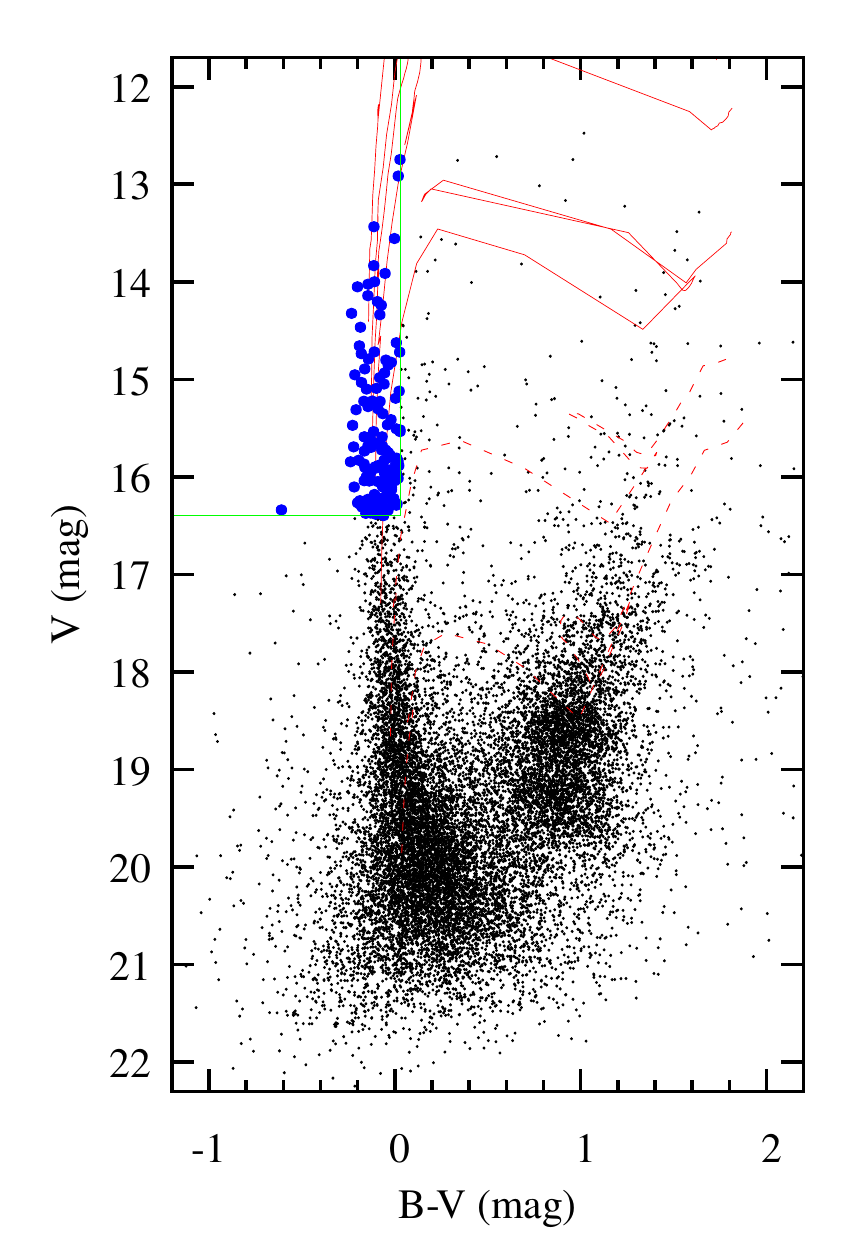}
    \includegraphics[height=0.23\vsize]{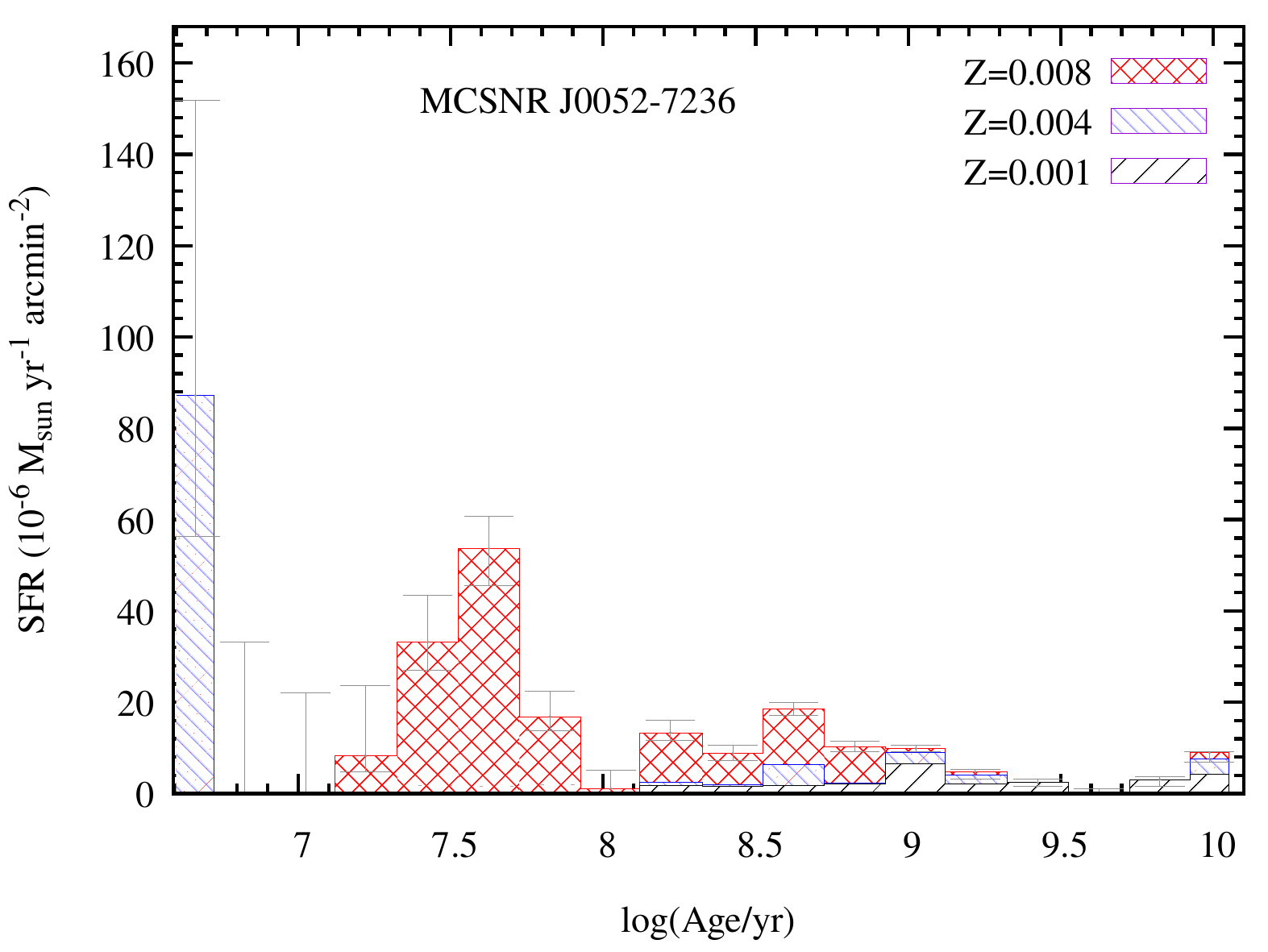}
    \includegraphics[height=0.23\vsize]{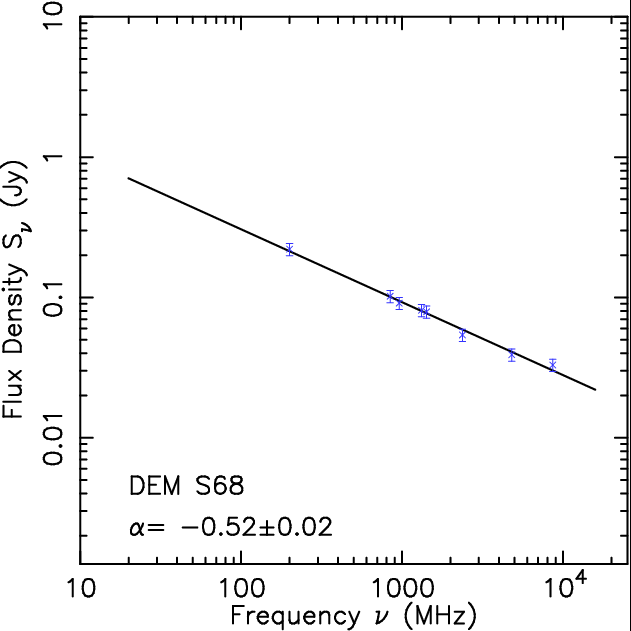}

    \includegraphics[height=0.225\vsize]{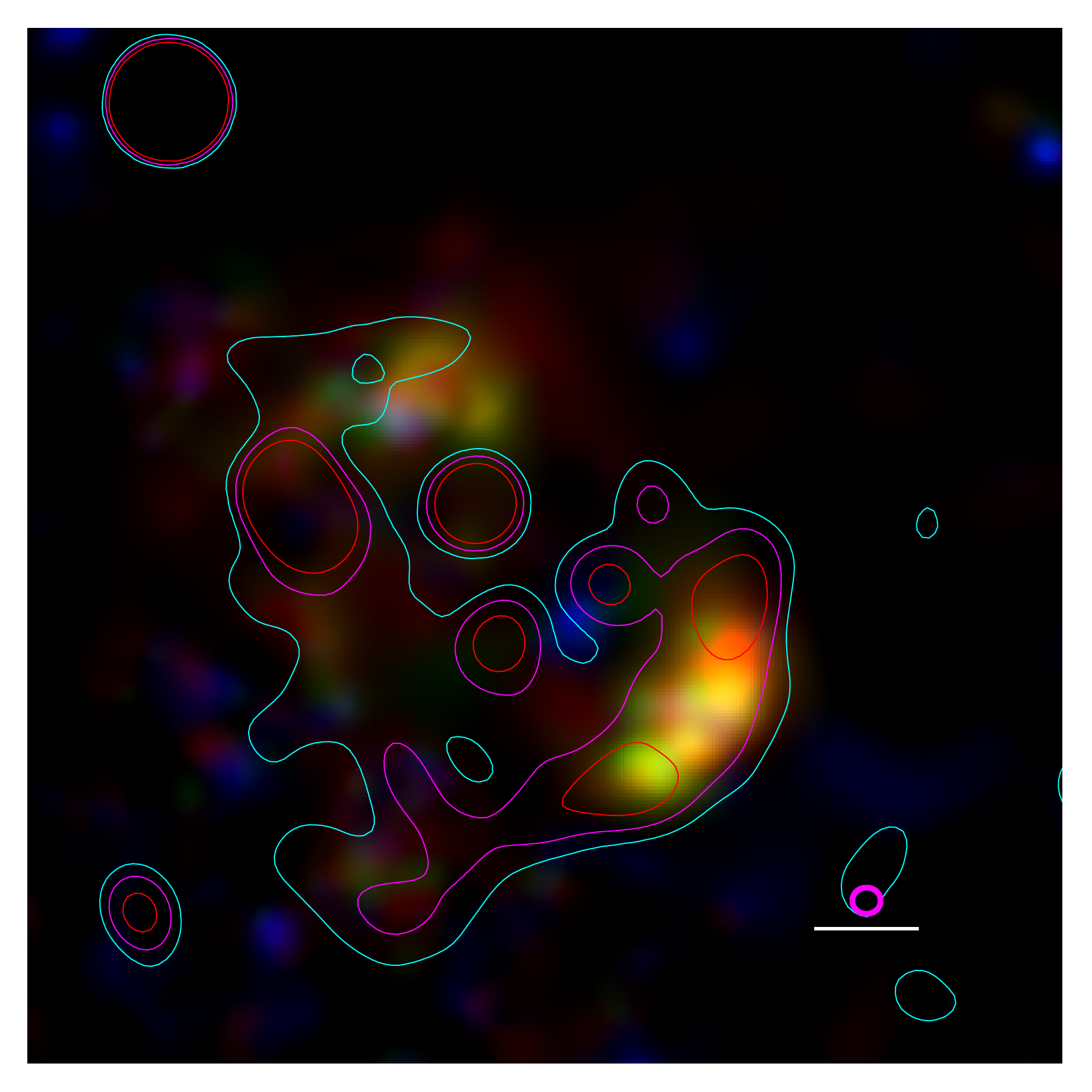}
    \includegraphics[height=0.225\vsize]{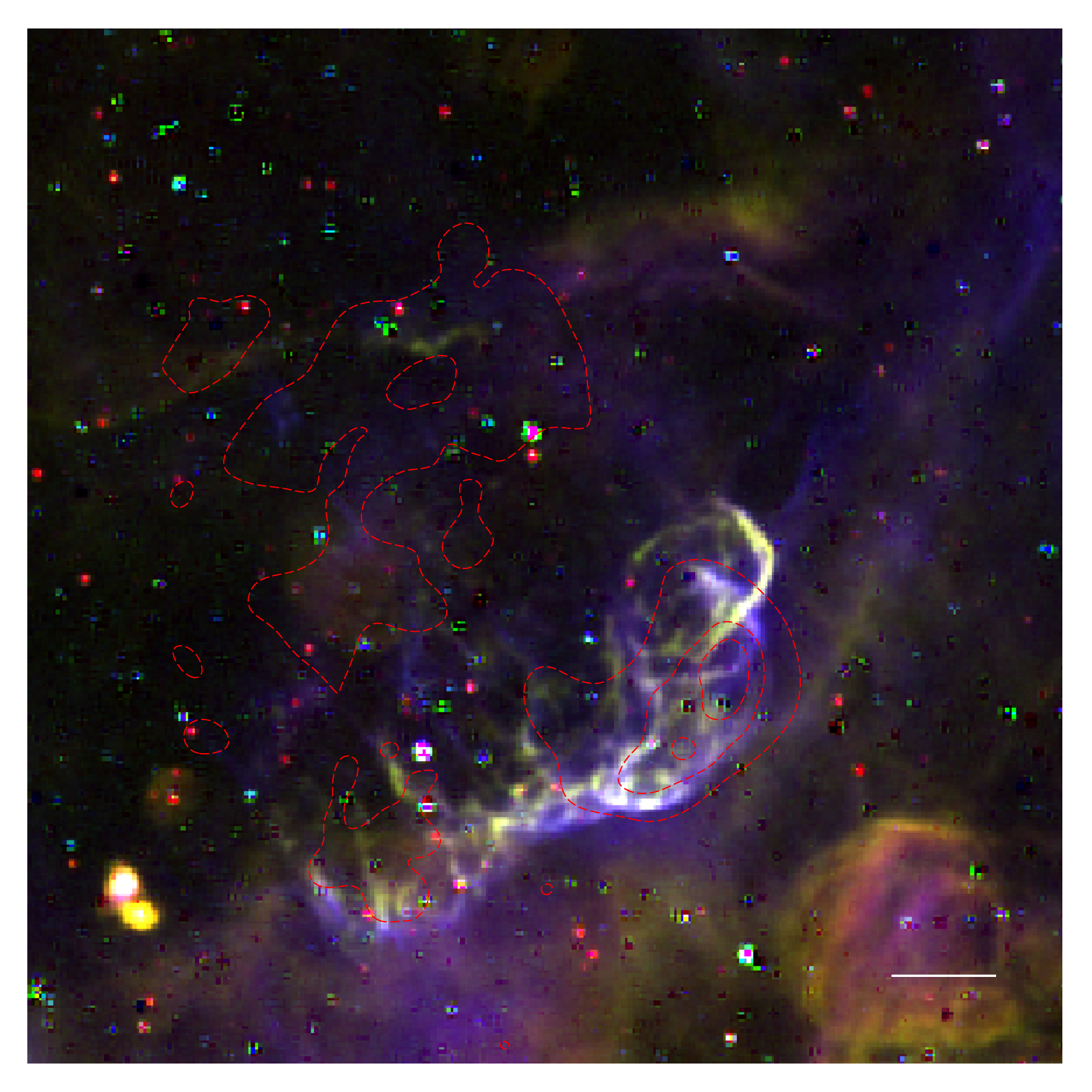}
    \includegraphics[height=0.225\vsize]{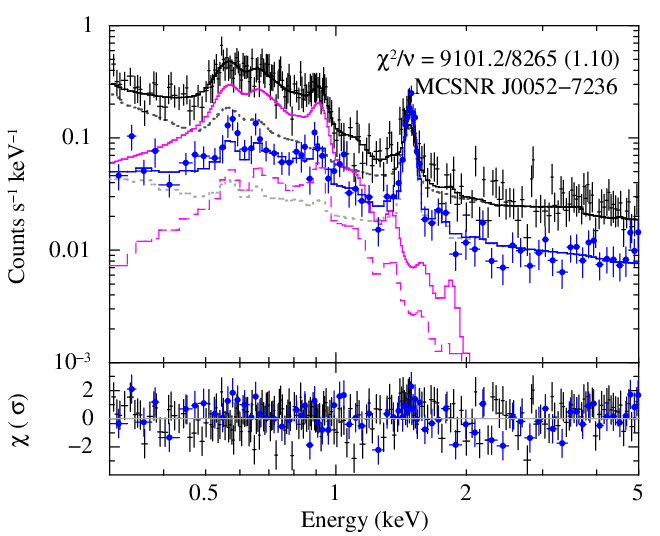}

    \caption{Same as Fig.\,\ref{fig_appendix_sfh0} for MCSNR~J0051$-$7321 (top part), with radio contour levels at 0.5, 1, and 1.5~mJy/beam, and for MCSNR~J0052$-$7236 with levels at 0.3, 0.5, and 0.8~mJy/beam (bottom part).}
  \label{fig_appendix_sfh2}
\end{figure*}

\begin{figure*}[t]
    \centering
    \includegraphics[height=0.23\vsize]{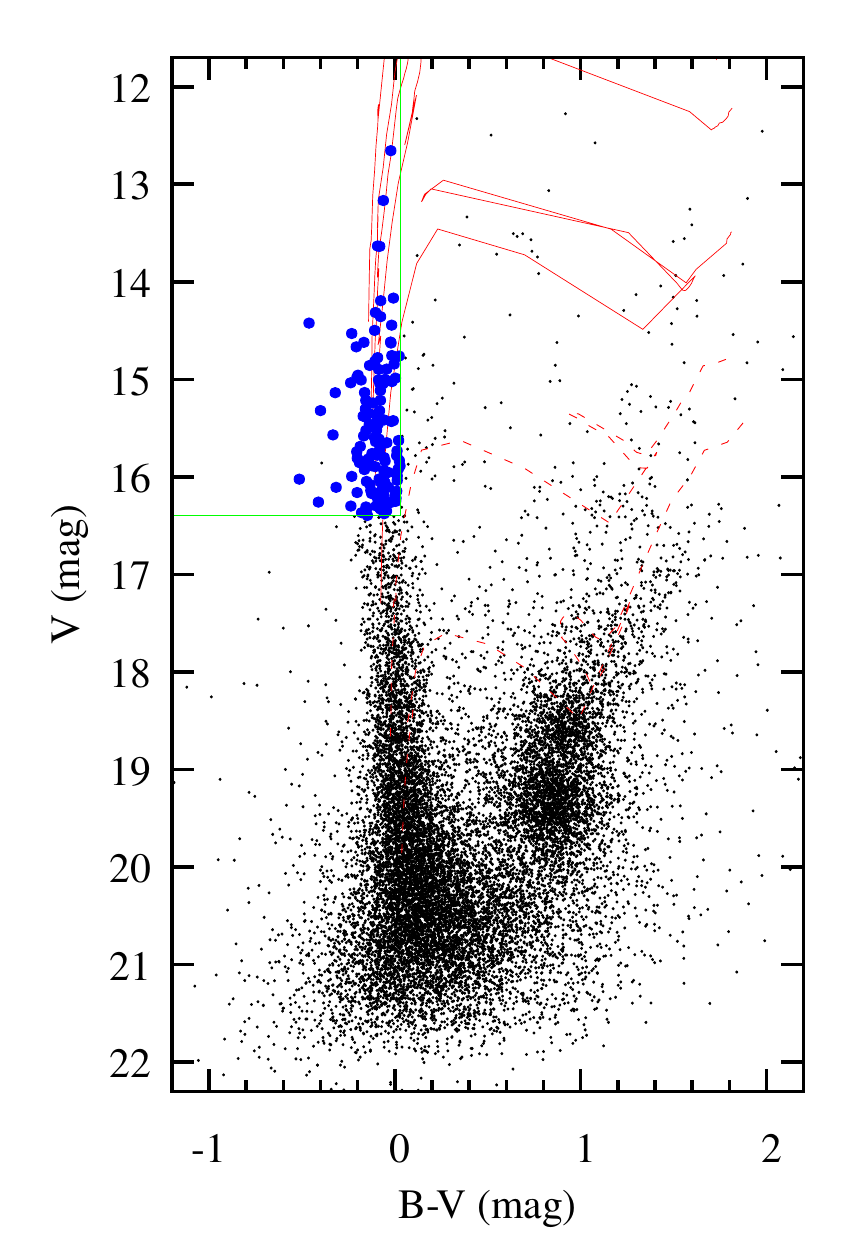}
    \includegraphics[height=0.23\vsize]{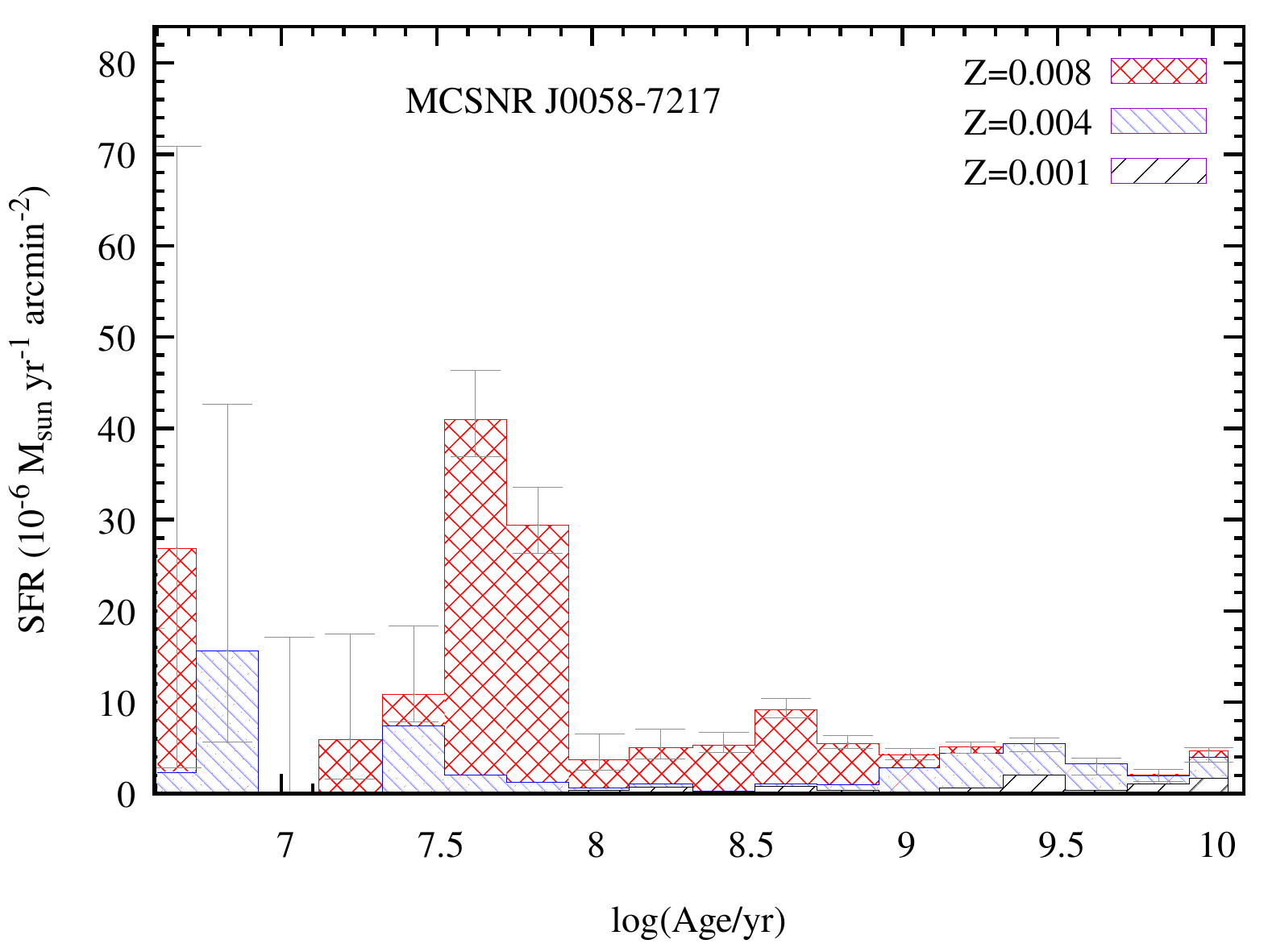}
    \includegraphics[height=0.23\vsize]{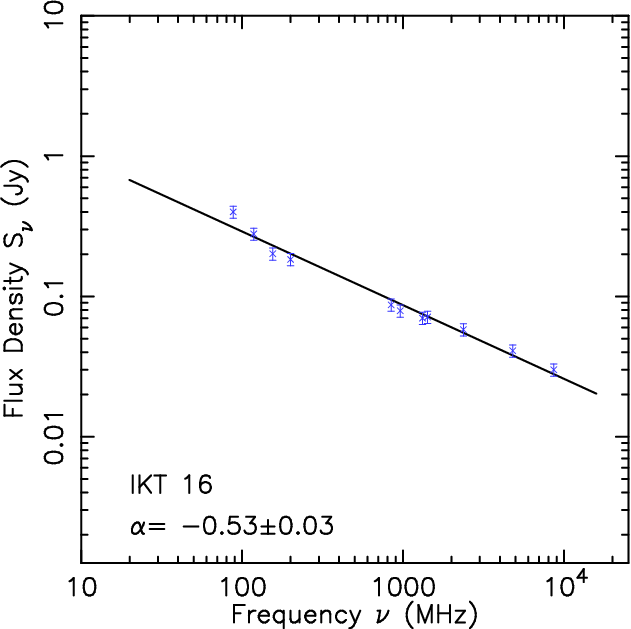}

    \includegraphics[height=0.225\vsize]{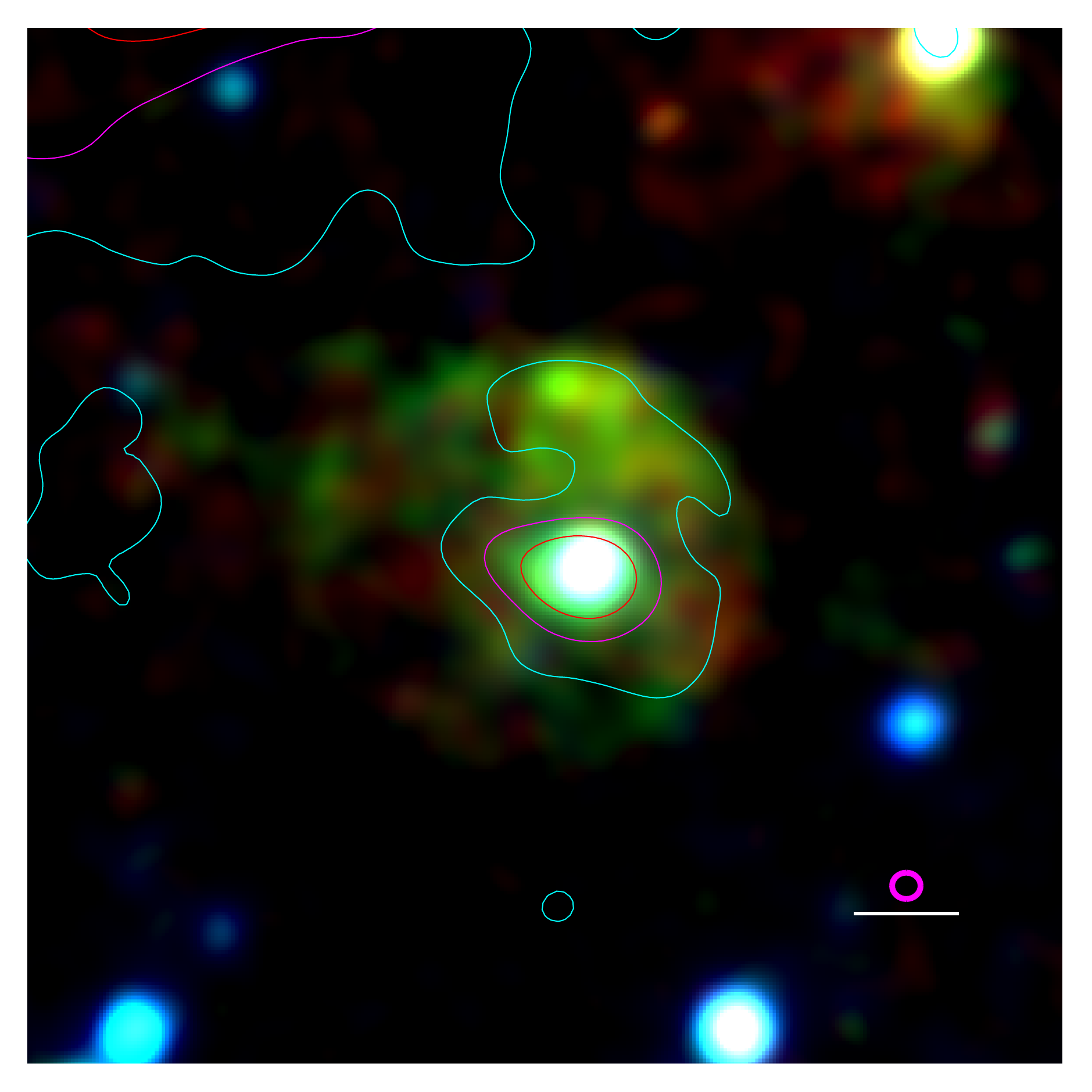}
    \includegraphics[height=0.225\vsize]{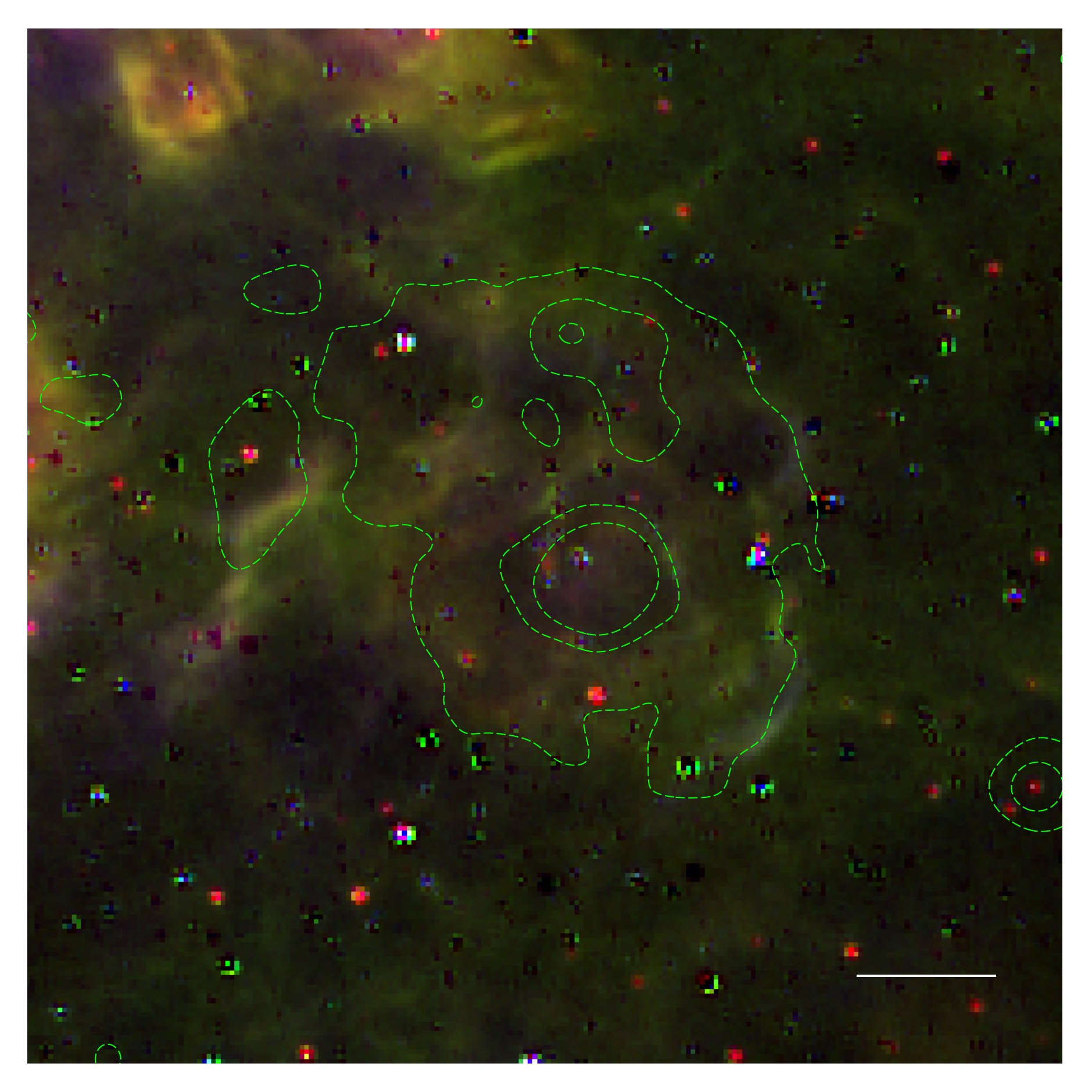}
    \includegraphics[height=0.225\vsize]{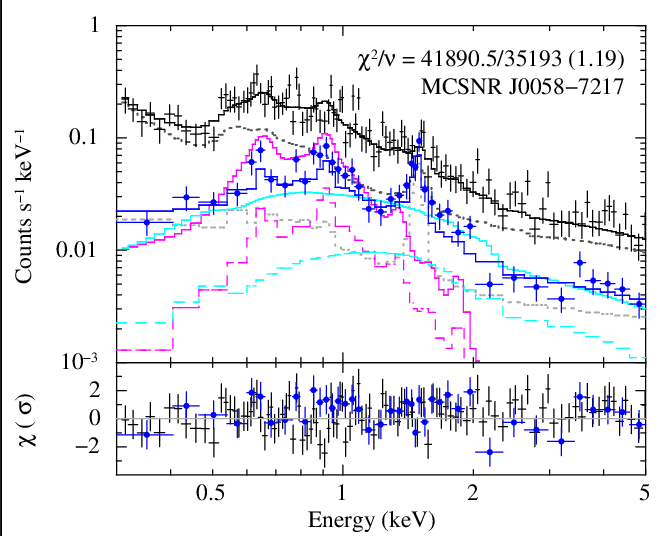}

    \vspace{1em}
    \hrule
    \vspace{1em}

    \includegraphics[height=0.23\vsize]{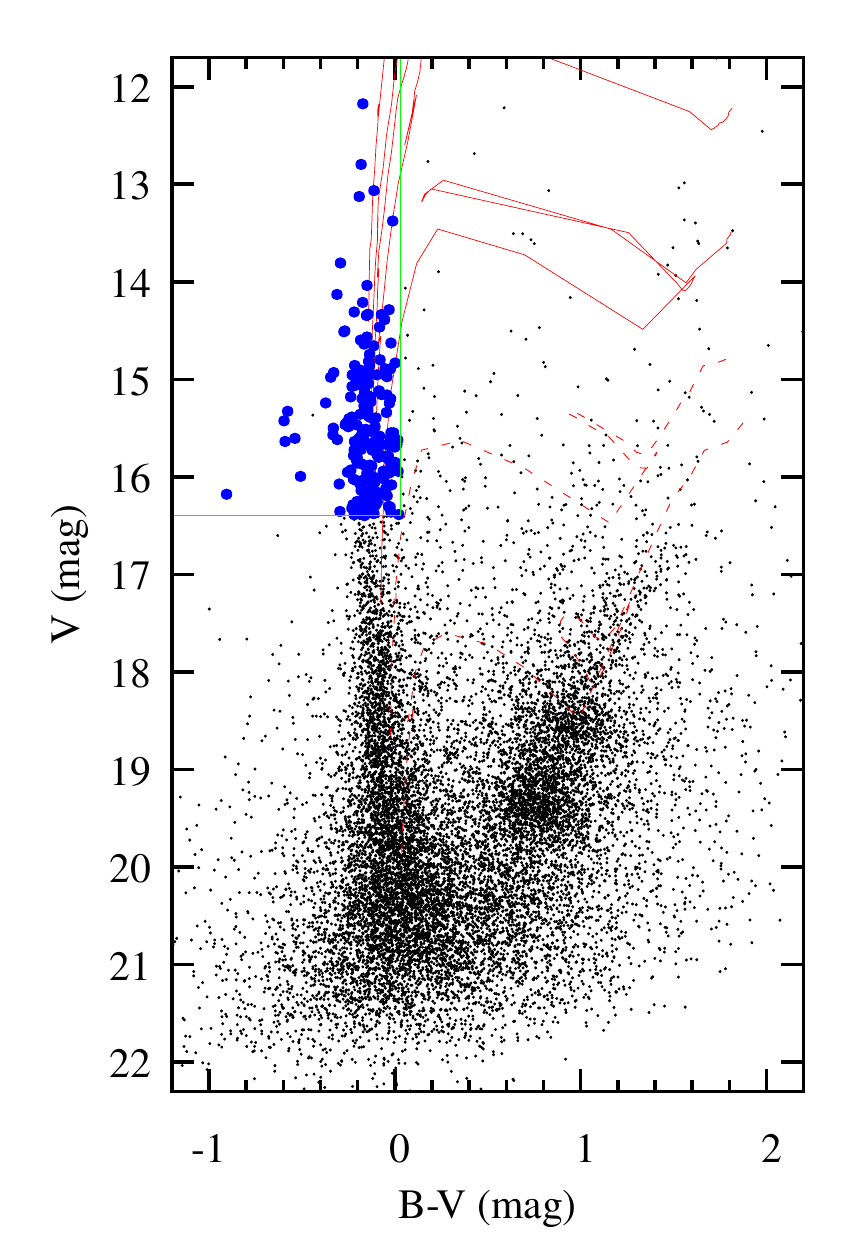}
    \includegraphics[height=0.23\vsize]{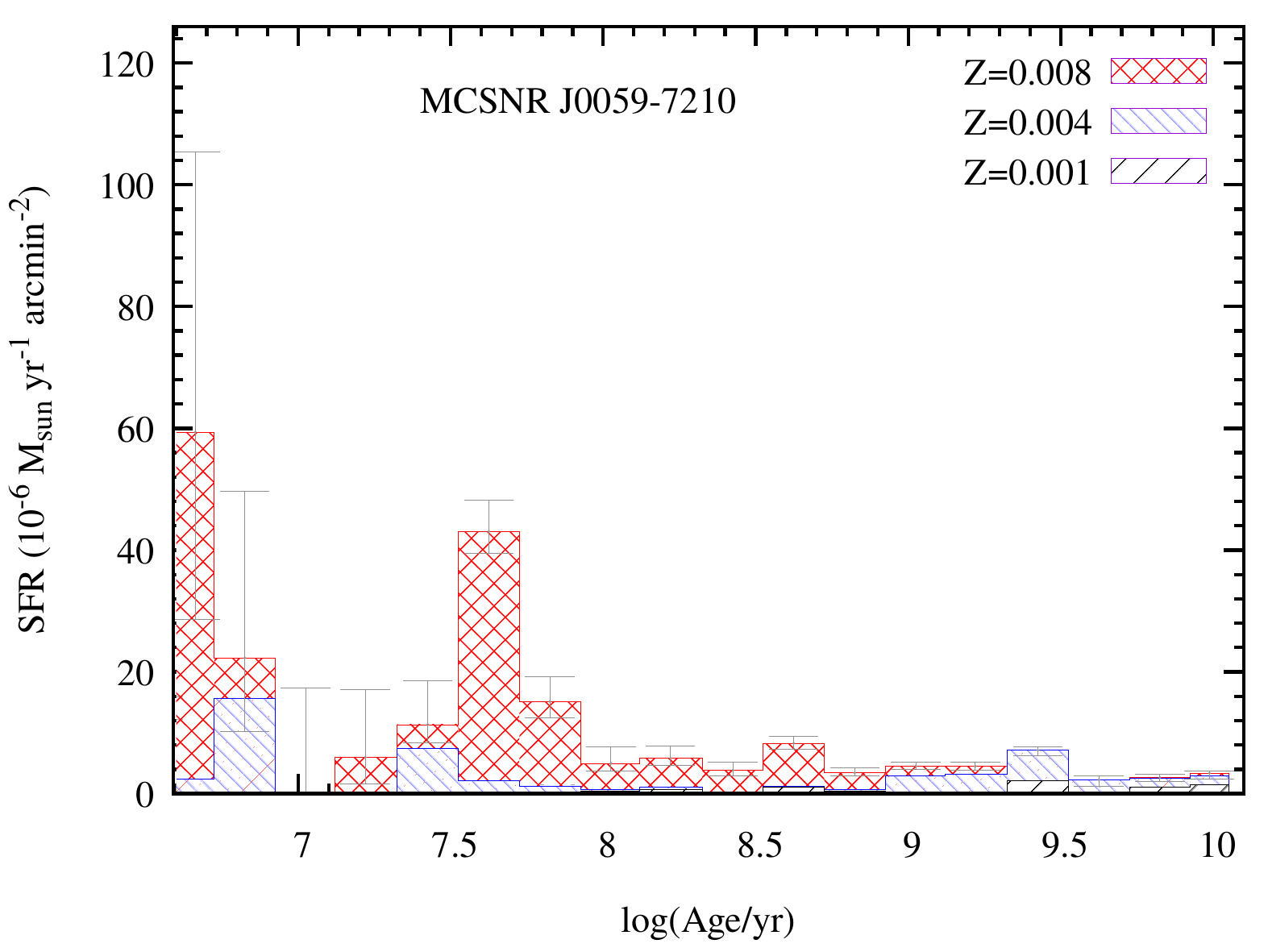}
    \includegraphics[height=0.23\vsize]{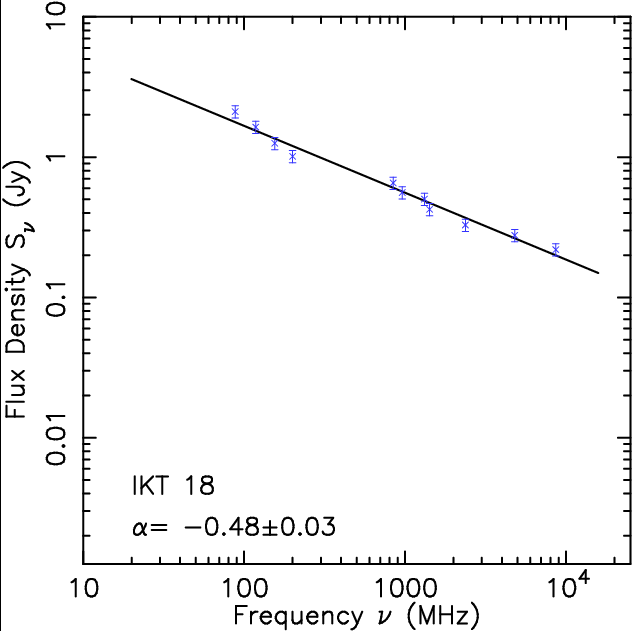}

    \includegraphics[height=0.225\vsize]{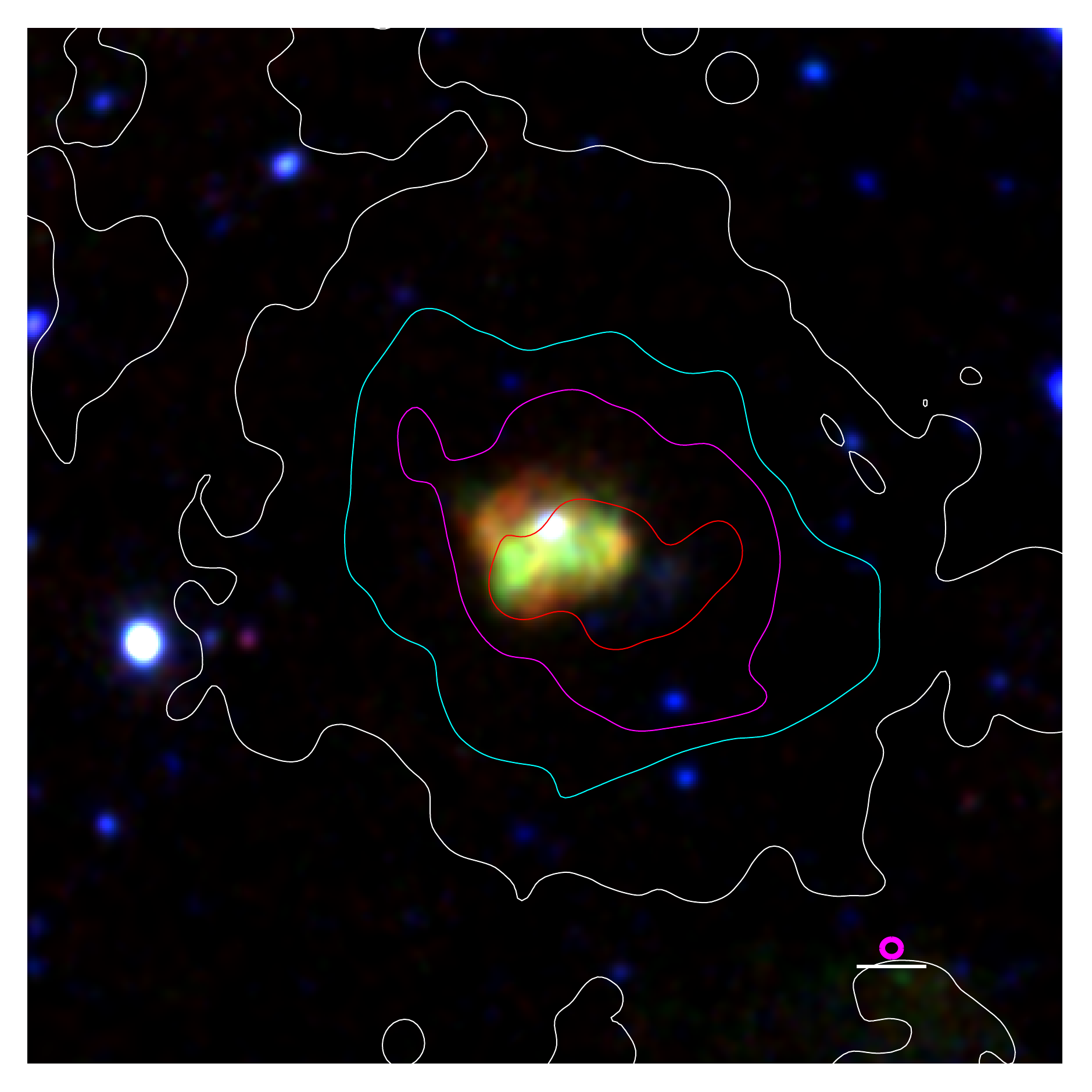}
    \includegraphics[height=0.225\vsize]{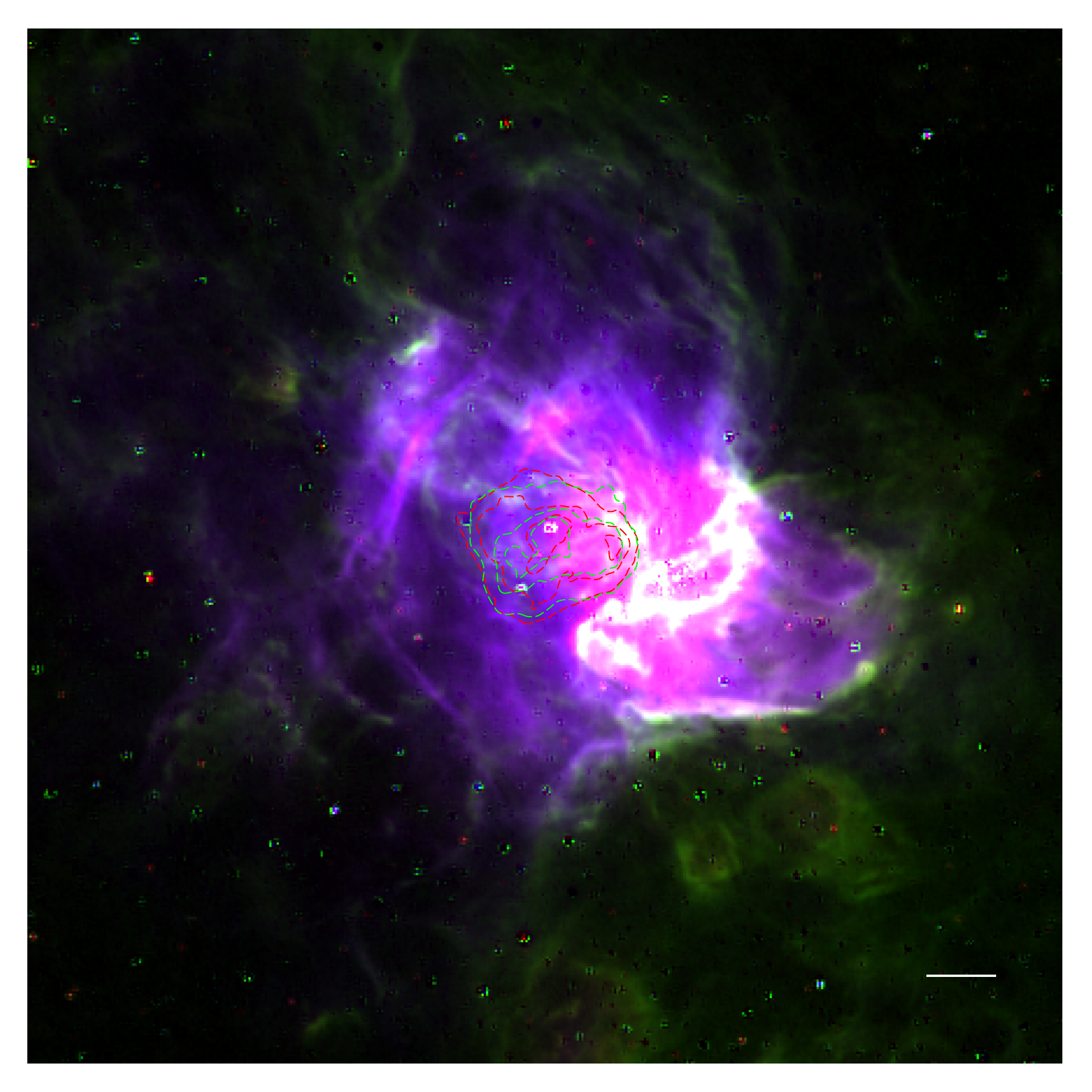}
    \includegraphics[height=0.225\vsize]{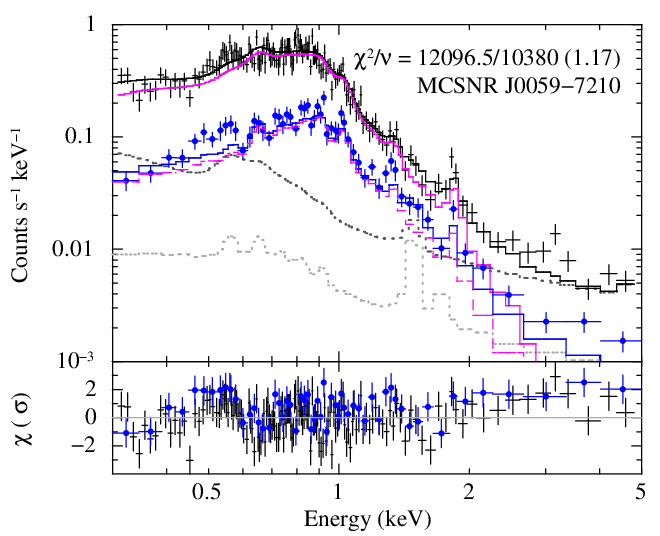}

    \caption{Same as Fig.\,\ref{fig_appendix_sfh0} for MCSNR~J0058$-$7217 (top part), with radio contour levels at 0.15, 0.6, and 1.5~mJy/beam, and for MCSNR~J0059$-$7210 with levels at 0.15, 0.8, 2, and 8~mJy/beam (bottom part).}
    
  \label{fig_appendix_sfh3}
\end{figure*}

\begin{figure*}[t]
    \centering
    \includegraphics[height=0.23\vsize]{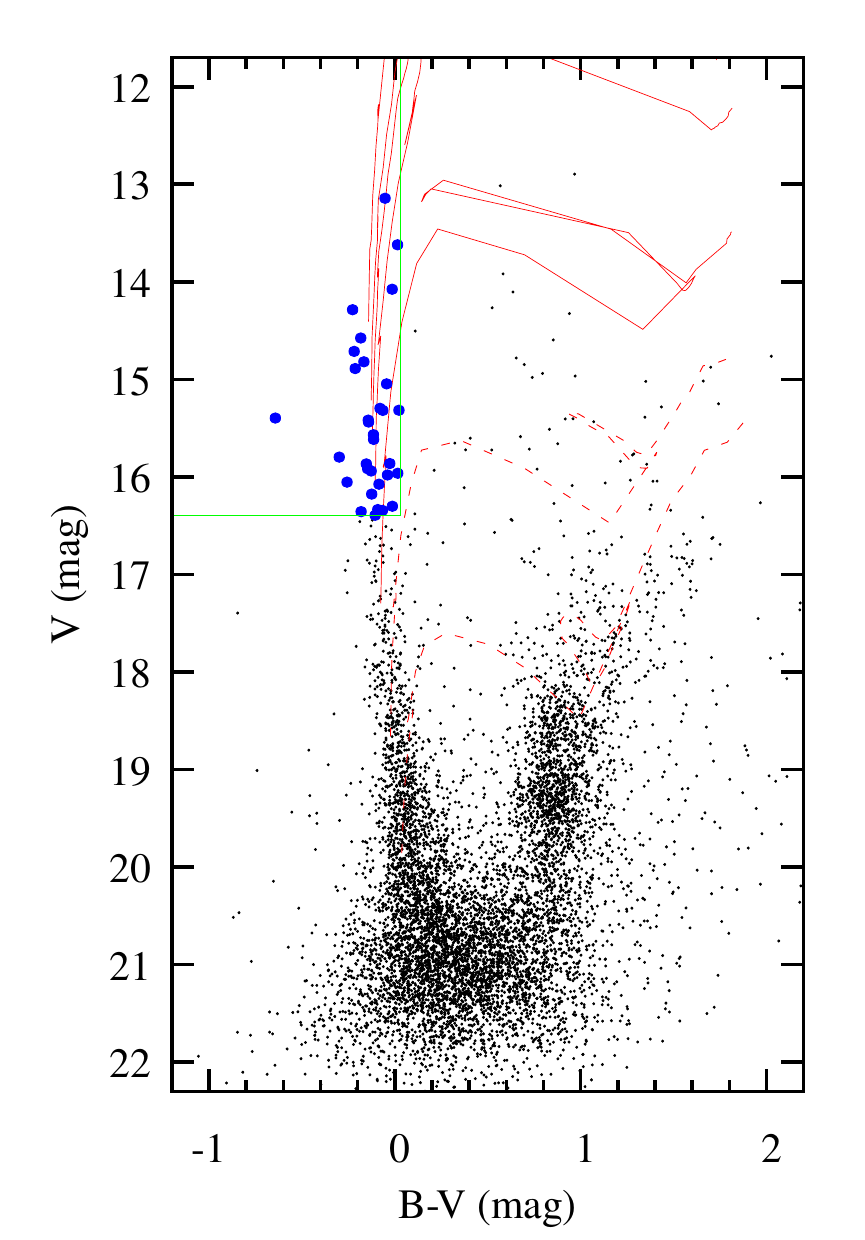}
    \includegraphics[height=0.23\vsize]{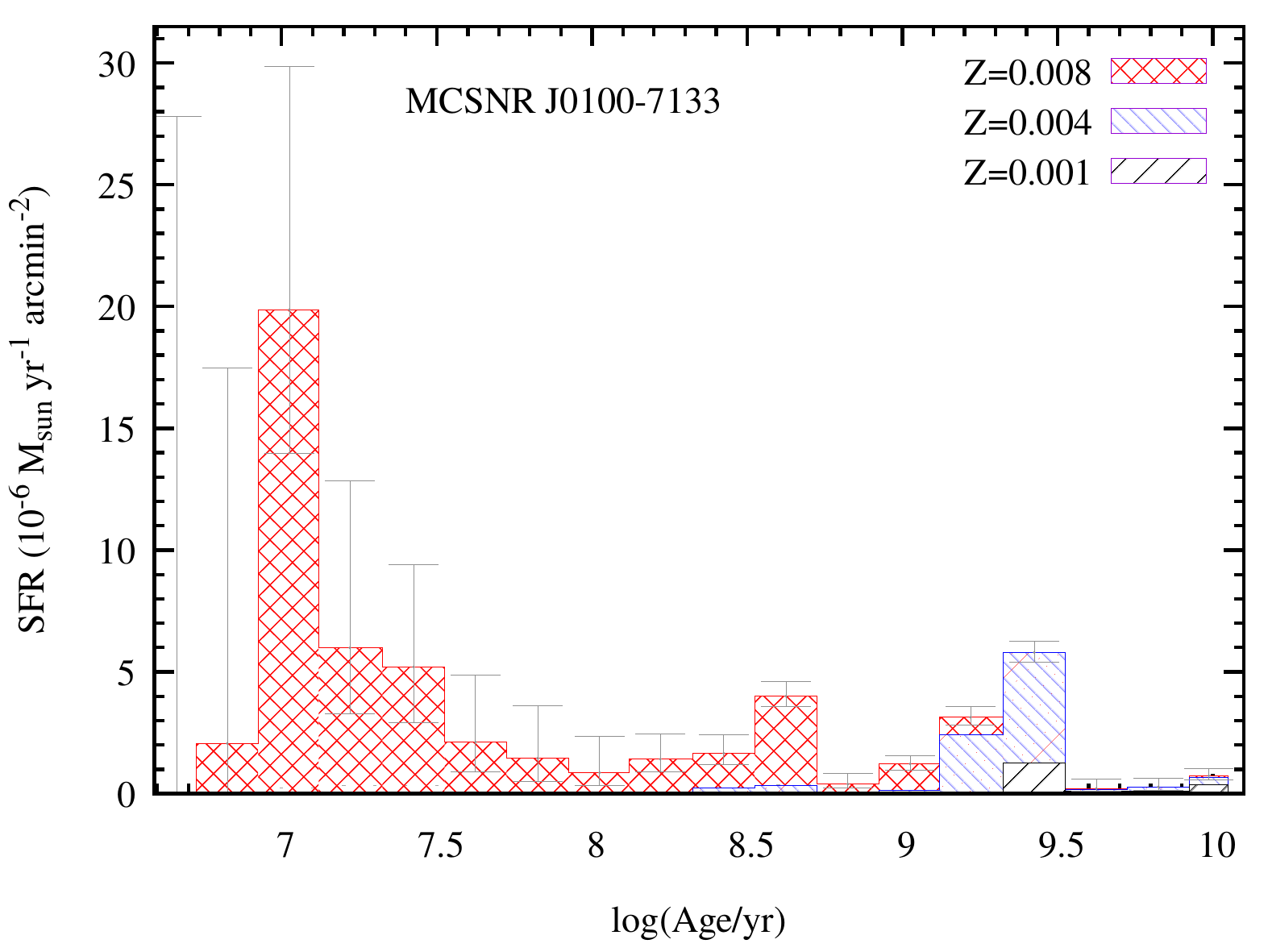}
    \includegraphics[height=0.23\vsize]{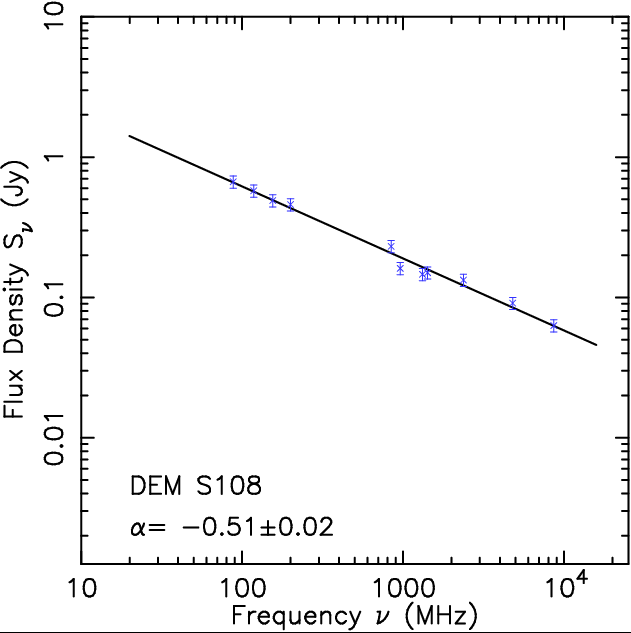}

    \includegraphics[height=0.225\vsize]{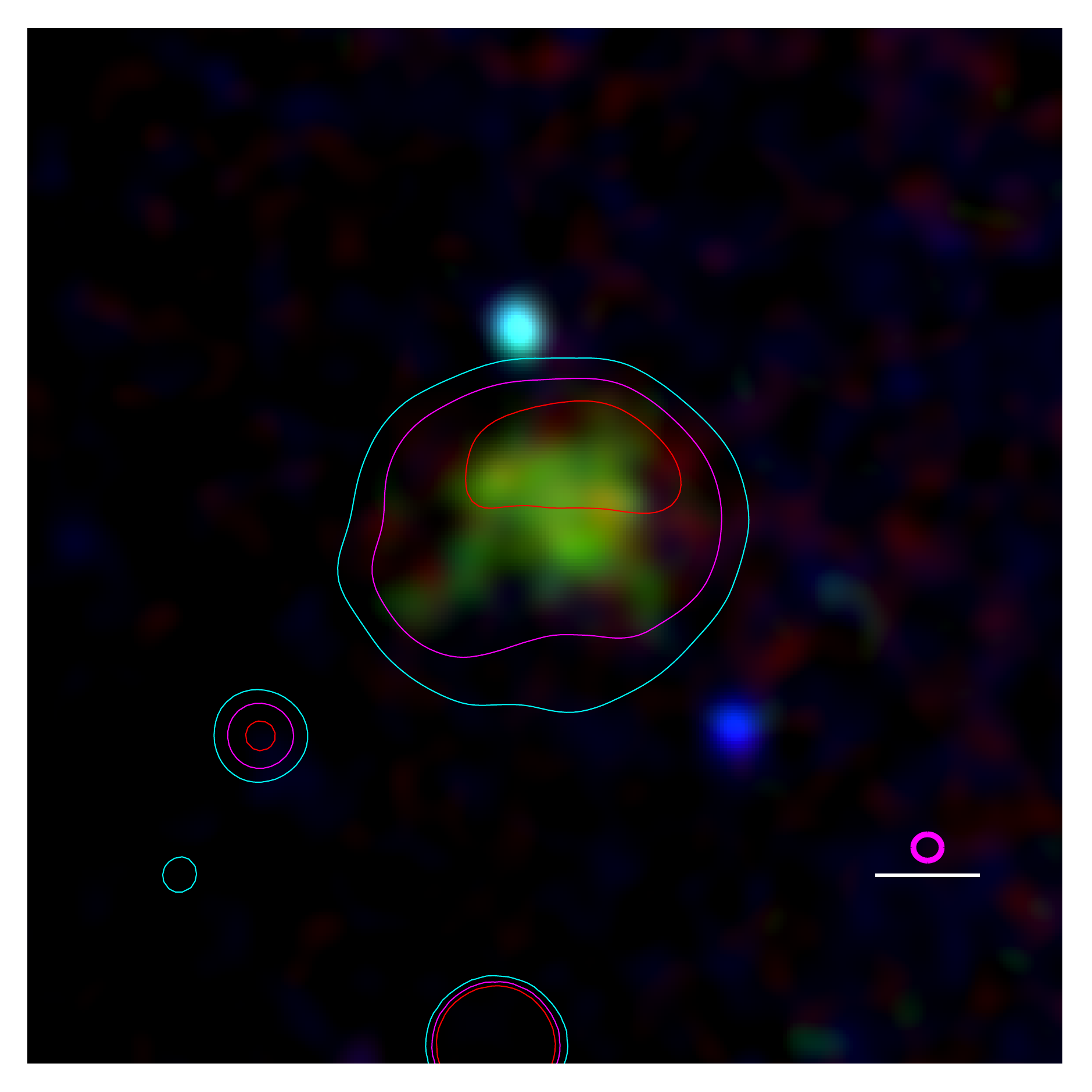}
    \includegraphics[height=0.225\vsize]{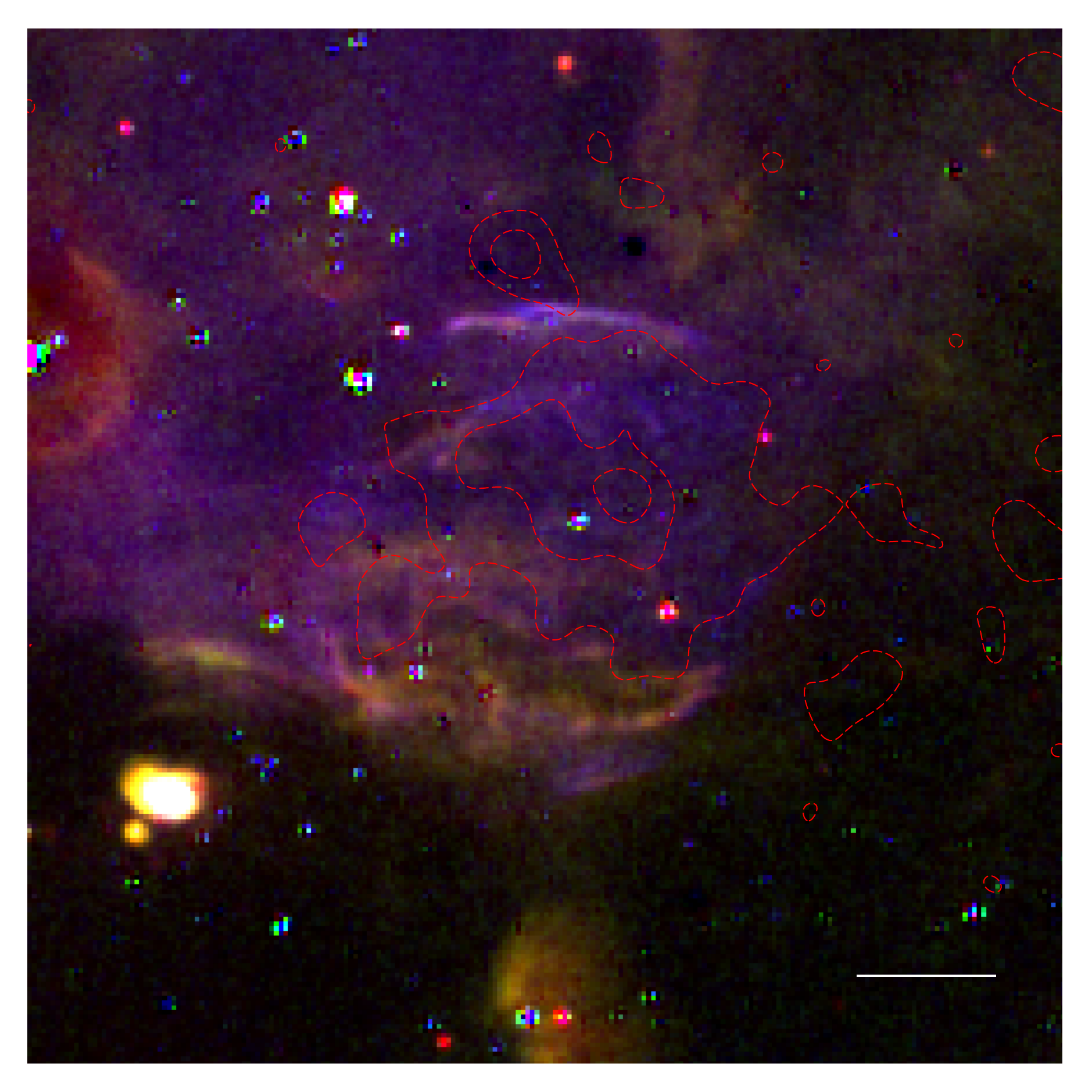}
    \includegraphics[height=0.225\vsize]{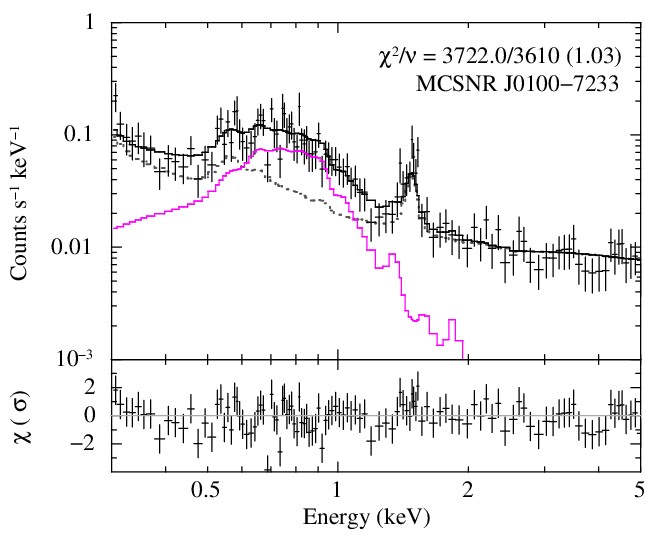}

    \vspace{1em}
    \hrule
    \vspace{1em}

    \includegraphics[height=0.23\vsize]{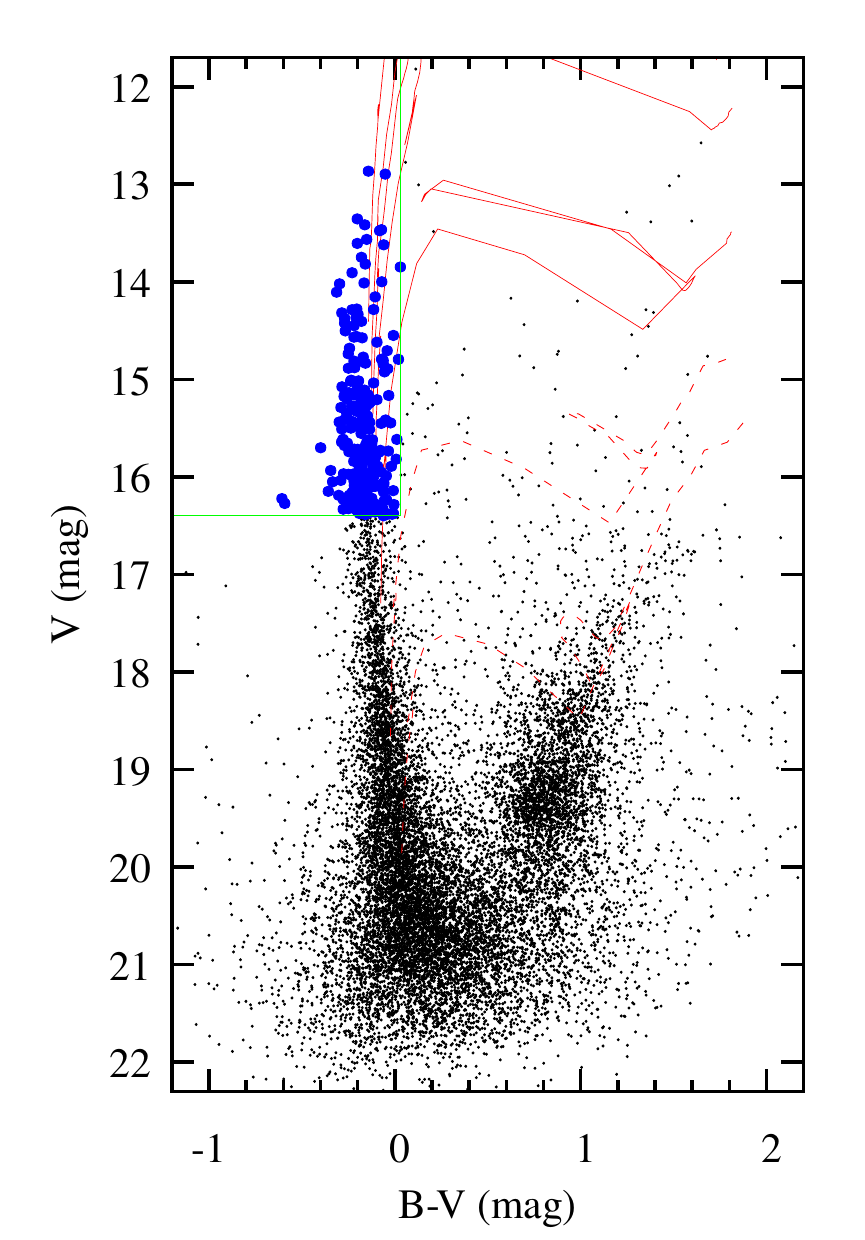}
    \includegraphics[height=0.23\vsize]{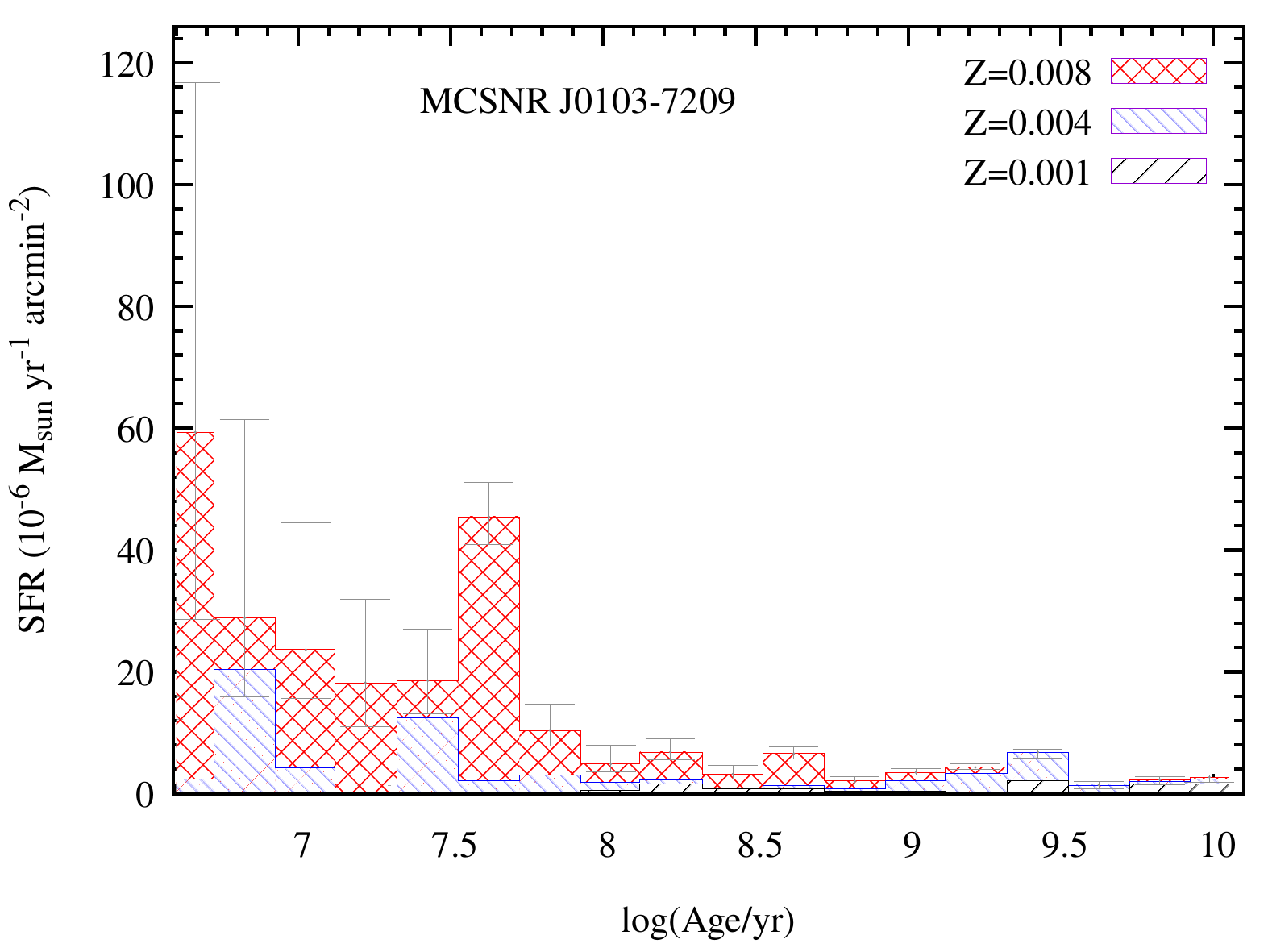}
    \includegraphics[height=0.23\vsize]{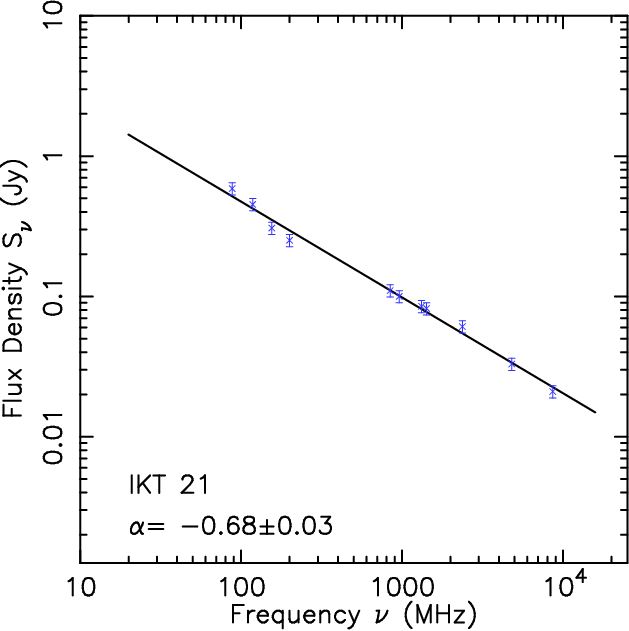}

    \includegraphics[height=0.225\vsize]{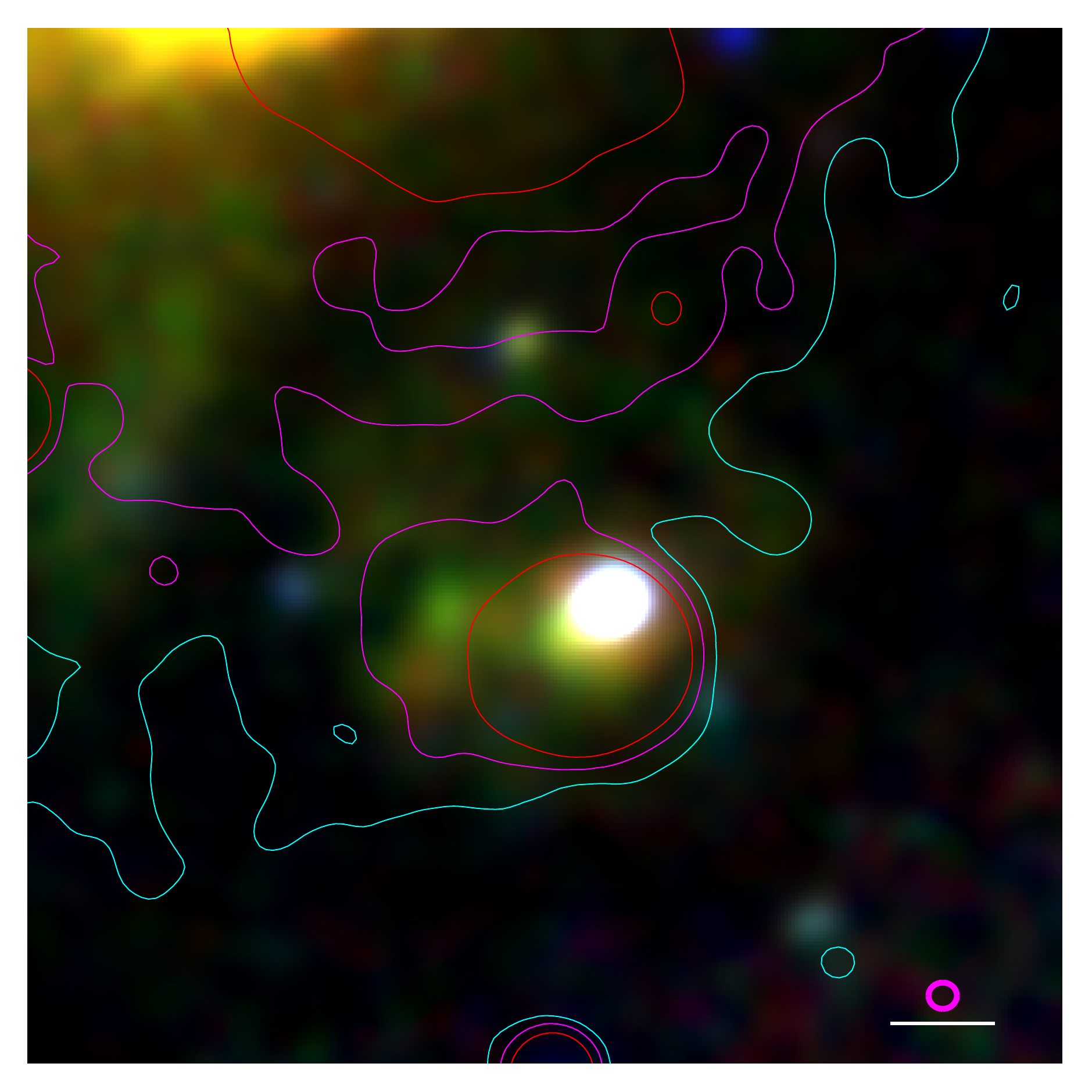}
    \includegraphics[height=0.225\vsize]{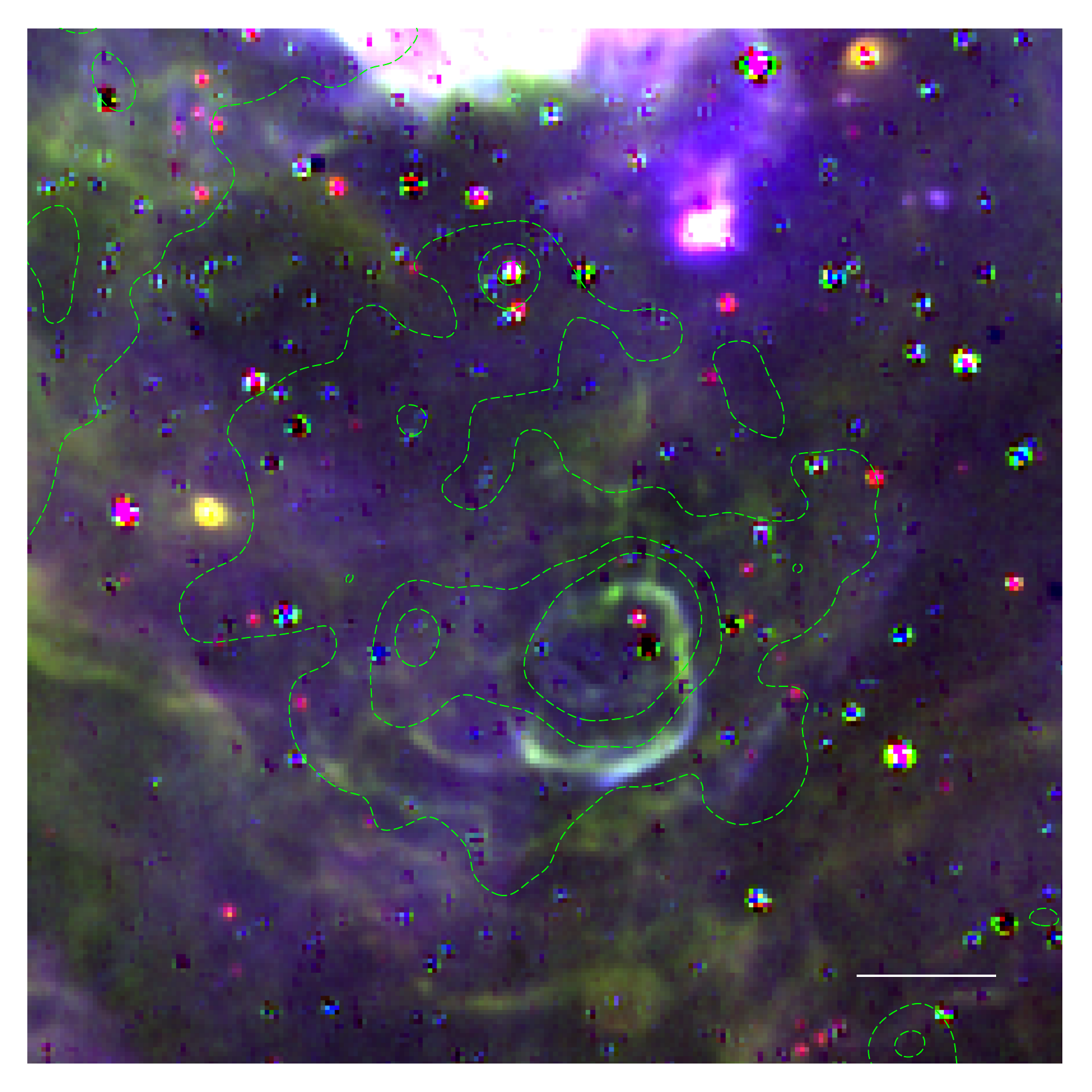}
    \includegraphics[height=0.225\vsize]{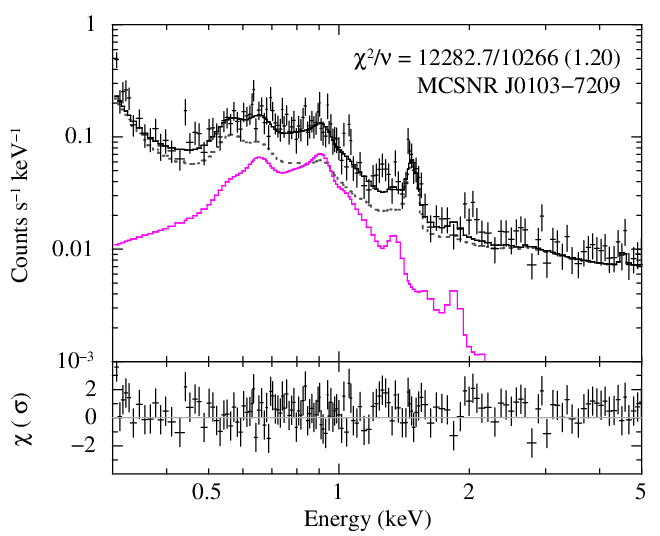}

    \caption{Same as Fig.\,\ref{fig_appendix_sfh0} for MCSNR~J0100$-$7133 (top part), with radio contour levels at 0.3, 0.8, and 1.5~mJy/beam, and for MCSNR~J0103$-$7209 with levels at 0.15, 0.3, and 0.6~mJy/beam (bottom part).}
  \label{fig_appendix_sfh4}
\end{figure*}

\begin{figure*}[t]
    \centering
    \includegraphics[height=0.23\vsize]{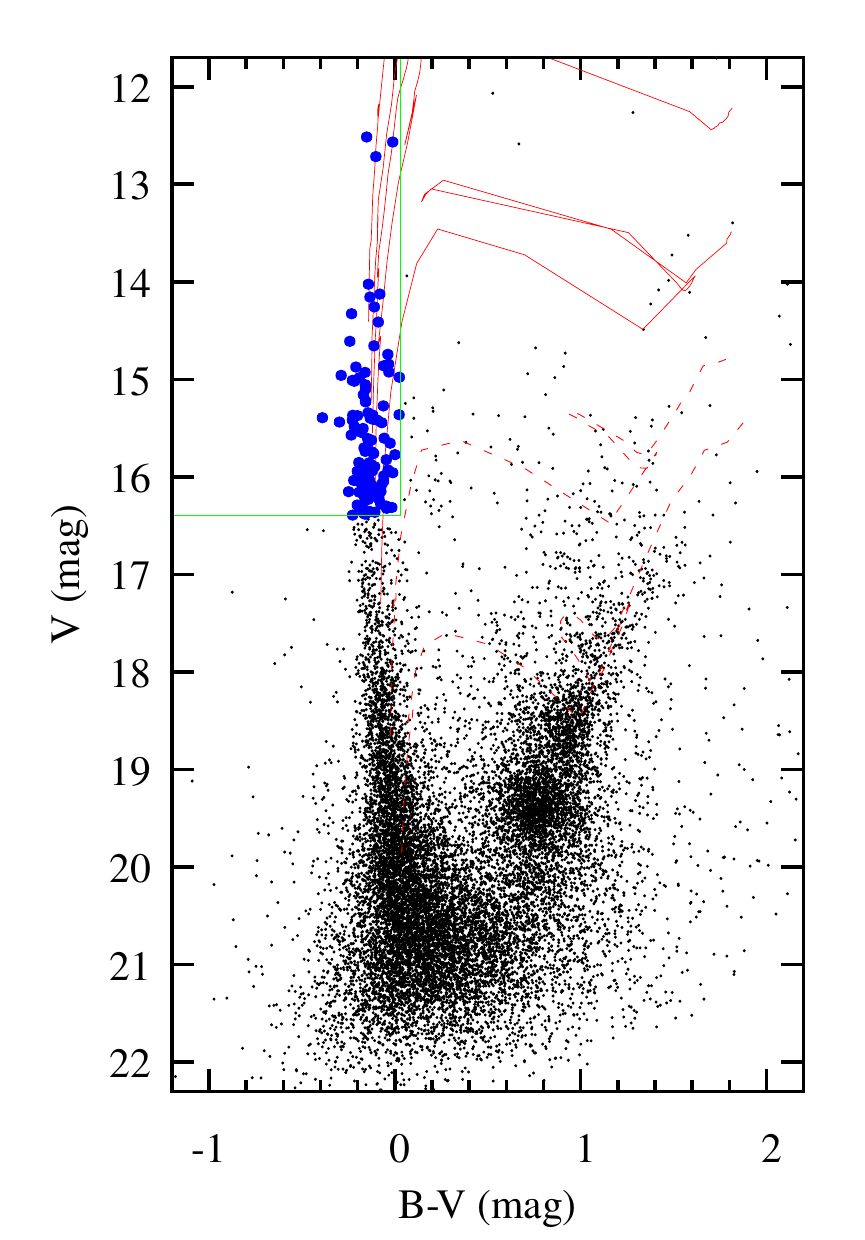}
    \includegraphics[height=0.23\vsize]{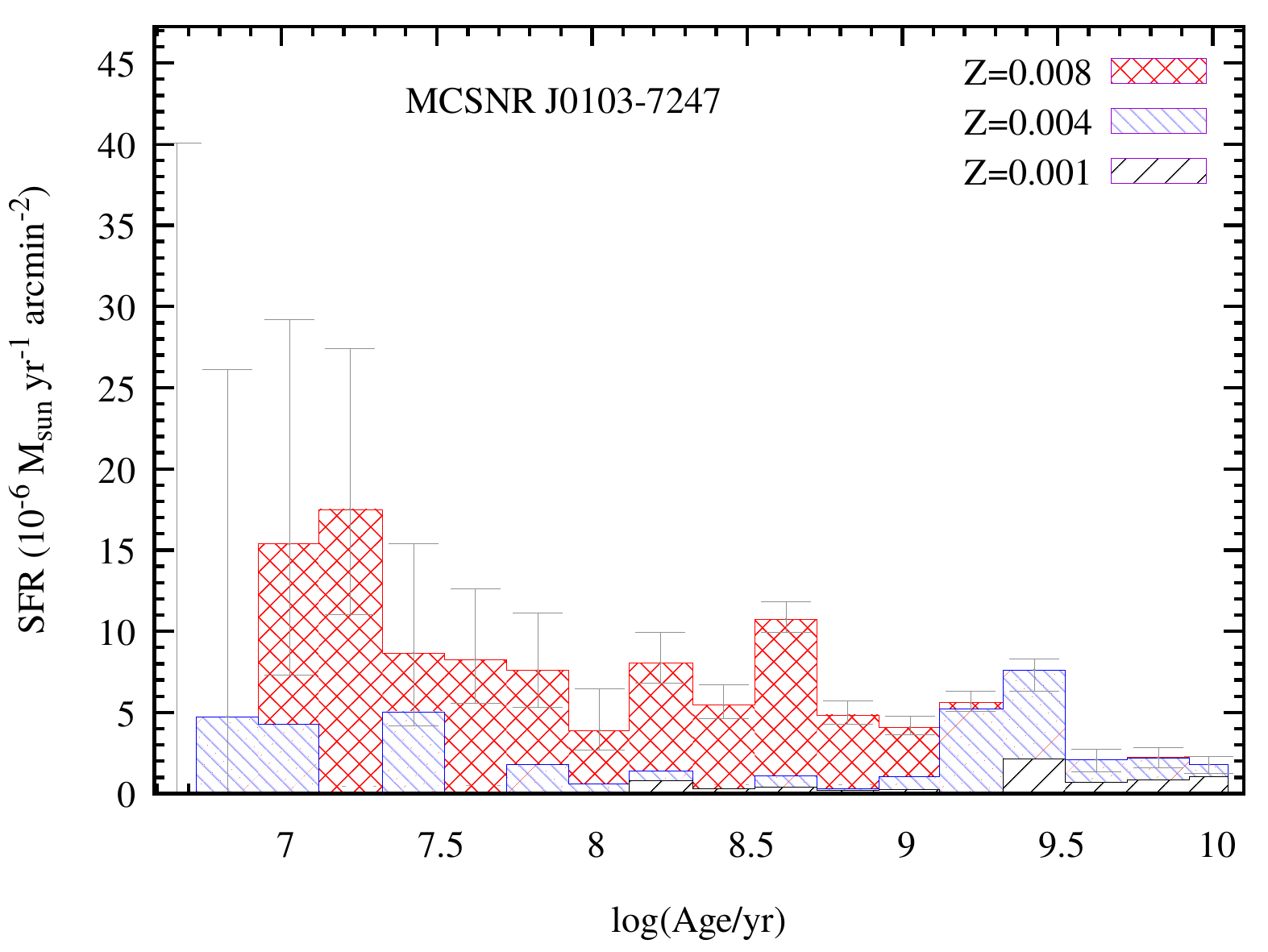}
    \includegraphics[height=0.23\vsize]{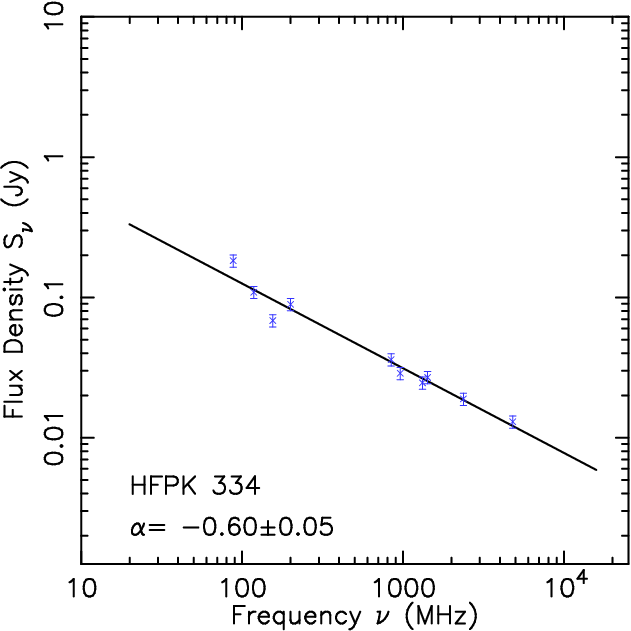}

    \includegraphics[height=0.225\vsize]{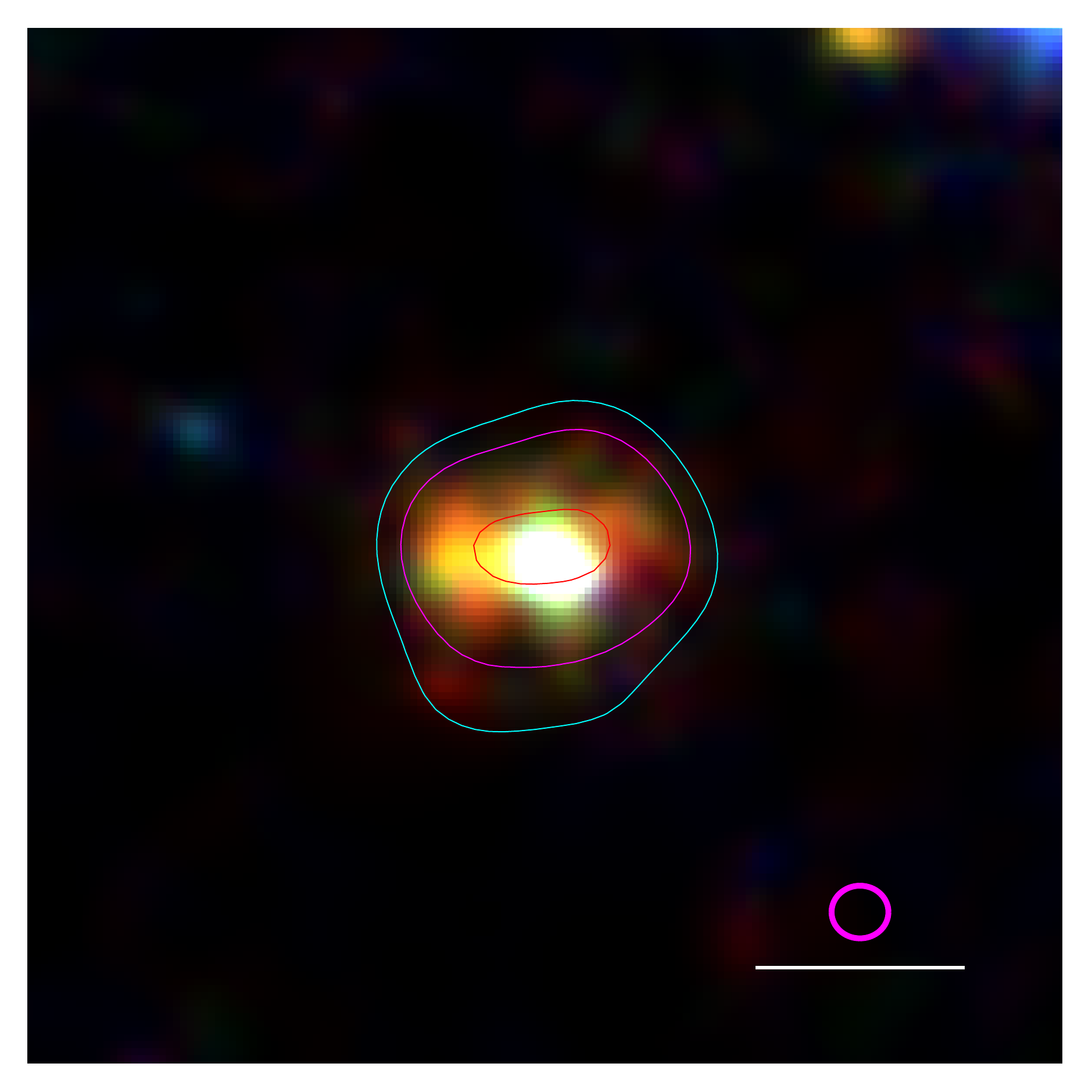}
    \includegraphics[height=0.225\vsize]{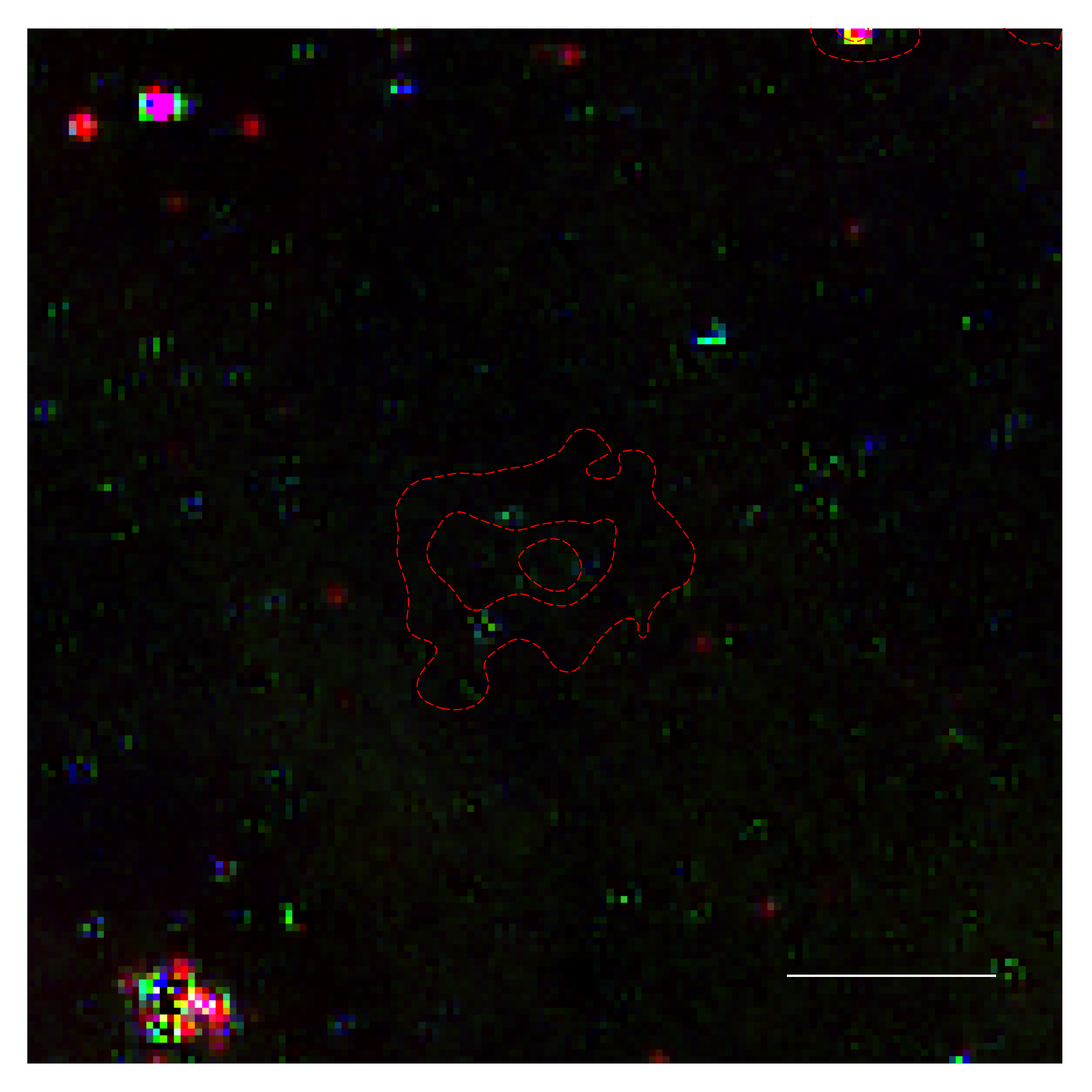}
    \includegraphics[height=0.225\vsize]{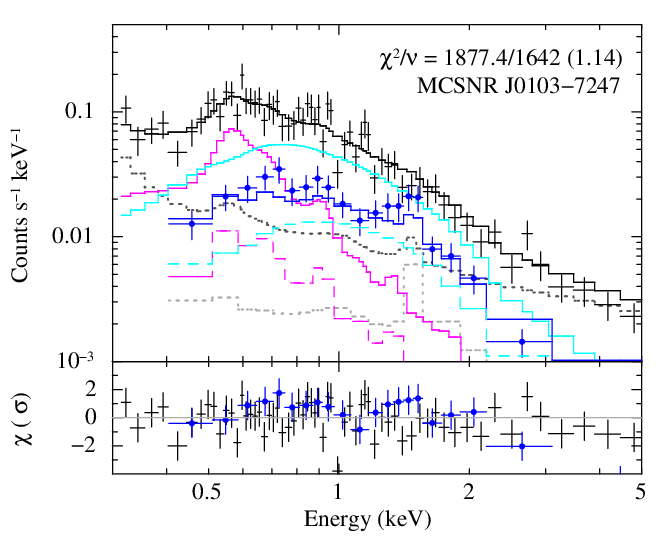}

    \vspace{1em}
    \hrule
    \vspace{1em}

    \includegraphics[height=0.18\vsize]{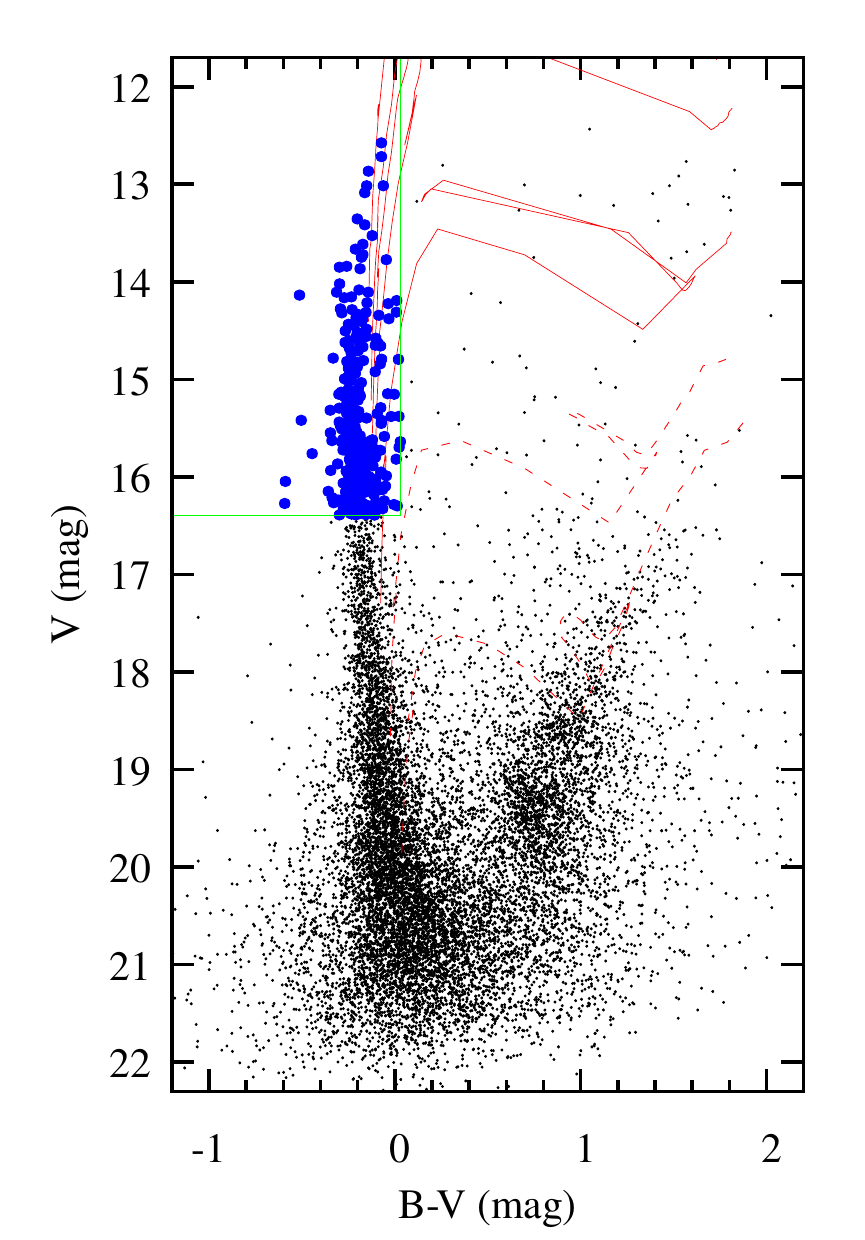}
    \includegraphics[height=0.18\vsize]{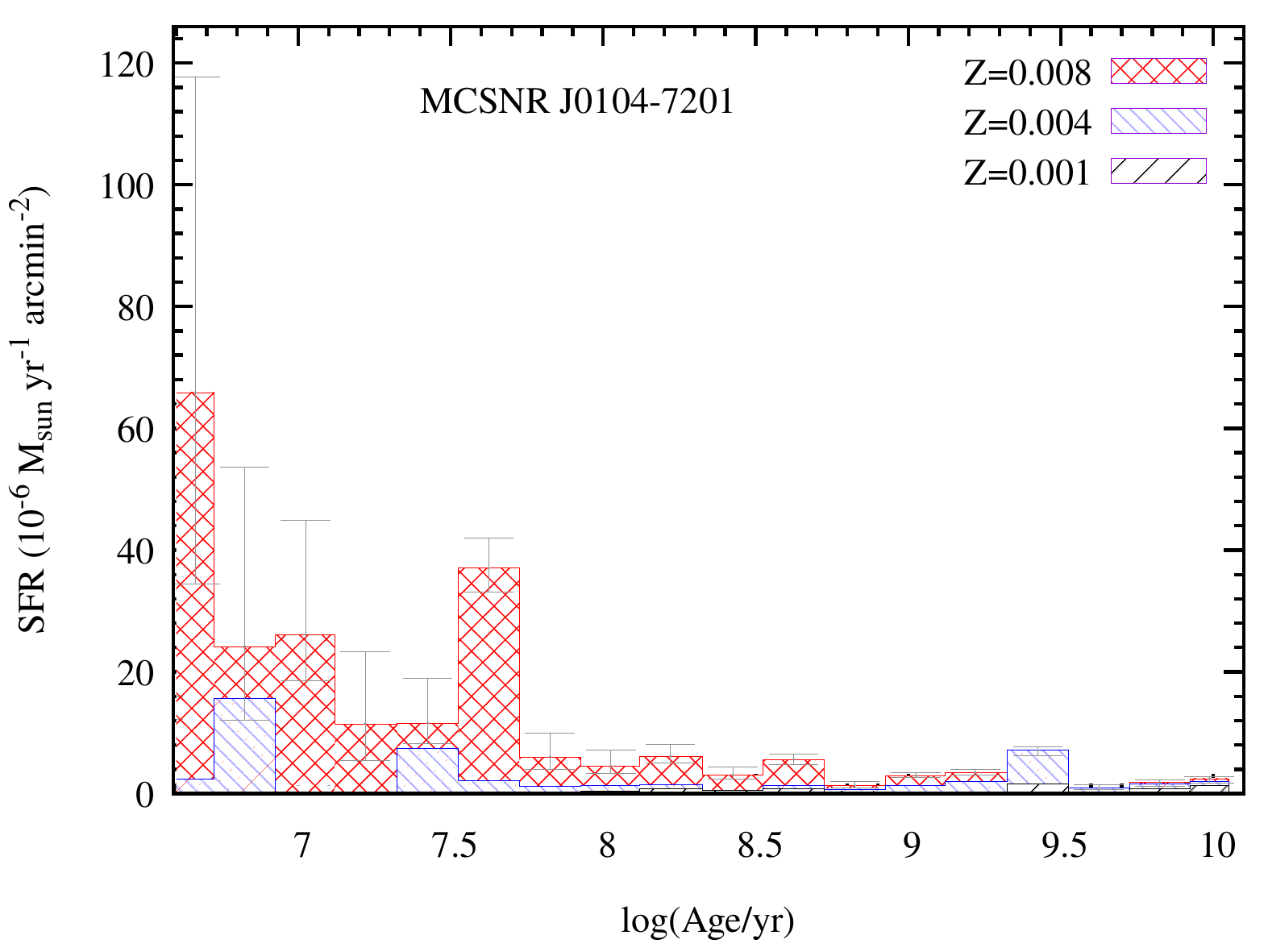}
    \includegraphics[height=0.18\vsize]{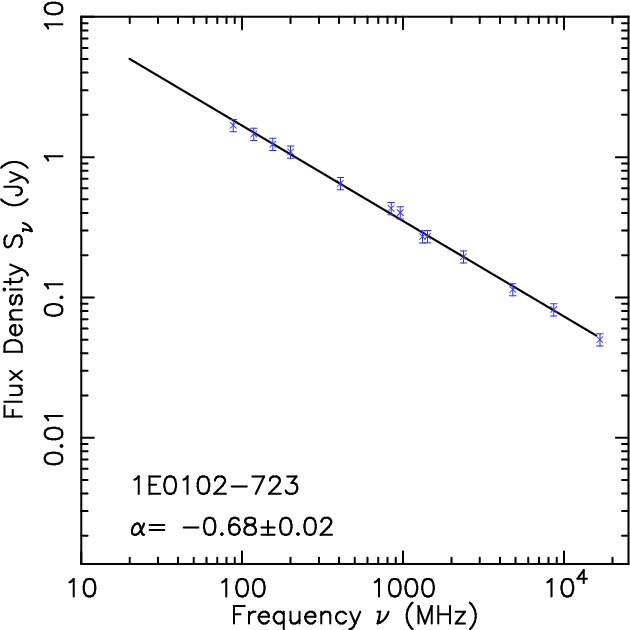}
    \includegraphics[height=0.18\vsize]{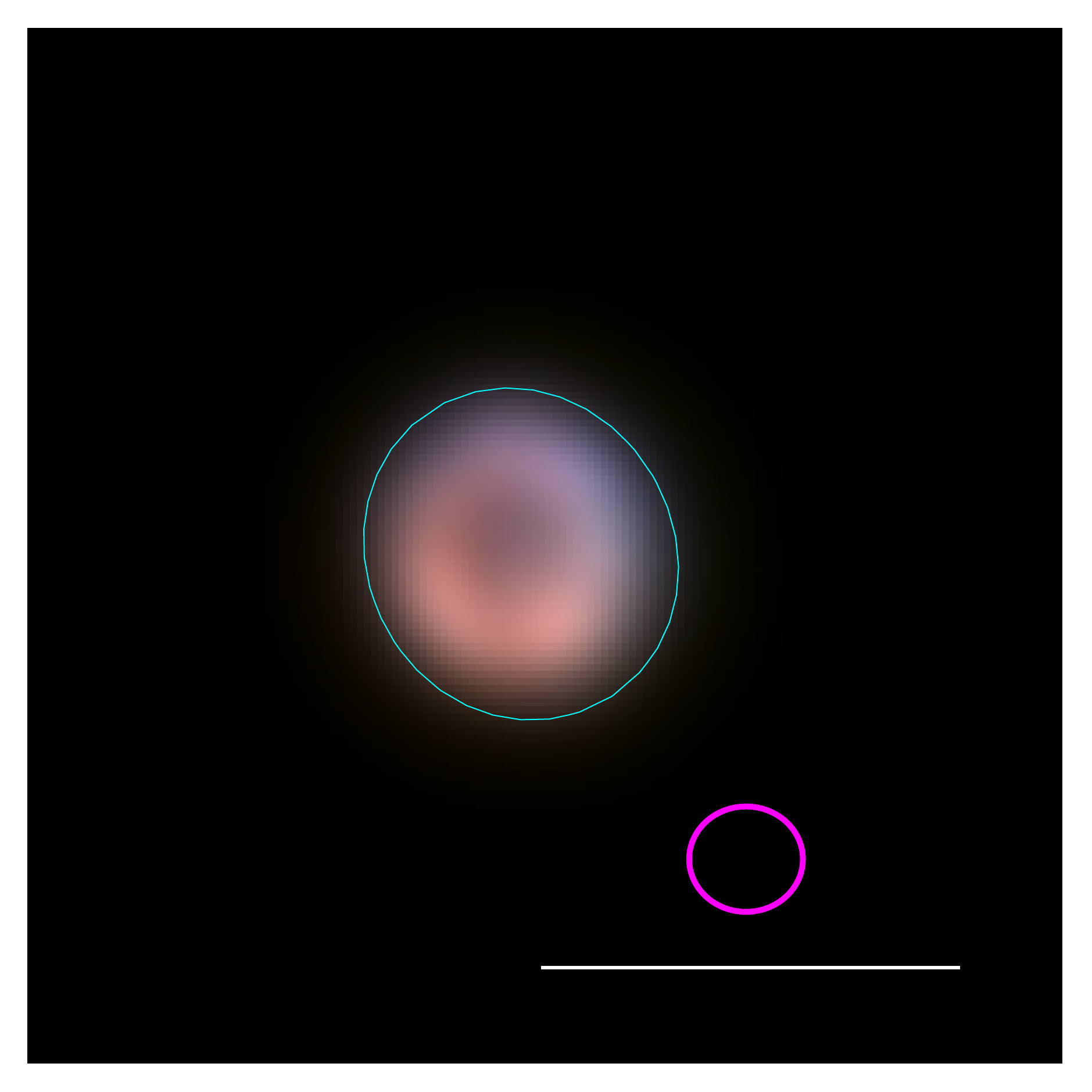}

    \includegraphics[height=0.225\vsize]{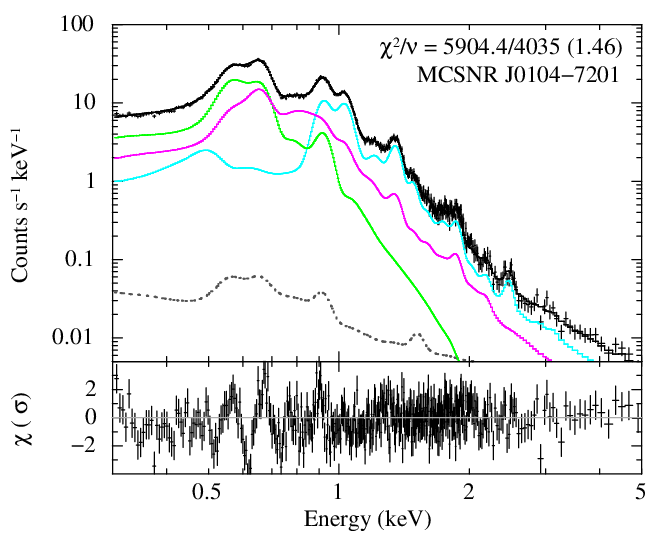}    
    \includegraphics[height=0.225\vsize]{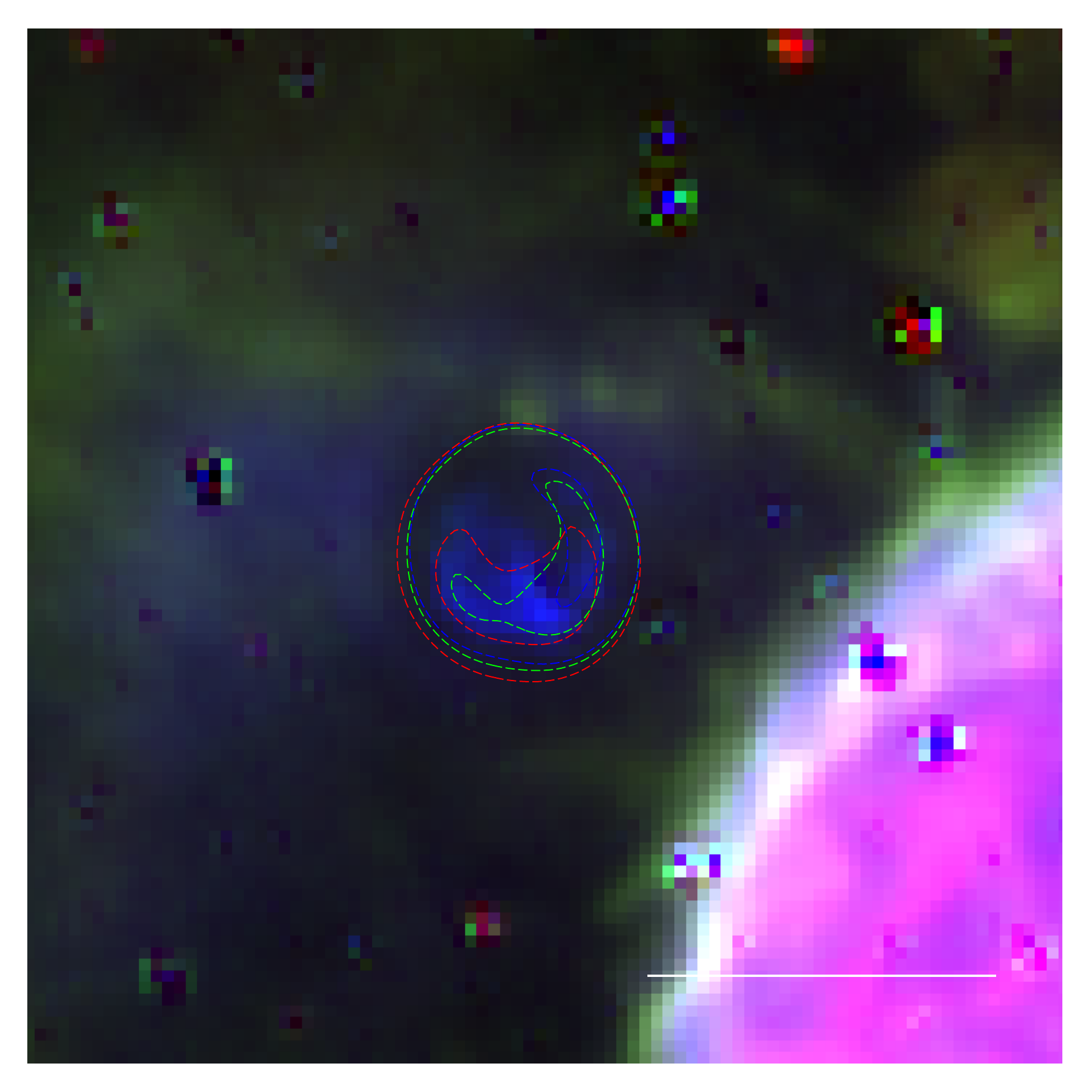}
    \includegraphics[height=0.225\vsize]{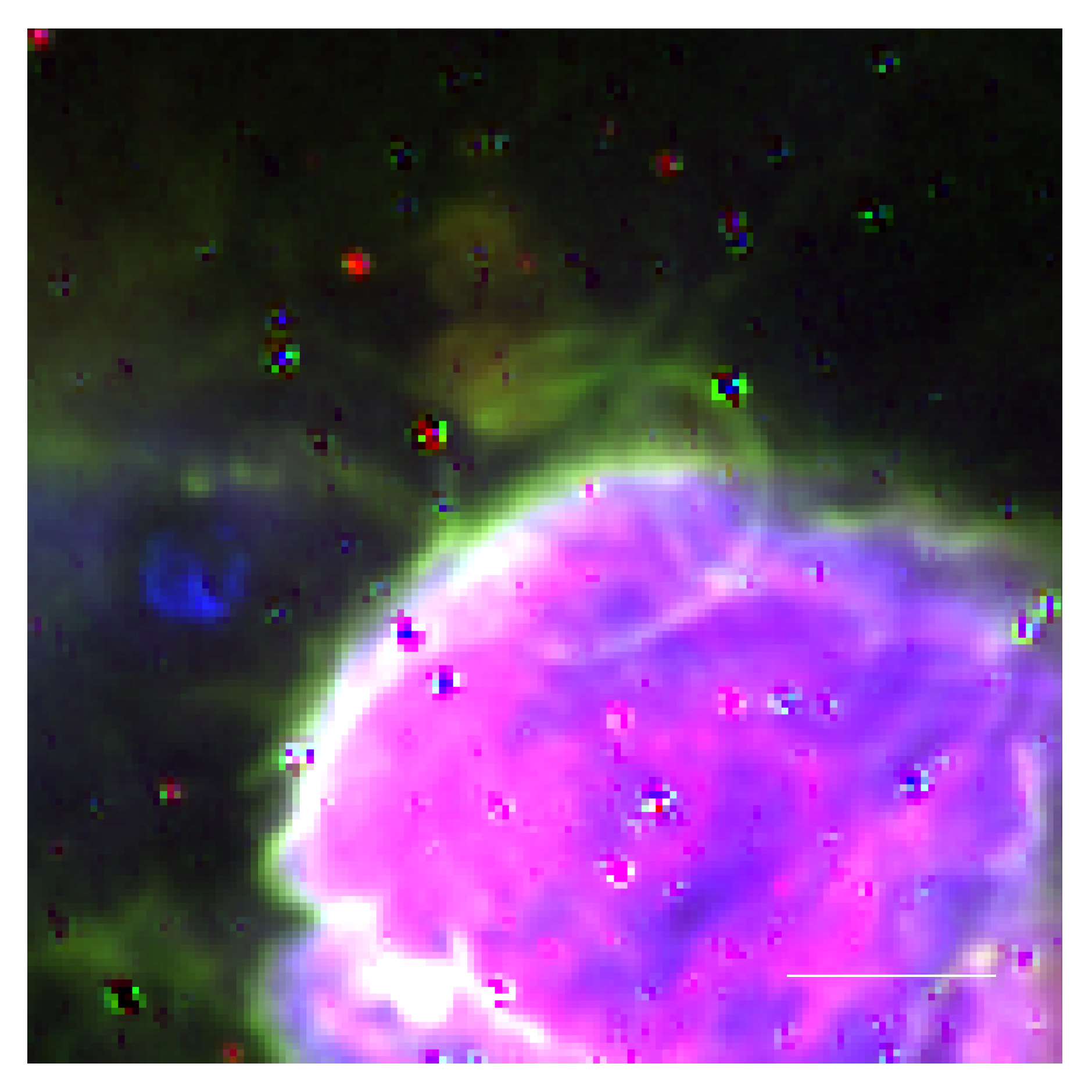}
    
    \caption{Same as Fig.\,\ref{fig_appendix_sfh0} for MCSNR~J0103$-$7247 (top part), with radio contour levels at 0.3, 0.6, and 1.5~mJy/beam, and for MCSNR~J0104$-$7201 with one level at 20~mJy/beam (bottom part). At the bottom right we show the MCELS image of MCSNR~J0103$-$7201, identified by its H$\alpha$ shell. Its CMD and SFH are essentially the same as that of the neighbouring J0104$-$7201. }

  \label{fig_appendix_sfh5}
\end{figure*}

\begin{figure*}[t]
  \centering

    \includegraphics[height=0.23\vsize]{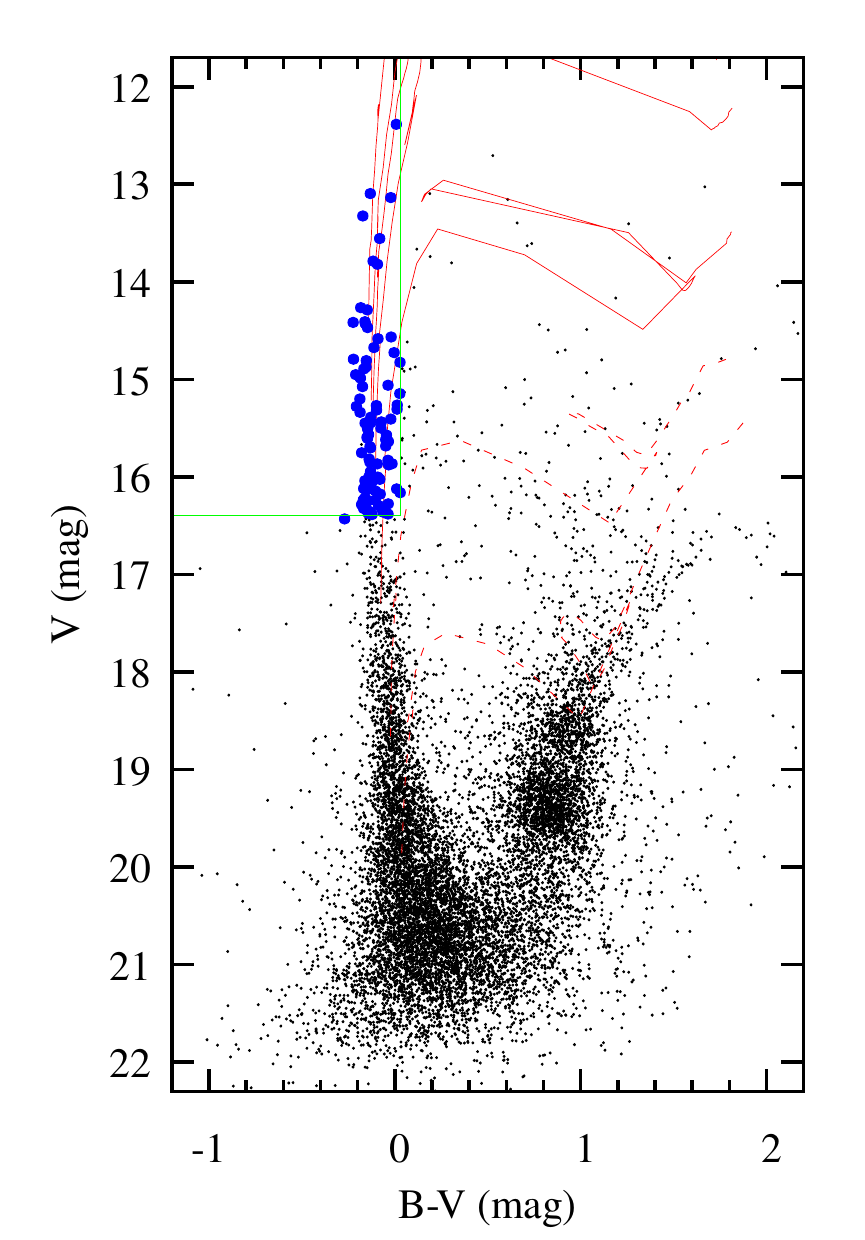}
    \includegraphics[height=0.23\vsize]{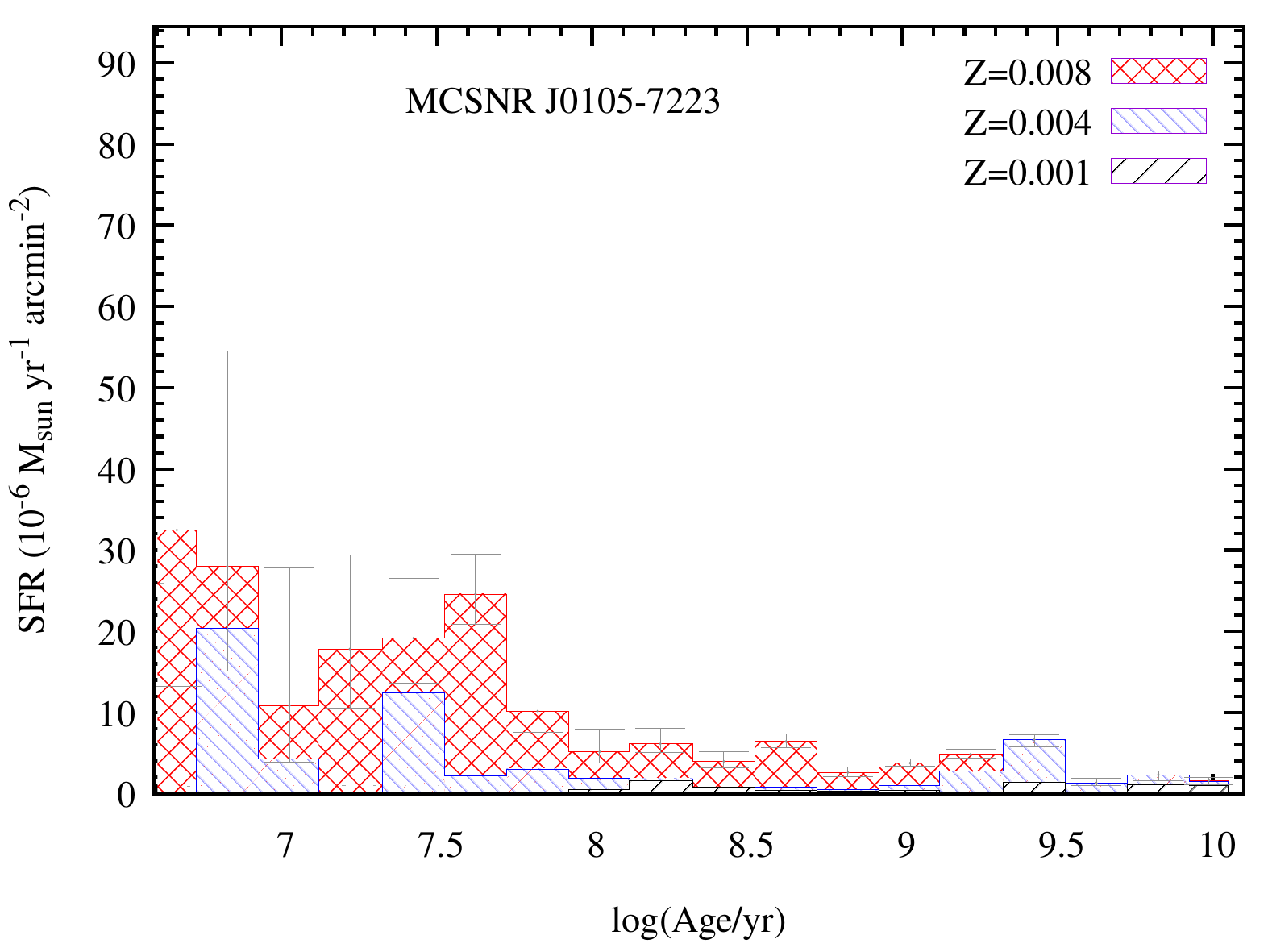}
    \includegraphics[height=0.23\vsize]{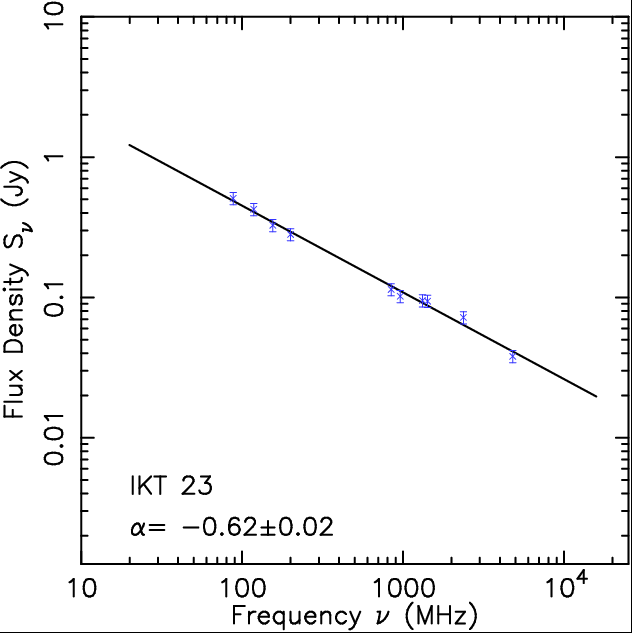}

    \includegraphics[height=0.225\vsize]{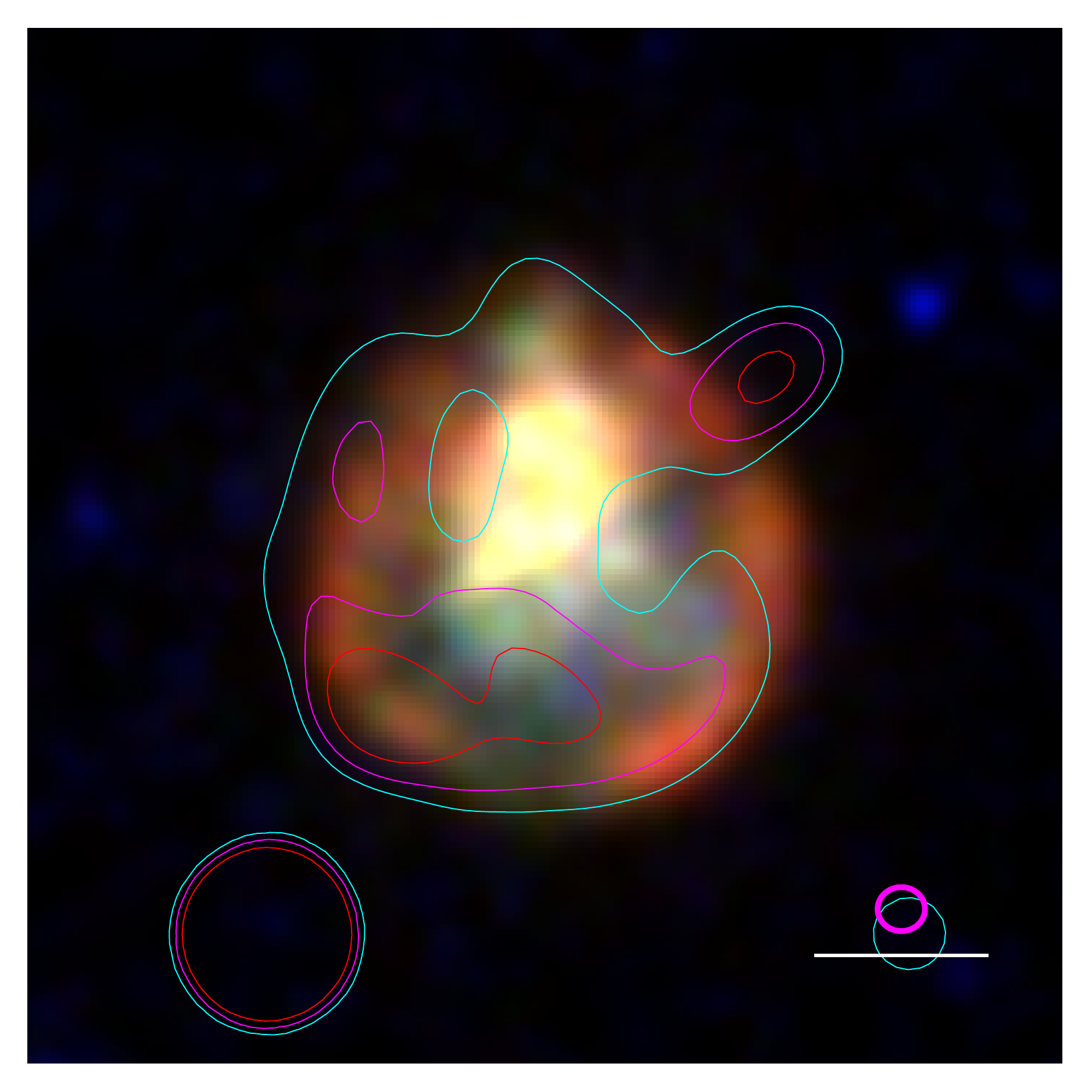}
    \includegraphics[height=0.225\vsize]{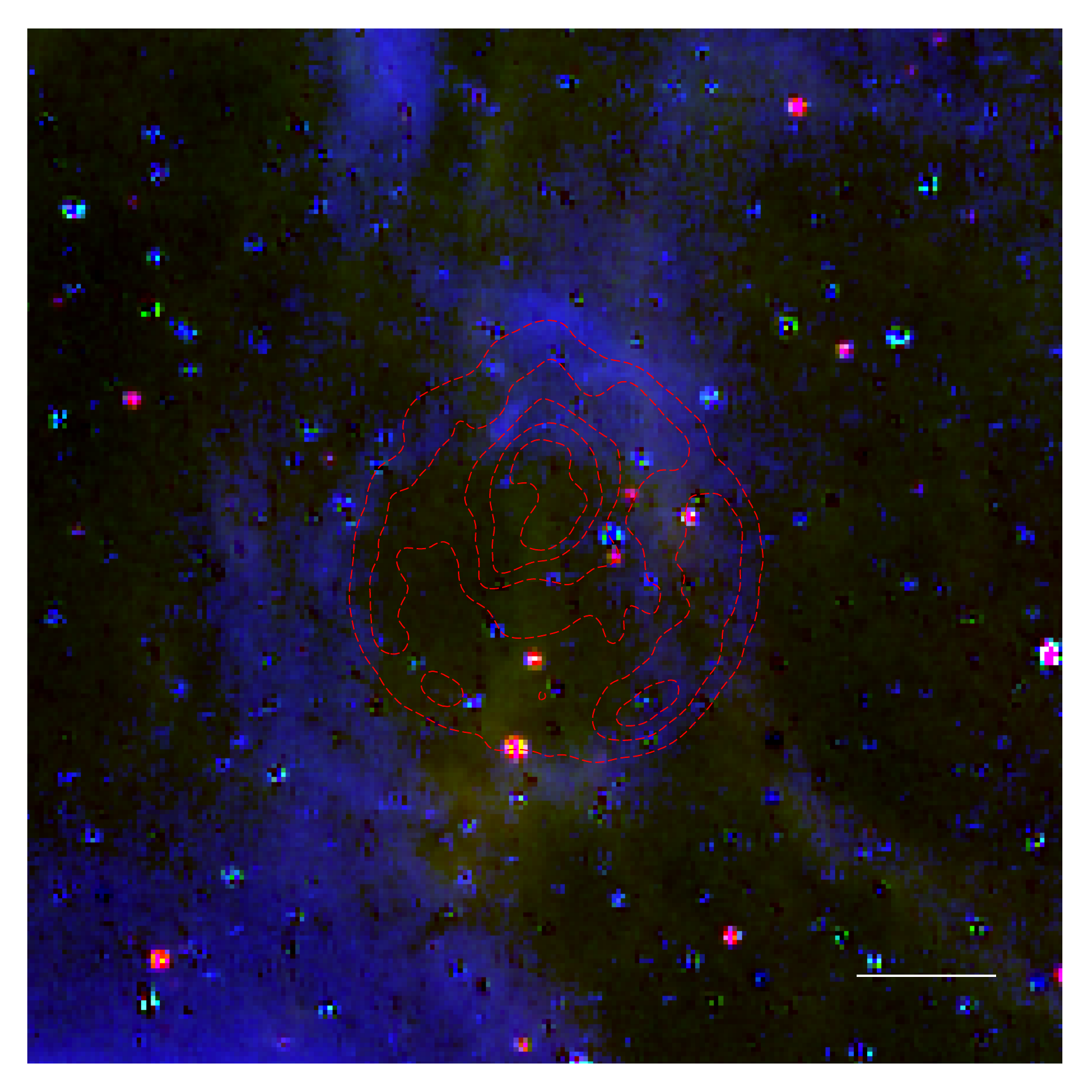}
    \includegraphics[height=0.225\vsize]{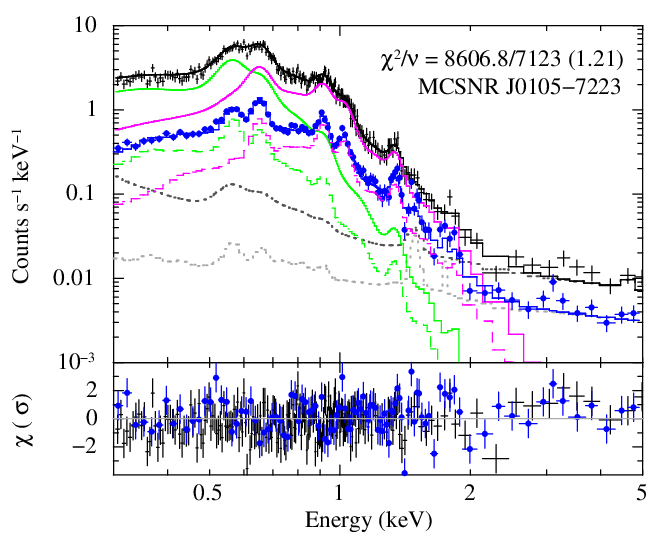}

    \vspace{1em}
    \hrule
    \vspace{1em}

    \includegraphics[height=0.23\vsize]{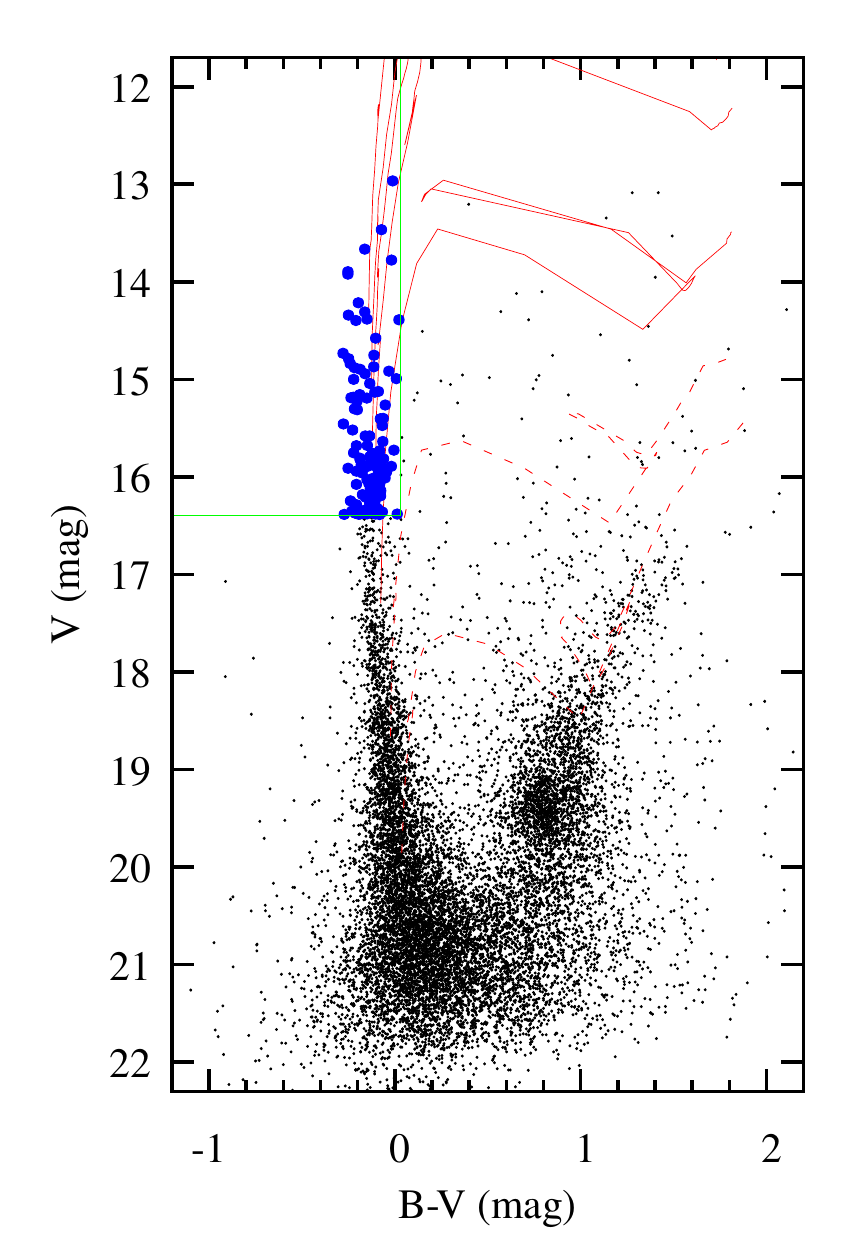}
    \includegraphics[height=0.23\vsize]{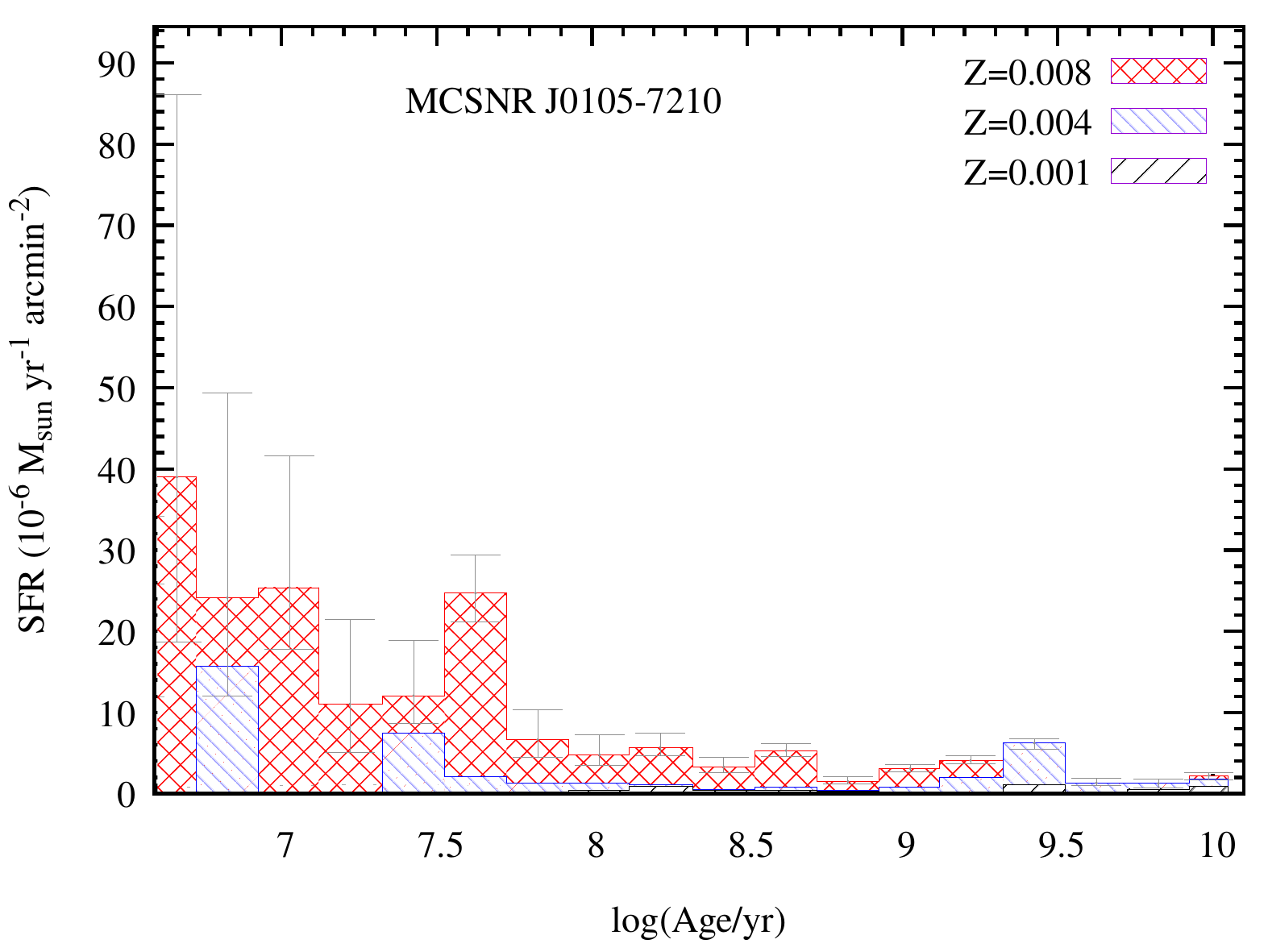}
    \includegraphics[height=0.23\vsize]{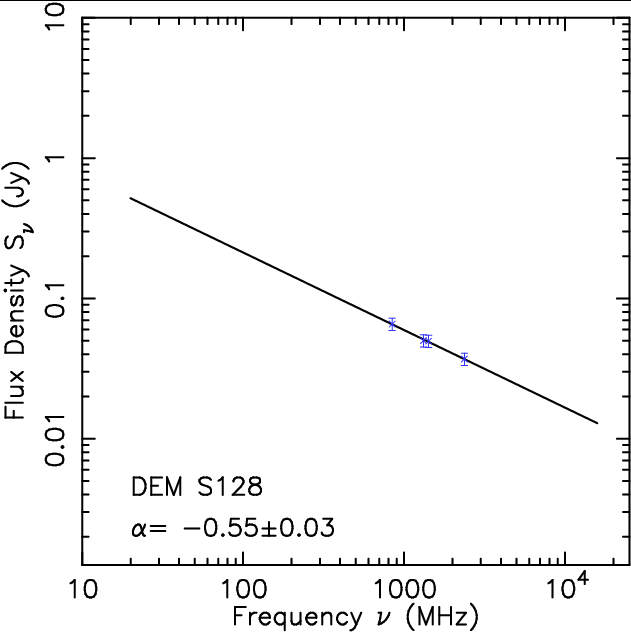}

    \includegraphics[height=0.225\vsize]{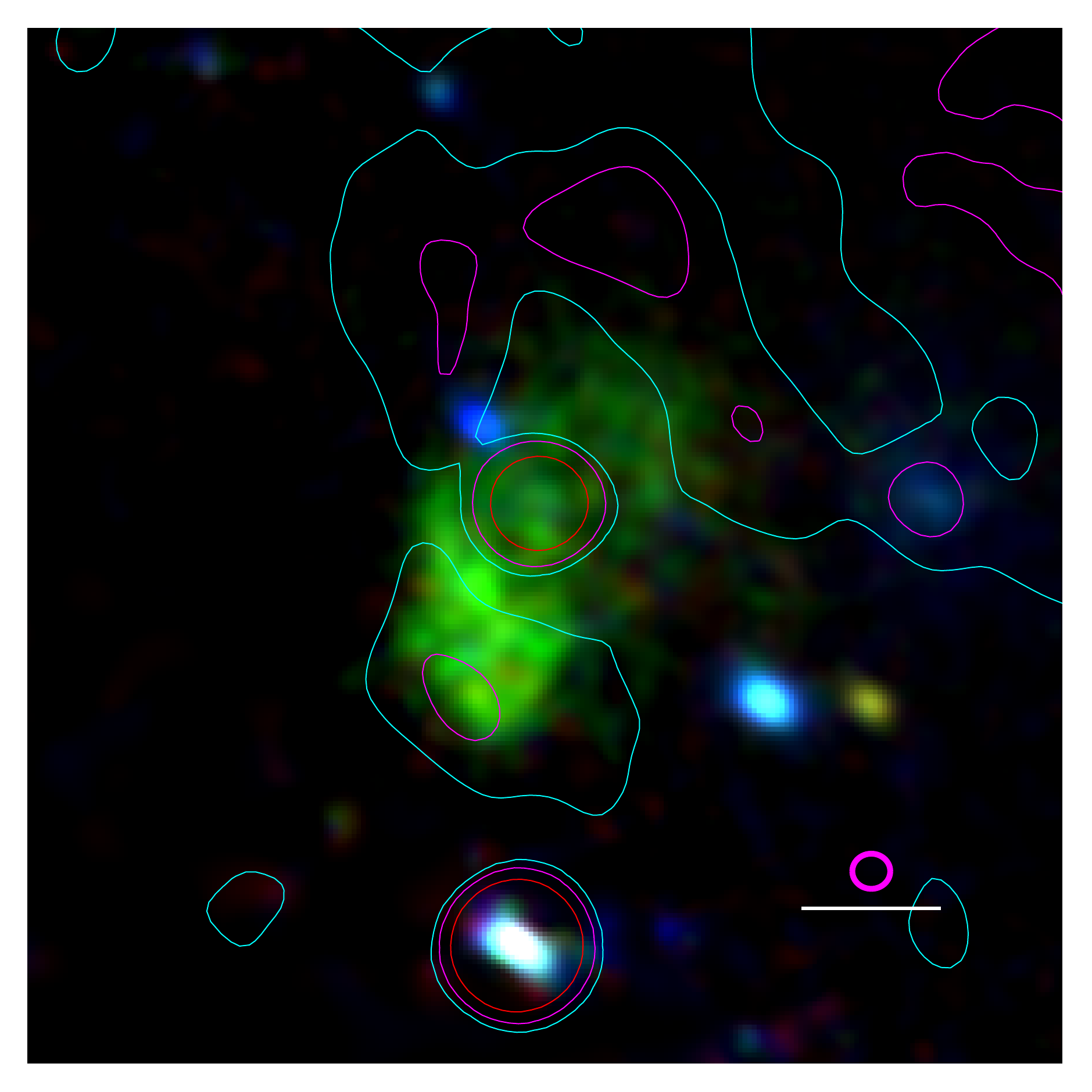}
    \includegraphics[height=0.225\vsize]{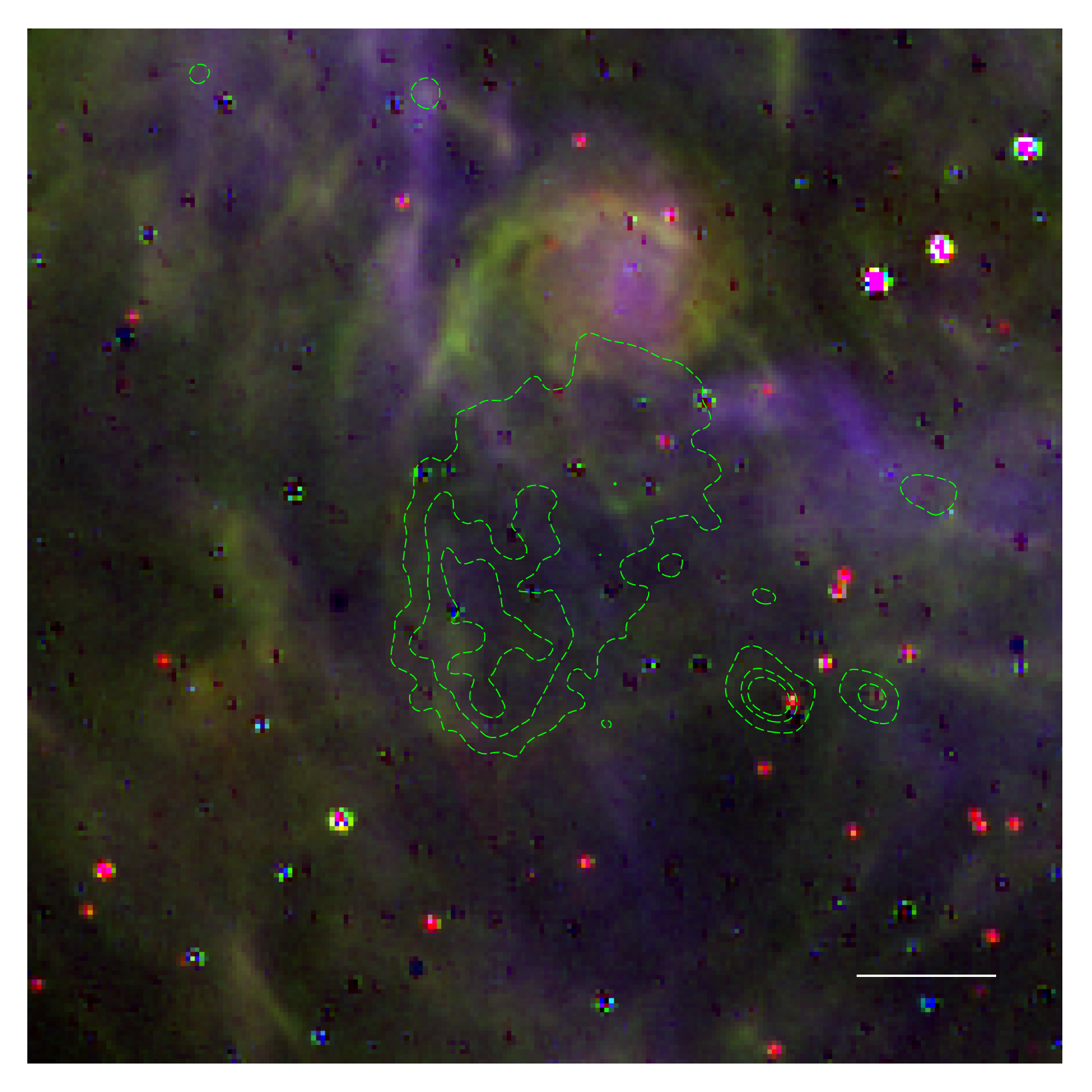}
    \includegraphics[height=0.225\vsize]{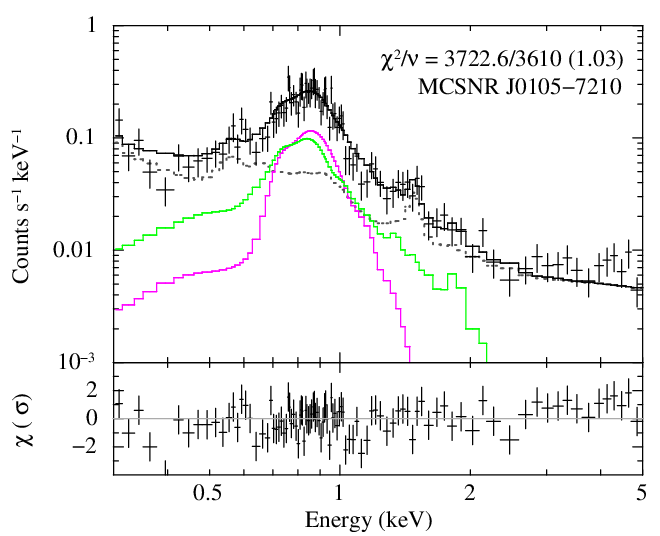}

    \caption{Same as Fig.\,\ref{fig_appendix_sfh0} for MCSNR~J0105$-$7223 (top part), with radio contour levels at 0.3, 0.5, and 0.8~mJy/beam, and for MCSNR~J0105$-$7210 with levels at  0.1, 0.2, and 0.5~mJy/beam (bottom part).}
  \label{fig_appendix_sfh6}
\end{figure*}

\begin{figure*}[t]
    \centering

    \includegraphics[height=0.23\vsize]{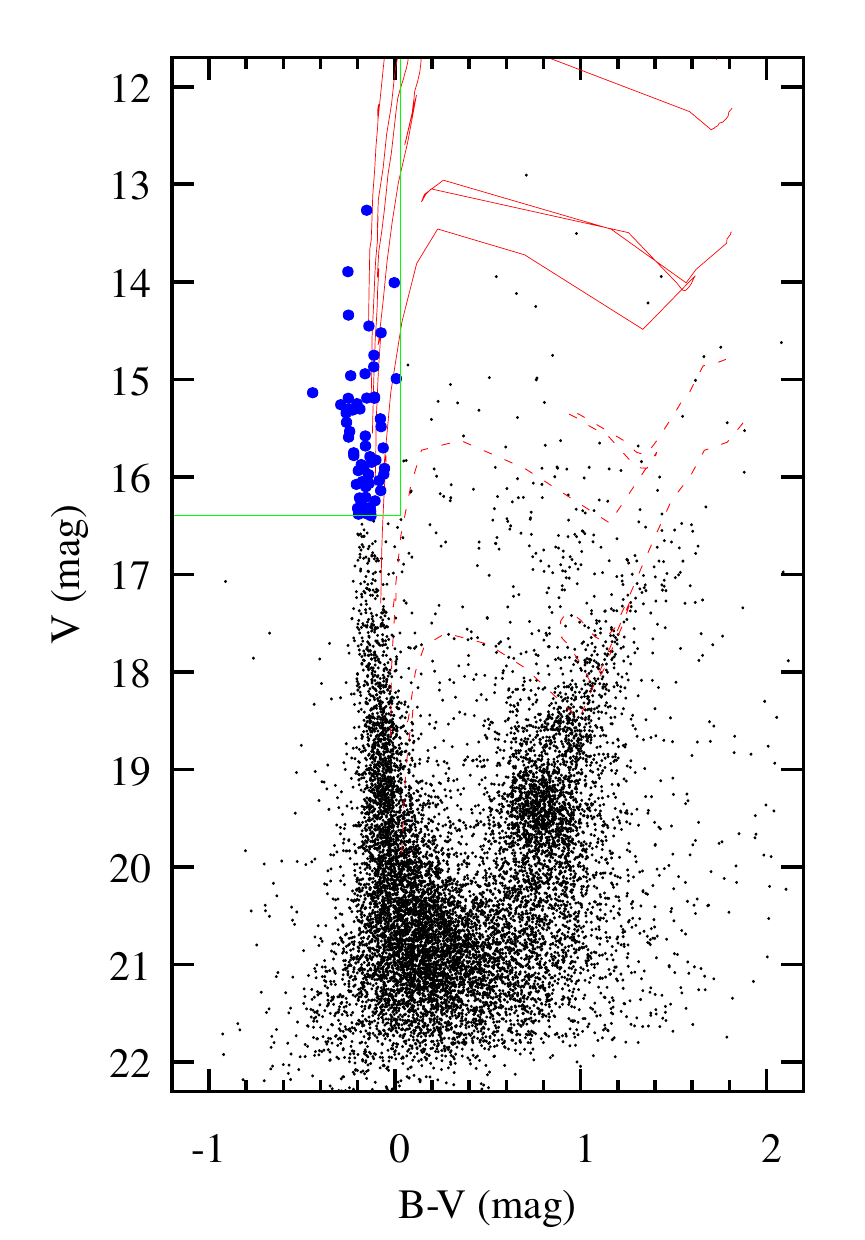}
    \includegraphics[height=0.23\vsize]{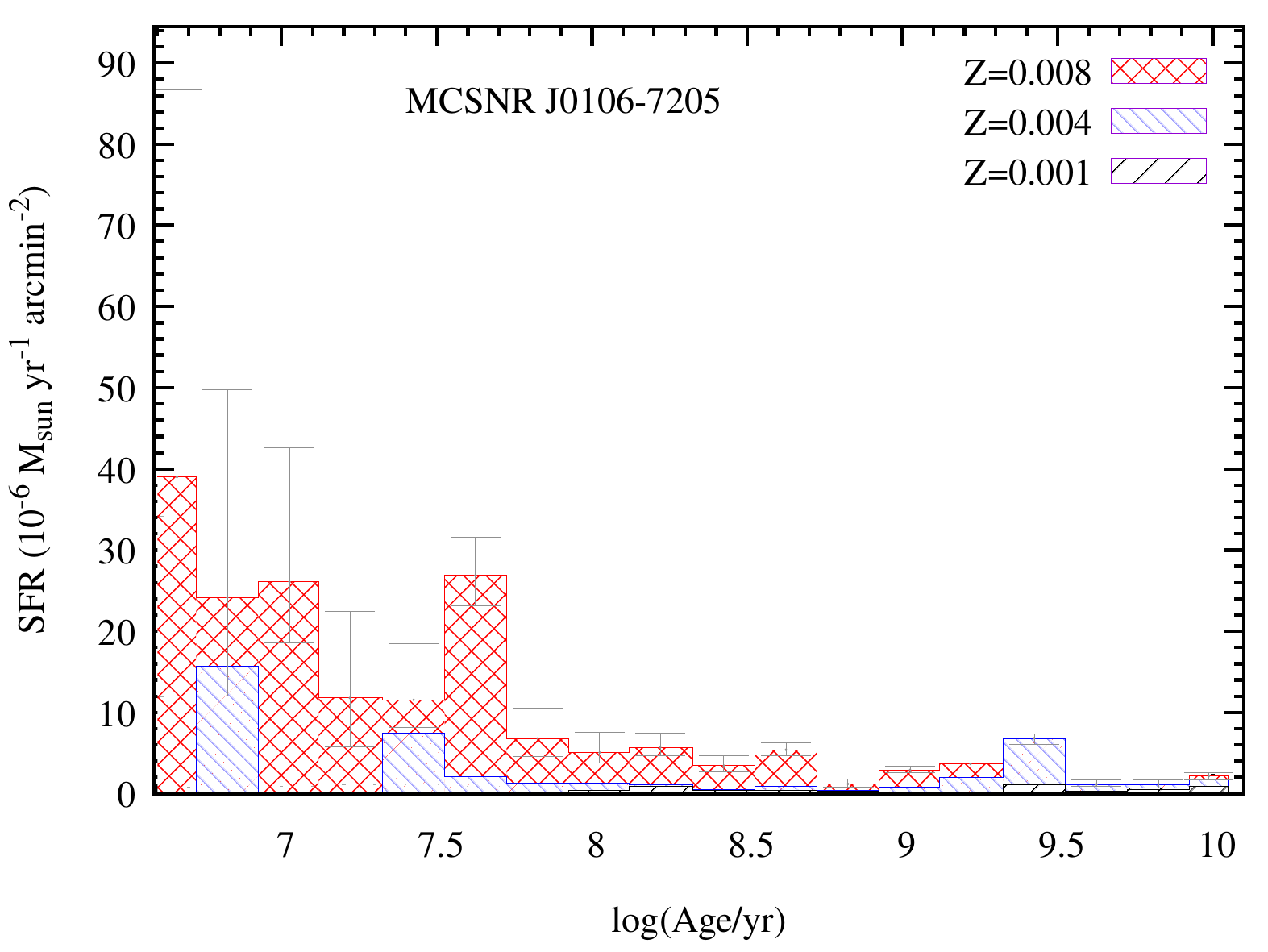}
    \includegraphics[height=0.23\vsize]{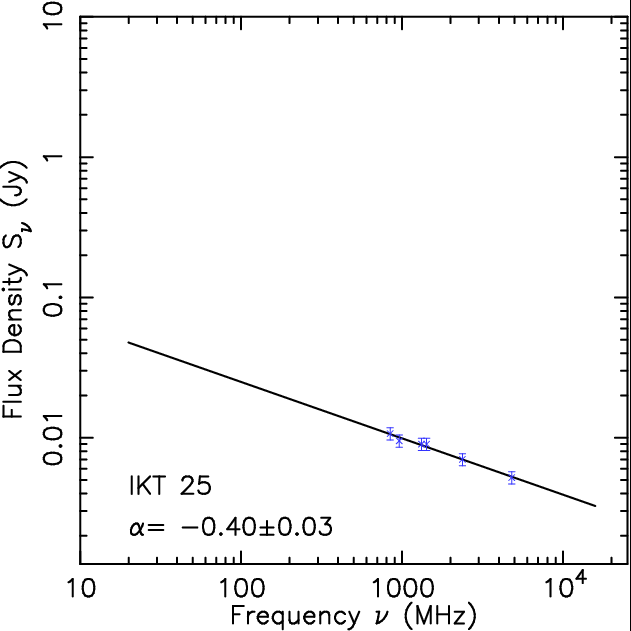}

    \includegraphics[height=0.225\vsize]{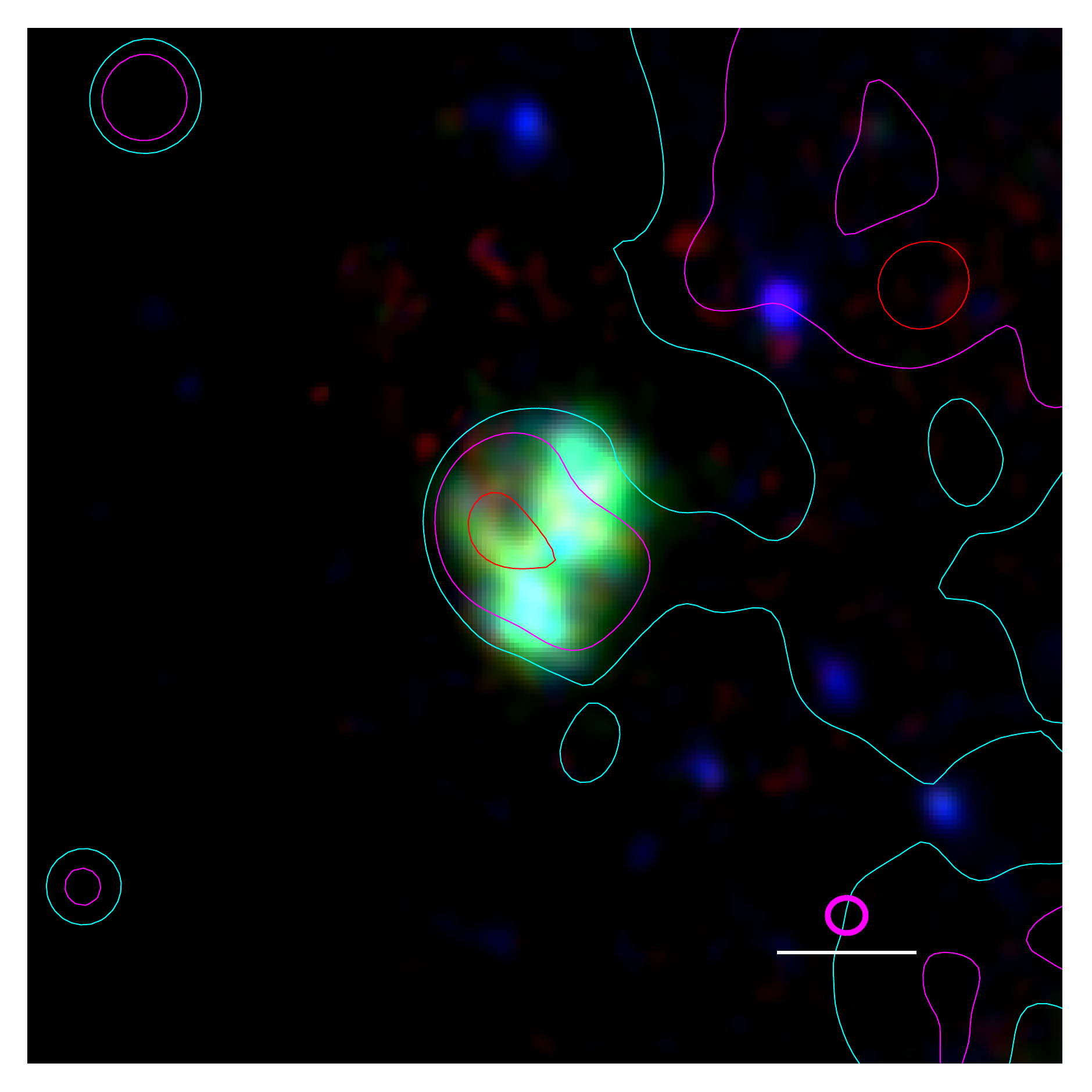}
    \includegraphics[height=0.225\vsize]{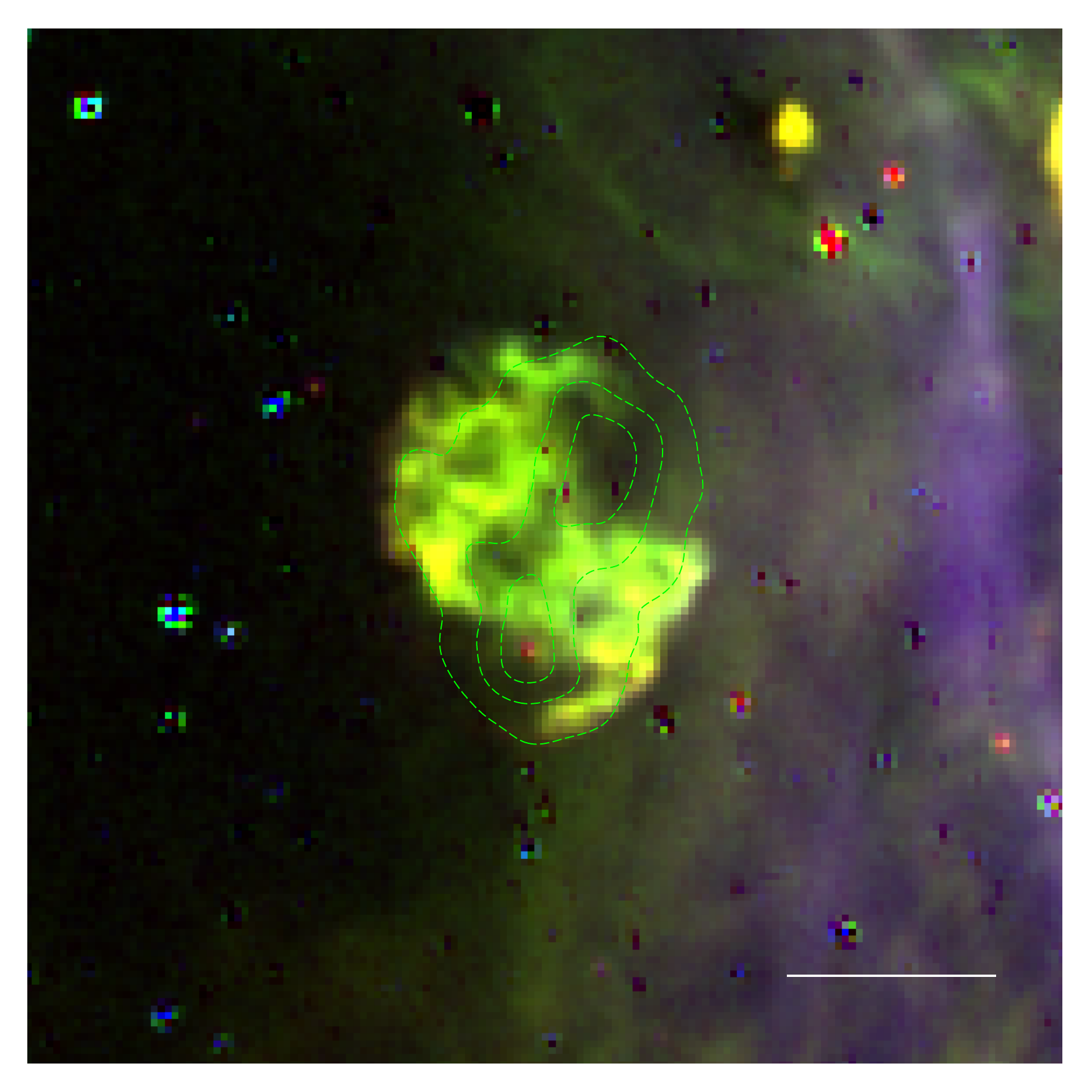}
    \includegraphics[height=0.225\vsize]{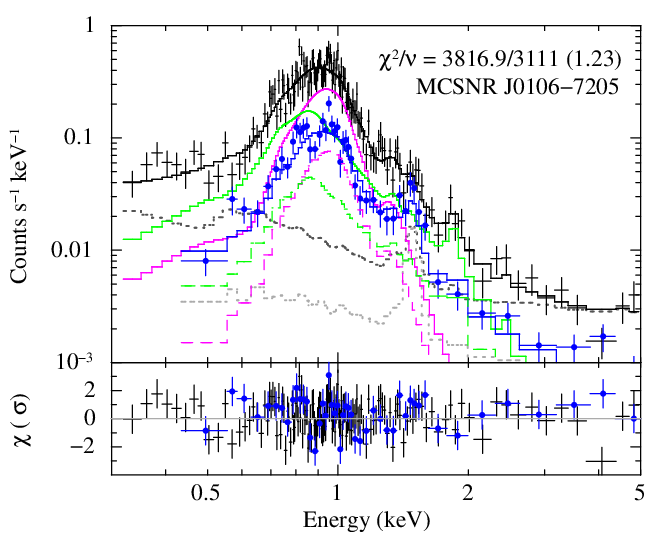}

    \vspace{1em}
    \hrule
    \vspace{1em}

    \includegraphics[height=0.225\vsize]{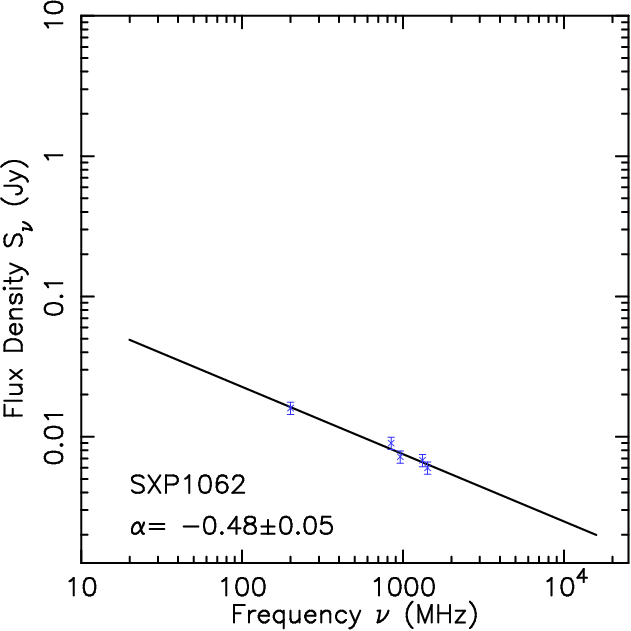}
    \includegraphics[height=0.225\vsize]{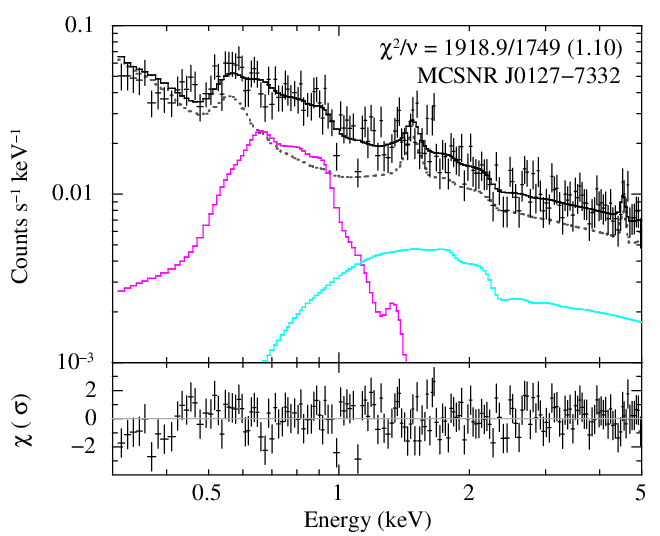}
    
    \includegraphics[height=0.24\vsize]{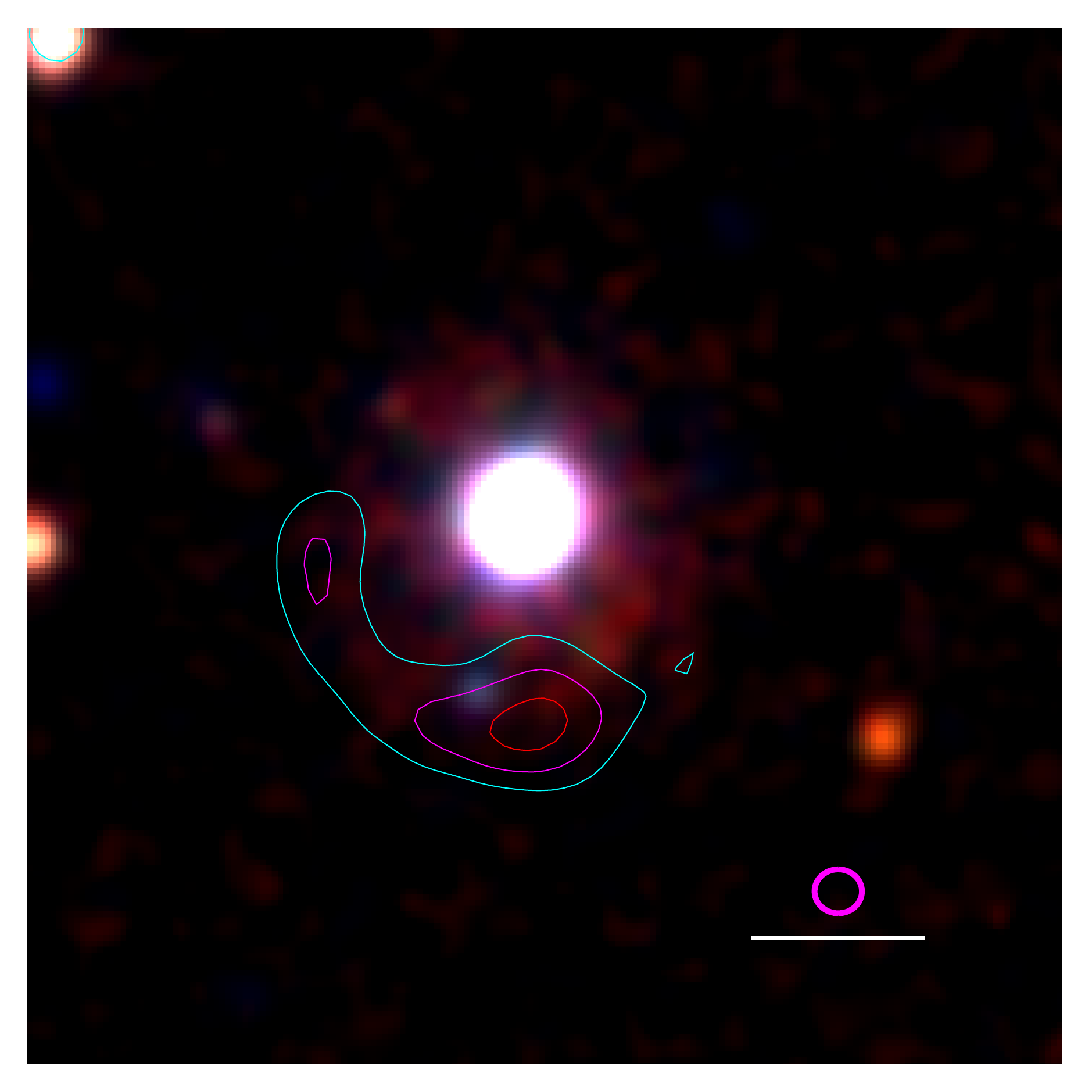}
    \includegraphics[height=0.24\vsize]{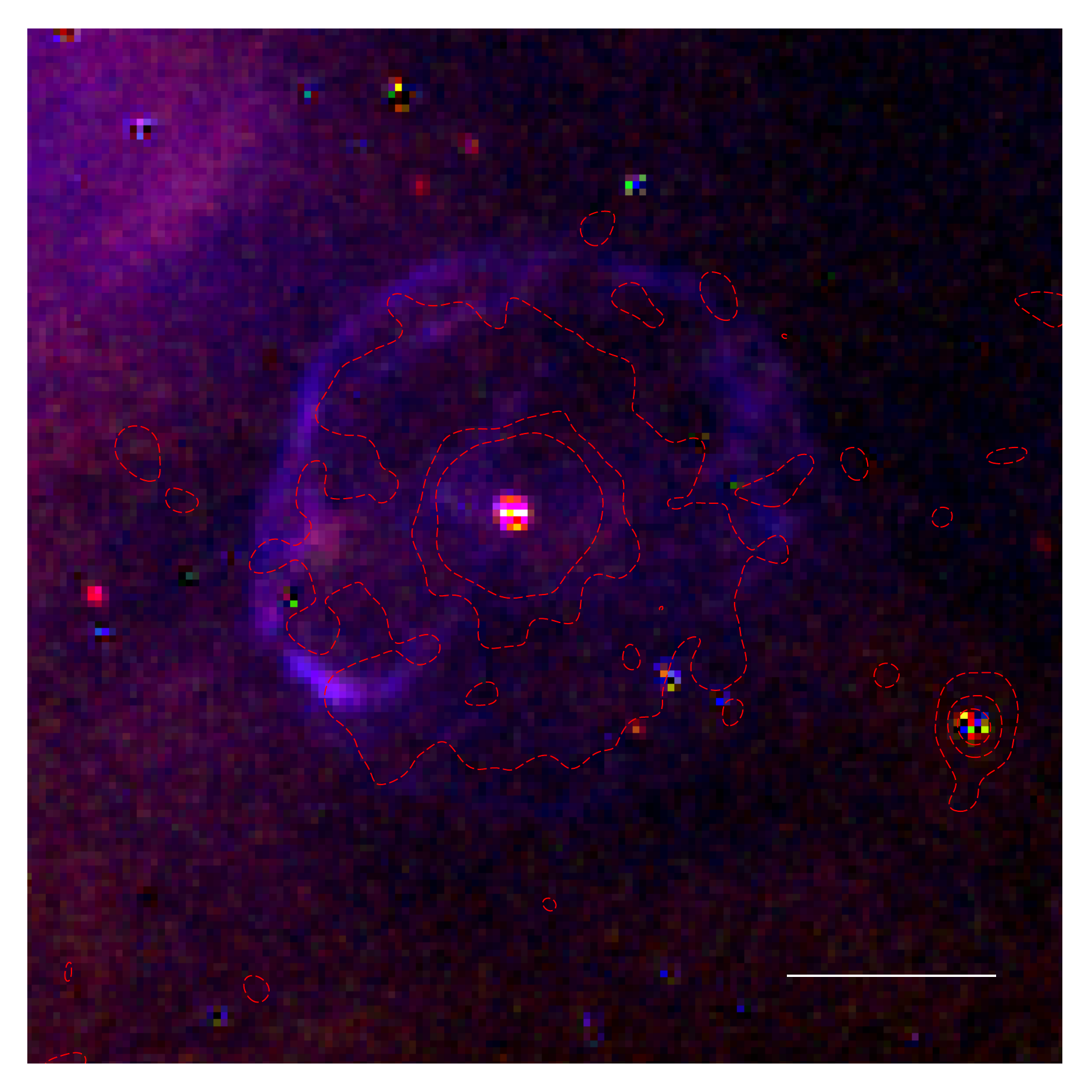}

    \caption{Same as Fig.\,\ref{fig_appendix_sfh0} for MCSNR~J0106$-$7205 (top part), with radio contour levels at 0.1, 0.2, and 0.5~mJy/beam, and for MCSNR~J0127$-$7332 with levels at 0.1, 0.15, and 0.2~mJy/beam (bottom part). For the last one, there are neither CMD nor SFH, as it was outside the area covered in the MCPS \citep{2004AJ....127.1531H}.}  
  \label{fig_appendix_sfh7}
\end{figure*}

\begin{figure*}[t]
    \centering

    \includegraphics[height=0.225\vsize]{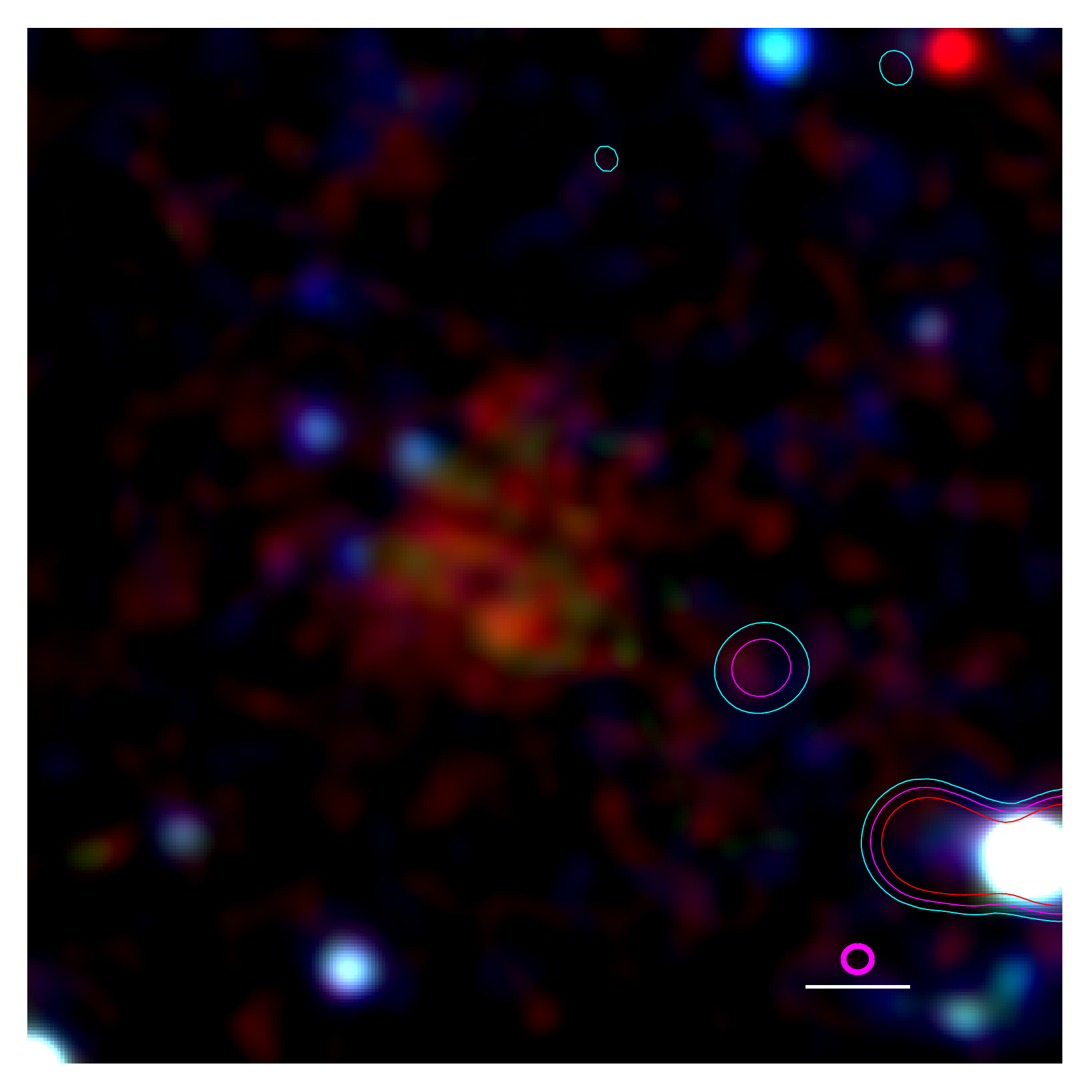}
    \includegraphics[height=0.225\vsize]{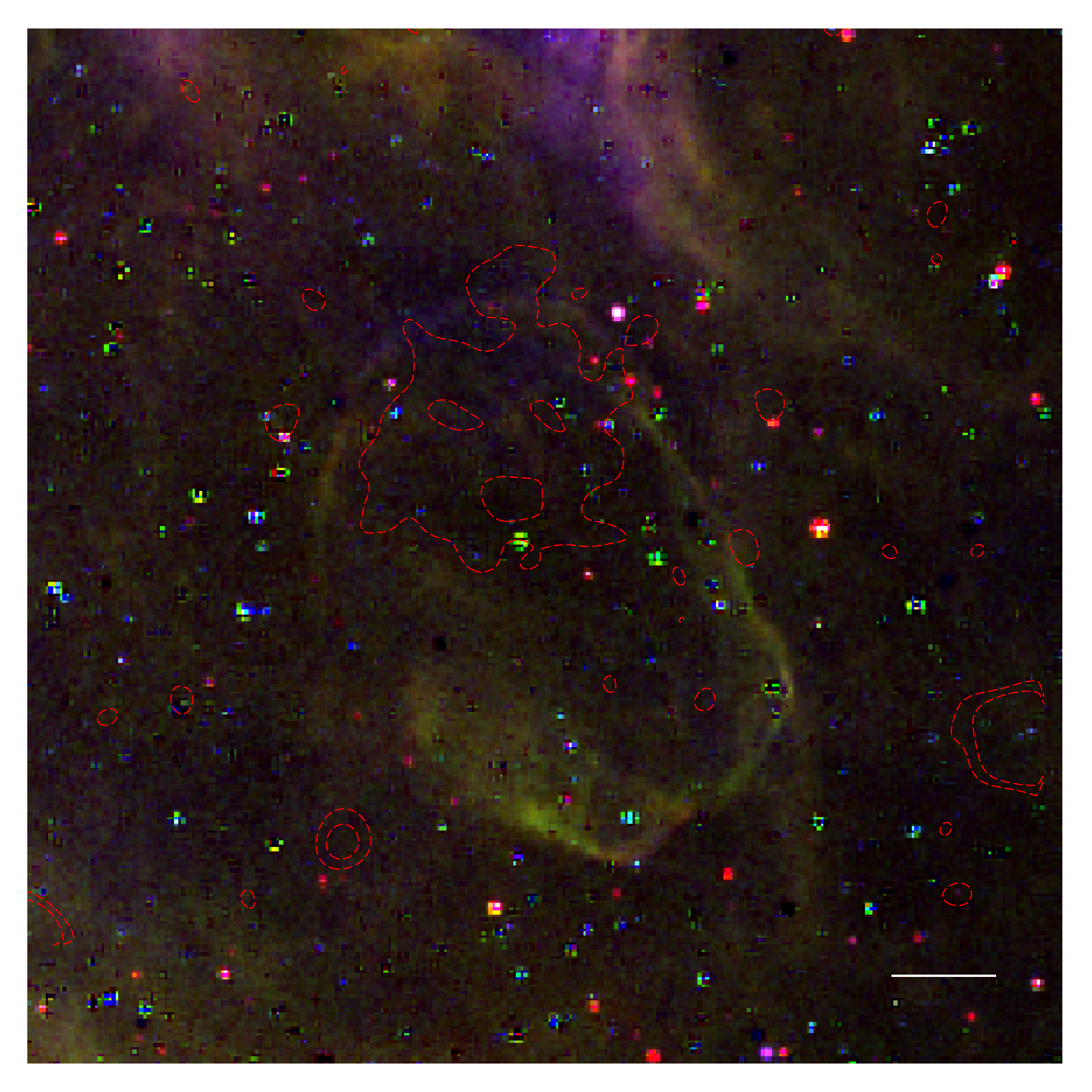}
    \includegraphics[height=0.225\vsize]{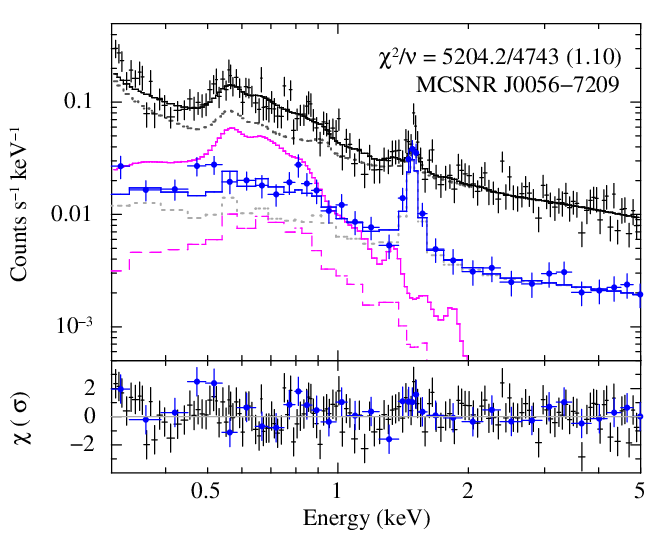}

    \vspace{1em}
    \hrule
    \vspace{1em}
    
    \includegraphics[height=0.225\vsize]{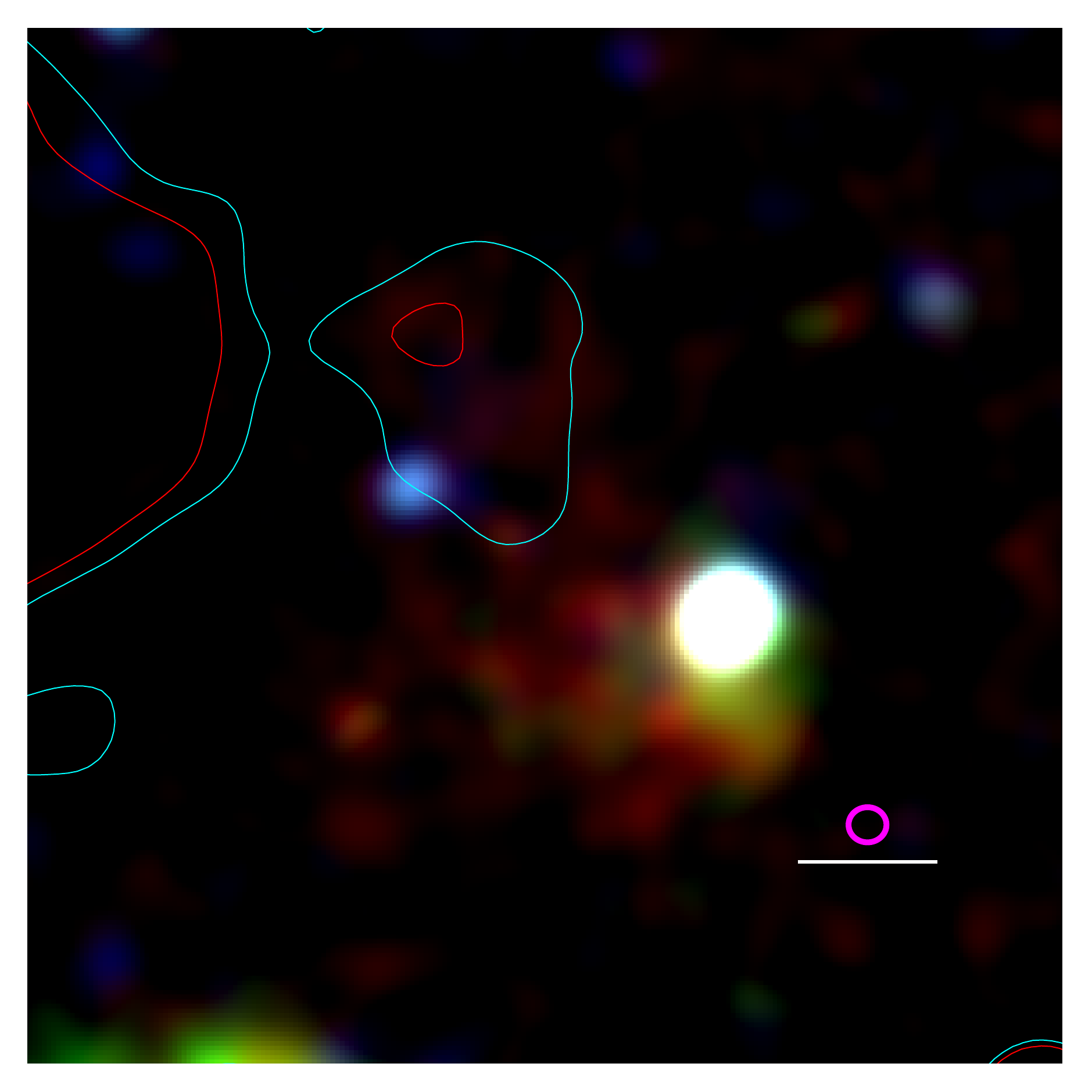}
    \includegraphics[height=0.225\vsize]{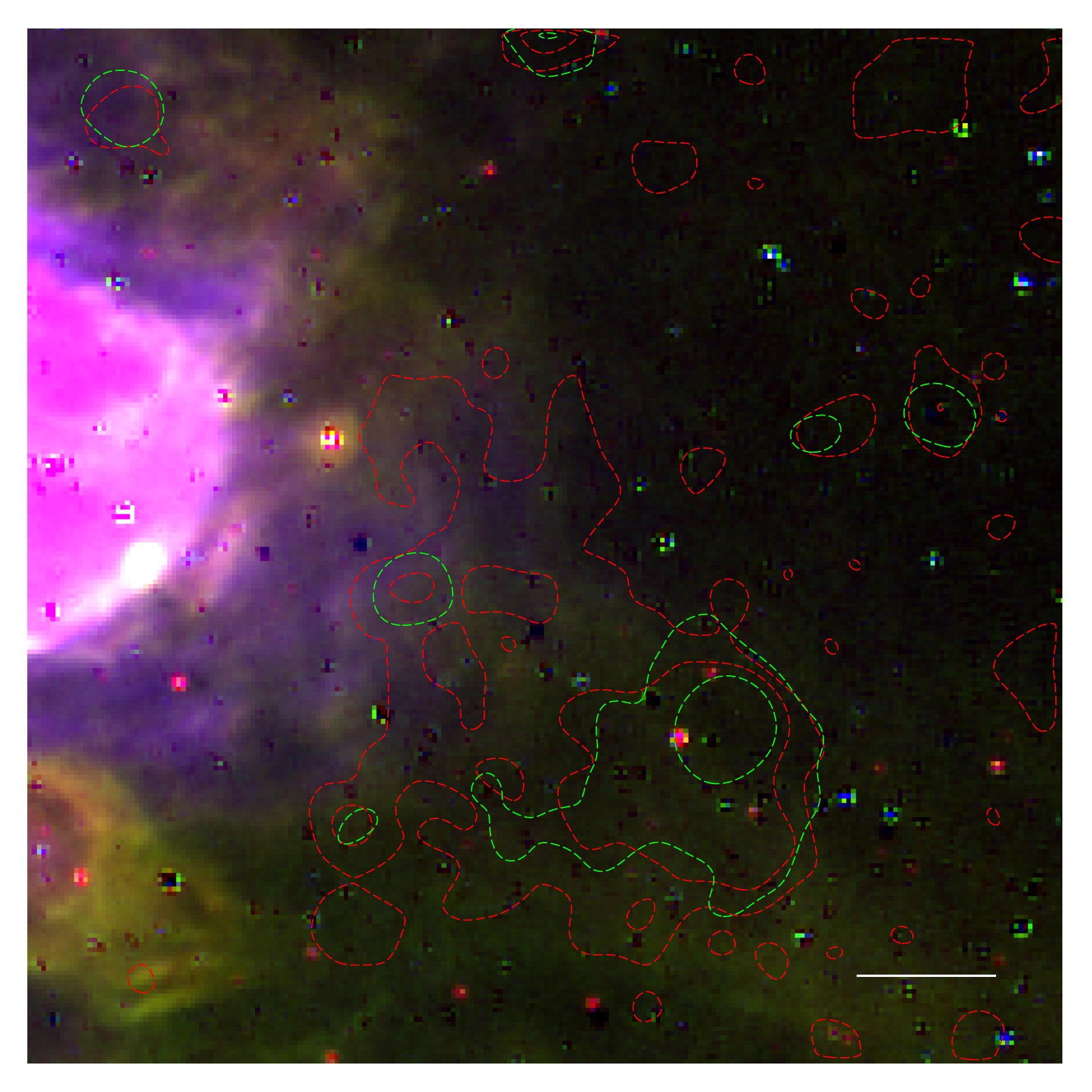}
    \includegraphics[height=0.225\vsize]{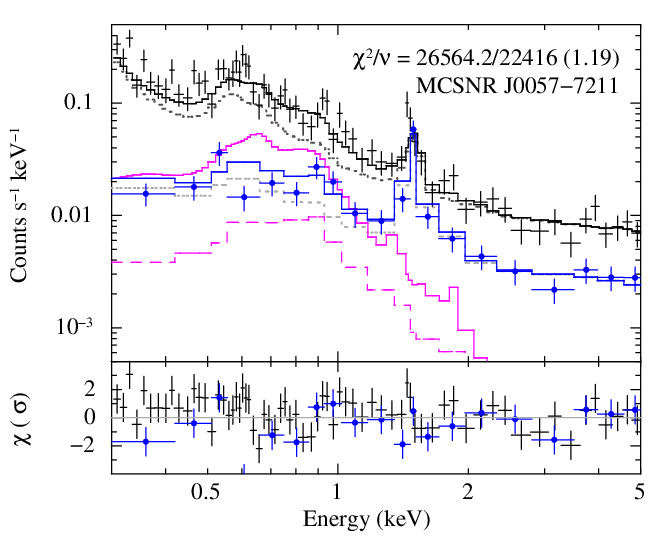}

    \vspace{1em}
    \hrule
    \vspace{1em}
    
    \includegraphics[height=0.28\vsize,angle=0]{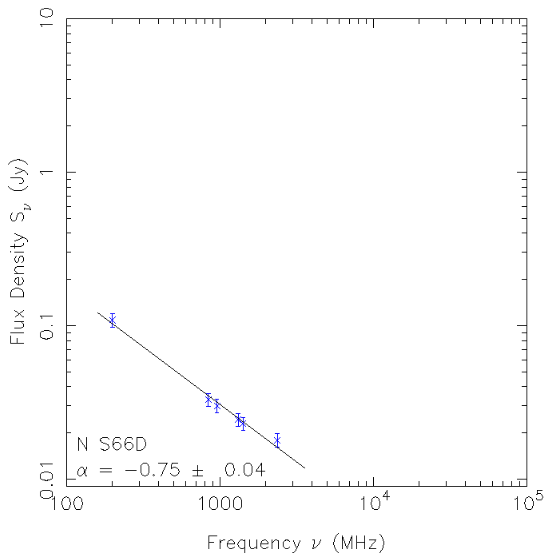}
    \hspace{1.2em}
    \includegraphics[height=0.28\vsize,angle=0]{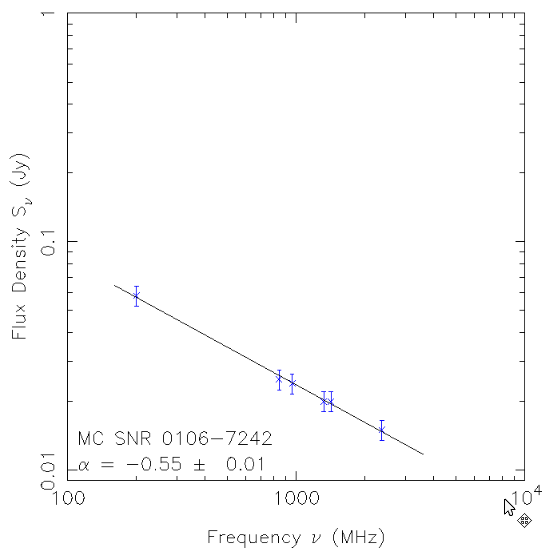}

    \caption{\textit{Top row\,:} X-ray image of the newly confirmed MCSNR~J0056$-$7209 with radio contours, levels at 0.3, 0.8, and 2~mJy/beam; MCELS image with X-ray contours; and X-ray spectrum. \textit{Middle row\,:} Same as above for the newly confirmed MCSNR~J0057$-$7211. The radio contours are at the 0.3 and 0.5~mJy/beam level. \textit{Bottom row\,:} Radio spectra for the newly confirmed MCSNR~J0057$-$7211 (N S66D) and MCSNR candidate J0106$-$7242.}  
  \label{fig_appendix_candidates}
\end{figure*}


\end{document}